\documentclass[journal=chreay,manuscript=review]{achemso}
\usepackage[version=3]{mhchem} 
\usepackage[bookmarks,bookmarksopen]{hyperref}
\usepackage{graphicx}
\usepackage{dcolumn}
\usepackage{bm}
\def\ref#1{$^{#1}$}

\DeclareRobustCommand*{\citen}[1]{%
  \begingroup
    \romannumeral-`\x 
    \setcitestyle{numbers}%
    \cite{#1}%
  \endgroup
}

\newcommand{\jcp}{J. Chem. Phys.}
\newcommand{\aap}{Astron. Astrophys.}
\newcommand{\aaps}{Astron. Astrophys. Supplt.}
\newcommand{\apj}{Astrophys. J.}
\newcommand{\apjs}{Astrophys. J., Suppl. Ser.}
\newcommand{\apjl}{Astrophys. J. Lett.}
\newcommand{\pasp}{Publ. Astron. Soc. Pac}
\newcommand{\araa}{Ann. Rev. Astron. Astrophys.}
\newcommand{\mnras}{Mon. Not. R. Astron. Soc.}
\newcommand{\pra}{Phys. Rev. A}
\newcommand{\qjras}{Q. J. R. Astron. Soc}
\newcommand{\nat}{Nature}
\newcommand{\jqsrt}{J. Quant. Spectrosc. Radiat. Transf.}

\newcommand{\HH}    {\mbox{H$_2$}}           
\newcommand{\CCO}    {\mbox{CO}}           
\newcommand{\DD}      {\mbox{D$_2$}}         
\newcommand{\HHHp}    {\mbox{H$_3^+$}}       
\newcommand{\NHH}    {\mbox{NH$_2$}}       
\newcommand{\NHHH}    {\mbox{NH$_3$}}       
\newcommand{\NNHp}    {\mbox{N$_2$H$^+$}}       
\newcommand{\HHO}  {\mbox{H$_2$O}}         

\newcommand{\HHOp}   {\mbox{H$_2$O$^+$}}      
\newcommand{\HHHOp}   {\mbox{H$_3$O$^+$}}      

\newcommand{\HCOp}  {\mbox{HCO$^{+}$}}       

\author{Evelyne Roueff} 
\email{evelyne.roueff@obspm.fr}
\affiliation{Laboratoire Univers et Th\'eories, Observatoire de Paris, 92190, Meudon, France}

\author{Fran\c{c}ois Lique} 
\email{francois.lique@univ-lehavre.fr}
\affiliation{LOMC - UMR 6294, CNRS-Universit\'e du Havre, 25 rue Philippe Lebon, BP 540, 76058, Le Havre, France}

\title[Molecular excitation in the Interstellar Medium]
  {Molecular excitation in the Interstellar Medium: recent advances in
collisional, radiative and chemical processes}

\abbreviations{IR,NMR,UV}

\begin{document}

\tableofcontents


\section{Introduction}
\label{sec:intro}
Our main knowledge of astrophysical environments relies on radiative spectra. Absorption studies require background sources which,
in astrophysical environments, are located hapharzadly although they may offer quite useful information. Sufficiently high spectral resolution  and sensitivity  also allow  to measure absorption spectra from different rotational levels, including the ground state, giving access to the full column density of the detected molecule. Astrophysical observations allow indeed to derive only column densities representing the integration of the abundances over the line of sight. Assuming an homogeneous cloud, one can then convert these column densities in  abundances after estimating the thickness of the absorbing/emitting region.  Alternatively, emission spectra 
arise after a proper excitation of the emitting level. The emitted radiation is directly linked to the population of the upper level but  may  also depend on the population of the lower level and  on the 
properties of the  environments located between the source and the observer. As an example, the presence of dust particles induces additional absorption, scattering which may considerably modify the observed emitted spectrum. Such effects are resulting from radiative transfer effects and will not be discussed in this review. We only consider specific microscopic processes which may lead to excitation. Collisional excitation is the basic process and its efficiency relies on the composition of the medium providing the density of the main perturbers and the temperature which establishes the degree of excitation, and consequently drives the intensities   of the radiated emission.  Nevertheless, other excitation mechanisms may be at work such as radiative or chemical pumping. 

 Detection of several transitions of a given species allows to draw a so-called excitation diagram where the $y$-axis stands for the logarithm of the column density $N(j)$ derived 
 for each considered level  $j$ divided by its statistical weight $g_j$ and the $x$-axis reports the energy (often expressed in Kelvin).
 The corresponding points are aligned when thermodynamical equilibrium or local thermodynamic equilibrium (LTE) is fulfilled and the temperature is simply linked to the slope of the curve: $$ N(j) = g_j \times \frac{N(0)}{g_0} \times exp[-(E_j - E_0)/kT],  $$ where $g_j$ ($g_0$)  is the statistical weight of level j (0)
 and $N(0)$ the column density of the lowest energy level.
 Astrophysicists introduce so-called excitation temperatures allowing to describe  abundances (column  densities)  as a function of the energy of the involved levels in a similar expression where $T_{ex}$ replaces the kinetic temperature.  This relation holds at least for 2 levels, but often for more and allows to obtain a simple derivation of the column densities of the various levels.
  Under low density conditions, such as those present in the ISM, the excitation temperature is most frequently below the kinetic temperature and the conditions are labelled as subthermal.  This is easily seen for a 2-levels system ($u$ for upper and $l$ for lower), where collisional
  de-excitation competes with spontaneous emission expressed by the Einstein $A$ coefficient  (s$^{-1}$), which allows to describe the population of the excited level, $x_u$, by the following formula
 $$  x_u = \frac{ g_u}{g_l} \times  x_l\times  exp[-(E_u - E_l)/kT_{ex}] , $$ where 
  $$T_{ex} = T  \times \frac{1}{1+ [\frac{kT}{(E_u-E_l)}~ \times ~\frac{A_{u\to l}}{ n~k_{u \to l}}]} .$$  This simple expression, where $n$ stands for the perturber density and $k$ for the collisional de-excitation rate of the $u \to l$ transition,  is obtained if one neglects
 background absorption  and induced emission. It shows that the excitation temperature is always smaller than the kinetic temperature $T$  and that it equals the kinetic temperature for high densities when the second factor in the denominator becomes negligible, i.e. when collisional de-excitation  dominates radiative quenching. The critical density is defined as the ratio between the Einstein coefficient (in s$^{-1}$) and the collisional de-excitation rate coefficient in cm$^3$ s$^{-1}$. More complicated expressions arise when absorption, induced emission are taken into account and are found in all textbooks  (as for example in Lequeux \cite{lequeux:05}), and the excitation temperature terminology is extended accordingly.  

 In a general way however, LTE is rarely fulfilled. Then, knowledge of the collisional rate coefficients involving the most  abundant species, i.e. H, H$_2$, He and electrons is required to solve the coupled differential equations describing the evolution of the different excited states of each molecule. These equations are non-linear in nature as the radiation source terms depend  themselves on the populations of the individual levels involved in the emitted transitions. Different techniques of resolution are available such as the Monte-Carlo approach\cite{bernes:79} or the Large Velocity Gradient (LVG)  approximation  used in Ref. \citen{vandertak:07}. 
 Goldsmith and Langer \cite{goldsmith:99}
 introduce an additional multiplicative optical depth correction factor  $C_{\tau}= \frac{\tau}{1-e^{-\tau}} $, where $\tau$ is the optical depth of the transition. Such a procedure improves considerably the interpretation of the observations. 
 Specific radiative transfer effects 
 may also occur in anomalous excitation conditions as for example in maser or anti maser transitions
 \cite{elitzur:94,daniel:13}. 
 
\begin{figure}
\includegraphics[width=8. cm]{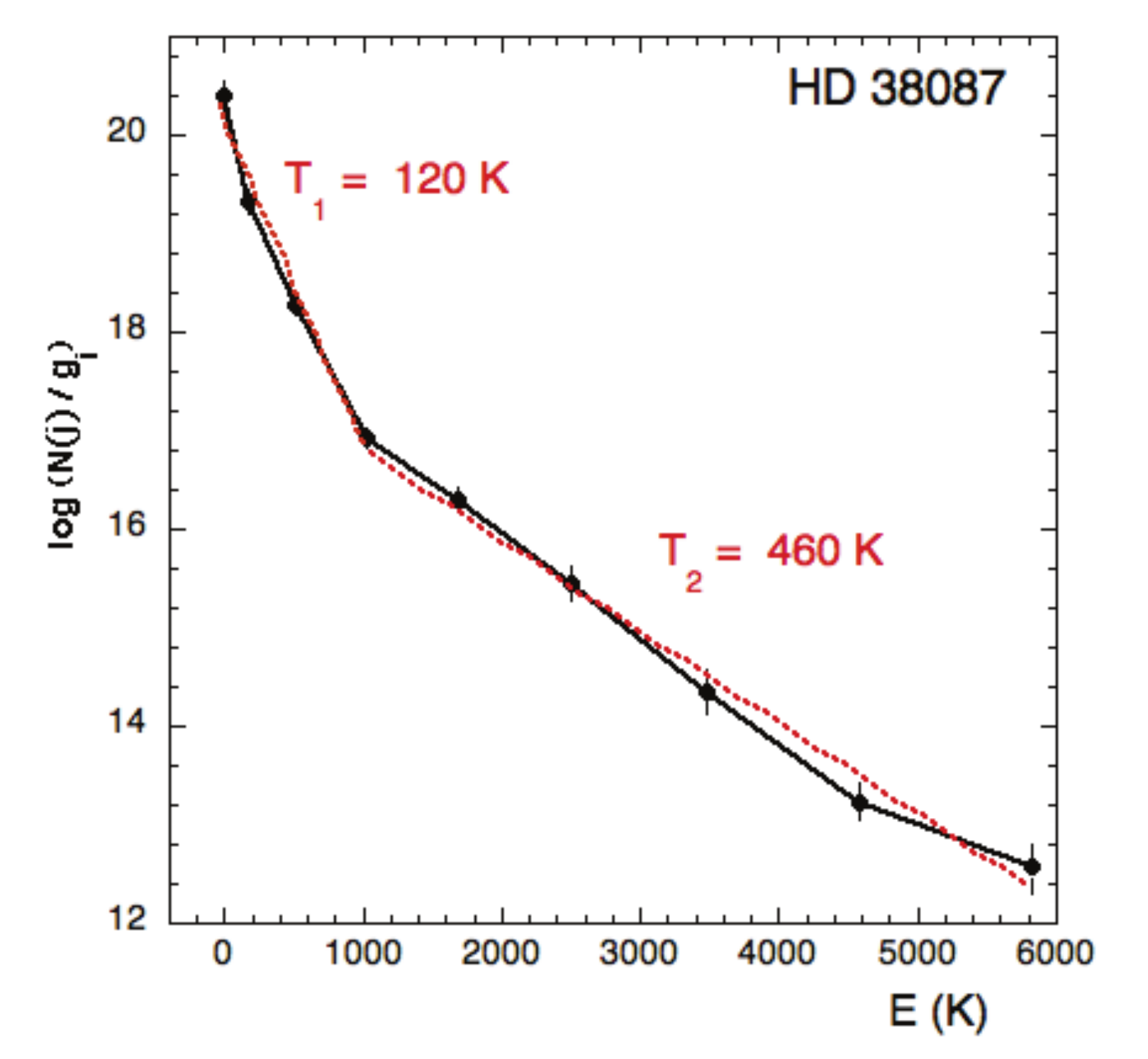}
\caption{log($N(j)$/$g_j$) as a function of the energy of level $j$ expressed in K. $N(j)$ values of  {\HH} (in cm$^{-2}$) are taken from Ref. \citen{jensen:10}. The 2 linear fits are also shown with the corresponding derived excitation temperatures.
 }
\label{fig:excitation}
\end{figure}

  Figure \ref{fig:excitation} displays an example taken from {\HH} observations  performed towards HD~38087 (a blue bright giant star in the constellation Orion) with FUSE and reported in Ref. \citen{jensen:10}.  Two excitation temperatures are derived from low ($120 \pm 4$~K) and higher ($460 \pm 11$~K) excitation levels. The physical relevance of such "temperatures" is not warranted.
 
 An additional example of departure from LTE conditions may be found in the analysis of far infrared carbon monoxyde PACS spectra where curved rotational diagrams may be explained by a single isothermal component \cite{neufeld:12}.

Our view of the interstellar medium (ISM) has been considerably improved in recent years thanks to new observational facilities. As far as spatial observatories are concerned, the far ultraviolet domain has been made accessible with FUSE (Far Ultraviolet Spectroscopic Explorer), the STIS (Space Telescope Imaging Spectrograph) and COS (Cosmic Origins Spectrograph) instruments on board of the Hubble Space Telescope with various spectral resolutions and sensitivities. At infrared and submillimeter wavelengths, SPITZER, SOFIA and Herschel  have allowed to probe in great details dark clouds including both low and high mass star formation regions and revealed the chemical complexity of these environments. Many new species have been detected such as water which is specifically discussed  in this volume and a bunch of molecular ionic hydrides, amongst other new molecular species, which revealed unexpected aspects of chemical complexity. Last but not least, ground based telescopes such as APEX (Atacama Pathfinder EXperiment), CSO (Caltech Submillimeter Observatory), the 30m telescope and the Plateau de Bure interferometer of IRAM (Institut de RadioAstronomie Millimetrique), have extended their spectral resolution bandwidth and sensitivity capabilities and the ALMA (Atacama Large Millimeter Array) era is now in front of us with both high
 spatial and spectral resolution with already very impressive and successful science verification preliminary results. 
 
 There are presently more than 170 observed gas phase molecular species including saturated molecules, radicals, different isomers, positive and negative molecular ions that have been successfully detected in the interstellar and circumstellar space as listed on  the Cologne Database for Molecular Spectroscopy  (CDMS)  \cite{cdms,muller:01,muller:05}.  This inventory does not include the isotopic substitutions by D, $^{13}$C, $^{15}$N, $^{34}$S, etc... Whereas already many isotopologues have been detected and identified, much more will be available  in the near future thanks to the increased sensitivity and spectral resolution of the different instruments and will shed a new light on the connection between the ISM and solar system objects where the isotopic ratio can also be studied. This specific aspect is discussed by Charnley and Milam in this volume. However, despite very similar spectroscopic properties, transposing excitation properties from the main isotopologue may lead to serious errors and one should realize that each isotopologue requires a specific  study \cite{dumouchel:12NH}.
 The potential field of investigations is thus almost unlimited and  we apologize in advance for overlooking some publications.
 We principally  emphasize the theoretical frame of collisional excitation studies in Section~\ref{sec:col} as both  theoretical tools and computational facilities have been improved very significantly. Data currently used in the interpretation of astrophysical spectra are available on database web sites such as the LAMDA database 
 \cite{lamda,schoier:05} and BASECOL  
 \cite{basecol,Dubernet:13}. We thus will focus our review on more recent material. Amongst other possible excitation processes, radiative and chemical pumping are also potentially at work.  These mechanisms deserve dedicated studies 
 only if the basic collisional excitation processes are understood. We describe the general frame of such possibilities  and discuss some illustrative examples in Section \ref{sec:rad}. Section \ref{sec:conclusion} summarizes our conclusions 
 and emphasizes some outlook for future studies.

\section{Collisional excitation}
\label{sec:col}

The experimental and theoretical study of collision-induced ro-vibrational energy transfer has received much attention during the past 40 years \cite{bernstein1979atom,Flower:07book,Paterson:12}. This has been motivated by the development of combined molecular beam and laser spectroscopy experiments along with fully-quantum mechanical treatments of the collision dynamics based on accurate potential energy surfaces (PES). 
However, up to the beginning of the present century, only  few molecular systems of astrophysical interest have been the object of detailed studies. 

In cold (T~$\sim$~10~K) molecular interstellar clouds, the most abundant species are H$_2$ and He at a level of 20\% in numerical density, 
which should be taken into account in the astrophysical modeling of observational spectra. 
He, as a closed shell atom with two electrons, is sometimes considered as a reasonable template of molecular Hydrogen 
\cite{green74,schoier:05,lique08b}.

In warmer regions such as  diffuse, translucent clouds (T~$\sim$~70~K) and Photon Dominated Regions (PDRs) (T~$\sim$~200~K), the abundances of atomic and molecular hydrogen are comparable and collisions by H should also be taken into account. Hence, in this review, we will focus on collisional excitation studies of interstellar molecules by H, He and H$_2$. One should recall that  ro-vibrational excitation of molecules by electronic collisions may also be of astrophysical interest  for molecular ions or when molecules possess a significant 
permanent dipole moment. However, the fractional electronic abundance is thought to be at maximum about 10$^{-4}$, which hardly compensates the differences in corresponding rate coefficients. 
We do not include this  topic in the present review which involves different theoretical methods of studies.

The formal quantal close-coupling theory of rotational excitation has been first established by Arthurs and Dalgarno \cite{arthurs:60}.
 Quantitative  investigations of interstellar molecules excitation began then in the early 1970's with the work of Townes and co-workers\cite{Townes:69}, Miller and co-workers\cite{Augustin:74} and Green and co-workers \cite{green74,green:76}. These studies were essentially limited by the computational resources and the approximate models employed to describe the interaction potentials. However, they were quite successful in qualitatively explaining the observed anomalous excitation of molecules such as H$_2$CO\cite{Townes:69} or SiO \cite{Watson:80}. Since these early works, the level of dynamical theory has not dramatically improved but the huge increase of computational resources and the significant progresses in {\it ab initio} quantum chemistry have led to a spectacular improvement of accuracy and a large increase of studied systems. 

During the 1970's, 1980's and 1990's, the theoretical calculations of inelastic collisional rate coefficients were essentially performed by Dalgarno and co-workers \cite{Chu:75,Lepp:95}, Green and co-workers \cite{green74,green78,bieniek83,green91,green93,green94,Neufeld:94}, Flower and co-workers \cite{flower1:81,flower2:83,danby86,danby88,jaquet92,Flower:99a} and Monteiro and co-workers \cite{monteiro84,monteiro86}. They were mainly focused on the calculations of collisional data involving the most abundant molecules (H$_2$, CO, H$_2$O, HCN, NH$_3$, HCO$^+$, ...). 

However, at the beginning of the present century, the funding by the European Community of the "Molecular Universe" research training network has stimulated the study of molecular processes for a number of key interstellar molecules, both experimentally and theoretically. 
Since then, a large number of new collisional systems have been investigated, 
which considerably improved  and extended the number of computed systems, mainly with He as a collider. 
 Tables \ref{tab1} and \ref{tab2} present the most recent studies dedicated to interstellar molecules. 
Most recent theoretical efforts involve molecular {\HH}, including its para ($j=0$) and ortho ($j=1$) forms.

\begin{table*}
\caption{Rotational excitation collisional studies  by He, {\HH}  and H of linear molecules (sorted by family and electronic symmetry).}
\label{tab1}
\begin{tabular}{|c|c|c|c||c|c|c|c|}
\hline
 & \multicolumn{3}{c||}{collisional partner} & & \multicolumn{3}{c|}{collisional partner}  \\ \hline
Molecule & He & H$_2$ & H & Molecule & He & H$_2$ & H \\ \hline
H$_2$($^{1}\Sigma_g^+$) & Ref. \citen{Balakrishnan:99} & Ref. \citen{Quemener:09} & Ref. \citen{lique:12:H2} & HD($^{1}\Sigma^+$) & Ref. \citen{Roueff:00}  & Ref. \citen{Flower:99c} & Ref. \citen{Flower:99c} \\ \hline
D$_2$($^{1}\Sigma^+$) & & & Ref. \citen{lique:12D2} &C$_2$($^{1}\Sigma_g^+$) & Ref. \citen{najar:08} & Ref. \citen{najar:09} & ... \\ \hline
CO($^{1}\Sigma^+$) & Ref. \citen{Cecchi:02} & Ref. \citen{Flower:12CO} & Ref. \citen{Shepler:07} &  CS($^{1}\Sigma^+$) & Ref. \citen{Lique:06CS,Lique:07} & ... & ... \\ \hline
HCN($^{1}\Sigma^+$) & Ref. \citen{Dumouchel:10} & Ref. \citen{benabdallah:12} & &HNC($^{1}\Sigma^+$) & Ref. \citen{Dumouchel:10} & Ref. \citen{Dumouchel:11} & ... \\ \hline
 HCl($^{1}\Sigma^+$) & Ref. \citen{Lanza:12}  & ... & ... & HF($^{1}\Sigma^+$) & Ref. \citen{Reese:05} & Ref. \citen{Guillon:12} & ...   \\ \hline
AlF($^{1}\Sigma^+$) & Ref. \citen{gotoum:12}  & ... & ... &    PN($^{1}\Sigma^+$) & Ref. \citen{Tobola:07} & ... & ...   \\ \hline
HC$_3$N($^{1}\Sigma^+$)  & Ref. \citen{Wernli:07,Wernli:07a} & Ref. \citen{Wernli:07,Wernli:07a} & ... &  HCP($^{1}\Sigma^+$) & Ref. \citen{Hammami2008} & Ref.\citen{Hammami:08} & ...    \\ \hline
SiS($^{1}\Sigma^+$) & Ref. \citen{Vincent:07,Tobola:08} & Ref. \citen{klos08SiS} & ... & OCS($^{1}\Sigma^+$) & Ref. \citen{green78} & Ref. \citen{green78} & ... \\ \hline
  C$_3$($^{1}\Sigma^+$) & Ref. \citen{Benabdallah:08} & ... & ...& & & & \\ \hline \hline
CH$^+$($^{1}\Sigma^+$) & Ref.\citen{Turpin:10} & ... & ... & SiH$^+$($^{1}\Sigma^+$) & Ref. \citen{Nkem:09} & ... & \\ \hline
HCO$^+$($^{1}\Sigma^+$) & Ref. \citen{Buffa:09} & Ref. \citen{Flower:HCOp} & ...  &DCO$^+$($^{1}\Sigma^+$) & Ref.\citen{Buffa:12} & Ref. \citen{Pagani:12} & ... \\ \hline
HCS$^+$($^{1}\Sigma^+$) & ... & Ref. \citen{monteiro84} &   ... &CO$^+$($^{2}\Sigma^+$) & & & Ref. \citen{andersson:08} \\ \hline \hline
CN$^-$($^{1}\Sigma^+$) & ... & Ref. \citen{Klos:11} & ... &C$_2$H$^-$($^{1}\Sigma^+$)  & Ref. \citen{dumouchel:12} & ... &...  \\ \hline \hline
CN($^{2}\Sigma^+$) & Ref. \citen{lique:10CN,lique:11CN} & Ref. \citen{kalugina12} & ... & C$_2$H($^{2}\Sigma^+$) & Ref. \citen{Spielfiedel:12} & ... & ... \\  \hline \hline
O$_2$($^{3}\Sigma_g^-$) & Ref. \citen{lique:10} &  Ref. \citen{Kalugina:12O2}  & ... &SO($^{3}\Sigma^-$) & Ref. \citen{lique:05,Lique:06SO,Lique:06SO_2} & Ref. \citen{lique:07b} & \\ \hline
NH($^{3}\Sigma^-$) & Ref. \citen{tobola:11,dumouchel:12NH} & ... & ... & ND($^{3}\Sigma^-$) & Ref. \citen{dumouchel:12NH} & ... & ... \\ \hline
C$_4$($^{3}\Sigma_g^-$) & Ref. \citen{Lique:10C4} & ... & ...   & ... & & &  \\ \hline \hline
OH($^{2}\Pi$) & Ref. \citen{klos:07OH} & Ref. \citen{offer:94} & ... & SH($^{2}\Pi$) & Ref. \citen{klos:09} & ... & ...  \\ \hline
NO($^{2}\Pi$) & Ref. \citen{klos:08NO,lique:09} & ... & ... & & & &  \\ \hline 
\end{tabular}
\end{table*}

\begin{table*}
\caption{Rotational excitation collisional studies by He and  {\HH}  of  polyatomic bent molecules (sorted by family and electronic symmetry).}
\label{tab2}
\begin{tabular}{|c|c|c||c|c|c|}
\hline
 & \multicolumn{2}{c||}{collisional partner} & & \multicolumn{2}{c|}{collisional partner}  \\ \hline
Molecule & He & H$_2$  & Molecule & He & H$_2$  \\ \hline
H$_2$O($^1A_1$) & Ref. \citen{yang:07,yang:13h2o} & Ref. \citen{Dubernet:09,Daniel:10,Daniel:11} & HDO($^1A_1$) & ... & Ref. \citen{Faure:12}  \\ \hline
D$_2$O($^1A_1$) & ... & Ref. \citen{Faure:12} & H$_2$CO($^1A_1$) & ... & Ref.\citen{Troscompt:09} \\ \hline
 NH$_3$($^1A_1$) & Ref. \citen{machin:05,yang:08} & Ref. \citen{maret09}  &NH$_2$D($^1A_1$) & Ref.\citen{Machin:06} & ...  \\
\hline
ND$_2$H($^1A_1$) & Ref.\citen{Machin:07} & Ref.\citen{wiesenfeld11} &ND$_3$($^1A_1$) & Ref. \citen{yang:08} & \\ \hline
SiC$_2$($^1A_1$) & ... & Ref. \citen{Chandra:00}  &C$_3$H$_2$($^1A_1$) & ... & Ref. \citen{Chandra:00}  \\ \hline \hline
CH$_3$OH($^1A'$) & Ref.\citen{Rabli:10a,Rabli:11} & Ref.\citen{Rabli:10b} & CH3CN($^1A_1$) & Ref. \citen{Green:86CH3CN} & ...  \\ \hline
CH$_3$COOH($^1A'$) & Ref.\citen{Faure:11} & ...  & HNCO($^1A'$) & Ref. \citen{HNCO} & ...   \\ \hline
\hline
CH$_2$($^3B_1$) &  Ref. \citen{Ma:12} & ... & CH$_3$($^2A_2''$)  & Ref. \citen{Dagdigian:11} &   \\ \hline \hline
H$_3^+$($^1A_1'$) & ... & Ref. \citen{Hugo:09} &  HOCO$^+$($^1A'$) & Ref.\citen{Hammami:07} & ...  \\ \hline
\end{tabular}
\end{table*}

\subsection{Methods}

\subsubsection{Theory}

The computation of collisional inelastic rate coefficients usually takes place within the Born-Oppenheimer approximation for the separation of electronic and nuclear motions. Scattering cross sections are thus obtained by solving  the motion of the nuclei on an electronic potential energy surface (PES), which is independent of the masses and spins of the nuclei. 
Recent studies have demonstrated that computational techniques employing advanced treatments for both electronic and nuclear motion problems can rival with experimental measurements \cite{Carty:04,lique:10CN} in terms of the achieved accuracy. 

\subsubsection{Potential energy surfaces}

The potential energy surfaces (PES) must be accurate since the dynamical calculations at typical interstellar collisional energies are very sensitive to the PES quality in the range of intermediate intermolecular distances. The most accurate treatments are  based on modern methods of {\it ab initio} quantum chemistry \cite{Werner:12}. The process of rotational excitation of a molecule colliding with He or H$_2$ relates usually to systems in their electronic ground state (such as van der Waals complex), as temperatures are generally low (T~<~300~K) in the ISM. In addition, closed shell  systems are adequately represented by a single electronic configuration, which allows the use of mono-configurational methods like coupled clusters \cite{Hampel92,knowles:93,knowles:00} or perturbative methods{\cite{Werner:96,celani:00,Jeziorski:94}. The coupled clusters  approach, usually (partially spin-restricted) coupled cluster with the single, double and perturbative triple excitations [(R)CCSD(T)], will be preferably used for the determination of non-reactive PES owing to its high accuracy (of the order of a few cm$^{-1}$). 
For some open shell radicals, that are not correctly described by a single electronic configuration, the PES calculations requires the use of {\it ab initio} methods based on interaction configuration. Configuration interaction methods like MRCI \cite{Werner:88,knowles:88}, which is currently the most accurate method, will then be used to describe all geometries that could be explored by the nuclei during the collisional process. Such approach will generally also be used for molecule - atomic hydrogen interactions since the open-shell character of H usually implie the use of interaction configuration methods. 

Very recently, new {\it ab initio} methods have been developed. The standard coupled clusters methods lead to quite accurate results for small systems but some limitations reside in the CPU cost of the generation of the
 interaction potential. For instance,
standard {\em ab initio} methods, in spite of their accuracy, are computationally expensive and especially   
demanding when the molecular system contains more than four/five atoms.
The explicitly correlated (R)CCSD(T)-F12x (x=a,b) methods recently developed by Adler {\em et al.} and Knizia {\em et al.}~\cite{adler:07,Knizia:09} are much less expensive in computational time and remain an accurate technique for mapping the short and long range multi-dimensional PES of single configurational electronic states. The validity of this technique has been checked for the generation of the PES of the O$_2$(X$^3\Sigma_g^-$)--He model system \cite{Lique:10C4} and may be used for the determination of  PES involving larger polyatomic systems as it was recently done for the C$_4$--He collisional system\cite{Lique:10C4}.  

 
In all the methods described above, the quality of the result is also determined by the choice of atomic orbitals to describe the molecular orbitals and the electronic configuration. The chosen atomic orbitals basis set, from which are built the molecular orbitals, must be large enough to correctly represent the correlation energy and not too extended so that the computation time remains acceptable. 
The  augmented correlation-consistent valence triple zeta (aug-cc-pVTZ), quadruple zeta (aug-cc-pVQZ) or quintuple zeta (aug-cc-pV5Z) basis sets of Dunning and co-corkers~\cite{kendall:92,woon:93,woon:94} are usually well adapted to interaction PES calculations.
The quality of the results can be increased by the addition of bond functions in the middle of a van der Waals bond \cite{Tao:92}. The majority of calculations describing van der Waals interaction involves bond functions. However, It should be mentioned, that bond functions have a tendency to increase Basis Set Superposition Error (BSSE) and to alter the electrostatic energy. For these reasons they should be used with caution.
The standard quantum chemistry methods described above are implemented in several widely used numerical codes (MOLPRO\cite{molpro}, GAUSSIAN\cite{g09} or SAPT2008\cite{SAPT2008}).

An additional important step related to the interaction PES determination is the building of its analytic  representation in order to adequately perform
the dynamical calculations.  Great 
  care should be taken in order to maintain the accuracy of an {\it ab initio} PES in its analytical representation
through elaborate fitting techniques. 
The PES are generally expanded over angular functions (spherical harmonics or symmetric top wave functions)  as described in Ref. \citen{arthurs:60} and Ref. \citen{green:75} where the angular functions are products of associated Legendre functions $P_{lm}$.
Alternatively, new fitting methods such as the Reproducing Kernel Hilbert Space (RKHS) method\cite{ho:96} have been recently developed to obtain the analytic representations. These methods allow to get as many radial expansion coefficients as needed for the scattering calculations compared to the "classical expansion" over associated Legendre functions that limit the number of coefficients to the number of angular geometries taken into account in the {\it ab initio} calculations.   Alternative fitting strategy has been suggested \cite{Faure:11,valiron:08,Wernli:07} but describing the different fitting strategies is beyond the purpose of this review and is a research topic in itself.
Generally, the long range values derived from the {\it ab initio} calculations are not very accurate as they result from the (small) difference of two large numbers. Then, one should keep in mind the importance of  carefully  extending   the analytical PES  values to the long range part which can be more precisely derived from perturbation calculations as a 
$1/R $ expansion, where $R$  represents the distance between the centers of charge of the two interacting systems. The long range part is then described by the electrostatic, induction and dispersion terms contributing  to the total interaction energy of the complex  
with the proper angular and radial dependences \cite{Stone:96}.
Small unphysical irregularities of magnitude $\simeq$ 1 cm$^{-1}$ at long range can significantly affect the dynamical calculations. These effects are even more crucial in the field of cold and ultra-cold collisions.


Finally, we also stress out  that   calculations involving H$_2$ (para- and ortho-H$_2$), as a perturber,  can be very CPU time consuming, both in terms of quantum chemistry and scattering calculations. It is possible occasionally  to obtain an accurate estimate of the collisional rate coefficients with para-H$_{2}(j=0)$ (the most abundant collisional partner in the cold molecular ISM) 
from scattering calculations that do not include coupling with $j>0$ levels of H$_{2}$ \cite{lique08b}. Indeed, for collisions at low/moderate temperatures, the probability of rotational excitation of H$_{2}$ is low (the energy spacing between the $j=0$ and $j=2$ levels in para-H$_{2}$ is 510~K) and it is possible to restrict H$_2$ to its lowest rotational level.  In this case, the interaction PES is obtained from an average over the angular motions  of the  H$_{2}$ molecule. 
The resulting PES can then be obtained from a limited number of geometries \cite{lique08b} and the scattering calculations are  equivalent to molecule-structureless atom calculations. This approximation has been found successful when comparing with the full calculations \cite{Dumouchel:11} in the case of SiS\cite{lique08b} or HNC\cite{Dumouchel:11}.  It can be very useful for heavy  or complex molecules where exact calculations with H$_2$ are tedious and time consuming.

\subsubsection{Scattering Calculations}

Once the PES has been determined, the collisional excitation cross sections and   rate coefficients are derived from the solution of the nuclear Schr{\" o}dinger  equations within a given PES. The computation of rotational excitation cross sections is generally performed by using methods based on a time-independent quantum formalism, called "close-coupling" (CC), where the various rotational wave functions are solutions of coupled second order differential equations. The quantal formalism has been introduced by Arthurs and Dalgarno \cite{arthurs:60}  for collisions between a rigid rotor and a spherical atom, and extended somewhat later by Green \cite{green:75}  for collisions between two rigid rotors. This  approach, implemented in several numerical codes (MOLSCAT\cite{molscat:94}, MOLCOL\cite{molcol} or Hibridon\cite{hibridon}) is currently the most precise method as far as all coupling terms are taken into account. The computing time typically varies as $N^3$, $N$ being the number of channels considered (open and closed). Then, when $N$ becomes too large  (high energy collisions with many degrees of freedom and / or low spectroscopic rotational constants), approximations may be introduced. The "coupled-states" (CS) approximation where the Coriolis coupling is neglected during the collision   can be used for heavy systems \cite{Tobola:07,lique08SiS,Troscompt:09}  and preserves a good accuracy in many cases. When the number of rotational states to consider is very large, the infinite order sudden (IOS) approximation that neglects the structure of the molecule, becomes helpful. It usually provides the correct order of magnitude of the rate coefficients in reasonable CPU time.
The CC method, as well as the CS and  IOS approximations have then been applied to rigid  symmetric and asymmetric tops in collisions with a spherical atom and/or a rigid rotor \cite{Green:76top,Green:79top,Garrison:76,Phillips:95,Rist:93}.
Further extension to open shell molecules ($^2\Sigma$ and $^2\Pi$), rotationally excited in collisions with a spherical atom, has been introduced by Alexander \cite{alexander1:82} and Alexander \cite{alexander2:82}, respectively.  These cases are part of the basis routines implemented in the Hibridon public code.
 
Multiple energy surfaces are involved for collisions between open-shell radicals of  $^2\Pi$ electronic ground state like OH, CH or NO and, for example,  a spherical atom, and non-Born-Oppenheimer terms occur. Indeed, the $^2\Pi$ electronic level is split into a lower (labelled $F_1$) and upper (labelled $F_2$) spin-orbit manifold \cite{herzberg}.  When a radical of  {$^2\Pi$}  electronic symmetry interacts with a spherical structureless target, the doubly-degenerate $\Pi$ electronic  surface is split into two distinct PES,  of $A'$  and  $A''$ symmetries\cite{alexander:00}.
Then, in the scattering calculations  it is more appropriate \cite{alexander:85} to introduce  the average  
\begin{equation}
V_{sum}=\frac{1}{2}\left( V_{A''}+V_{A'} \right )
\end{equation}
and the half-difference  
\begin{equation}
V_{dif}=\frac{1}{2}\left( V_{A''}-V_{A'} \right )
\end{equation}
of these two PES.  In the pure Hund's case (a) limit (when the spin-orbit term is much larger than the rotational constant),  
$V_{sum}$   is  principally responsible for inducing inelastic collisions within a given spin-orbit manifold, and $V_{dif}$   allows inelastic collisions between the two spin-orbit manifolds.

$^3\Sigma$  ground state molecules (such as O$_2$, NH, SO ...) are better described through Hund's case (b) \cite{herzberg} and the scattering equations involving such radicals have been first derived by considering a pure Hund's case (b) description of the energy levels \cite{corey:83} and has then been extended to the intermediate coupling as done in Ref. \citen{lique:05}.

To the best of our knowledge, the CC formalism for the interaction between an open-shell radical and a non spherical perturber (H$_2$ in a $ j>0$ 
level) has only been considered for OH($^2\Pi$)  by Offer and van Dishoeck \cite{offer:92}.

\paragraph{Hyperfine structure} Hyperfine structure is usually resolved in  high resolution millimeter astrophysical spectra involving the first rotational quantum numbers. Then, it is necessary to include the corresponding additional levels involved and possible couplings in dynamical calculations as well. 
The quadrupole hyperfine splitting is usually most prominent and occurs for nuclei with a nuclear spin equal or larger than 1 (D, N, Cl, ...). The corresponding pattern is well known for several commonly observed molecules ({\NHHH}, {\NNHp}, HCN, HCl, ....).
A recent review on the possible theoretical approaches to take into account the hyperfine structure in the dynamical calculations was done by Faure and Lique \cite{Faure:12HFS}. As exact CC calculations are extremely heavy   due to the 
large number of coupled hyperfine states, 
 approximate methods are used to provide  hyperfine resolved rate coefficients despite exact theory has been established \cite{Stutzki:85}.
The hyperfine scattering problem  can be significantly simplified \cite{alexander:85hyp} if the hyperfine levels are considered as degenerate.
The collisional cross sections and rate coefficients between hyperfine levels can
then be obtained from the scattering matrices between rotational levels
(i.e. computed without  rotation - nuclear spin coupling terms) using the recoupling method of Alexander and Dagdigian
\cite{alexander:85hyp}. Such a procedure is usually justified and requires careful handling of the scattering matrix elements
as done in Daniel et al.\cite{Daniel:04}. 
The derivation of hyperfine collisional excitation rate coefficients from pure rotational excitation\cite{Neufeld:94} rate constants is 
however less rigorous. 
The IOS approximation allows to derive the hyperfine branching ratios from the rotational rate coefficients and such an approximation is adequate, as shown by Faure and Lique \cite{Faure:12}
, except at very low temperatures ($\simeq$10~K). This approach overrules the randomizing approximation \cite{alexander:85,Guilloteau:81}, where the hyperfine  rate coefficients are expressed as:
$$k_{jF \rightarrow j'F'}=\frac{(2F'+1)}{(2j'+1)(2I+1)}\times k_{j\rightarrow j'} , $$ 
where $I$ is the nuclear spin.
An alternative point of view is introduced by Keto and Rybicki \cite{keto:10} who recommend two approximate methods which are designed
as hyperfine statistical equilibrium (HSE) and proportional approximation. In the HSE methods, hyperfine levels within each rotational level are assumed to be populated in proportion to their statistical weights whereas the second approximation assumes that the collisional rates between the individual hyperfine levels are proportional to the total rate between their rotational levels and the statistical degeneracy of the final hyperfine level of the transition which is equivalent to the randomizing approximation described above. At the light of more recent studies, we find that these approaches 
are less reliable than the recoupling method or the IOS approximation.

\paragraph{Vibrational excitation}
Rovibrational excitation of interstellar molecules has also been the object of theoretical studies. Indeed, over the last two decades, several linear molecules were found in vibrationally excited states, particularly  in circumstellar envelopes \cite{agundez:12,Cernicharo:99,Highberger:00}. 
Many additional geometries are required for the PES calculations in order to overcome the rigid rotor approximation and to include the vibrational motions as well. Calculations of vibrationally inelastic rate coefficients employing the CC or CS approach have been  performed for the stretching mode of vibration \cite{krems:02,Flower:00} where the inclusion of the harmonic approximation
for the vibrational wave function is  discussed.
However, due to the large number of ro-vibrational states to be considered, the Vibrational Close Coupling Rotational Infinite Order Sudden
(VCC-IOS) method \cite{parker:78} has been settled: the vibrational motion of the molecule is described at a CC level whereas  the rotational motion is solved within the IOS 
 approximation.  
 A recent attempt to introduce coupling between rotation and bending at the CC level is reported for collisions between HCN and He \cite{stoecklin:13}.

Alternatively,  calculations of the (vibrationally) inelastic rate coefficients can be done using quasi-classical trajectory (QCT) method \cite{Abrines:66,Bonnet:04,Mandy:09}. The QCT method combines the use of classical mechanics, to treat the scattering process, but the quantization of the reactants is taken into account. Quantization is simulated by means of a "binning" procedure, which involves allocating the final states to discrete values of the corresponding quantum numbers. However, the QCT method is only  valid as long as the classical mechanics that underpins it. When the collision energy decreases, the cross-sections may fail to satisfy detailed balance, which is a symptom of the breakdown of the method. Then, this method may not be very adapted to the calculations of collisional data for cold ISM.

\subsubsection{Experiments}

Experimental studies of rovibrational energy transfer due to collisions are numerous (e.g. review of Ref.\citen{Schiffman:95,Dagdigian:96,Smith:11}). However, most of the work has been done at room temperature and with collisional partners (Ar, Ne, N$_2$ ...) that are not fully relevant for astrophysical applications. Indeed, the use of heavier collisional partners than He or H$_2$ may simplify the experimental setup. Then, measurements of inelastic rate coefficients of astrophysical species are sparse. The determination of collisional rate coefficients at physical conditions relevant for the interstellar medium is  challenging both in terms of temperature and collisional partners.

There are two essential ingredients for carrying out a state-resolved scattering experiment: the preparation of the initial state and detection of the final states populated in the collision process. Double resonance (DR) methods  allow to meet these requirements. In such experiments, one source of radiation, the "pump",  perturbs the rotational thermal distribution of the sample molecules and a second source of radiation, the "probe",  is used to follow the subsequent time evolution until   equilibrium as collisions redistribute molecules amongst the rotational levels.
Microwave-microwave DR  methods in rotational energy transfer studies were pioneered by Oka and coworkers \cite{daly:70} who were able to derive some propensity rules for the $\Delta j$ induced collisions of ammonia by a variety of perturbers, including He and {\HH} at room temperature. 
Later, this technique has been extended to  ground-breaking infrared (IR) - IR DR experiments by Brechignac and co-workers \cite{Brechignac:80}  at room temperature and 77~K for the CO--H$_2$ system. A continuous IR probe system has been settled, yielding total quenching rate coefficients through rotational energy transfer from selected states. 
Even more recently, DR experiments involve a tunable pulsed IR pump laser to prepare molecules in a single rotational level of an otherwise unpopulated excited vibrational state and a second either IR or tunable ultra violet (UV) probe laser to monitor the fate of the state-selected subset of excited molecules \cite{Orr:95}. This method has been coupled to a CRESU (Cin\'etique de R\'eaction en Ecoulement Supersonique Uniforme) apparatus where temperatures as low as 7~K for NO \cite{James:98} and 15~K for CO \cite{Carty:04} can be achieved. A tunable IR laser was used to promote molecules to selected rotational levels in CO ($v = 2$) and NO ($v = 3$). The fate of these molecules is derived by following  the fluorescence induced by a tunable UV probe laser.  
Rate coefficients for collisional removal from the initially excited rotational level $j$ to all other $j'$ levels in the same vibrational state, and  state-to-state rate coefficients for the transfer of molecules from $j$ to various final rotational levels $j'$ are obtained in such experimental studies. In both cases the main gas, and therefore effective collision partner, was He, but such experiments with H$_2$ as the bath gas are feasible. As  several theoretical calculations \cite{krems:02,Lique:07} show that  pure rotational excitation is almost independent on the vibrational state considered, then one can expect that the data provided by Ref. \citen{James:98} and Ref. \citen{Carty:04} could be applied to values involving molecules in their ground vibrational state.

Additionally, rotational energy transfer effects have been studied through spectroscopic line-broadening measurements \cite{Delucia:88,Willey:89,Beaky:96,Willey:00} under very low temperature conditions. 
Although pressure broadening  and rotational energy transfer cross sections are closely connected, as they are expressed in terms of the same $S$-matrix  within the impact approximation\cite{benreuven:66},  
they do not involve the same combination of matrix elements. 
Indeed, the pressure broadening cross sections can be related to the total inelastic state-to-state cross sections and to a pure elastic term via the optical theorem \cite{Baranger:58}.
Therefore, to extract cross sections and rate coefficients for rotational energy transfer from pressure-broadening data, theoretical scattering calculations should be carried out. 

Another  experimental method of studying  collision-induced rotational energy transfer involves molecular beams. 
No absolute values are obtained in such experiments, which represents a significant  shortcoming. As a general rule,
scaling is performed with a theoretical result involving a cross section of relatively large magnitude in order
to minimize possible errors.
For collisions involving ground molecular electronic states, a supersonic jet source is a very effective means of initial state preparation since all molecules are cooled into the lowest state, especially for molecules with large rotational constants.
The crossed molecular beam experiments reported by ter Meulen \cite{meulen:97} serve as examples. Molecules were prepared in a single initial state by electrostatic state selection. This beam of molecules (NH$_3$, OH, or D$_2$CO) was crossed with a second beam of He or H$_2$. Relative state-to-state cross sections were determined by measuring the distribution of molecules over final states using either laser-induced fluorescence or resonance-enhanced multiphoton ionization. Another recent example of such experiments is provided by the work of Yang et al. \cite{Yang:10,Yang:11} on the inelastic scattering of H$_2$O in collisions with He atoms and H$_2$ molecules. The extracted differential cross sections were compared with the results from full CC quantum calculations and excellent agreement was found. 
Finally, very recently \cite{Chefdeville:12}, crossed beam measurements of inelastic transitions of CO by collision with H$_2$ have been performed up to kinetic energies as low as $\simeq$ 4 cm$^{-1}$, opening the way to a detailed validation of theoretical calculations at low temperature. 

\subsection{H$_2$, CO and H$_2$O molecules as benchmark systems}

Among the collisional systems that have been studied, three molecules have been the object of extensive investigations, 
due to their large abundances and crucial role in astrochemistry  
as well as their relevance for theoretical and experimental studies.
H$_2$, CO and H$_2$O molecules are indeed the first, second and third most abundant molecules in the ISM and 
the corresponding inelastic collisions have been the object of both theoretical and experimental  studies during the last fifty years. 

\subsubsection{H$_2$}

The large rotational energy level spacings of H$_2$($^{1}\Sigma_g^+$) involved in  rotational transitions make this molecule well suited for state resolved crossed beam scattering experiments, as well as detailed quantum scattering calculations.

Collisions between hydrogen molecules and helium atoms have been the object of extensive theoretical studies. Indeed, the PES involved in H$_2$--He collisions, as well as the H$_2$--He scattering are relevant in quantum chemistry as the simplest case between a molecule / closed-shell atom. 
Then, the H$_2$--He collisional system has been widely studied from both the theoretical and experimental viewpoint (See Ref. \citen{Tejeda:08} and reference therein) with more emphasis in the vibrational relaxation problem than in the pure rotational decays. 

Two recent theoretical studies \cite{Flower:98b,Balakrishnan:99} have been published using the the so-called MR (Muchnick-Russek) PES\cite{Muchnick:94}. 
Flower et al.\cite{Flower:98b} present the results of quantum mechanical calculations of cross sections and hence rate coefficients for ro-vibrational transitions in ortho- and para-H$_2$, induced by collisions with He. Ro-vibrational levels up to $\simeq$ 15000 cm$^{-1}$ were included in the calculations, and rate coefficients are available for temperatures up to 6000~K. Comparison is made with previous calculations of rate coefficients for pure rotational transitions within the vibrational ground state and with measurements of the rate coefficient for vibrational relaxation  at both high and low temperatures. Agreement was found to be good.

Balakrishnan et al. \cite{Balakrishnan:99} also describe quantum-mechanical scattering calculations of ro-vibrational transitions in H$_2$ induced by collisions with He using the MR PES. Rate coefficients for rotational and vibrational transitions in ortho- and para-H$_2$ were presented in the temperature range 10 $\le$ T $\le$ 5000~K.

A new PES of the H$_2$--He system  has been published more recently by Boothroyd et al.\cite{boothroyd:03} and used in a theoretical scattering study  by Tejeda et al. \cite{Tejeda:08} that also performed   experiments between 22 and 180~K. 
State-to-state cross sections and rates were calculated at the CC level employing both the MR and Boothroyd et al. PES. The fundamental rates $k_{j=2 \to j'=0}$ and $k_{j=3 \to j'=1}$ for H$_2$--He collisions were assessed experimentally on the basis of a master equation describing the time evolution of rotational populations of H$_2$ in the vibrational ground state. These populations are measured in the paraxial region of a supersonic jet of H$_2$:He mixtures by means of high-sensitivity and high spatial resolution Raman spectroscopy. Good agreement between theory and experiment was found for the $k_{j=2 \to j'=0}$ rate derived from the MR-PES and the experimental $k_{j=3 \to j'=1}$ rate was compatible within experimental errors with both PES calculated values. 

Being the simplest neutral molecule-molecule system, the H$_2$--H$_2$ system has also served as a prototype for accurate calculations of tetratomic PES as well as for accurate quantum dynamics treatment of diatom-diatom collisions. 
Several PES of the H$_4$ system are available in the literature \cite{Boothroyd:91H4,Aguado:94,Boothroyd:02,Hinde:08}. They have been included in several studies of rotational energy transfer in H$_2$--H$_2$ collisions within the rigid rotor approximation at thermal energies \cite{Danby:87,Flower:98}.
As a light molecule, one can question the validity to treat  {\HH}  as a rigid rotor. Inclusion of the radial dependence is discussed both in quantum time independent  \cite{Flower:98,Quemener:09}, quantum time dependent  \cite{Lin:03,Otto:08} and semi-classical \cite{Zenevich:99} approaches. Finally, a full dimensional quantum mechanical treatment of  {\HH} with para and ortho- {\HH}  has been performed over a wide range of energies \cite{santos:11,santos:13}. It is found that state to state transitions are very specific in rovibrationnally excited molecules.
Inelastic transitions that conserve the total rotational angular momentum of the diatoms and that involve small changes in the internal energy are found to be highly efficient. The effectiveness of these quasiresonant processes increases with decreasing collision energy.

From the experimental point of view, Mat\'e et al. \cite{Mate:05} carried out measurements of rotational excitation in collisions between two ground state para-H$_2$ molecules and reported rate coefficients for temperatures between 2 and 110~K. They generally found good agreement between their measured data and the theoretical results. Theoretical calculations for the vibrational quenching were also found to be in reasonable agreement with experimental data \cite{Audibert:75}. 

The H$_3$ system can be considered as a prototype of polyatomic systems. 
The corresponding PES has been extensively studied and we only refer to the two most recent calculations  by Boothroyd et al. \cite{Boothroyd:91} and Mielke et al. \cite{Mielke02} and used 
both for inelastic and reactive collisions, including isotopic variants. 
The  inelastic rotational and rovibrational excitations have mainly been studied theoretically as the corresponding experiments are 
challenging. In a pure inelastic process, only $\Delta j =2$ collisions can take place, restricting collisions between either para or ortho forms of  \HH. Quantum mechanical CC calculations have been performed for the first  50 levels of para and ortho-H$_2$  and the corresponding collisional rate coefficients are conveniently expressed as temperature dependent analytical expressions as discussed in Ref. \citen{wrathmall:07} updating older values computed with the older surface \cite{lebourlot:99,flower:00b}. The relative importance of H, He and {\HH}  collisions is discussed in Ref \cite{wrathmall:06}.

  Due to the breakdown of the homonuclear symmetry, this $\Delta j =2$ selection rule does not hold anymore for HD-H collisions, which have been studied by Flower and Roueff\cite{Flower:99, Flower:99c}. 
 $\Delta j = 1$ collisions can also take place within {\HH} if a reactive channel  occurs, which corresponds to an exchange of atomic H during the collision. 
 The full quantal CC study, including both reactive and non reactive processes has recently been revisited
 by  Lique et al. \cite{lique:12:H2} with the most recent PES for the first 9 rotational levels of  {\HH} and the corresponding rate coefficients are derived between 300 and 1500~K.  The agreement with experimental results \cite{Schulz:65}
 is a nice illustration of the validity of the PES and the dynamical treatment, as displayed in Figure \ref{fig4}.

\begin{figure}
\includegraphics[width=8. cm]{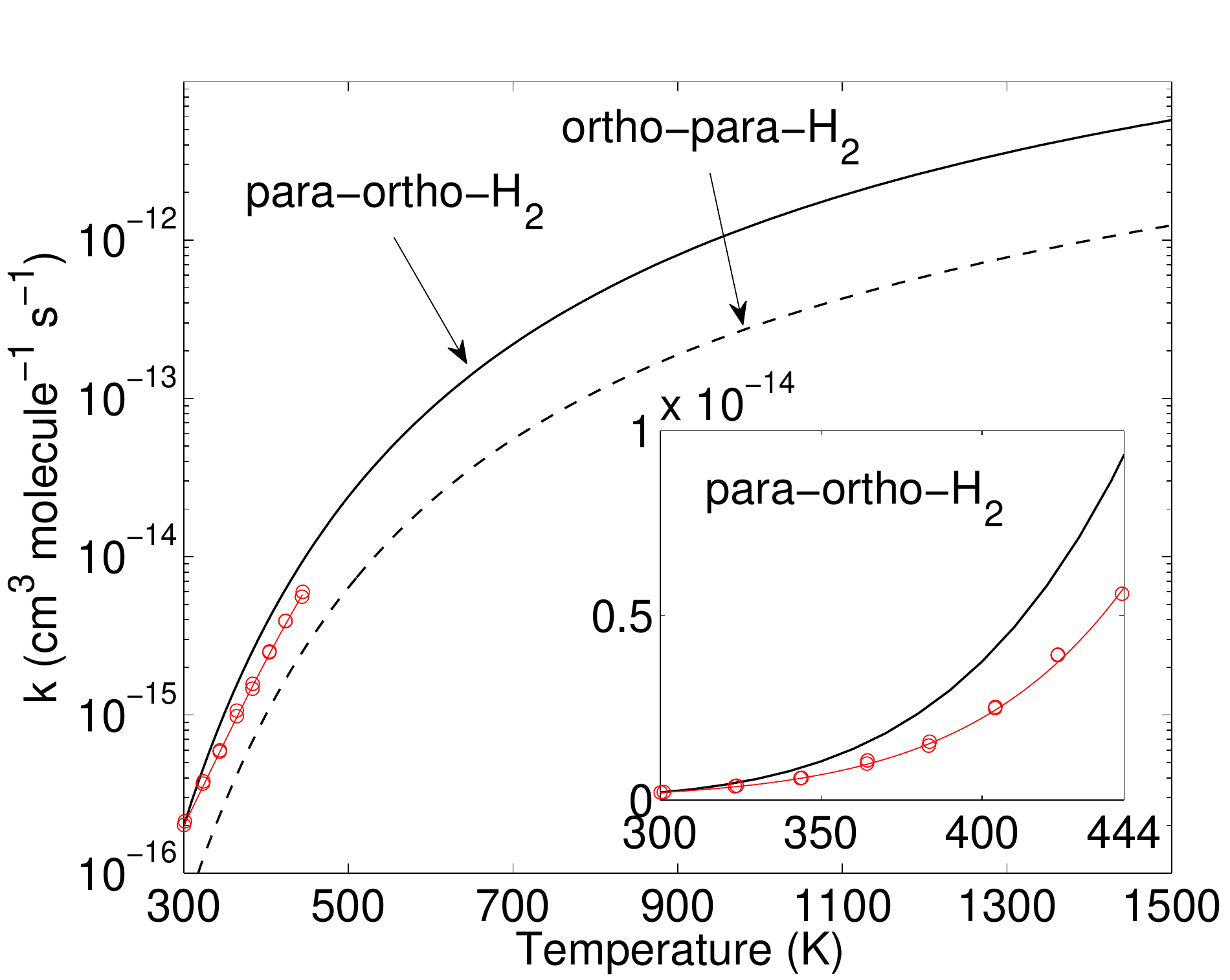}
\caption{Temperature dependence of the rate coefficients for the
  para--ortho-H$_2$ and ortho--para-H$_2$ conversion. The line with
  circles indicates the experimental results of Schulz and Le Roy
   \cite{Schulz:65}. Reprinted with permission from Ref. \citen{lique:12:H2}. {Copyright 2012  American Institute of Physics.}}
\label{fig4}
\end{figure}

The H+H$_2  \rightarrow$ H$_2$+H  reactive collisions (principally its isotopic analogs) 
has been extensively studied as a prototype of  a reaction involving a triatomic
system, both experimentally and theoretically.  
In order to make short a long story, the present status
is an undebatable agreement between experimental and quantum mechanical results, from
state-resolved cross sections to thermal rate coefficients (see
Ref. \citen{Aoiz05} for a review). 

\subsubsection{CO}

Since its discovery in interstellar space more than 40 years ago, carbon monoxide (CO) is extensively and conveniently observed. CO($^{1}\Sigma_g^+$) is indeed the second most abundant molecule in the universe after H$_2$ and 
is the principal mass tracer of molecular gas in galactic and extragalactic sources. Collisional excitation of rotational levels of CO occurs in a great variety of physical conditions and emission from levels with very high rotational quantum numbers have first been identified in circumstellar environments \cite{cernicharo:96} with ISO (Infrared Space Observatory). An impressive bunch of excited rotational lines is also present in protostars and in low and high mass star forming regions \cite{manoj:13,garcia:13} as found in the recent Herschel spectra.
Rovibrational transitions of carbon monoxide and isotopologues are then valuable diagnostic probes  of densities  in moderate to high densities and represent one major contribution to the cooling. 

Collisional rotational excitation of CO by H, He, and H$_2$ was early studied by Green and Thaddeus \cite{green:76} using approximate PES
based on the electron gas model approximation. 
Calculations of  collisional excitation of CO by  He  on an accurate He--CO PES was  initiated only at the end of the 1990's. 
This collisional system has then been the subject of numerous theoretical studies, making it a good candidate for benchmark calculations. An accurate PES have been reported by Heijmen et al. \cite{Heijmen:97b}. Cecchi-Pestellini et al.\cite{Cecchi:02} reported a large set of CC and IOS approximation calculations using this PES. 

Experimental studies have also been devoted to this system either through crossed molecular beams with resonance enhanced multi photon ionization detection \cite{Antonova:99} or double resonance experiments \cite{Smith:04,Carty:04}. 
A satisfactory agreement is obtained, when compared to the recent calculations of Yang et al. \cite{Yang:05} as shown in Figure \ref{fig:CO}. 

Inclusion of the dependence of the CO bond length has been performed in the PES computed by Kobayashi et al. 
 \cite{kobayashi:00}, allowing study of vibrational quenching of CO by He.  Rovibrational relaxation has then been computed using a complete CC approach and very satisfactory agreement is obtained both with experimental measurements reported by Wickham-Jones  et al. \cite{jones:87} and CS computations \cite{peterson:05}.

\begin{figure}
\includegraphics[width=14. cm]{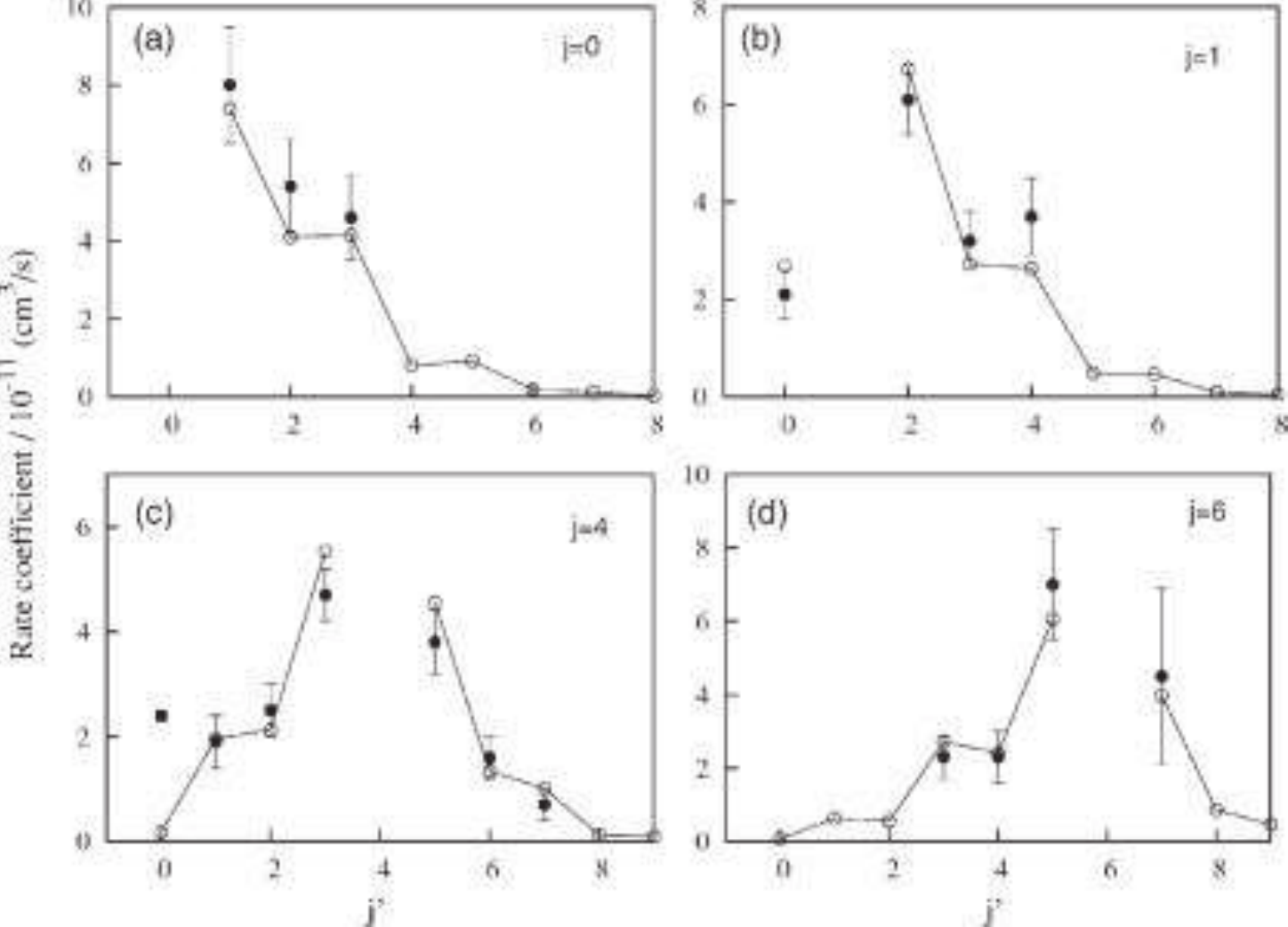}
\caption{State-to-state rate coefficients for rotationally inelastic collisions of CO ($\nu = 2$) with He atom at 63 K: (a) $j = 0$, (b) $j = 1$, (c) $j = 4$, and (d) $j = 6$. Lines with open circles: Yang et al. \cite{Yang:05}; solid circles with error bar: measurements of Carty et al.,  Ref. \citen{Carty:04}. Reprinted with permission from Ref. \citen{Yang:05}. Copyright 2005 American Institute of Physics. }
\label{fig:CO}
\end{figure}

Collisions with {\HH} are however more astrophysically relevant. Jankowski and Szalewicz \cite{Jankowski:98} have used the symmetry adapted perturbation theory (SAPT) to compute a four-dimensional PES of the CO--H$_2$ complex for purpose of comparison with van der Waals experimental spectra \cite{McKellar:98}. 
Based on this PES, Flower \cite{Flower:01} and  Mengel et al. \cite{Mengel:01} have calculated an extensive set of inelastic rate coefficients using a CC approach that agree reasonably well with those computed on an older and less accurate PES \cite{Schinke:85}. However, in contrast to previous works, these authors found that the inelastic rates with $\Delta j=1$  are larger than those with $\Delta j=2$  for collisions with para-H$_2$. This result reflects the crucial importance of the PES, whose inaccuracies are one of the main sources of error in collisional rate calculations. 

Jankowski and Szalewicz \cite{Jankowski:05}   have subsequently refined their PES study by using the CCSD(T) level of theory and by including the dependence of the PES compared to the H$_2$ intermolecular distance and obtained a much better agreement both with the van der Waal complex spectrum and  the second virial coefficient. 
Wernli et al. \cite{wernli06} computed new CO--H$_2$ inelastic collisional rotational excitation rate coefficients from this new PES. Their results were restricted to low temperatures (T < 70~K) because the new PES calculations are found to affect the rotational rate coefficients only at low temperature. Yang et al. \cite{yang:10b} have subsequently extended the temperature range and the number of rotational levels involved in a quantal CC approach and recovered the previous results of Wernli et al \cite{wernli06}.
Differences with the results of Flower \cite{Flower:01} range from a few percent to a factor of 2. 
It is interesting to note that for collisions with para-H$_2$, inelastic rates with $\Delta j=1$  are larger than those with  $\Delta j=2$ while the reverse applies for ortho-H$_2$. As Kobayashi et al. \cite{kobayashi:00} also computed the CO-H$_2$ PES, including radial dependence of CO, Flower \cite{Flower:12CO} has then very recently calculated cross-sections and rate coefficients for the rovibrational excitation of CO by ortho- and para-H$_2$ collisions using the quantal coupled channels method, including CO rotational levels in the $v = 0, 1$ and 2 vibrational manifolds. These results provide reliable values of the rate coefficients for rovibrational excitation of CO for temperatures up to 1000~K.
Extrapolation techniques have further been used by Neufeld \cite{neufeld:12} to extend the collisional excitation rate coefficients of Yang et al. \cite{yang:10b} rates both in temperature (up to 5000~K) and in the number of energy levels.

Finally, Chefdeville et al. \cite{Chefdeville:12} report on crossed-beam experiments  and quantum-mechanical calculations performed on the CO($j=0$) + H$_2$($j=0$) $\to$ CO($j=1$) + H$_2$($j=0$) system. The experimental cross sections show resonance structures in the threshold region in qualitative agreement with the theoretical values.  Despite the PES is shown to reproduce the van der Waals spectroscopy to a high precision, some additional  refinement in the PES and/or in the scattering experiments is required for a  quantitative agreement.  

CO--H collisions may occur at the edge of molecular clouds and the pioneer quantal results of  Green and Thaddeus \cite{green:76}, obtained with an approximate PES, have been widely used in the astrophysical literature. 
It has been recognized recently that a reactive channel is involved in the HCO system and corresponding PES have been specifically computed to account for the HCO spectroscopy \cite{Bowman:86,Keller:96} and the C + OH reaction \cite{Zanchet:06}.
In 2002, Balakrishnan et al. \cite{Balakrishnan:02} reported rate coefficients for rotational and vibrational transitions in CO induced by H atoms using quantum-mechanical scattering calculations and the CO--H interaction potential of Keller et al. \cite{Keller:96} Rate coefficients are presented for temperatures up to 3000~K. Differences by a factor of 30 were found for low temperature rate coefficients compared to earlier results of Green and Thaddeus \cite{green:76}. The discrepancies are attributed to the differences in the details of the interaction potentials, especially the long-range part to which the low-temperature rate coefficients are most sensitive. In particular, Balakrishnan et al. \cite{Balakrishnan:02} reported a large cross section for the fundamental $j= 0 \to j'=1$ rotational transition. This result is in contradiction with that  obtained using the earlier PES of Bowman et al. \cite{Bowman:86} for which the cross section is quite small and which is consistent with an expected homonuclear-like propensity for even $\Delta j$ transitions. 
In order to solve this contradiction, Shepler et al. \cite{Shepler:07} performed CC rigid-rotor  calculations for CO($v = 0, j = 0$) rotational excitation by H  on four different PES whereas  two of them were computed using state-of-the-art quantum chemistry techniques at the CCSD(T) and MRCI levels of theory. They found that the cross sections for the $j= 0 \to j'=1$, as well as other odd $\Delta j$ transitions are significantly smaller compared to even $\Delta j$ transitions using the CCSD(T) and MRCI surfaces. 
 This system has been revisited  by Yang et al. \cite{yang:13co} who used two different recent potential surfaces   \cite{Shepler:07}  and performed CC and CS calculations of the rotational (de)excitation of CO by H. De-excitation cross sections satisfy even {$\Delta j$} propensity rules due to the nearly symmetric character of CO. The rate coefficients obtained with the two potentials agree well for temperatures above about 150K, but discrepancies by a factor of 2 are found for lower temperatures. The temperature dependence of the collision rates has an undulatory behavior due to the resonances in the cross sections. Collisions with H are more important than those involving Helium in mostly atomic regions due to the ten percent abundance ratio of both atoms. In molecular regions however, collisions with molecular hydrogen are largely predominant over those involving hydrogen atoms. He collision rates have a similar order of magnitude than those with molecular hydrogen and  their contribution should be included as the abundance ratio 
is twenty percent. The critical densities of the first J= 1-0, 2-1, 3-2, 4-3 rotational transitions are of the order of 2 $\times$ 10$^3$, 8 $\times$ 10$^3$, 3 $\times$ 10$^4$, 7 $\times$ 10$^4$  cm$^{-3}$ at 10 K. The values increase with the excitation state as the A Einstein emission coefficients vary approximately as J$_{upper}^3$ whereas the collision rates stay within a same order of magnitude for $\Delta$ J =1 de-excitation collisions. 

\subsubsection{\HHO}

Water is the third most abundant molecule and is valuable  both respective to the chemistry and cooling budget of the warm molecular gas.  
 After the detection of a water  highly excited maser transition at 23.235 GHz by Cheung et al \cite{cheung:69} in different galactic sources, the fundamental submillimeter water transitions were  first unambiguously detected by the ISO and SWAS satellites \cite{benedettini:02} and the WISH (Water in star-forming regions with Herschel with informations available at \\
  \verb+ http://www.strw.leidenuniv.nl/WISH/+) key program is specifically devoted to its study and diagnostics. Hence,  rotational excitation collisions of {\HHO} ($^1A_1$) molecules with He and {\HH} have been the object of numerous and detailed studies.

The first water collisional rate coefficients were calculated using He as a collisional partner in the early 1980's by Green \cite{Green:80}. One decade after, Green et al. \cite{green93} have provided improved inelastic rate coefficients for temperatures  ranging from  20 to 2000~K. Several other PESs have been obtained since these studies using different techniques.  The SAPT approach has been used by Hodges et al \cite{hodges:02} and Patkowski et al. \cite{patkowski:02} whereas Calderoni  \cite{calderoni:03} reported a potential on Valence Bond calculations. Stancil and collaborators \cite{yang:07,yang:13h2o} have used the Patkowski {\it{et al.}} PES and computed  quenching of rotationally excited {\HHO}  in collisions with He. Rate coefficients are reported from 0.1 to 3000 K. The present state-to-state coefficients  were found to be in good agreement with previous results obtained by Green \cite{green93} but significant differences were obtained at lower temperatures.

Collisions with {\HH} were also determined \cite{Phillips:95,Phillips:96} making use of a 5D PES described in Phillips et al. \cite{Phillips:94}. This study showed that considering either ortho- or para-{\HH} as a collisional partner could lead to substantial differences in the magnitude of the rate coefficients. However, due to limited CPU resources, Phillips et al. \cite{Phillips:95} considered a limited number of  {\HHO} energy levels, and results are only  available for a reduced range of temperatures. 

At the beginning of the present century, new and extensive theoretical and experimental works on the  {\HHO--\HH} collisional system were started. Dubernet and co-workers \cite{Dubernet:02,Dubernet:03} have extended the calculations of Phillips et al. \cite{Phillips:95} to lower temperatures (to cover the  5--20~K temperature range) using the same PES. A new highly accurate PES for the H$_2$O--H$_2$ system has been subsequently computed\cite{Faure:05,valiron:08} including the full nine dimensional problem. The H$_2$O--H$_2$ interaction PES has been obtained by performing both rigid-rotor and non-rigid-rotor calculations using CCSD(T) approach with moderately large but thoroughly selected basis set. The resulting surface was further calibrated using high precision explicitly correlated CCSD(T)-R12 calculations on a subset of the rigid-rotor intermolecular geometries. The accuracy of the PES is estimated to be about 3 cm$^{-1}$ in the van der Waals minimum region of the interaction. The inclusion of this new PES leads to significantly different collisional excitation rate coefficients, especially for collisions with H$_2(j=0)$ where difference up to a factor of 3 could be found \cite{Dubernet:06}.

This new PES has then be extensively used to provide a full set of data for the collisional excitation of water by molecular hydrogen.
Dubernet and coworkers \cite{Dubernet:09,Daniel:10,Daniel:11} computed inelastic rate coefficients for transitions between a large number of H$_2$O and H$_2$ levels for temperature ranging from 5 to 1500~K. 
Hence, they provided  to the astrophysical community all the data that are useful to interpret the water observations from   the Herschel satellite. If the rate coefficients with para-H$_2$ and ortho-H$_2$ differ significantly at low temperatures, Daniel et al. \cite{Daniel:11} found that the ortho-H$_2$ to para-H$_2$ ratio of para-H$_2$O rate coefficients tends to unity when  temperature increases for all transitions (See Fig. \ref{H2O}). 
Such a finding  allows to restrict the calculations of rate coefficients with a single form of {\HH} at high temperatures.

\begin{figure}
\includegraphics[width=14. cm]{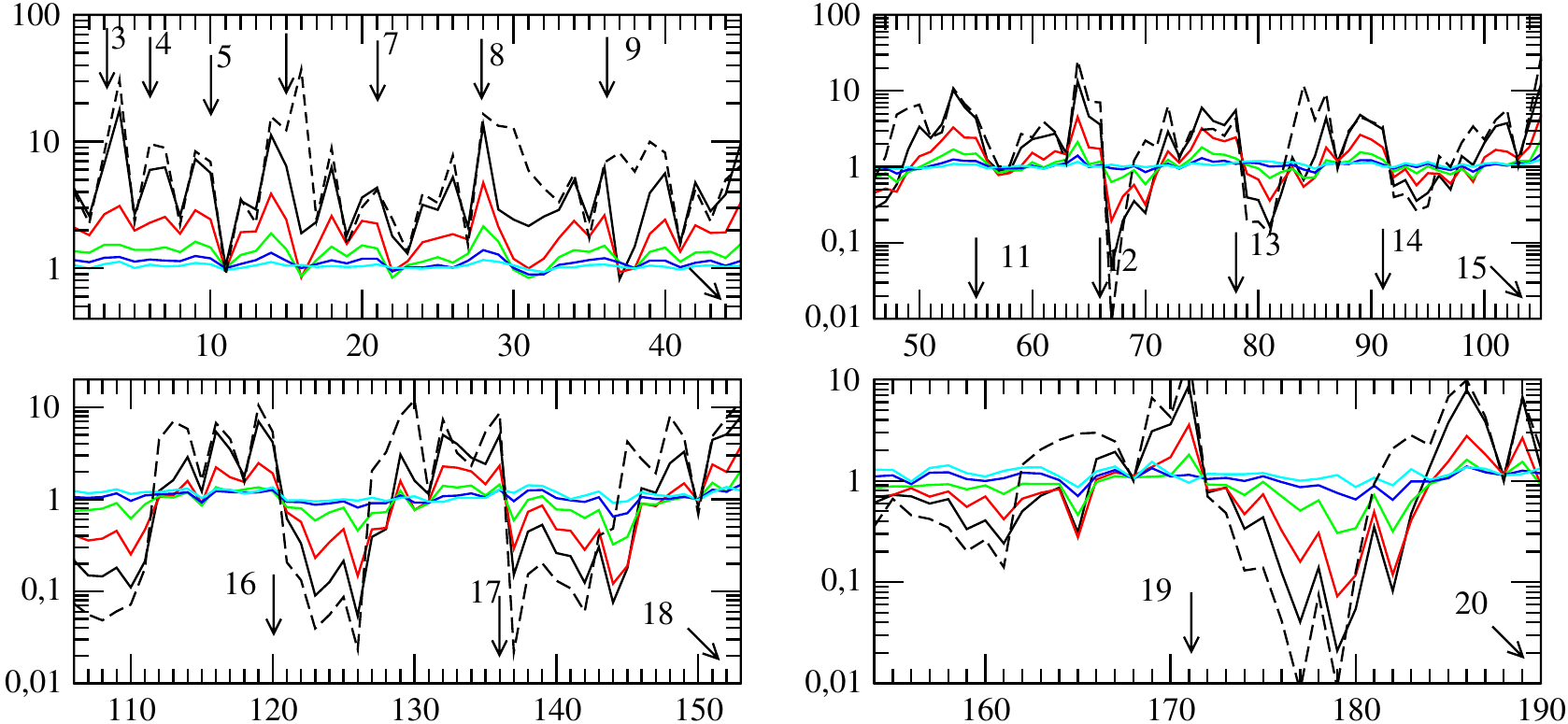}
\caption{Ratios (ortho-H$_2$/para-H$_2$) of thermalized de-excitation rate coefficients from the 1st to the 190th de-excitation transition of para-H$_2$O and for the following temperatures: T~=~20~K (back broken line), T~=~100~K (black), T~=~200~K (red), T~=~400~K (green), T~=~800~K (blue), and T~=~1600~K (cyan). The abscissae indicate the labeling of the de-excitation transitions as given in Ref. \citen{Daniel:10}. Reprinted with permission from Ref. \citen{Daniel:11}. Copyright 2011 ESO.  }
\label{H2O}
\end{figure}

Using the same PES and making use of laboratory measurements for the vibrational relaxation of water and QCT calculations \cite{Faure:05,Faure:07}, ro-vibrational rate coefficients for the first five vibrational states were computed by Faure and Josselin \cite{Faure:08}. 

In complement to  theoretical calculations, Yang et al. \cite{Yang:10,Yang:11} have measured relative rotationally resolved state-to-state differential and integral cross sections for H$_2$O collisions with H$_2$ and He.  A crossed beam machine working at  moderate  collisional energies is combined with velocity map imaging detection. 
 For the H$_2$O-He and H$_2$O-H$_2$ inelastic scattering systems, the state-to-state differential  and integral sections were  found to be in good agreement with full CC quantum calculations (See Fig. \ref{H2O_exp}) after  normalization to the theoretical value of the total quenching cross section of the  1$_{11}$ level of \HHO.

\begin{figure}
\includegraphics[width=8. cm]{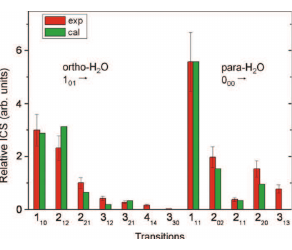}
\caption{Comparison of experimental relative integral cross sections with quantum mechanical calculations for collisions of ortho- and para-H$_2$O with para-H$_2$. 
The experimental uncertainty is $\simeq$ 20\%. Reprinted with permission from Ref. \citen{Yang:11}. Copyright 2011 American Institute of Physics. }
\label{H2O_exp}
\end{figure}

The {\it ab initio} H$_2$O--H$_2$ PES has  also been tested by computing the bound rovibrational levels of the H$_2$O --H$_2$ dimer for the first total angular momenta \cite{Avoird:11}. Theoretical rovibrational levels  agree also very well with the high-resolution van der Waals spectrum reported by Weida and Nesbitt \cite{Weida:99}. 

The considerable work on  water excitation represents an illustrative example of the combined efforts of theoreticians and experimentalists to provide the best possible data to the astronomical community. The impact of the accuracy of the H$_2$O collisional rate coefficients on the astrophysical modeling has  recently been addressed by Daniel et al. \cite{Daniel:12}. They performed non-local non-LTE (Local Thermodynamic Equilibrium) excitation and radiative transfer calculations aiming at comparing the line intensities predicted for H$_2$O when making use of the different collisional rate coefficients. In the absence of radiative pumping by dust photons, it was found that the results based on the quantum and QCT rate coefficients sets will lead to line intensities which are of the same order of magnitude. 

The extensive studies dedicated to  H$_2$, CO and H$_2$O rovibrational excitation also illustrate the present ability of performing accurate and extensive calculations on collisional excitation of molecules. The generally good agreement that is obtained with experimental measurements confirms the accuracy achieved in $ab$ $initio$ theoretical PES calculations. 
It should be kept in mind that such studies involve many collaborations over a long period of time so that the choice of the systems to study should be carefully debated.

\subsection{Other recent results}

In this section, we only review the recent results provided during the last decade and we mainly focus on the interstellar molecules that are frequently observed. We refer the reader to table \ref{tab1} for a more extensive list of available data.

\subsubsection{CN / HCN / HNC}

The cyano radical (CN) molecule is   widely distributed in the interstellar medium (ISM). It was the second interstellar molecule to be identified in diffuse gas, thanks to its visible absorption features, going back to the 40's \cite{mckellar:40,adams:41}. Since then, CN has also been detected through its rotational millimeter emission spectrum  in PDRs  \cite{fuente:95}, Ultra Compact HII regions \cite{Hakobian:11} and prestellar cores \cite{hilyblant:08}. With its open shell structure (the ground electronic state is of $^2\Sigma^+$ symmetry) and high dipole moment, CN also allows to probe the magnetic field through Zeeman spectrum and is a good tracer of high density gas. 
HCN and HNC, closely related molecules, are also powerful probes of high density gas  due to their large dipole moments and large abundances, since, at the opposite of molecules like CO or CS, they do not seem to deplete on grain surfaces in the denser cold areas of prestellar cores \cite{hilyblant:08}. 

Scaling of CO or CS excitation rate coefficients to CN has been tentatively performed \cite{Truong87,fuente:95} but the validity of such a procedure is questionable since CN has a specific rotational energy  diagram resulting from the open shell structure of the ground electronic state. The presence of the nuclear spin of nitrogen further induces an hyperfine structure pattern.
The first realistic CN rate coefficients have only been provided during the last three years. The CN--He PES has been computed at the CCSD(T) level using large atomic basis set and accurate CN--He rate coefficients taking into account the fine and hyperfine structure of CN have been calculated in a quantal CC approach by Lique et al.\cite{lique:10CN} and Lique and K{\l}os \cite{lique:11CN}. Rate coefficients between the lowest 41 fine  structure levels of CN in collisions with He have been calculated for temperatures ranging from 5 to 350~K. These calculations were extended to the CN hyperfine structure using a recoupling approach. The CN--He rate coefficients were then compared with  room temperature experimental rotational excitation (with unresolved fine and hyperfine structure) rate coefficients of Fei et al. \cite{fei:94} (See Fig. \ref{fig:CNexp}) who measured the rotational transfer within the excited $v=2$ vibrational level through the double resonance technique. An excellent agreement was found between the two sets of data as shown in Figure \ref{fig:CNexp}. As rotational transfer is assumed to be independent on the vibrational excitation, the comparison is encouraging.

\begin{figure}
\begin{center}
\includegraphics[width=8.0cm,angle=0.]{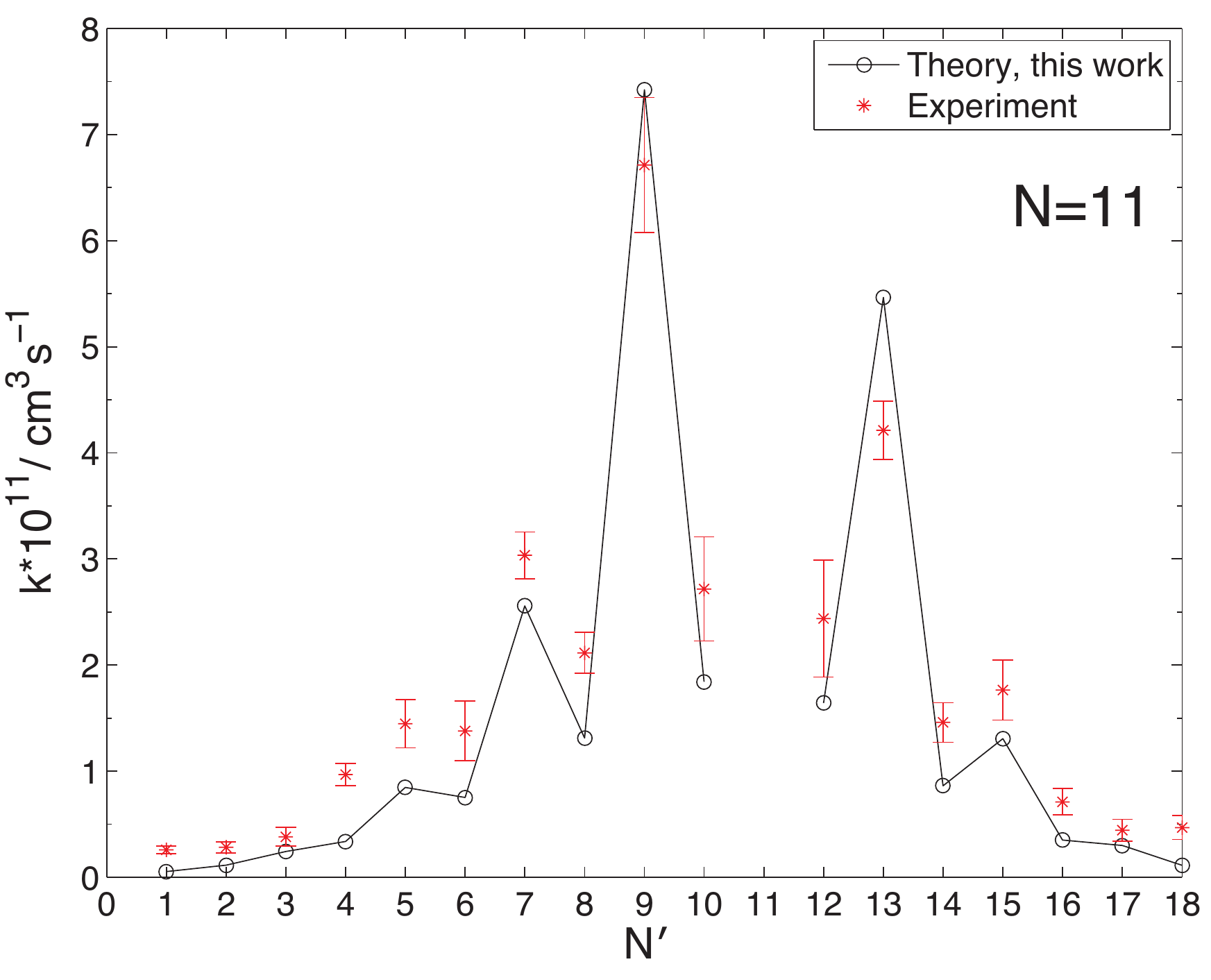}
\includegraphics[width=8.0cm,angle=0.]{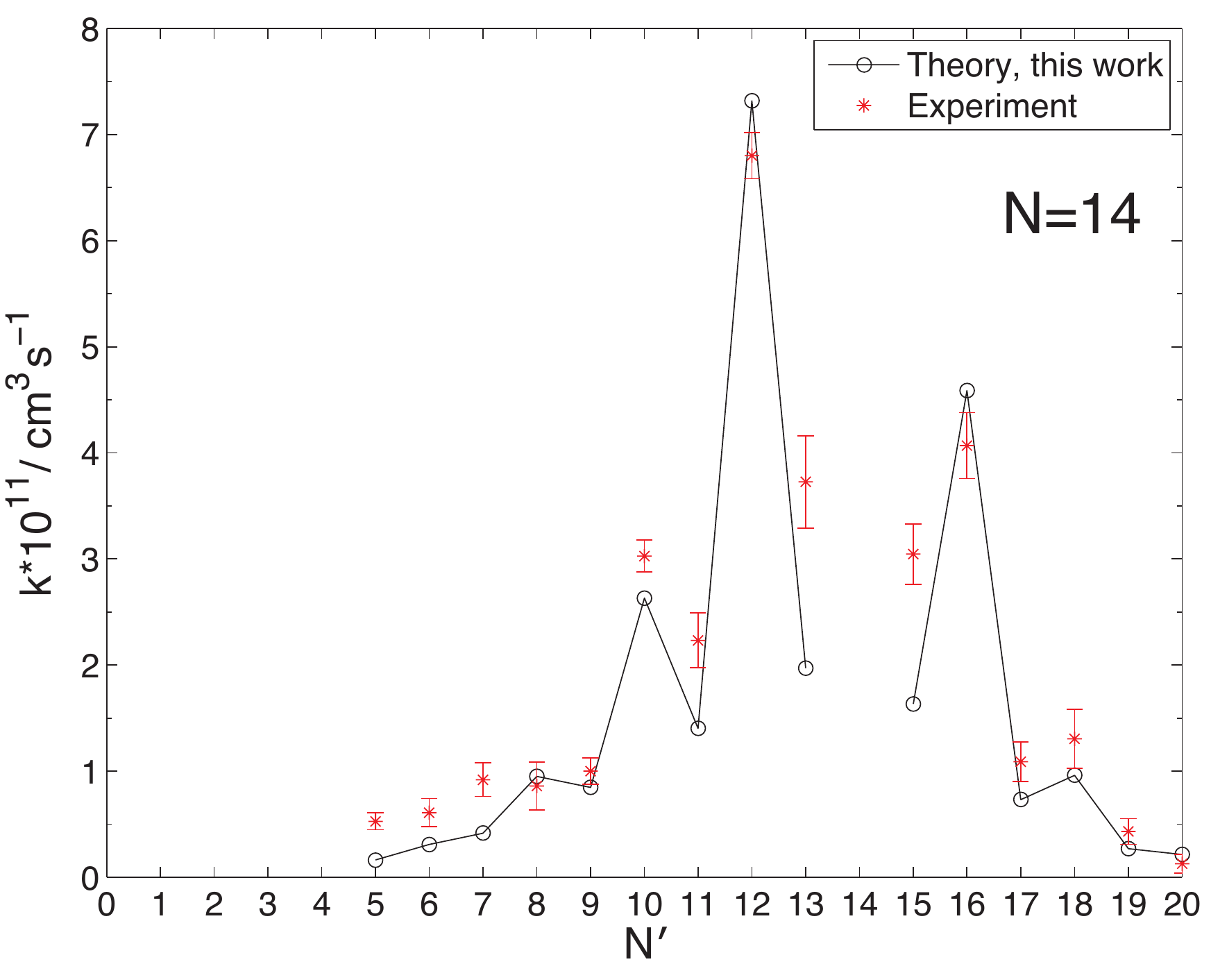}
\caption{CN - He system. Inelastic rate coefficients at 300~K out of the $N=11$ and $N=14$ level, summed and averaged over spin fine-structure: Comparison between theoretical (Ref. \citen{lique:10CN}) and experimental results (Ref. \citen{fei:94}). Reprinted with permission from Ref. \citen{lique:10CN}. Copyright 2010 American Institute of Physics.} 
\label{fig:CNexp}
\end{center}
\end{figure}

The CN--He calculations have  been further extended to the CN--{\HH} system both in terms of  PES and  inelastic cross sections calculations  by Kalugina et al. \cite{kalugina12}.  A new, four dimensional (4D) highly-correlated RCCSD(T) PES has been calculated and used in CC quantum scattering calculations to investigate rotational energy transfer in collisions of CN with H$_2(j=0)$. 
Fine and hyperfine state-resolved rate coefficients were obtained for temperatures ranging from 5 to 100~K.

A strong propensity rules in favor of $\Delta j = \Delta N$ transitions was found for fine structure resolved rate coefficients  both 
for CN--He and CN--H$_2$ collisional systems. For hyperfine structure resolved rate coefficients, the collisions exhibit strong propensity rules in favor of $\Delta j = \Delta F$ transitions for $\Delta j = \Delta N$ transitions (see Fig. \ref{fig:CN}).

\begin{figure}
\begin{center}
\includegraphics[width=8.0cm,angle=0.]{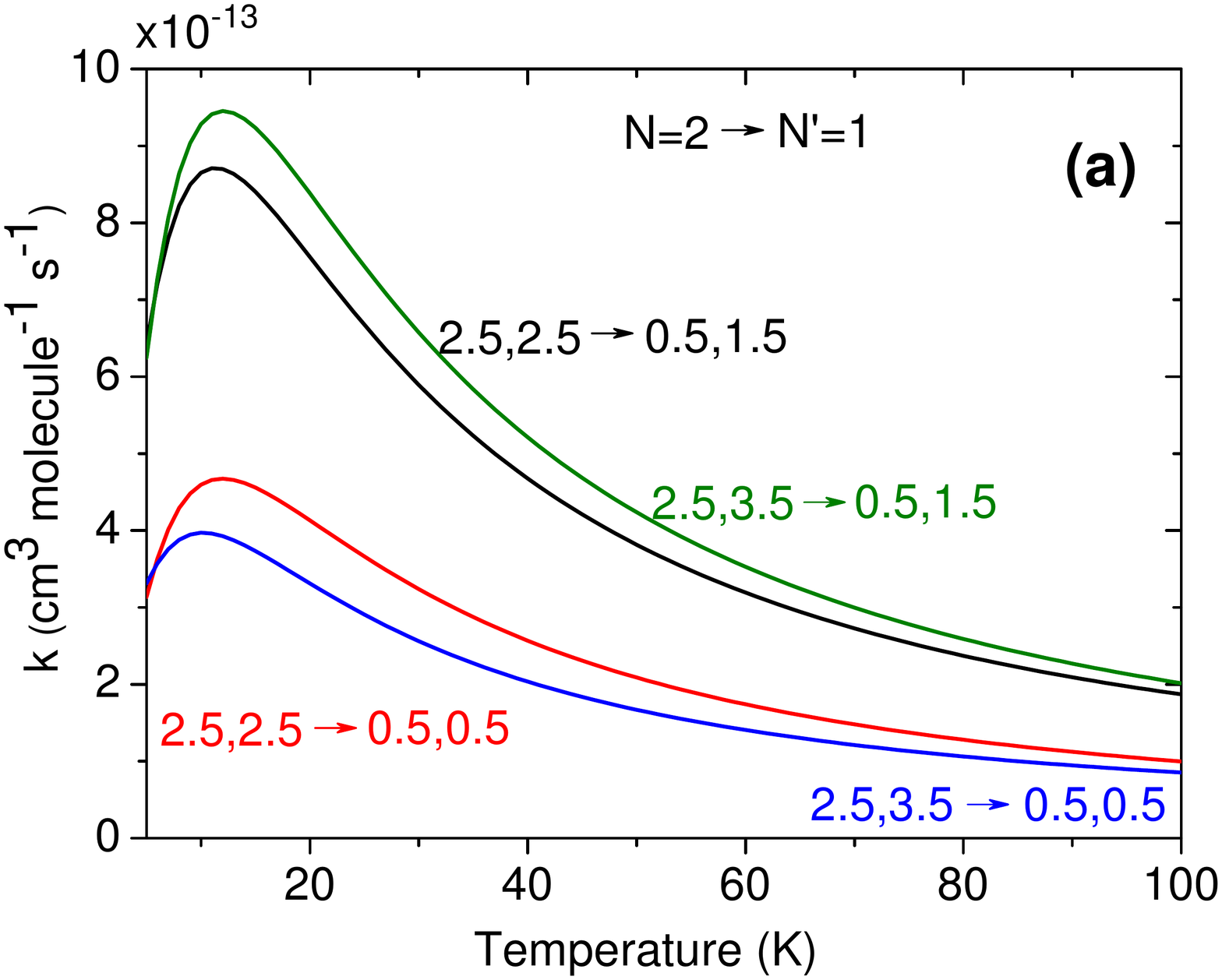}
\includegraphics[width=8.0cm,angle=0.]{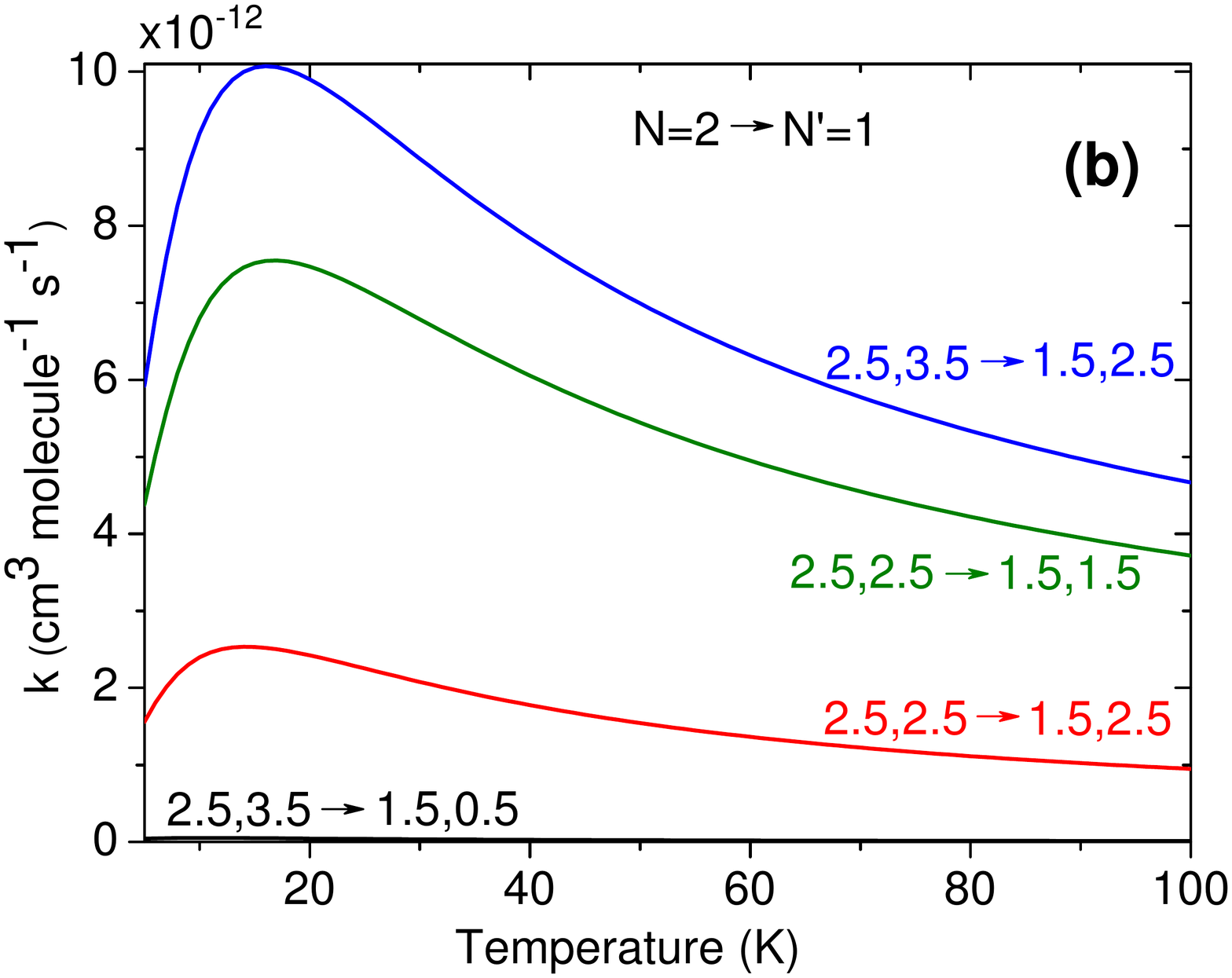}
\caption{Temperature variation of the hyperfine resolved CN--H$_2(j=0)$ rate coefficients for $N=2, j=2.5, F \to N'=1, j', F'$
transitions. Panel (a) corresponds to $\Delta j \ne \Delta N$ transitions. Panel (b) corresponds to $\Delta j = \Delta N$ transitions. Reprinted with permission from Ref. \citen{kalugina12}. Copyright 2012  Oxford University Press} \label{fig:CN}
\end{center}
\end{figure}

Calculation of collisional excitation rate coefficients for the HCN molecule have been amongst the first calculations dedicated to interstellar
applications. 
The rotational and hyperfine excitation of HCN by He [as a substitute for H$_{2}$] has been studied by Green and Thaddeus\cite{green74} and Monteiro et al.\cite{monteiro84,monteiro86}. Indeed, due to the nuclear quadrupole of the nitrogen ($^{14}N$) nucleus with a spin $I=1$, each rotational level $j$ (except $j = 0$) of HCN is split into three well separated hyperfine levels. The HCN hyperfine levels pattern has been found out of thermal equilibrium in a number of cases \cite{Walmsley:82,Turner:97}.
In particular, the three hyperfine transitions arising from the $j=1 \rightarrow j'=0$ rotational transition exhibit a wide variety of relative intensities differing significantly from the ''LTE'' predictions. 
The need for considering explicitly the hyperfine coupling in rotational collisional excitation was then pointed out by Guilloteau and Baudry \cite{Guilloteau:81}. The detection in cold dark interstellar clouds of the isomeric species HNC at a similar and even larger abundance level \cite{hirota98}, despite a significant lower stability, has been widely debated in the astrochemical community \cite{schilke92,talbi:96}. It should be stressed that the corresponding observations have been modeled by assuming identical excitation rate coefficients for both isomers, as no information was available for HNC.

In 2010, Lique and co-workers\cite{Sarrasin:10,Dumouchel:10,Dumouchel:11,benabdallah:12} decided to reevaluate  HCN and  compute  HNC collisional excitation rate coefficients . Two new  2D \textit{ab initio} PES for HCN--He and HNC--He has been built and the corresponding rotational excitation rate coefficients were calculated for a large number of rotational levels ($j = 0-25$) for a wide range of temperatures (5 to 500~K) \cite{Dumouchel:10}. These studies have shown unambiguously that the collisional excitation rates of HCN and HNC by He are quite different as discussed in Sarrasin {\it et al.} \cite{Sarrasin:10}. A strong propensity rule for even  $\Delta j$ transitions is found in the case of HCN--He whereas transitions with odd  $\Delta j $ is favored for HNC--He. These behaviors result directly from a larger (odd) anisotropy of the PES in the HNC--He case. These results were promptly included in the observational modeling studies and modified considerably the determination of the relative abundance of HNC, compared to the previous estimates. The abundances of HNC and HCN were then found almost equal, in much better agreement with astrochemical expectations.

These authors then investigated the HCN/HNC--{\HH}  systems and computed the corresponding  four dimensional \textit{ab initio} PES.  HCN--H$_2$ (para-H$_2$) and HNC--H$_2$ (para- and ortho-H$_2$) (de)excitations rate coefficients were also calculated using a CC approach for a smaller number of rotational levels ($j=0-10$) and for temperatures ranging from 5 to 100~K. The trends and propensity rules are similar to those found for He when para-- and ortho-{\HH} is involved as shown in  Fig. \ref{fig:HCN_HNC}. However absolute values of the collisional rate coefficients are significantly different for para and ortho-\HH. The hyperfine structure of HCN was introduced subsequently by using the recoupling technique.  
Inclusion of the full sample of collisional excitation rate coefficients due to He and {\HH} confirmed the previous finding when only He was included \cite{Sarrasin:10}.  This example is a very nice illustration of interdisciplinary collaborations  leading to a spectacular
improvement  of understanding of both the physical  and chemical context.

\begin{figure}
\begin{center}
\includegraphics[width=8.cm,angle=0.]{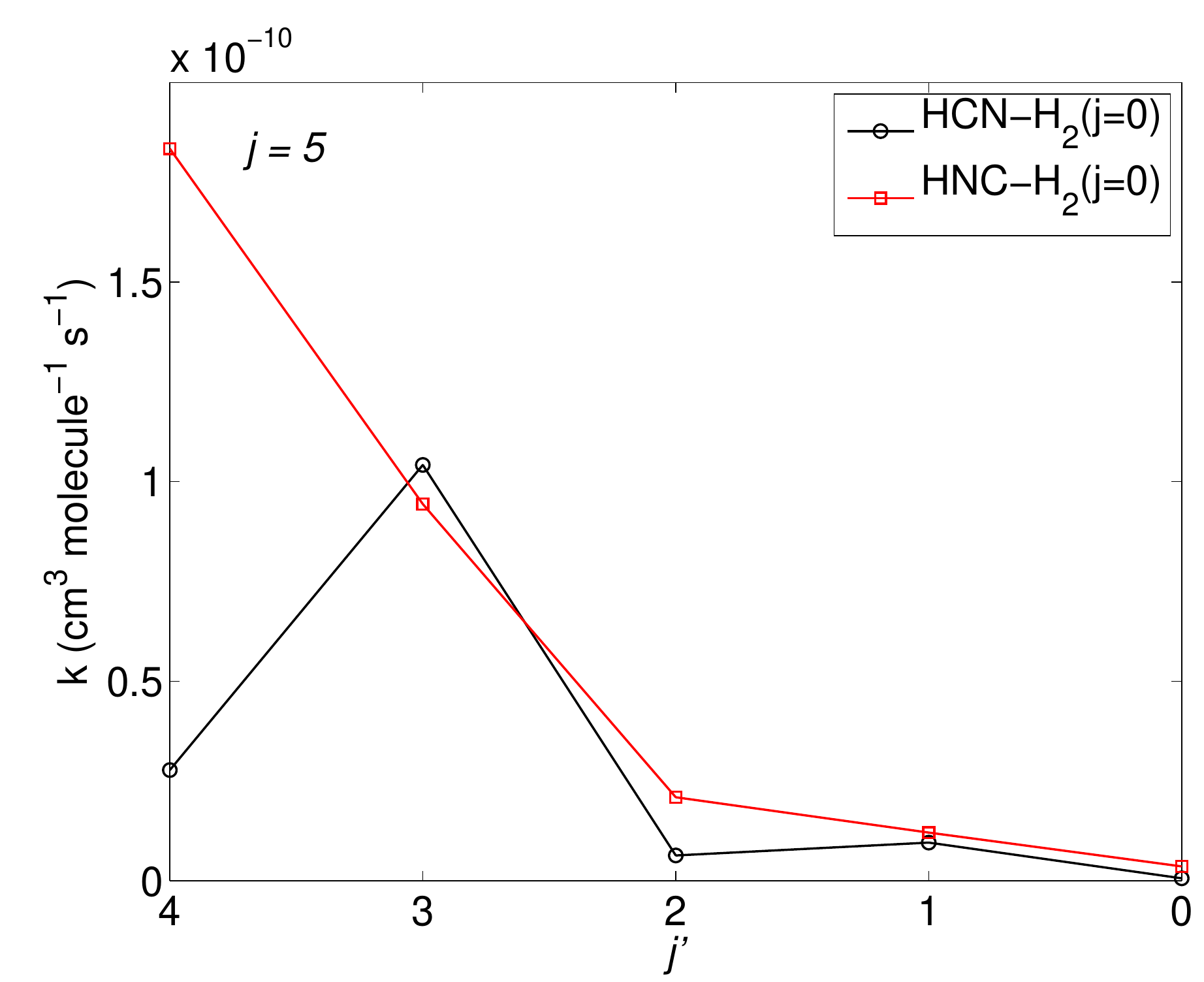}
\caption{Propensity rules for transition out of initial $j$=5 states of the HCN and HNC molecule with para-H$_2(j=0)$ collisions for T~=~50~K. Reprinted with permission from Ref. \citen{benabdallah:12}. Copyright 2012  Oxford University Press.}
\label{fig:HCN_HNC}
\end{center}
\end{figure}

\subsubsection{CS / SiO / SiS / SO / SO$_2$}

Sulfur bearing species are relatively ubiquitous and are valuable diagnostics of  dark clouds, star forming regions, PDRs as well as tracers of shocked regions (along with the SiO molecule).
Recent detailed calculations of two dimensional PESs, involving CS($^1\Sigma^+$)--He\cite{Lique:06CS}, SiO($^1\Sigma^+$)--He\cite{Dayou:06} and SiS($^1\Sigma^+$)--He\cite{Vincent:07} have been performed with highly correlated coupled clusters {\it ab initio} methods. Standard quantal CC dynamical calculations have
been subsequently achieved for pure rotational excitation by the same authors. 
The results of CS and SiO updated older results provided by Turner \cite{turner:92} whereas the SiS -- He system was studied for the first time. The intramolecular dependence of CS and SiS has been further introduced in the PES by Lique and Spielfiedel\cite{Lique:07} and Tobo{\l}a et al.\cite{Tobola:08} respectively, in order to compute also ro-vibrational excitation by He. The corresponding dynamical calculations have been performed at the VCC-IOS  approximation level
as the number of coupled channels is very large for an heavy molecule like CS or SiS.  


The SiS--He calculations has then been extended both in terms of the PES and of the inelastic collisional cross sections, to the SiS--H$_2$ system \cite{lique08SiS,klos08SiS}. A new 4D {\it ab initio} PES has been computed at the CCSD(T) level. 
The authors calculated the collisional rate coefficients involving the first 41 rotational levels of the SiS molecule in its ground vibrational state in collision with para- and ortho-H$_2$. Dynamical calculations of pure rotational (de)excitation were performed within the CS approximation (validated by comparison with CC calculations).  
State-to-state rate coefficients were calculated for temperatures ranging from 5 to 300~K. The authors found that the SiS collisional de-excitation rate coefficients are dependent on the initial rotational level of the H$_2$ molecule: the rate coefficients involving collisions with H$_2(j = 0)$ are significantly lower than the rate coefficients involving collisions with H$_2(j = 1)$  and H$_2(j = 2)$ (see Fig. \ref{fig:SiS}). Such a behavior is similar  to that obtained previously for CO--H$_2$ collisional rotational de-excitation rate coefficients \cite{wernli06}. 

\begin{figure}
\begin{center}
\includegraphics[width=15.cm,angle=0.]{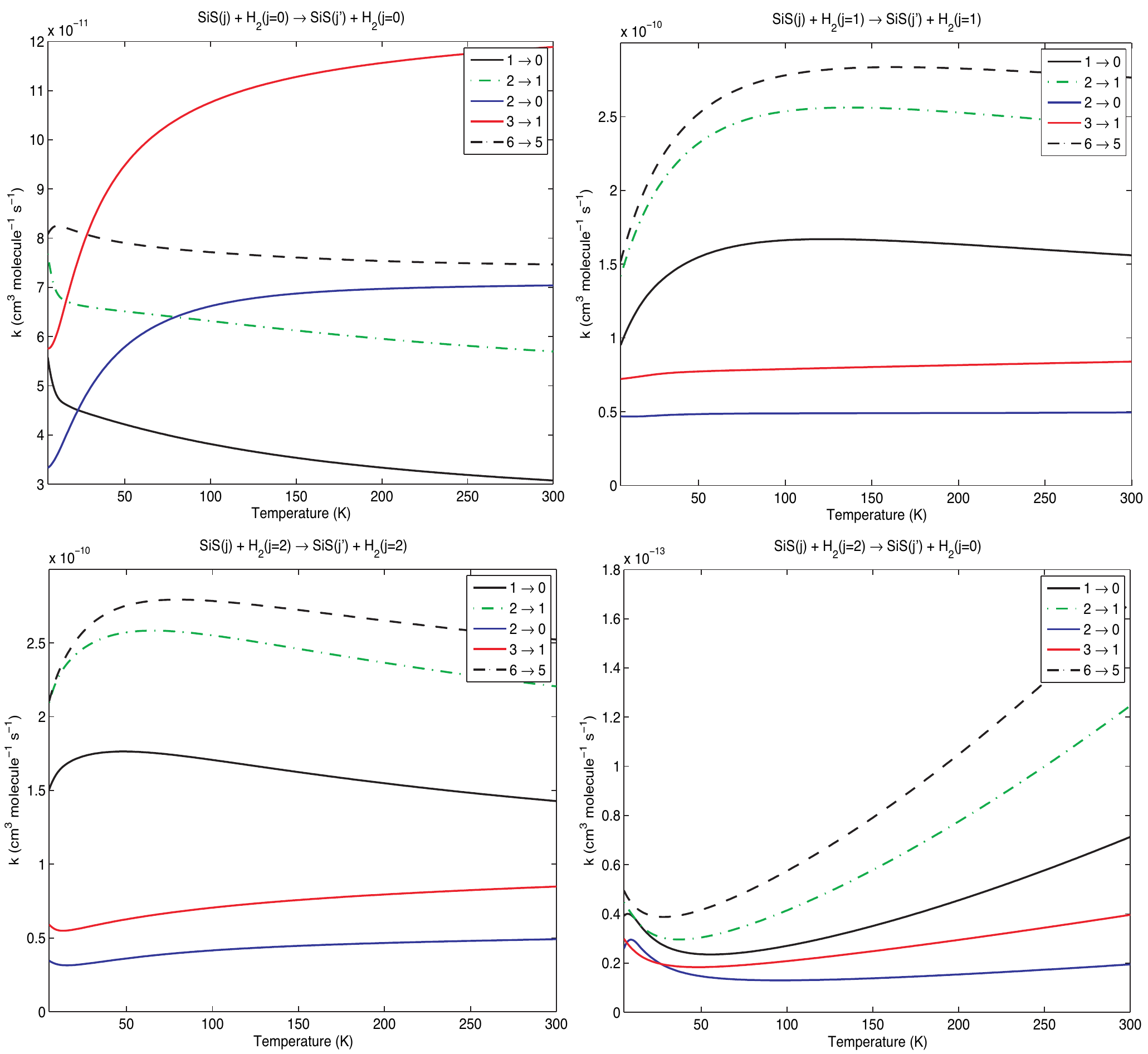}
\caption{Temperature variation  of some relevant collisional rotational de-excitation rate coefficients of SiS by para- and ortho-H$_{2}(j=0, 1, 2)$. Reprinted with permission from Ref. \citen{klos08SiS}. Copyright 2008 Oxford University Press.}
\label{fig:SiS}
\end{center}
\end{figure}

As SO has a ($^{3}\Sigma^{-}$) ground electronic state, the calculations of the corresponding collisional rotational excitation require additional care. 
Indeed, the rotational levels are split by spin-rotation coupling and the fine structure levels of the molecule 
may be described in the intermediate coupling scheme \cite{corey:86}. In 2005, Lique et al.\cite{lique:05} computed a new RCCSD(T) SO--He PES and calculated the  SO--He rate coefficients by using a purely quantum CC approach. 
The quantal coupled equations were solved in the intermediate coupling scheme using the MOLSCAT code\cite{molscat:94} , which was modified to take into account the additional fine structure coupling. 
Rotational and fine structure excitation rate coefficients between the first 31 levels (highest energy of 88.1 cm$^{-1}$) of SO were provided for temperatures ranging from 5 to 50~K. The results differ significantly from those obtained previously by Green\cite{green94} who used the CS--{\HH} PES and  a combination of CS and  IOS approximations. 

The impact of these new SO collisional rate
coefficients  on the interpretation of cold dark clouds observations was also performed and checked  satisfactorily  against other density diagnostics \cite{lique:06}. 
The new collision rate coefficients allow to constrain the physical conditions in 

The fine structure resolved SO--He rate coefficient calculations were extended to higher temperatures and to ro-vibrational transitions (always taking into account the fine structure) using an hybrid approach combining the accuracy of the full CC method and the less time consuming approximations (IOS, IOS scaling relationship)\cite{Lique:06SO,Lique:06SO_2}.

SO--H$_2$ rate coefficients have also been provided in 2007 by Lique et al.\cite{lique:07b} for temperatures ranging between 5 and 50~K using the same approach than for SO--He\cite{lique:05} as only collisions with spherical H$_2(j=0)$ were taken into account.

SO$_2(^1A_1)$,  which is detected with the Herschel Space Observatory, has also been the object of recent studies. The 5D PES of  the SO$_2$--{\HH} has been computed by Spielfiedel et al \cite{spielfiedel:09} and the collisional rate coefficients of the 31 first energy levels have been computed  for temperatures between 5 and 30~K with para-{\HH} as a perturber. 
In 2011, Cernicharo et al.\cite{Cernicharo:11} extended these full quantum scattering calculations to collisions of SO$_2$ with para and ortho-H$_2$. 
As in the cases of CO and SiS, rate coefficients with para and ortho-H$_2$ differ by a factor of two on average, and the largest values of the rate coefficients involve rather  ortho-H$_2$. The rate coefficients were carefully compared to those obtained by Green\cite{green95}, who used the approximate IOS approximation in the collisional treatment and considered He as good representative of \HH. Large differences, exceeding a factor ten for some transitions, were found,  and the rate coefficients involving H$_2$ are always larger than those obtained with He. No particular propensity rule was found, but the value of the rate coefficients decrease for increasing $|\Delta j|$ and $|\Delta k_a|$. 

The new collisional rates of SO$_2$ have also been used to predict the behavior of the SO$_2$ transitions in the centimeter and millimeter domain \cite{Cernicharo:11}. The 12.26 GHz transition connecting the 1$_{11}$ level to the 2$_{02}$ metastable level should be detectable in absorption against the cosmic background radiation field for a wide range of densities and the 3$_{13}$ $\to$ 4$_{04}$ transition at 383.8 GHz is found  as a potential maser.  These predictions have indeed been confirmed by observations towards dark clouds \cite{Cernicharo:11}, which help constraining the abundance of SO$_2$ in these environments. 

\subsubsection{NH$_3$ / NH}

Ammonia was the first polyatomic molecule detected towards the galactic center thanks to its centimeter inversion transitions \cite{Cheung:68}. It was immediately recognised as a potential thermometer since  the level energies involved in these  inversion transitions can only be connected via collisions. In addition, as they arise from quite different level energies but take place  in a narrow
wavelength window close to 1.3 cm, they  can be probed by a single telescope and detector, i.e. at a same spectral and spatial resolution. Detection of  rotational transitions was  {achieved  significantly later as the corresponding frequencies lie in the submillimeter and far infrared. 
 The transitions  are detected in emission in hot core regions and are also seen in absorption from the foreground envelopes \cite{persson:10,hily10}.
Whereas NH \cite{meyer:91}, NH$_2$\cite{vandishoeck:93}, have been detected respectively in the visible and submillimeter  absorption two decades ago in diffuse clouds and Sgr B2(M), these molecules are now detected in various sources thanks to the Herschel HIFI instrument  and allow to constrain the nitrogen chemistry. 
Study of ammonia inelastic collisions in presence of He has been investigated by Green \cite{Green:76top} who used an electron gas model PES and introduced the quantal dynamical formalism for the excitation of  symmetric top molecules  with additional inversion motion. 
Consideration of the PES involving {\HH} has been subsequently performed by Danby et al.  \cite{danby86,danby88} with the self consistent field approximation and long range perturbation theory.  Then, these authors derived the corresponding collisional excitation rate coefficients and provided the necessary data to achieve temperature diagnostics. Experiments on broadening cross sections obtained at very low temperatures \cite{willey:02} further 
agreed with these predictions and showed that cross sections involving molecular Hydrogen were at least five times larger than those obtained with He. 
Thanks to the computation of a new PES obtained for NH$_3$--He collisional system by Hodges and Wheatley\cite{Hodges:01}, Machin and Roueff\cite{machin:05} computed quantal CC calculations of rotational excitation rate coefficients of NH$_3$ through collisions with He which were compared to the previous results of Green \cite{Green:76top} and some 
experimental cross sections given at a well defined relative kinetic energy by Schleipen and ter Meulen \cite{schleipen:91}.
Significant differences were obtained for the fundamental ortho transition of ammonia.
The same PES was subsequently used for isotopologues of NH$_3$, as will be discussed later. 
The  NH$_3$--{\HH} PES was also reexamined at a very high level of accuracy within the coupled-cluster CCSD(T) level with a basis set extrapolation procedure.  NH$_3$--para-H$_2(j=0)$ rates coefficients were then obtained for all transitions involving ammonia levels with $j \le 3$ and for kinetic temperatures in the range 5-100 K\cite{maret09}. 
The results are in good agreement with previous calculations of Danby et al \cite{danby88} and display significant differences from those obtained for Helium as a perturber. 
The  NH$_3$--para-H$_2(j=0)$ calculations were  extended to the  ND$_2$H  -- para-H$_2(j=0)$ system by Wiesenfeld et al. \cite{wiesenfeld11} and the calculations involving the other isotopologues should also been undertaken. 
The NH($^{3}\Sigma^{-}$) -- He collisional system has been investigated by Tobola et al. \cite{tobola:11} who computed the collisional rotational and fine structure excitation rate coefficients of NH by He in a quantal CC approach. The results are validated through comparison with room temperature measurements by Rinnenthal and Gericke\cite{rinnenthaltully:02}. The hyperfine resolved rate coefficients were then obtained by using the recoupling technique involving both nitrogen and hydrogen nuclear spins \cite{dumouchel:12NH} and corresponding values are reported for a range of temperatures between 5 and 150~K. Fine structure resolved rate coefficients present a strong propensity rules in favor of $\Delta j = \Delta N$ transitions, as expected from theoretical considerations. The $\Delta j= \Delta F_1 = \Delta F$ propensity rule is observed for the hyperfine transitions, $F_1$ and $F$ stands for the hyperfine level due to the coupling with the nuclear spin of H and N, respectively. 
ND--He rate coefficients were also computed and the results will be discussed in the isotopologues section of this review. 

\subsubsection{O$_2$ / OH / NO }

Oxygen compounds apart from CO and H$_2$O are also detected and we focus on O$_2$, predicted to be the oxygen reservoir in cold dark cloud chemical models and on the OH and NO radicals. 
With a symmetric $X^3\Sigma_g^-$  ground electronic state, O$_2$ can not be detected  via electric dipole transitions and observational searches have been performed via  its magnetic dipole rotational transitions. In addition, one has to avoid the oxygen abundant terrestrial atmosphere so that its discovery in the interstellar medium requires spatial instrumentation.  Finally, O$_2$ has been detected very recently with the ODIN  \cite{larsson:07} and Herschel \cite{goldsmith:11,liseau:12} satellites at a very low abundance level. The main formation channel is supposed to be the OH + O $\to$ O$_2$ + H reaction which has been
studied in the laboratory down to 70~K \cite{carty:06}. Numerous theoretical studies have also been devoted to this system 
with different dynamical approximations and 
the most recent study of Lique et al. \cite{Lique:09O2} predicts an efficient formation of O$_2$ at low temperatures.
As far as collisional excitation is concerned, O$_2$--He rate coefficients were calculated only in 2010\cite{lique:10}. As for the SO molecule, the quantal coupled equations were solved in the intermediate Hund's coupling  representation of the energy levels. Collisional excitation rate coefficients  involving the lowest 36 fine structure levels of the O$_2$ molecule were determined for  temperatures ranging from 1 to 350~K. 
The values of collisional rate coefficients depend slightly on $N$, the rotational quantum number and a propensity rule
for  F-conserving transitions  ($\Delta j = \Delta N$  in pure Hund's b coupling case) is found.
Very recently, Kalugina et al.\cite{Kalugina:12O2} published a new 4D PES for the O$_2$--H$_{2}$ system neglecting the reactive channels that open only at very high temperatures. Calculations of fine structure resolved rate coefficients are in progress.

OH and NO are both open shell radicals with a $^2\Pi$ ground state electronic structure. OH has been first detected thanks to its  
$\Lambda$ doubling transitions at 18 cm where hyperfine structure is resolved. Subsequently, rotational transitions have also been detected in the far infrared and sub millimeter windows with the ISO and Herschel satellites. Its presence is strongly linked to water formation/destruction and it is widely observed under various physical conditions. As an heavier molecule,
 rotational transitions of NO lie in the millimeter window and have been first detected by Liszt and Turner \cite{liszt:78} in the galactic
 center. Further detections have then been reported in dark cold clouds \cite{gerin:92}. Its presence is linked to OH via the N + OH  $\to$ NO + H reaction.


The OH--{\HH}  system has been thoroughly studied in Flower's group \cite{dewangan:87,offer:94} who provided rotational, fine and hyperfine structure excitation rate coefficients in collisions with para and ortho \HH. These values are still considered as state of the art and compare satisfactorily to experimental studies.
OH--He and NO--He fine structure resolved rate coefficients were computed by K{\l}os et al.\cite{klos:07OH,klos:08NO}. The calculations were performed using the CC and CS approaches, respectively. For OH, the collision rate coefficients within a spin-orbit  ladder ($F_1 \to F_1$ or $F_2 \to F_2$)  are of the same order of magnitude than those involving spin-orbit  changing collisions ($F_1 \to F_2$ or $F_2 \to F_1$). 
The situation is opposite for NO--He induced transitions where collision rate coefficients within a spin-orbit ladder are about 2 orders of magnitude larger than the others.
These differences can be explained by comparing the $V_{sum}$ and $V_{dif}$ terms of the two PES.
The OH--He calculations were validated by a detailed comparison with the crossed beam experiments of Schreel et al. \cite{Schreel:93} and of Kirste et al.\cite{Kirste:10} (See Fig. \ref{fig:OH}). 
\begin{figure}
\begin{center}
\includegraphics[width=8.cm,angle=0.]{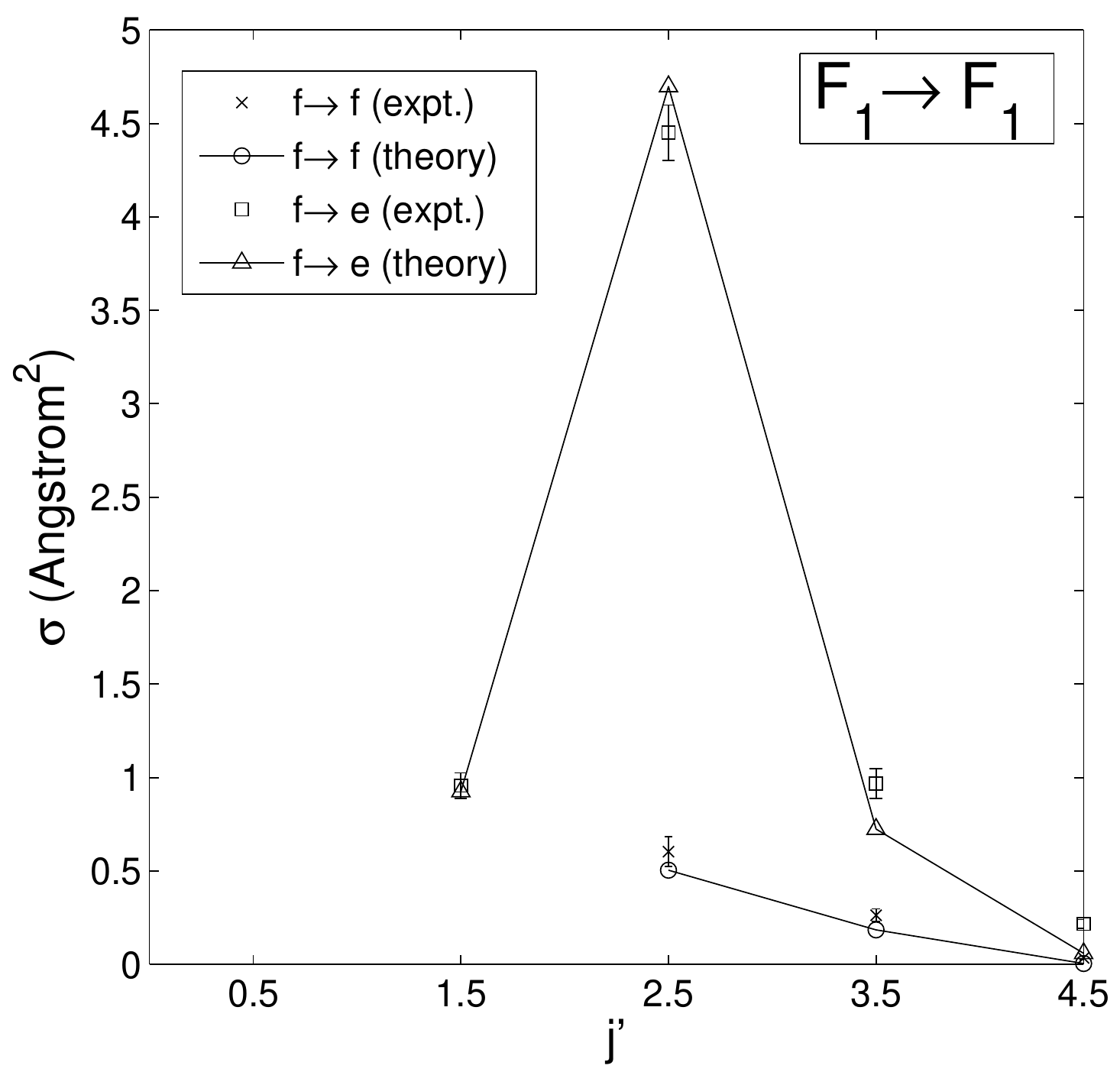}
\includegraphics[width=8.cm,angle=0.]{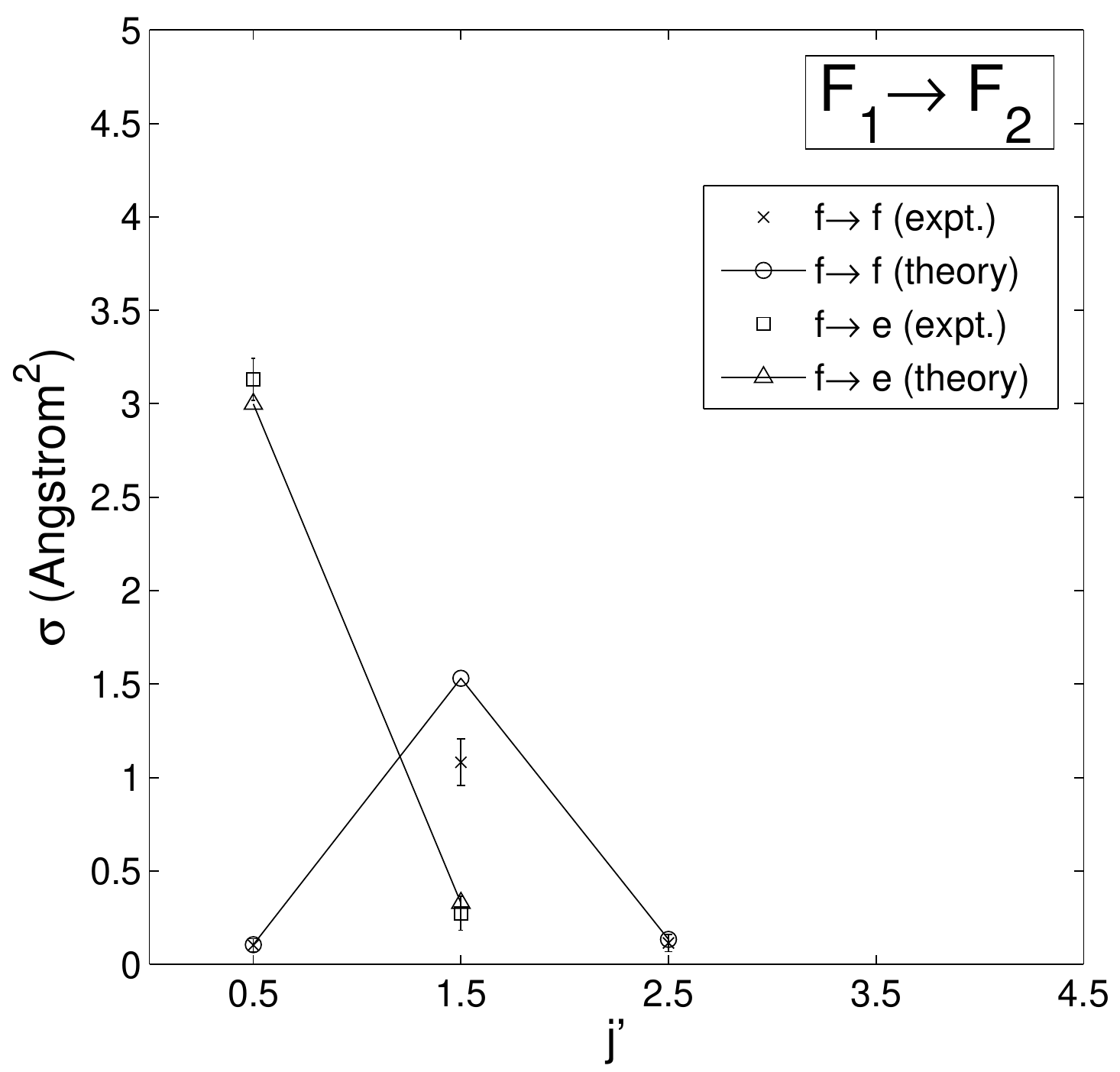}
\caption{OH -- He system. Comparison between  theoretical values \cite{klos:07OH} and experiments  \cite{Schreel:93}
for a collision energy of 394 cm$^{-1}$. (a) 
Integral cross sections involving spin-orbit conserving transitions out of the $N=1$, $j=1.5$ ($F_1$), {$f$} level [$e/f$ label is related to parity as defined by Brown et al. \cite{brown:75}].
 (b) Integral cross sections involving  spin-orbit changing transitions ($F_1\to F_2$) out of the  $N=1$, $j=1.5$ ($F_1$), {$f$}. 
 }
\label{fig:OH}
\end{center}
\end{figure}

The NO--He rate coefficients calculations of K{\l}os et al.\cite{klos:08NO} have subsequently been extended to the hyperfine level transitions within a statistical approach \cite{lique:09}. These rate coefficients have further been introduced in radiative transfer codes and allow to derive the actual values of NO column densities. 

Fine structure resolved NO--He rate coefficients were also measured by James et al. \cite{James:98} using the CRESU technique. State-to-state rate coefficients for rotational energy transfer 
have been measured for six temperatures : 295, 149, 63, 27, 15 and 7~K. 
The comparison with theoretical calculations is very satisfactory for all considered temperatures.

\subsubsection{C$_2$ / C$_2$H / C$_3$ / C$_4$ / HC$_3$N}

Small carbon chain molecules are present in a variety of interstellar environments and circumstellar envelopes of carbon stars and constitute a non-negligible reservoir of carbon. Some of them have been first detected in space and their identification resulted from a combination of quantum $ab$ $initio$ calculations and a good knowledge of the astrophysical environment. It is remarkable that such an extraterrestrial spectroscopy
has been sometimes achieved before or at the same time as the full experimental analysis 
\cite{guelin:78,cernicharo:91a, cernicharo:91b,guelin:97}. Two main hypothesis of production are suggested, one related to the  building of carbon chains from successive addition of carbon, the other resulting from the destruction of larger carbon compounds related to dust such as PAHs, amorphous carbon, etc...

Calculations of inelastic collision rate coefficients involving carbon chains are challenging because the ground and the first excited electronic states are often close in energy, as in the case of C$_2$. In addition,  small carbon clusters are generally "floppy" in their electronic ground state: the frequency of the bending modes may be  as low as 100 cm$^{-1}$ and the resulting spectrum is hardly expressed with the usual spectroscopic constants involving a good decoupling between the various spectroscopic modes.
The symmetric C$_2$ molecule is not detectable through radioastronomy but has been detected in absorption in the red band Phillips  electronic system in front of bright stars up to rotational quantum number values of 12 \cite{adamkovics:03,iglesias:11,welty:13}.
The rotational excitation and de-excitation of  C$_2$($^1\Sigma_g^+$) by He and H$_2$ has been investigated by Najar et al.\cite{najar:08,najar:09}. Due to the multiconfigurational character of the ground electronic state of the two complexes, new {\it ab initio} PES for the C$_2$--He and C$_2$--H$_2$ systems were computed using MRCI method. Inelastic scattering cross sections of C$_2$ by He and para-H$_2(j=0)$ were calculated using both the CC approach (for low collisional energies) and the IOS approximation yielding, after Boltzmann thermal averaging, to C$_2$--He and C$_2$--H$_2$ rate coefficients up to temperature of 300 and~1000 K, respectively. C$_2$--He and C$_2$--H$_2$ rate coefficients were compared and significant differences were found between the two sets of data, questioning  the  prediction of closely related values based on the SiS case \cite{lique08b}.

Collisional rotational excitation of the C$_{2}$H ($^{2}\Sigma^{+}$) radical by He was investigated by Spielfiedel et al.\cite{Spielfiedel:12}. A new PES is computed at the RCCSD(T) level. This radical is isoelectronic of CN and the hyperfine structure is
due to the magnetic dipole of the hydrogen. 
State-to-state collisional excitation cross sections linking the 25 first  fine structure levels of C$_{2}$H were calculated for energies up to 800 cm$^{-1}$ which yields, after thermal averaging, to corresponding  rate coefficients up to T~=~100~K. The exact spin splitting of the energy levels was taken into account. It was shown that the rate coefficients for $\Delta j= \Delta N$ transitions are much larger than those for $\Delta j \ne \Delta N$ transitions and that  $ \Delta j = \Delta F$ propensity rules hold for the hyperfine transitions. 

Rotational (de-)excitation of C$_3$($^1\Sigma^+$) by collisions with He was studied in 2008 by Ben Abdallah et al.\cite{Benabdallah:08} for low  kinetic energies, in order to restrict the study within the ground vibrational level. An excited bending mode is indeed occurring at an energy about $\simeq$ 60 cm$^{-1}$ above the ground state. The PES has been calculated using CCSD(T) approach. Full CC quantum scattering calculations for collision energies up to 50 cm$^{-1}$ were performed in order to deduce the rate constants for rotational levels of C$_3$ up to $j = 10$, covering the temperature range 5-15~K. 

C$_4(X ^3\Sigma^-)$--He collisional excitation rate coefficients were computed by Lique et al.\cite{Lique:10C4}. After a careful validation of the RCCSD(T)-F12x (x=a/b) methods for the generation of multi-dimensional PES for scattering calculations, the authors studied the interaction of the carbon rich interstellar species C$_4$ with atomic He using a new {\it ab initio} PES (See Fig \ref{fig:C4}).

\begin{figure}
\begin{center}
\includegraphics[width=10.cm,angle=0.]{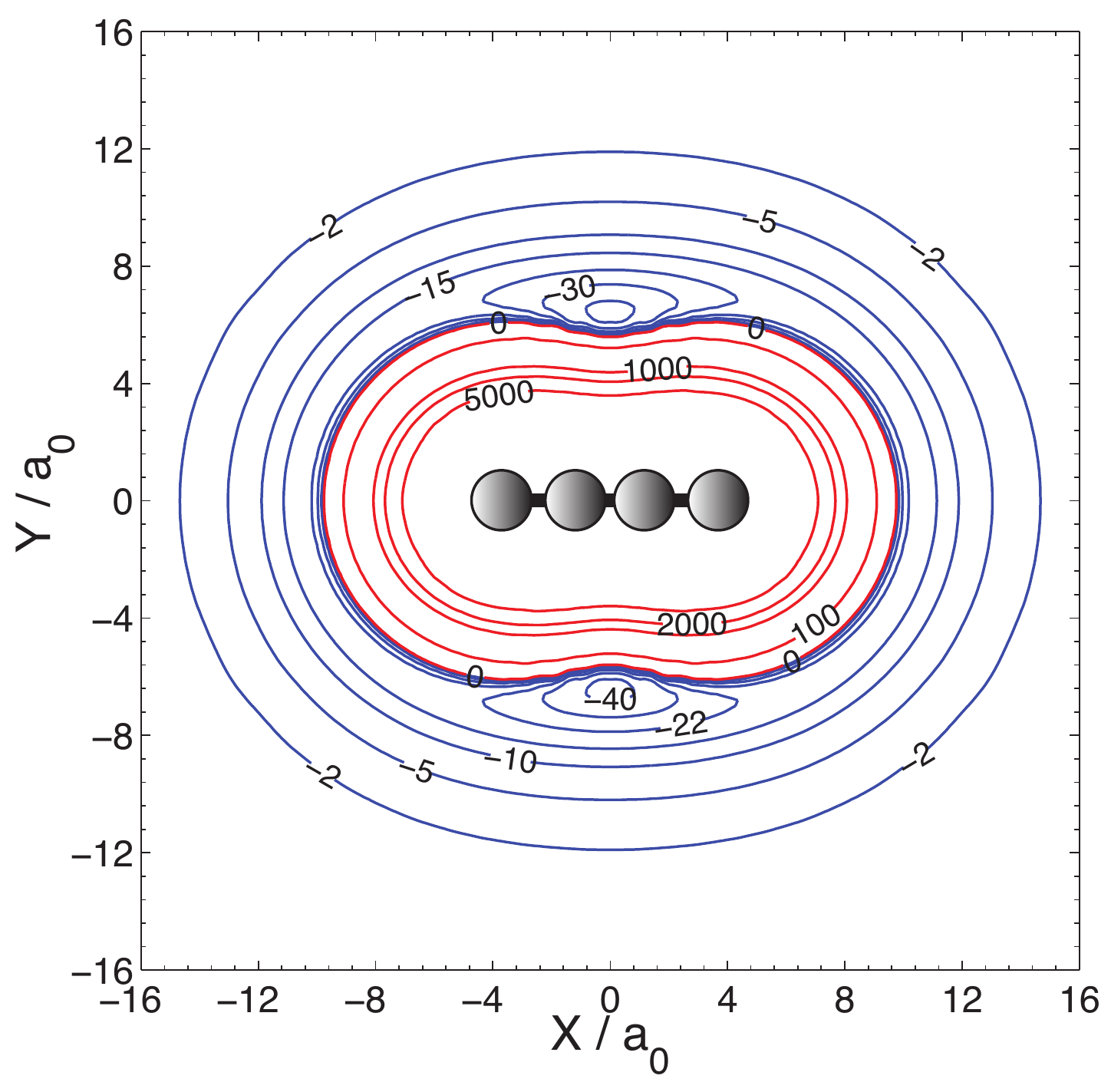}
\caption{2D-contour plot of the aug-cc-pVTZ/RCCSD(T)-F12a PES for the C$_4$--He $^{3}A"$ state correlating to the $\mathrm{C_{4}}$( X $^{3}\Sigma_g^-$) + $\mathrm{He}$($^{1}S$) asymptote. Energies are in cm$^{-1}$. Reprinted with permission from Ref. \citen{Lique:10C4} with permission of  RSC Publishing.}
\label{fig:C4}
\end{center}
\end{figure}

Collisional excitation cross sections of the fine-structure levels of C$_4$ by He were calculated at low collisional energies using a CC approach. Corresponding rate coefficients were calculated up to 50~K. As far as propensity rules between fine-structure levels are concerned, collisional excitation cross sections within a same fine structure ladder are favored, especially for high-$N$ rotational levels.

Finally, rate coefficients for rotational excitation of HC$_3$N($^1\Sigma^+$) by collisions with He atoms and H$_2$ molecules were computed for  temperatures in the range 5-20~K and 5-100~K, respectively\cite{Wernli:07}.
The data were obtained from extensive quantum and quasi-classical calculations using new accurate PES calculated at the CCSD(T) levels. For the PES calculations, the authors used a cautious sampling strategy, building a spline interpolation in a first step, and postponing the troublesome angular Legendre expansion to a second step. This procedure allows to avoid oscillations in the numerical fit of the PES over Legendre expansions.
Unfortunately, in this work, the calculations involving ortho-H$_2$ were subject to an erratum \cite{Wernli:07a} and 
only  rate coefficients with He and para-H$_2$ are available.

\subsubsection{Complex Organic Molecules : H$_2$CO / HCOOCH$_3$ / CH$_3$OH}

Complex organic molecules (COMs) have received a lot of attention as they are found in many interstellar environments,
not only in the so-called hot cores \cite{wang:09} or hot corinos \cite{sakai:12} but even in low mass protostellar environments \cite{caux:11}. Understanding the associated spectroscopy requires major efforts as the microwave spectrum contains many transitions diluted over large spectral bandwidths which represent the "weeds" \cite{delucia:06} in observational spectra  which have to be cleaned and/or removed to recover the "flowers", the new potential molecular carriers. Collisional excitation may not only involve rotational modes but also more subtle torsional degrees of freedom which are linked to the variation of orientation and bond distances of the interacting species. Their detection also opens the way to even search   for prebiotic molecules, a  fascinating area, and linked to astrobiology, which is mainly related to the discovery of extraterrestrial planets.

Formaldehyde (H$_2$CO) has been discovered in the early days of radioastronomy thanks to its 6 cm "anomalous" absorption (1$_{11}$ - 1$_{10}$) by Palmer et al. \cite{palmer:69}, which directly motivated Townes and Cheung \cite{Townes:69} to invoke a specific pumping mechanism.  H$_2$CO has been extensively observed both in galactic and extragalactic sources. The relatively large dipole moment and abundance of the molecule make the rotational lines relatively easy to observe from ground-based observations. H$_2$CO is the simplest and often most abundant polyatomic organic molecule containing oxygen observed in space. So, intensive work has been early devoted to the calculation of its collisional excitation rate coefficients where different levels of approximation have been introduced \cite{Townes:69,Augustin:74,Garrison:76}. 

In 2009, Troscompt et al.\cite{Troscompt:09} provided the first fully quantum calculations of rate coefficients for the rotational excitation of the ten lowest levels of ortho-H$_2$CO by collisions with H$_2$ molecules for kinetic temperatures between 5 and 100 K, using a highly accurate PES. Cross sections are obtained from extensive, fully converged, quantum-mechanical scattering calculations. Scattering calculations were carried out for H$_2$ molecules in both para and ortho rotational levels.
The new rate coefficients were shown to differ significantly from those available previously in the literature. Moreover, the values involving para- and ortho-H$_2$ results may also differ significantly such as for the 1$_{11}$ - 1$_{10}$ transition (See upper panel of Fig \ref{fig:H2CO}). 
The collisional pumping mechanism invoked to explain the anomalous absorption of H$_2$CO\cite{Townes:69} was reinvestigated \cite{Troscompt:09b}.
The importance of the ortho/para form of {\HH} is emphasized as can be guessed from Fig. \ref{fig:H2CO}.

\begin{figure}
\begin{center}
\includegraphics[width=10.cm,angle=0.]{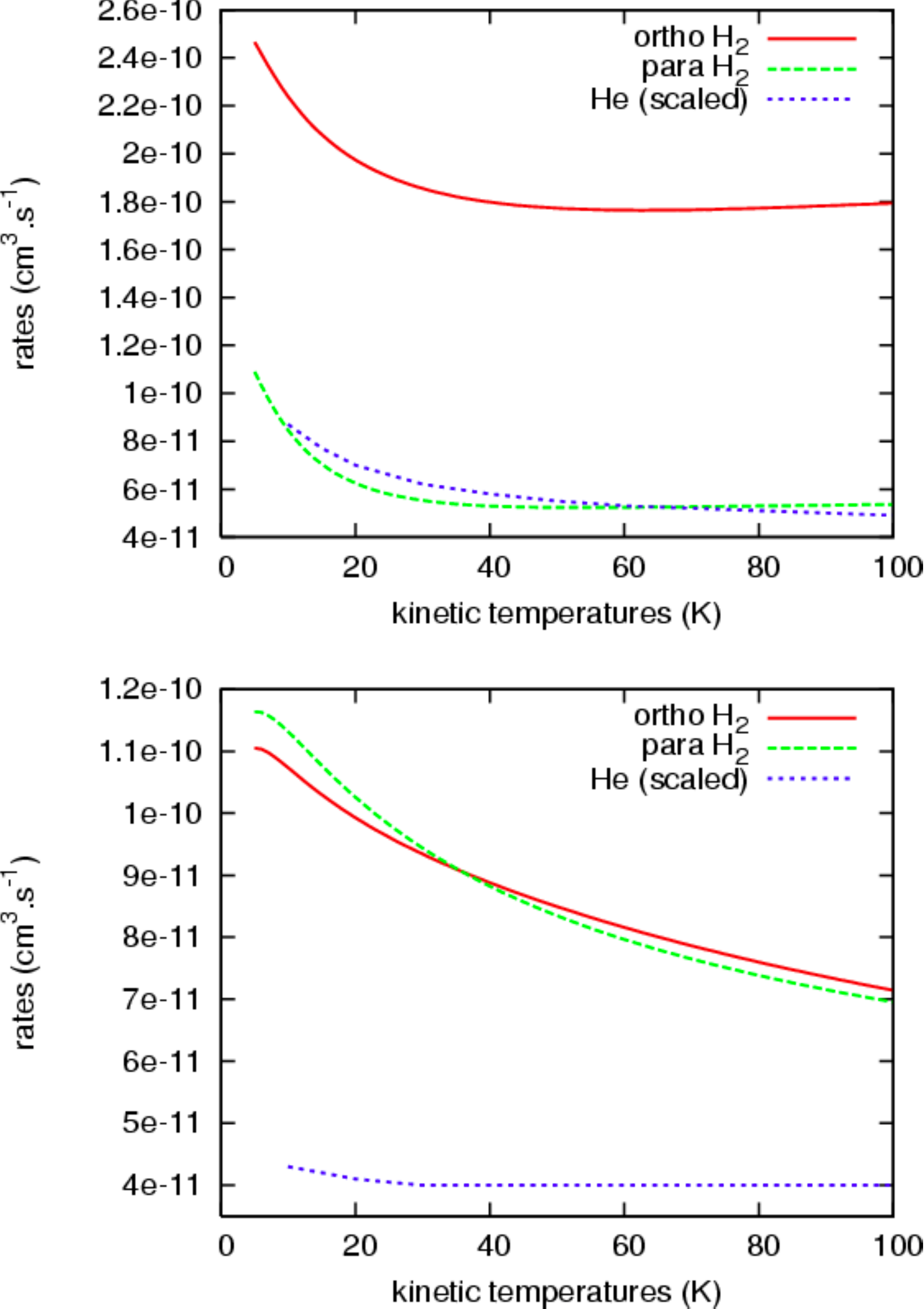}
\caption{H$_2$CO. Rate coefficients for the rotational transitions $1_{10} \to 1_{11}$ (upper panel) and  $2_{12} \to 1_{10}$ ( lower panel) as functions of temperature for (scaled) He (Green 1991), para-H$_2 (j = 0)$ and ortho-H$_2 (j = 1)$. Reprinted with permission from Ref. \citen{Troscompt:09}. Copyright 2009 ESO.}
\label{fig:H2CO}
\end{center}
\end{figure}

Flower and co-workers have put many efforts to investigate  the excitation of methanol (CH$_3$OH) by He \cite{pottage:01,pottage:02,pottage:04,Rabli:10a} and molecular Hydrogen \cite{pottage:04a,Rabli:10b}. 
New PESs have been calculated with progressive inclusion of the internal torsional motion of the molecule and corresponding dynamic calculations have been performed at the quantal CS and CC levels. 
The calculations have been recently extended to rotationally and torsionally inelastic transitions\cite{Rabli:11} between levels of the torsional manifolds $v= 0$, 1 and 2 of methanol using He as collisional partner, which represents 406 levels of A and E-type methanol up to temperatures of 400~K.   The authors found that the torsional inelasticity is more pronounced for transitions between the excited states $v= 1$ and 2. The torsionally inelastic cross-sections are smaller, on average, than those involving  pure rotational inelastic transitions, by typically an order of magnitude. 
In the case of methanol in its ground torsional state excited by \HH, rotational transitions involving 256 levels of A and E-type methanol in its ground torsional state have been computed for para-{\HH} and  100 levels of A and E-type methanol for ortho-{\HH} up to 200~K.
These calculations are, to the best of our knowledge, the only quantum calculations available for torsionally inelastic collision of a polyatomic molecule. The corresponding rates have been further introduced in a MHD shock model \cite{flower:10,Flower:12extr} which predicts that the $A^+~7_0 \to 6_1$  (44.07 GHz) and $E~4_{-1} \to 3_0$  (36.17 GHz) transitions are the most intense transitions of the two methanol types. These transitions have further been detected in a class I methanol maser source, in which the emission appears to be related to an outflow
along the line of sight and an associated shock wave \cite{voronkov:10}. Both population inversion and anti-inversion are consistent with the collisional propensities favoring transitions in which $k$, the projection of $j$ on the symmetry axis of the molecule remains unchanged, as recognized by Walmsley et al. \cite{walmsley:88}.

Finally, collisional excitation of methyl formate HCOOCH$_3$\cite{Faure:11} has also received recent interest \cite{Faure:11}. Methyl formate is an asymmetric top. Since the lowest rotational excitation energy of methyl formate is only $\simeq 0.4$ cm$^{-1}$, HCOOCH$_3$ is rotationally excited even in the coldest regions of the ISM. 
A PES for helium interacting with methyl formate has been computed using high-level electronic structure methods. The interaction energies have been computed on a three dimensional grid. The analytical fit derived from the individual point values ensures a correct long range analytic expansion involving angular and radial dependence. Cross sections have then been derived through full quantal CC calculations up to 100 cm$^{-1}$. As a rule, cross sections involving transitions corresponding to $\Delta k_a = 0$ with the same final value of $j$ are found  larger. Also, within each $k_a = 1$ doublet, the lower level is favored as in the case of formaldehyde. This propensity rule however, disappears for other values of $k_a$. It is also remarkable that largest cross sections are obtained for   $\Delta j = 2$ transitions, in contrast to radiative selection rules.  Corresponding rate coefficients remain to be derived.

\subsubsection{Cations : CH$^+$ /  SiH$^+$ / HCO$^+$ / N$_2$H$^+$ / HOCO$^+$ }

Molecular ions, despite a low fractional abundance compared to \HH,  are a very important component of the ISM as they allow 
coupling with the ambient magnetic field and contribute to the overall stability of interstellar clouds. Their presence results from the impinging UV photons 
in illuminated regions (diffuse clouds, PDRs) whereas cosmic rays, heavy particles ejected form supernovae, are able to ionize most neutrals in dark cloud regions at an ionization rate of the order of 10$^{-17}$ - 10$^{-16}$ s$^{-1}$. Molecular ions allow then to derive the electronic fraction. They also  may appear as protonated forms of symmetric abundant molecules \cite{herbst:77}, such as H$_3^+$, N$_2$H$^+$, HOCO$^+$, offering a way to infer the abundance of these undetectable species. 
As far as molecular physics is concerned, the long range interaction between ions and neutral is dominated by  the induction 
energy term resulting from the polarization of He and molecular Hydrogen by these molecular ions.

Most of the detected cations are closed shell systems and do not react with molecular hydrogen, which would destroy them very efficiently. There are however some outstanding exceptions such as CH$^+$, OH$^+$, H$_2$O$^+$ which are found in a variety of environments. 
Most of the collisional excitation studies of  these reactive  cations have started   with He  as collisional partner. 

The CH$^+$($^1\Sigma^+$)--He system has received recent attention almost simultaneously by Turpin et al. \cite{Turpin:10}    
 and Hammami et al. \cite{Hammami:09}. 
Both calculations of the PES were performed using a CCSD(T) method and the dynamical calculations were performed using the quantal CC   approach which is required due to the large rotational energy transfers involved in the target species. The rate coefficients were provided for   rotational levels up to $j = 5$  and  temperatures lower than 500~K.
The isovalent system SiH$^+$--He has been further studied by Nkem et al \cite{Nkem:09}, but this ion has (yet) not been detected in the ISM.

HCO$^+$,   isoelectronic of the CO molecule, is the first cation found in the cold ISM and its identification is a beautiful example of successful interdisciplinary efforts
\cite{buhl:73, herbst:74,Kraemer:76}.   It is also ubiquitous and is closely linked to CO chemistry. First computations have been performed by Monteiro where the PESs are obtained mainly using the configuration interaction approach for He \cite{monteiro84} and para-{\HH} \cite{monteiro:85} . Collisional excitation rate coefficients were derived up to $j=4$ and for temperatures lower than 30~K. Very recently, new rate coefficients of rotational excitation  of  HCO$^+$($^1\Sigma^+$) induced by collisions with helium were obtained by Buffa et al. \cite{Buffa:09}, for temperatures ranging from 10 to 80~K. The calculations are based on a new PES for the HCO$^+$--He interaction. A comparison between these new results and those of Monteiro\cite{monteiro84} results at 10~K shows that the new rates are somewhat larger, with a difference ranging from a few to 50\%. The H/D substitution effect on the rate coefficients will be discuss at the end of this section. 
Flower \cite{Flower:HCOp}, on the other hand, extended previous dynamical calculations  of  Monteiro \cite{monteiro:85} by using the same PES for collisions
with para-\HH. 
Rate coefficients are derived  for the 21 first rotational levels up to T~=~400~K. 

Diazenylium, the N$_2$H$^+$ ion, is also isoelectronic of CO, extensively observed and its extended hyperfine structure allows
to check in a narrow range of  frequencies  the LTE approximations of its hyperfine components. Green\cite{Green:75N2Hp} computed a PES  representing the N$_2$H$^+$--He system and derived corresponding 
rotational excitation rate coefficients at a moderate convergence level.  Daniel et al.\cite{Daniel:04,Daniel:05} have 
computed a new 2D PES calculated at the CCSD(T) level obtained with a large orbital basis set and derived  dedicated rotational excitation  rate coefficients. The differences between earlier rotational results by Green\cite{Green:75N2Hp} and new calculations by Daniel et al.\cite{Daniel:05} span the range from a few percent to 100\%. 
The cross sections between hyperfine levels are obtained by introducing the recoupling of S matrix elements and inlude the  2 nuclear spins of nitrogen\cite{Daniel:04}. The authors conclude that the only well defined propensity rule is $\Delta F = \Delta F_1 = \Delta j$ (see Fig. \ref{fig:N2Hp})

\begin{figure}
\begin{center}
\includegraphics[width=10.cm,angle=270.]{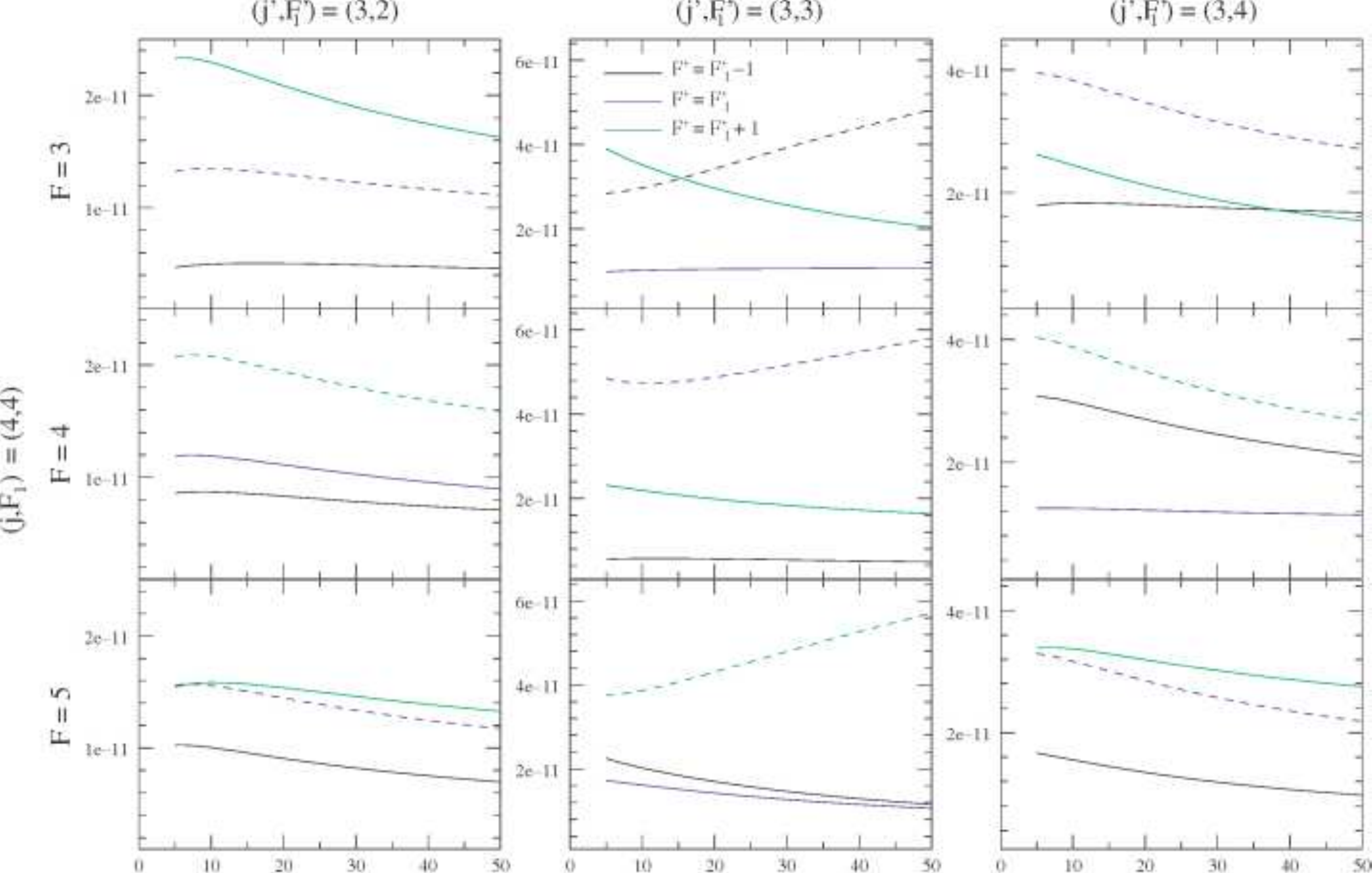}
\caption{N$_2$H$^+$ -- He. Hyperfine rate coefficients (in cm$^3$s$^{-1}$) as a function of temperature (K) from the $j, F_1 = (4, 4)$ states and $F= 3, 4, 5$. The rate coefficients in dashed lines are the rates expected to be of highest magnitude according to regular propensity rules from the behaviour of Wigner-6j coefficients. Reprinted with permission from Ref. \citen{Daniel:05}. Copyright 2005 Oxford University Press}
\label{fig:N2Hp}
\end{center}
\end{figure}

Finally, we mention the study of Hammami et al.\cite{Hammami:07} dedicated to the non linear ion   HOCO$^+$ collisionally
excited by  He collisions. Due to the large number of low lying rotational levels in HOCO$^+$, CS approximation is reasonably fulfilled and 
the authors adopt then this dynamical approximation  to compute corresponding inelastic cross sections for the 25 first rotational levels at low energies. Rate coefficients are derived for temperatures ranging from 5 to 30~K. 

Apart from the case of HCO$^+$, all these studies involve He as a perturber and a representative of molecular \HH. 
As the polarizabilities of H$_2$ and He are different, the long range expansions of the corresponding PESs are obviously different and the use of these data to mimic {\HH} collisions is questionable. An additional difficulty arising in the presence of  {\HH} is the availability of a reactive channel with no barrier, as can be expected for the CH$^+$--{\HH} system, which changes considerably the PES and 
would imply to consider both reactive and nonreactive channels in a single unified theoretical treatment. 

\subsubsection{H$_3^+$}
{\HHHp} deserves an additional comment. After a 20 years search, Geballe and Oka \cite{geballe:96} finally succeeded to detect  {\HHHp} via infrared absorption transitions in molecular clouds. Thanks to the improvement of infrared detector technology, it is now widely observed towards diffuse and translucent clouds \cite{mccall:98} and dense molecular clouds \cite{mccall:99}.
Its detection towards the Galactic center \cite{geballe:06,geballe:12} allows to probe excited rotational levels up to the $3_3$    
metastable level. These observations lead  Oka and Epp \cite{oka:04} to propose empirical formulae for the corresponding excitation collisional rate coefficients. The \HHHp--{\HH} system is a nice example where reactive channel may occur and different mechanisms involve pure inelastic, hoping, exchange of protons which should be disentangled. Bowman and colleagues \cite{xie:05} built a global PES for  the H$_5^+$  $\to$  {\HHHp} + {\HH}  system whereas Aguado et al. \cite{aguado:10} also produced a new accurate and full dimensional PES of H$_5^+$. Rate coefficients for ortho/para conversion of H$_3^+$ due to H$_2$ collisions have been computed by G\'omez-Carrasco et al \cite{gomez:12} based on dynamically biased statistical models. Alternatively Hugo et al \cite{Hugo:09} used a microcanonical approach using the Langevin model and the laws of conservation for mass, energy, angular momentum and nuclear spin following previous studies by Park and Light \cite{park:07}. Corresponding rate coefficients were given up to temperatures of 50~K. 


\subsubsection{Anions : CN$^-$ / C$_2$H$^-$}

With the recent discovery of negative molecular ions in the ISM, gas-phase ion-molecule chemistry has gained additional interest. 
The detected molecular anions have a closed shell structure, are linear and give rise to a repulsive long range interaction
in the presence of neutral He and \HH.
In 2010, the CN$^-$($^1\Sigma^+$) anion has been discovered in the carbon star envelope IRC +10216 \cite{Agundez:10}. The fractional abundance of CN$^{-}$ has been derived thanks to the knowledge of collisional excitation rate coefficients for the CN$^{-}$ - para-{\HH} ($j=0$)   system. The rate coefficients provided in this paper were obtained from preliminary calculations based on an  CN$^{-}$ -- H$_{2}$ PES averaged over H$_2$ orientations. These preliminary calculations  have been extended to take into account the rotational degrees of freedom of H$_2$ both in the PES and in the scattering calculations    \cite{Klos:11}.  These studies are the first calculations involving collisional excitation of an anion by
He , \HH. 
The calculations are based on a new reliable 5D {\it ab initio} PES of  the  CN$^-$--H$_{2}$ complex and expanded in a bispherical angular basis  \cite{Klos:11}. 
The authors applied 4D quantum CC scattering calculations to investigate rotational energy transfer in collisions of the CN$^-$ molecule with para-H$_{2}(j=0)$ and ortho-H$_2(j=1)$ molecules. 
The rate coefficients are there weakly dependent on the rotational level of H$_{2}$. 
The weak anisotropy of the PES obtained from different H$_2$ orientations explains this behavior. 

Then, the authors checked the possibility of estimating the collisional excitation rate coefficients of the anion from  the corresponding rate coefficients of a neutral  isoelectronic molecule.  CO   is isoelectronic with CN$^-$ and has the same ground electronic state $^1\Sigma^+$. Fig.~\ref{fig:CNm} compares, on a small data sample, the temperature variation of the de-excitation rate coefficients of CN$^-$ in collisions with para-H$_2(j_2=0)$ with those obtained recently for the neutral CO molecule in collision with para-H$_2(j_2=0)$ \citep{wernli06}.  
On can clearly see on Fig.~\ref{fig:CNm} that the CN$^-$ rate coefficients are significantly larger than those of CO. Indeed, the CN$^-$--H$_2(j_2=0)$ rate coefficients are up to a factor of ten larger than the one of CO--H$_2(j_2=0)$. The large difference can be explained by the shape of the PES. The well depth of the CO--H$_2$ PES\cite{Jankowski:05} is 93~cm$^{-1}$ whereas that of the CN$^-$--H$_2$ PES is 875~cm$^{-1}$,  which is about 9 times larger. 
Table \ref{tab:crit} displays the critical densities of  CN$^-$ for different transitions and temperatures. The values increase slightly with temperatures 

\begin{table*}
\caption{Critical densities for CN$^-$-- $normal$-{\HH}  collisions in cm$^{-3}$}
\label{tab:crit}
\begin{tabular}{cccccccc}
\hline
     &     &         \multicolumn{6}{c}{ T  (K)} \\
    \hline
   $j'$  &   $j'$    &   10      &        20      &         30     &       40     &        50       &      100 \\
   \hline  \hline
   1 &  0 &   6.24 10$^3$  & 6.88 10$^3$  &7.32 10$^3$   &  7.65 10$^3$   &  7.91 10$^3$  &   8.81 10$^3$  \\
   2   &1  &   3.71 10$^4$ &  4.00 10$^4$ &  4.2110$^4$ & 4.36 10$^4$  &  4.47 10$^4$ &   4.76 10$^4$ \\
   3  & 2  &   1.11 10$^5$  & 1.19 10$^5$  &  1.25 10$^5$   & 1.29 10$^5$  &  1.32 10$^5$   &  1.40 10$^5$ \\
   4  & 3 &    2.54 10$^5$ &   2.66 10$^5$  &  2.79 10$^5$  &  2.89 10$^5$   &  2.97 10$^5$   & 3.1510$^5$ \\
   5 &  4&     4.85 10$^5$  &5.01 10$^5$  &  5.22 10$^5$  & 5.41 10$^5$  &  5.57 10$^5$  & 6.01 10$^5$ \\
   6   &5  &   8.28 10$^5$ &   8.39 10$^5$  &  8.69 10$^5$  &  9.01 10$^5$  &  9.30 10$^5$   &  1.02 10$^6$ \\
   7 & 6  &   1.29 10$^6$  &  1.29 10$^6$  &  1.33 10$^6$   & 1.38 10$^6$   &  1.42 10$^6$   &  1.59 10$^6$ \\
   8  & 7  &   1.93 10$^6$ &   1.91 10$^6$  &  1.95 10$^6$  &  2.00 10$^6$   &  2.06 10$^6$   &  2.32 10$^6$ \\
   9  & 8   &  2.71 10$^6$   & 2.68 10$^6$   & 2.72 10$^6$  & 2.78 10$^6$   &  2.85 10$^6$    & 3.20 10$^6$ \\
  10  & 9   &  3.61 10$^6$  & 3.60 10$^6$  &  3.64 10$^6$   & 3.71 10$^6$  &   3.80 10$^6$  &  4.17 10$^6$ \\
  \hline
  \end{tabular}
  \end{table*}

\begin{figure}
\begin{center}
\includegraphics[width=8.cm,angle=0.]{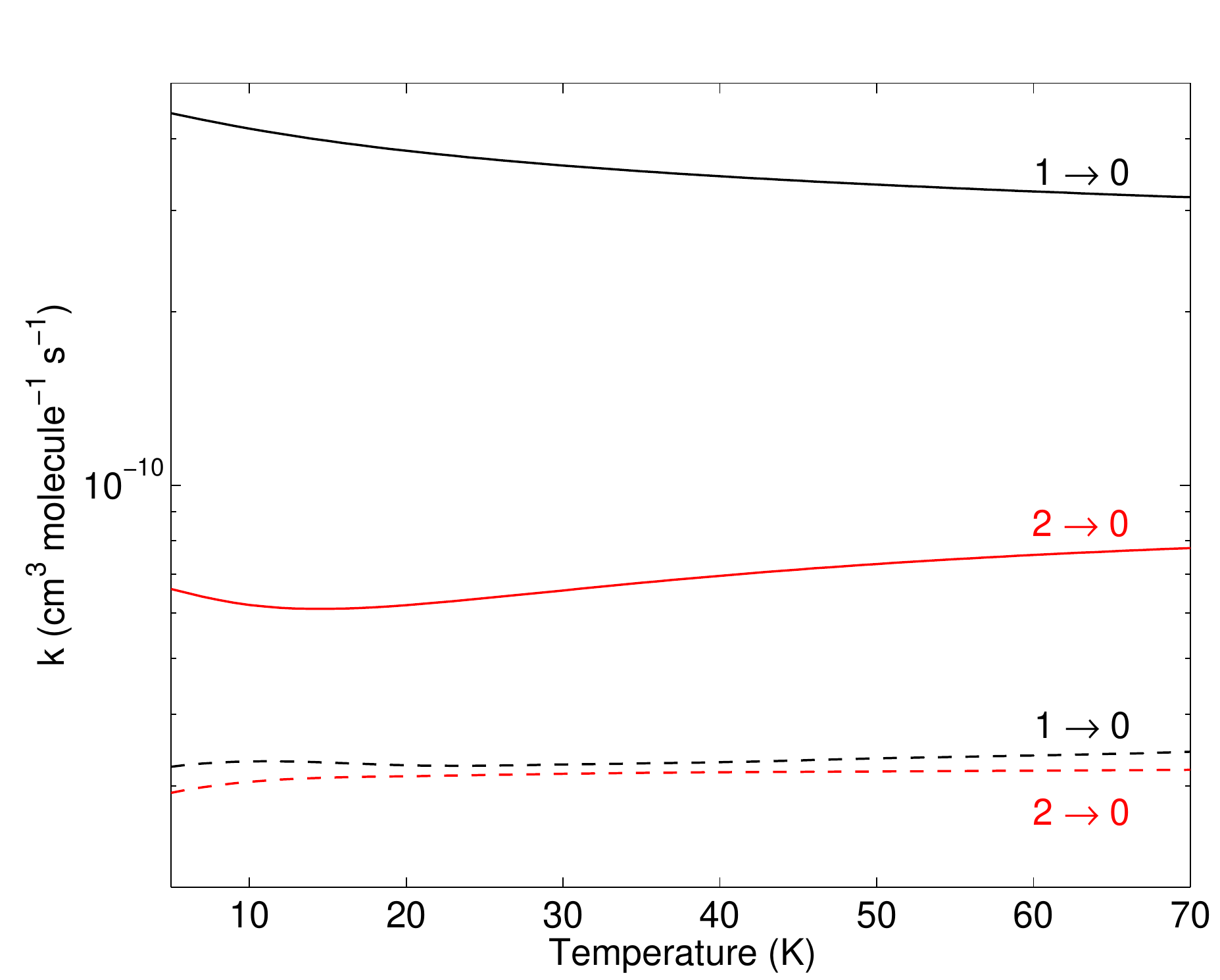}
\caption{Temperature variation of the CN$^-$ (solid lines) and CO (dashed lines) de-excitation rate coefficients for collisions with para-H$_2(j=0)$. Reprinted with permission from Ref. \citen{Klos:11}. Copyright 2011 Oxford University Press}
\label{fig:CNm}
\end{center}
\end{figure}

C$_2$H$^-$($^1\Sigma^+$) has also been considered recently in terms  of inelastic collision studies despite the molecule is not yet observed. From a new 2D PES, rotational excitation of the C$_2$H$^-$($^1\Sigma^+$) anion by collision with He has been investigated by Dumouchel et al.\cite{dumouchel:12}. Fully-quantum CC calculations of inelastic integral cross sections were done on a grid of collision energies large enough to insure converged state-to-state rate coefficients for the 13 first rotational levels of C$_2$H$^-$ and temperatures ranging from 5 to 100~K. A comparison with the spin-averaged He-rate coefficients of the related neutral C$_2$H molecule points out the relatively large difference between the anion and the neutral species, rate coefficients for C$_2$H$^-$ being larger by factor 2--5 than the corresponding rates for C$_2$H confirming the trend found K{\l}os and Lique for CN$^-$\cite{Klos:11}. 

As negative ions probably react rapidly with H$_2$ in an associative detachment mechanism, rate coefficients involving He may not be good representative of \HH.

\subsection{Isotopologues}

We first want to point out that any substitution of nucleus does not modify the electronic PES between 2 molecular species at first order unless strong non adiabatic coupling effects between rotation and electronic motion occur. The values corresponding to the different individual geometries for which the PES has been numerically computed remain identical. It is only at the step where fitting and decomposition of the PES over the various angular functions is performed that modifications may occur: the center of mass is not the same and the overall symmetry may have been destroyed in the substitution process. In addition, hyperfine structure will change and modify the spectrum. Rotational excitation of $^{13}$CO, C$^{18}$O, for example,  is often taken identical to that of $^{12}$CO in astrophysical modeling analysis and this approximation looks reasonable as the shift of the center of gravity is negligible and the transition energies 
of the three isotopologues are very close. 

The effect of H/D substitution is much more sensitive but has been rarely tested in detail. We describe below
the few examples where excitation of deuterated  molecules has been computed.
In addition to modify the zero point energy (ZPE) term \cite{Dubernet:06} and the transition energies,  inclusion of a deuterium nucleus in a symmetric molecule such as {\HH} , {\HHO} and {\NHHH} is significant
as it cancels the possible selection rules occurring for exchange between ortho and para species or E and A symmetries.
HD collisional excitation by He, H and {\HH} has been studied for rotational excitation \cite{Flower:99} and rovibrational  transitions as well  \cite{Flower:99c,Roueff:00}. Analytic functions of temperature have been derived
to express the different rate coefficients \cite{flower:00b} to derive the corresponding cooling functions. 
HDO rotational excitation excitation has also been considered  \cite{Faure:04,Faure:12},
as well as DCO$^+$\cite{Buffa:09,Buffa:12}, N$_2$D$^+$\cite{Daniel:05,Daniel:07}, NH$_2$D and ND$_2$H  \cite{Machin:06,Machin:07,maret09,wiesenfeld11}. 
On the other hand, H$_2$O--H$_2$ rate coefficients were compared with D$_2$O--H$_2$ rate coefficients in
Scribano et al.\cite{scribano10} and differences by up to a factor of 3 were observed. The use of state
averaged geometries by these authors allows to include the effect due to the change in  ZPE in a straightforward way. 
Very recently, Dumouchel et al.\cite{dumouchel:12NH} performed NH/ND rate coefficients calculations and compared them in great details in order to test the effect of H/D substitution in NH.
They found that significant differences exist between the two sets of data. The
temperature variation of the NH and ND rate coefficients differ at low
temperature and the ND--He rate coefficients are larger or smaller
depending on the transition. 
Taking into account the small difference in the reduced mass, they considered (and checked) that it doesn't significantly impact the magnitude of the rate coefficients. Surprisingly, they found that the different position of the center of mass has a strong influence on the values of the ND--He cross sections.  Within the Born-Oppenheimer approximation, the full electronic ground state potential is identical for the two isotopologues and depends only on the mutual distances of the three atoms involved. The only difference between the NH--He and ND--He PES is thus the position of the center of mass taken for the origin of the Jacobi coordinates. This small change in the position of the center of mass of the colliding system leads to a significantly different expansion of the interaction PES in terms of the Legendre polynomials. This effect has been also observed in the case of HCO$^+$/DCO$^+$ collision with He even if the difference was less than for NH/ND in collision with He.

Dumouchel et al.\cite{dumouchel:12NH} also found that the rotational structure has a significant impact on the magnitude of the rate coefficients. In the case of NH/ND molecules, the smaller energy spacings between ND levels compared to NH levels impact significantly the intensity of the cross sections.

In agreement with the previous work by Scribano et
al. \cite{scribano10} on H$_2$O/D$_2$O, 
it seems that, for light hydrides, one has to consider both the different structure and PES expansion caused by H/D substitution if one wants to determine accurate rate coefficients for the deuterated isotopologues. 

\section{Radiative and chemical excitation}
\label{sec:rad}

\subsection{Radiative effects}

Radiative excitation or ultraviolet pumping is the principal source of excitation in cometary environments  and can be directly observed through fluorescence emission spectra. Indeed, the bright solar radiation field located at distances of  several  Astronomical Units is impinging on the moving comet and, through a sequence of far VUV (Vacuum UltraViolet) absorption/ UV-Visible emission, populates the excited rotational/vibrational levels of the molecular ground states which can subsequently cascade via IR and millimeter transitions. The corresponding literature is multiple and is a research area in itself which we will not report here.  The occurrence of such pumping mechanisms is yet to be fully recognized in interstellar conditions where the ambient interstellar radiation field is much weaker and has a very specific wavelength dependence, extending from 91.2~nm (corresponding to atomic hydrogen ionization) up to the millimeter region. In addition, this field is very efficiently shielded by dust and molecular hydrogen in dense molecular clouds. Draine \cite{draine:78} has modeled the UV part of the interstellar radiation field (ISRF)  in connection with photoelectric effects on interstellar dust grains and provided an analytic expression which has been widely used for photodissociation calculations and is often referred as the "standard" ISRF value. However, the extrapolation of this radiation field to larger wavelengths is necessary to fully account for possible radiative pumping effects. An alternative model derived by Mathis et al. \cite{mathis:83} extends the evaluation of the ISRF up to larger wavelengths as a result of  infrared conversion by dust   of the impinging UV radiation and is another useful reference. It should be stressed that  both models agree within a factor of 2  in the VUV  range, as displayed in Fig. \ref{fig:rad}, which corresponds to a reasonable understanding.

\begin{figure}
\includegraphics[width=10.cm,angle=0.]{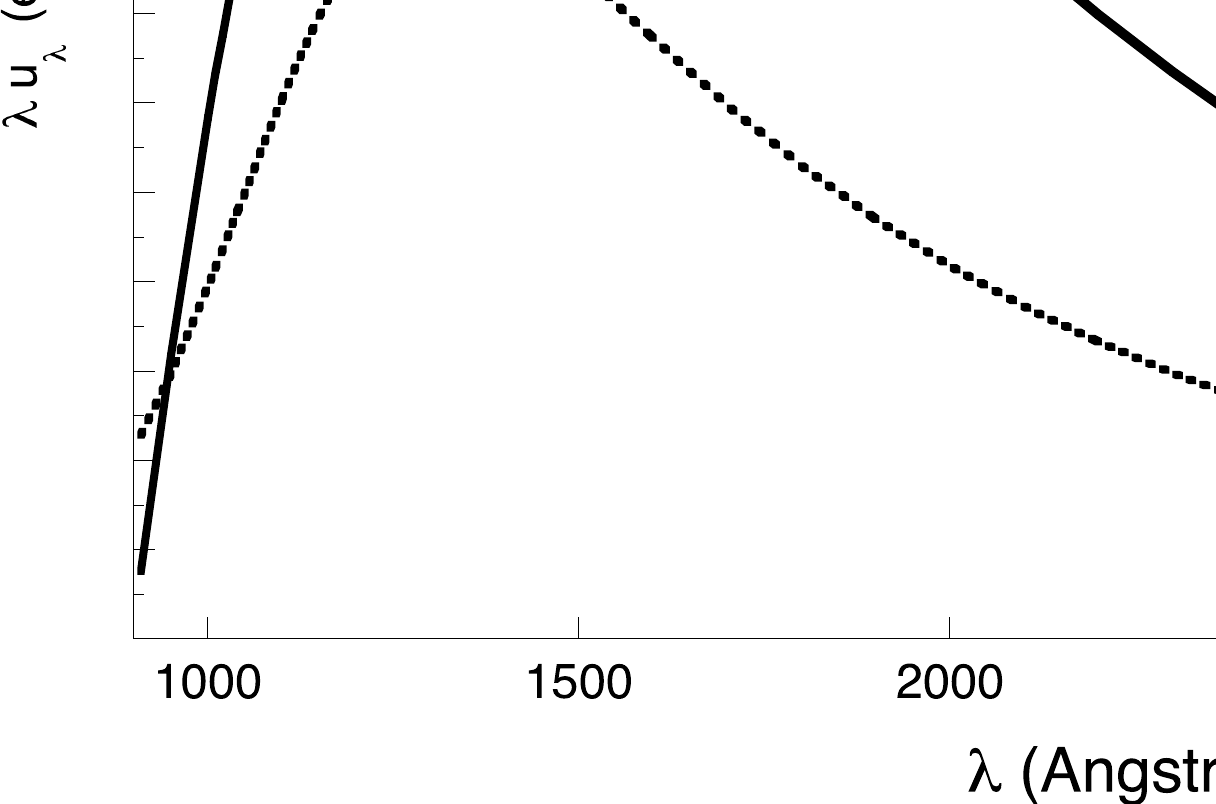}
\caption{Energy density (erg cm$^{-3}$ s$^{-1}$) of the standard interstellar radiation field. Draine (Mathis) model as  a full (dotted) line.}
\label{fig:rad}
\end{figure}

This later model allows to fully account for possible radiative pumping of infrared vibrational transitions. At even longer wavelengths, millimeter emission from dust grains and the cosmic blackbody radiation corresponding to a temperature of 2.7~K are at work.  It should also be kept in mind that the VUV range of the radiation field may be significantly enhanced in PDRs where a bright star is close to the region as in NGC7023, which is often modeled in the literature through an enhancement factor of the standard radiation field. However, this enhancement should be restricted to the VUV - Visible range corresponding to the maximum energy power of the illuminating star and not in the infrared where the VUV light is reprocessed by dust.  

The chemical physics associated studies (both theoretical and experimental) involve the knowledge of electronic and vibrational absorption~/~emission probabilities of the molecules in their different rotational~/~vibrational levels. The requested molecular data involve thus the detailed knowledge of
energy level transitions of the molecules and corresponding oscillator strengths or emission probabilities as well as the total radiative lifetime of the upper electronic levels involved. Both theoretical and experimental approaches can be used in order to determine these spectroscopic data. However, at the opposite of the collisional excitation field,  theoretical calculations of transition energies cannot reach the experimental remarkable accuracy,  
especially when excited electronic or ro-vibrational states are involved. However, theoretical calculations are very helpful to provide the associated electric dipole transition moments and   oscillator strengths.

The necessary information is spread in the spectroscopic literature 
where each molecule and associated transitions are more or less extensively studied.  As an example, we refer to the books of Lefebvre-Brion and Field \cite{lefebvre:86,lefebvre:04} devoted to diatomics. A systematic use of spectroscopic constants, transition strengths and Franck-Condon factors \cite{herzberg,herzberg:89,herzberg3} allows to generate extensive samples of energy transitions and emission probabilities. However, such a procedure may lead to serious errors as these expressions neglect all perturbation possibilities and 
are usually valid only for low rotational and vibrational quantum numbers.  Hsu and Smith \cite{hsu:77} provided
a review of spectroscopic information in the visible and ultraviolet region for diatomics more than 30 years ago and Kuznetsova \cite{kuznetsova:87} reported corresponding electronic transition strengths. More recently, Reddy and coworkers \cite{reddy:03} extended such systematic studies  but the corresponding  data are mainly used for the purpose of computing opacities in stellar atmospheres at a relatively poor spectral accuracy.
Finally, we recall the availability of spectroscopic databases of millimeter, submillimeter and infrared spectra such as JPL \cite{jpl,Pickett:98}, CDMS \cite{cdms,muller:05}, HITRAN \cite{hitran,Rothman:13} and GEISA \cite{geisa,Jacquinet:11}. 

\subsection{Chemical effects}

Reactive exothermic formation processes may also produce excited molecules as part of the energy release may be transferred in internal energy of the products. The fate of these excited
molecules depends on the relative time scales between production~/~chemical destruction~/~radiative decay~/~collisional decay. Dissociative recombination reactions  are usually very exothermic and may provide internal energy to the neutral products. Such a mechanism is known since a long time in planetary atmospheres where metastable nitrogen and oxygen atoms are efficiently produced from the dissociative recombination reaction of N$_2^+$ and O$_2^+$ in the atmosphere of Mars and Venus  \cite{fox:05,fox:12}.  Molecules resulting from photolysis of larger molecules are also subject to specific excitation. As an example, CN radicals result from photodissociation of HCN in comets \cite{fray:05} and in the envelope of the carbon star IRC +10216  \cite{lindqvist:00}.  CN photo-fragments are  formed principally in the first excited electronic state $A^2\Pi$ in various vibrational levels as found in the laboratory \cite{cook:00}. A small percentage of  fragments is also directly produced in rovibrational states associated with CN($B^2\Sigma^+$)($v=0,1$), leading to vibrationally excited CN in relatively cold regions.

Molecular excitation may also result from differential behavior of excited molecular levels in a reaction. Such effects have been observed   in dissociative recombination studies where  para- and ortho-{\HHHp}  levels are found to recombine differently \cite{varju:11,dohnal:12}  in experiments able to monitor the decay of different rotational populations of  \HHHp. 
Theoretical methods have also been developed to account for such rotational effects in some benchmark cases \cite{pagani:09}. 
The coupling of chemical  excitation to other excitation processes (collisional, radiative) is not   commonly achieved  for interstellar applications although the molecular hydrogen case is a noticeable exception, as will be discussed below. However, the possible role of chemistry is progressively accounted for
when other processes of excitation are failing, as shown in the following examples. The possible application field is huge and implies that numerous detailed state to state reaction rate coefficients are available. Such efforts are really relevant 
only if  all other possible excitation mechanisms are well understood.   Indeed,   studying  chemical reactions at a state to state level represents a theoretical and experimental challenge. 
Such achievements have been performed for atom - diatomic and some triatomic systems such as the F + {\HH} reaction
\cite{che:07,lique:08FH2}, Cl + {\HH} reaction \cite{Wang:08}  
or in photodissociation theoretical studies of  {\HHO}  towards H + OH \cite{zhou:13}
or {\NHHH} towards  {\NHH} + H \cite{ma:12nh3}.  Alternatively, the complexity of the corresponding astrophysical modeling 
increases as well. We describe some illustrative examples below.

\subsection{Examples}

\subsubsection{H$_2$}

Radiative excitation has been first invoked for explaining the rotational excitation of molecular Hydrogen. This issue is summarized in Ref. \citen{spitzer:76} where the first observations of molecular hydrogen via VUV absorption against bright stars by the Copernicus satellite are reported. Absorption is taking place in intervening interstellar clouds along the line of sight which  first have been named "diffuse", then "translucent"  \cite{snow:06} and containing comparable amounts of atomic and molecular hydrogen. 
The different excitation temperatures   found for the different rotational levels were interpreted as resulting from different mechanisms: low $j=0$ and $j=1$ levels reveal the kinetic temperature as only reactive collisions with protons and hydrogen atoms (involving a significant barrier) allow to transfer ortho ($j=1$) and para ($j=0$) levels. The typical value is 70~K, in good agreement with the expected temperature of these environments. Higher rotational levels are approximately described by a temperature of several hundreds K, which can be interpreted as resulting form radiative pumping effects due to the ultraviolet  ISRF. As the Hydrogen molecule has no dipole moment, radiative cascades within the ground electronic state  are very slow, resulting from quadrupole electric transitions \cite{wolniewicz:98} with a $\Delta j = 2$ selection rule. This issue has been revisited \cite{lacour:05} at the light of recent FUSE observations that have allowed to probe four lines of sight with larger extinctions in the 912-1200~{\AA} spectral window range, which have been subsequently labelled as translucent. Even more recently, Jensen et al \cite{jensen:10} provided an extended FUSE sample of more than twenty translucent clouds where excited rotational H$_2$ levels up to $j=8$ were reported (as shown previously in Fig.  \ref{fig:excitation}). It should also be pointed that infrared emission of rotationally excited {\HH}  is also detectable in translucent clouds with modern ultra sensitive space equipments such as SPITZER  \cite{ingalls:11}. 
Remarkably, additional information on {\HH} molecular excitation can be obtained from HST archival data at larger wavelengths, i.e. in the 1250-3000~{\AA} window, where absorption transitions from vibrationally excited {\HH} can take place.  Such an occurrence was first pointed out by  Federman et al. \cite{federman:95} for the cloud towards $\zeta$ Ophiuchi. Meyer et al.  \cite{meyer:01} subsequently reported the rich ultraviolet spectrum of vibrational excited {\HH} toward HD~37903 (HD 37903 is a bright UV-emitting star embedded in the L1630 molecular cloud, creating the reflection nebula NGC 2023) and raised the role of ultraviolet radiative pumping induced by the intense radiation field of this star, illuminating the bright reflection nebula NGC2023 in Orion. Finally, let us mention the reanalysis by Gnaci{\'n}ski\cite{gnacinski:11,gnacinski:13} of the HD~37903 and HD~147888 (a blue main-sequence star in the Ophiuchus constellation) lines of sight where both FUSE and HST available data allow to have a complete description of  {\HH} in these environments. It should be recalled that the derivation of accurate values of column densities requires very detailed analysis and is far from obvious. The main issue in the profile fitting method over the full spectrum is the determination of the unknown underlying continuum spectrum whereas the curve of growth method suffers from large uncertainties in the so-called square-root range, where unfortunately most rotationally excited absorption transitions are located. 

Despite these shortcomings, the resulting values provide a stimulating challenge to modelists
and a potential clue to the physical conditions of the environment.
The quantitative knowledge of spectral radiative properties involving {\HH} has indeed been greatly improved in the last twenty years. A complete $j$-dependent theoretical analysis of Lyman and Werner electronic  band systems of {\HH} has been performed by Abgrall et al \cite{abgrall:93,abgrall:93a,abgrall:93b} where non-adiabatic coupling between excited B, B', C and D electronic levels was taken into account to derive the energy level positions as well as the electronic wave functions, in order to compute radiative transition probabilities. These theoretical calculations have been successfully tested against the atlas of the VUV emission spectrum of \HH, recorded in Ref. \citen{roncin:95} and to the fluorescence emission laboratory spectra following electronic excitation \cite{liu:99,liu:03}. The uncertainty on the energy transitions is less than 1~cm$^{-1}$, as the values of the electronic ground state energy levels are taken from experiments \cite{dabrowski:84}. Spontaneous continuum radiation emitted from excited electronic levels has also been computed \cite{abgrall:00}, checked with experiments \cite{jonin:00} and astrophysical and planetary observations. This mechanism is controlling   photodissociation of interstellar \HH, as photons with energies larger than 13.6~eV are not available, preventing the direct photodissociation mechanism  to occur. 

Another unknown lies in the understanding of  {\HH} excitation resulting from its formation mechanism on grain surfaces. About 4.5~eV are available from the formation of  {\HH} resulting from recombination of two H atoms on the surface of the dust grains. In the absence of quantitative information, it was first assumed \cite{black:87} that the energy release is equivalently 
shared between kinetic energy, internal energy of the dust grain and internal energy of  \HH. This assumption allows to populate significantly vibrational and rotational energy levels which decay subsequently via the (slow) radiative electric
quadrupole transitions and collisions. 
The statistical equations describing the equilibrium of the rovibrational levels of {\HH} in its ground electronic state 
are solved by including all these processes, including chemical formation in a number of PDR models as summarized in Ref. \citen{roellig:07}. Recent experiments coupled to theoretical calculations on {\HH} formation on various surfaces seem to indicate that {\HH} could be formed in a narrow range of excited levels \cite{gough:96,creighan:06,lemaire:10,sizun:10}. Such scheme were  tested in astrochemical models \cite{islam:10,lebourlot:12}.  Observational evidence of the signature of this "formation pumping" is still debated \cite{burton:02,congiu:09} and new infrared spectroscopic facilities should help 
solving this issue.

In summary, one can reasonably state that radiative and collisional excitation processes are well understood for interstellar {\HH}. 
This fact has lead different groups to build PDR models incorporating these 
processes and check their basic features in a successful comparison study \cite{roellig:07}. The detailed description of these codes
is beyond the purpose of the present review. The main remaining uncertainty lies in the role of  {\HH} formation on dust grains for which many parameters are out of control (composition, temperature, ...).

\subsubsection{CO}

CO carbon monoxide is a useful mass tracer as its low frequency rotational transitions are seen ubiquitously. The so-called "X-factor" relies the molecular hydrogen column density $N(\HH)$ to the measure of CO ($j=1 \to j'=0$) integrated intensity,  W($^{12}$CO),  and has been first introduced by Solomon et al. \cite{solomon:87} who found a tight relation between the gas dynamical mass, as measured by the virial theorem and the CO luminosity. The integration is performed on the emission profile expressed in velocity unit  and the intensity is displayed in Kelvin so that W has units $ K ~km/s$.
$$X \equiv \frac{N(\HH)}{W(^{12}\CCO)} =  ~ ~ 3.0 \times 10^{20}~cm^{-2} ~K^{-1} ~ km^{-1} ~ s .$$ This quantity is  widely used in galactic and extragalactic studies for mass determination, despite the fact that it may not be as constant as previously thought \cite{bell:06,pineda:08,bolatto:13}. 
Understanding CO photodissociation is a major task as this $j=1 \to j'=0$ transition is mainly detectable at the edges of molecular clouds when the corresponding opacity is close to $\sim$ 1. Detailed studies cannot be undertaken without corresponding spectroscopic or laboratory studies. First attempts have been  performed with the LURE (Laboratoire
pour l'Utilisation du Rayonnement Electronique) and ACO synchrotron source \cite{letzelter:87} where predissociation of CO via  Rydberg states  was experimentally studied and further theoretically interpreted by Tchang-Brillet et al \cite{tchang:92}. Finally an extensive list of oscillator strengths and predissociation rates was provided  by Eidelsberg et al. \cite{eidelsberg:91},
which was further included in PDR models. 
The availability of VUV lasers and new synchrotron sources (SUPERACO, ALS, SOLEIL) with high spectral resolution facilities has been applied recently to the study of CO and its isotopologues \cite{cacciani:95,eidelsberg:06,chakraborty:08,eidelsberg:12}, which allows to resolve the narrow rotational structure of predissociating Rydberg levels of CO as shown in Fig. \ref{fig:photco}.

\begin{figure}
\begin{center}
\includegraphics[width=10.cm,angle=0.]{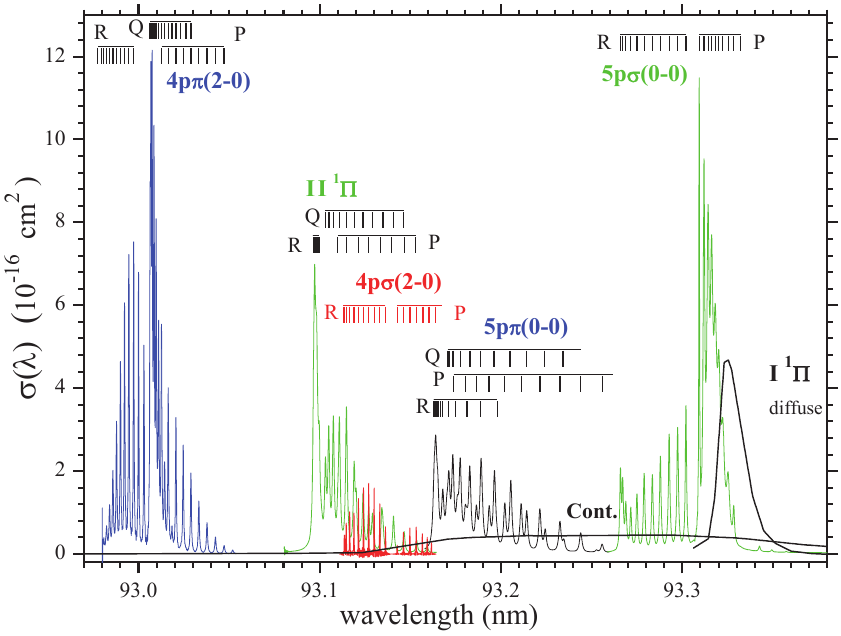}
\caption{Model absorption cross sections with rotational assignments of the
4p(2) and 5p(0) Rydberg complexes and $1\Pi$ valence states. Reprinted with permission from Ref. \citen{eidelsberg:12}. Copyright 2012 ESO.}
\label{fig:photco}
\end{center}
\end{figure}

Several PDR models include detailed photodissociation mechanisms of CO \cite{black:87, lebourlot:93, visser:09, roellig:13} coupled to collisional excitation although most recent detailed informations on $j$ dependent predissociation positions and widths are not yet fully included.

It has also been possible to detect CO up to very high rotational quantum numbers thanks to the opening of submillimeter region with Herschel and understanding the full rotational pattern opens new modeling constraints. Values up to $j=46$ have been observed with the Photodetector Array Camera and Spectrometer (PACS) in the spectra of 21 protostars in the Orion molecular clouds \cite{manoj:13}. The present interpretation involves   3 - 4 rotational temperature components, including high temperatures (T~$\sim$~2000~K) and low densities (n(\HH)~$\leq$~$\sim 10^6$~cm$^{-3}$) regions coming from shock heated molecular outflows. However, one may wonder if this high rotational excitation may not arise from 
other possibilities. Radiative excitation may take place through the VUV A-X band molecular system whose absorption spectrum lies around 1400 -  1500 \AA. However with a selection rule $\Delta j = 0 \pm 1$, excitation up to $j= 46$ is a real challenge.
Another source of excitation may be provided from a chemical channel: CO is a very stable molecule ($\sim$ 8 eV) so that its chemical formation involves large exothermicites. Most promising are the channels provided by dissociative recombination of 
{\HCOp} and the C + OH reaction. This last reaction has been theoretically studied by Bulut et al \cite{bulut:09} who find that no barrier is present and that significative vibrational excitation can be obtained in the reaction. Additional observations will help putting more constraints on the physical origin of CO excitation.

\subsubsection{OH / {\HHO} / \HHHOp}

Recent observational findings of anomalous excitation have been raised recently. Tappe et al \cite{tappe:08} report the discovery of extraordinary excited hydroxyl radical (OH) in SPITZER observations performed in the H 211 outflow. The detected transitions involve $j$ quantum numbers between 15/2 and 69/2, probably limited by the wavelength window of the instrument. The corresponding maximum energy is 28200~K above the ground level.  Such a temperature is obviously out of the range
of any kinetic relevance: The radiative decay of rotationally excited OH is of the order of 10 - 400~s$^{-1}$, which makes collisional excitation very ineffective at the low densities of the environment. The qualitative interpretation of these results  involves  formation of highly rotationally excited OH from the photodissociation of \HHO. Absorption of UV radiation in the {\HHO} (B - X) band produces OH in an excited state A$^2\Sigma$, which in turn yields highly excited ground state OH molecules via a non-adiabatic crossing between intersecting PESs \cite{vanharrevelt:00,Harich:00,zhou:13}. Nevertheless, the full quantitative analysis remains to be performed. A similar result  has been found by Najita $et ~al.$ \cite{najita:10} in the SPITZER spectrum of  the T Tauri star TWHya. These observations can be compared to the detection of strong water emission toward the protostellar object IRAS 4B in NGC 1333 \cite{watson:07,herczeg:12} where OH emission has been identified as well, but with $j$ upper quantum values less than 21/2.  These observations are interpreted as the signature of ice sublimation of grain mantles in the $\sim$~170~K warn surface layer, which  is also consistent with the weakness of OH emission in this particular source. 
Another puzzling result involving probably chemical effects is the recent observation of highly rotationally excited hydronium ion ({\HHHOp}) \cite{lis:12} with Herschel HIFI detected in absorption towards the galactic center Sagittarius B2. The highest excited level is the $j, k =11,11$ metastable level, corresponding to an energy larger than 1200~K.  The
interpretation takes advantage of the knowledge of the physical conditions available in the astrophysical source, which is undoubtedly complex. The source is well studied and consists of a rich molecular gas in the central molecular zone characterized by significant densities, high degree of turbulence as well as increased rates of cosmic and X-ray fluxes
 and it is remarkable that the $j,k=18,18$ metastable level of ammonia has been reported \cite{wilson:06} in the same source.
Whereas no complete analysis has been performed, in particular due to the fact that the corresponding collision rates are not available, there is a present assent that these observations could reflect the formation process which occurs through exothermic reactions between oxygen hydride ions and molecular hydrogen. The $\HHOp + \HH \to \HHHOp$ + H reaction has an exothermicity of 1.69 eV ($\sim$ 20000~K) and $\HHHp + \HHO \to \HHHOp + \HH $ of 2.81eV ($\sim$ 32000~K).
The distribution of this energy amongst the reaction products is not known but a significant amount of energy is available to populate excited {\HHHOp} levels. Metastable levels could be populated in such a process, and will then decay through collisions with He and  \HH. However, no information is yet available for such collisional rate coefficients.     

\subsubsection{CH$^+$}
Understanding CH$^+$ abundance and excitation is a major challenge. This molecular ion is observed ubiquitously in diffuse regions where carbon is photoionized but  the C$^+$ + {\HH}  reaction is endothermic by a significant amount of 0.4 eV and in addition, it reacts efficiently both with {\HH} and H, although recent studies showed that the reaction rate coefficients with H atoms decrease significantly below 60K \cite{plasil:11}. 
Then, all chemical modeling studies aiming to account for the presence of this molecular ion have introduced some additional energy input allowing to overcome the barrier.
Molecular shock waves 
\cite{forets:86}, non thermal chemistry and turbulent dissipation \cite{joulain:98,zsargo:03,godard:09} or the presence of some available vibrationally excited {\HH} \cite{agundez:10b} in PDRs are amongst the possibilities invoked. 
The recent introduction of radiative and chemical pumping in the excitation study of this ion deserves
a specific comment. Godard and Cernicharo \cite{godard:13} have built an excitation model where ultraviolet pumping and chemical pumping are introduced in addition to collisional excitation and spontaneous radiative decay. They showed that the intensities of the pure rotational transitions of CH$^+$ detected towards the Orion bar 
are reproduced with densities of about $5~\times 10^4$cm$^{-3}$, which is one order of magnitude below the value inferred from traditional  collisional excitation models. This work opens the road to further investigations for other 
molecules  and also 
emphasizes the need to better constrain the state to state chemistry at work in these environments.

\subsection{Role of the cosmic background radiation field}

\subsubsection{CN radical: a probe of the present cosmic background radiation field}

The observation of the interstellar CN radical is another example of radiative pumping effects and it was recognized very early that
the rotational excitation temperature obtained from the analysis of visible absorption transitions was of several K. The significance of this temperature has been recognized in Ref. \citen{field:66}, shortly after the detection of the 2.7 K black body cosmic radiation field by Penzias and Wilson \cite{penzias:65}, a black body relic radiation of the initial big bang explosion. The rotational pattern of this radical accidentally matches the energy range where the energy flux of the present cosmic background radiation field is maximum and the $N=0 \rightarrow N'=1$ rotational transition is directly pumped by the cosmic radiation field. This measurement has provided the most accurate value of the temperature of the cosmic background radiation field before the launch of the COBE satellite \cite{crane:86,meyer:89c,kaiser:90}, after subtraction of possible collisional effects \cite{black:91}.  Very high signal to noise spectra and careful detailed analysis \cite{ritchey:11} lead to a weighted mean value $T_R(CN) =  2.754 \pm 0.002$~K. A variant study \cite{krelowski:12} suggests an even slightly larger value. These determinations should be compared with the latest COBE value of the cosmic background radiation field \cite{fixsen:09} of $ 2.72548 \pm 0.002$~K and are definitively larger. The significance of these discrepancies is discussed in Ref. \citen{leach:12}, where it is not excluded that they may be due to an insufficient knowledge  of the corresponding oscillator strengths and some possible departure from a full decoupling of electronic, vibrational and rotational motions in CN. 

\subsubsection{Higher redshifts}

The cosmic background radiation field varies with the evolution time of the universe and the value of the corresponding
black body temperature, in the standard hot big bang model, can be expressed as $ T = T_0 (1 + z) $, where $T_0$ is the present value (corresponding to $z = 0$) and $z$ is the redshift, linked to the time evolution of the expanding universe. In distant objects corresponding to positive $z$  values, the transition wavelengths of any atom or molecule are displaced towards the red and can also be expressed in function of $z$ as, $\lambda = \lambda_0 (1+z)$, where $\lambda_0$ is the value of wavelength in the laboratory rest frame.  Then, VUV absorption transition are shifted in the visible / near infrared regions with increasing values of  $z$. Different atoms, ions and {\HH} molecules have 
been detected in front of distant quasars \cite{noterdaeme:08} in so-called high redshift damped Lyman $\alpha$ systems through VUV transitions shifted in the visible window from ground based telescopes. Very recently Srianand and co-workers \cite{srianand:08,noterdaeme:09,noterdaeme:10,noterdaeme:11} have detected the A-X electronic band system of CO in absorption from several rotationally excited ground state levels. As the observations involve diffuse regions, collisional effects are found negligible and the temperature of the exciting radiation follows the linear law given above for a range of redshifts between 1.7 and 2.8, corresponding to blackbody radiation temperatures in the 6 - 10~K temperature range. Their study leads to
the following expression $ T_{CMB}(z)=(2.725 \pm 0.02) \times (1+z)^{(1-\beta)}$~K with $\beta=-0.007 \pm 0.025 $, which supports the predictions of the standard cosmological big bang model. A study of emission rotational transitions of a variety of molecules in a galaxy lensing the quasar PKS 1830-211 at $z = 0.89$ \cite{muller:13} leads to a similar conclusion where the exponent value $\beta$ is further reduced to $ \beta=0.009 \pm 0.019$. This determination however has required a much more subtle disentangling of radiative/collisional effects through Monte Carlo Markov Chain (MCMC) analysis. 
Corresponding results are displayed in Fig. \ref{fig:redshift}.

\begin{figure}
\begin{center}
\includegraphics[width=8.cm,angle=0.]{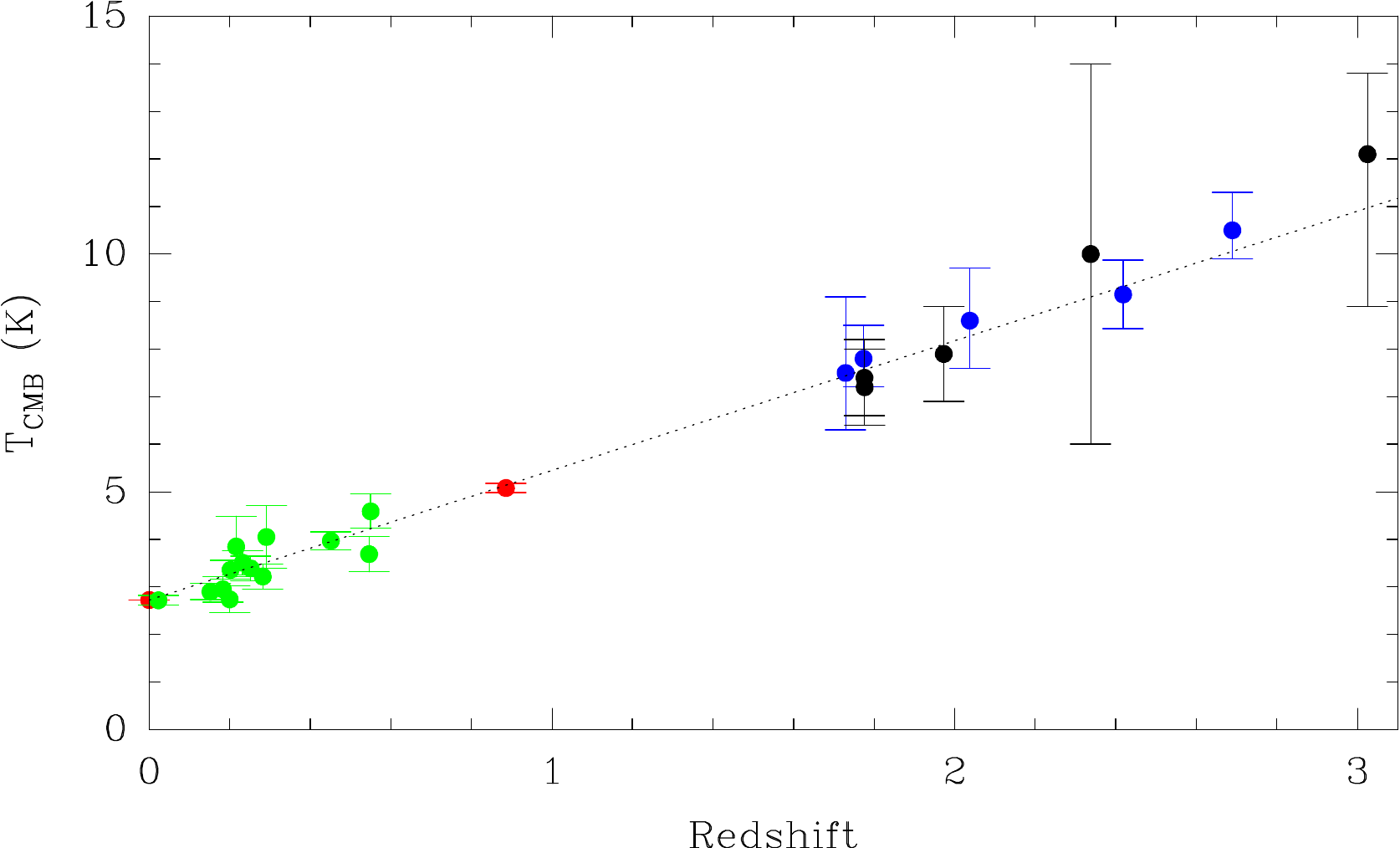}
\caption{Measurements of the CMB temperature as a function of redshift. Data points in green correspond to Sunyaev-Zeldovich (SZ)  measurements towards galaxy clusters, in black to C I and C II absorption studies, in blue to CO absorption and the value derived toward PKS 1830-211 absorption is marked in red. The dotted line corresponds to the law $T_{CMB}=T_0 \times (1+z)$. Reprinted with permission from Ref. \citen{muller:13}. Copyright 2013  ESO}
\label{fig:redshift}
\end{center}
\end{figure}

Molecular observations provide outstanding and unique evaluations of the temperature of the cosmic background thermal radiation at different redshifts and the present findings are in agreement with the standard cosmological model. Such a fundamental achievement can only be obtained if other excitation processes are well understood and emphasizes the extreme importance of accurate molecular collisional data for astrophysical studies.

\section{Discussion and outlook}
\label{sec:conclusion}

We have attempted to present an overview of recent concern and achievements related to physical mechanisms 
involved in molecular excitation.  The new observational facilities have provided a bunch of unexpected  results, extending from the presence of new species to the discovery of striking states of excitation of known molecules. 

A major part of this review is dedicated to the recent progresses accomplished in collisional excitation studies, which have been decisive to the success of  interpreting
these new observational data and constrain the physical conditions present in interstellar environments.  Such developments would not have been possible without a fruitful exchange between astrophysicists,  molecular experimental and theoretical physicists and quantum chemists. This review allows to acknowledge 
these efforts but should also serve to pursue the collaborations and prospect  future directions of investigation. 
This review provides also specific examples showing how risky it may be to use rate coefficients of one molecule to infer those involving an isoelectronic molecule (CO -  CN$^-$) or even an isomeric form (HCN -HNC). It is indeed not possible to predict the behavior and magnitude of collisional excitation rate coefficients without performing explicit calculations. However, some general trends can be extracted. For example, in the case of $^{2S+1}\Sigma$ electronic ground state radicals, a propensity in favor of $\Delta j = \Delta N$ holds generally whereas the $\Delta F = \Delta j$ propensity rules are obtained  for hyperfine structure excitation.



Computing and fitting a molecular PES,  performing subsequent dynamical calculations is a time consuming task and one may wonder about the sensitivity of astrophysical excitation models to molecular data, such as collisional rate coefficients and Einstein emission coefficients. Answering such an interrogation is not straightforward.  However, it is well know that CPU time dedicated to the computations on the PES and/or dynamical calculations can be greatly reduced if the expected accuracy is of 30-50\% for the collisional excitation rate coefficients. Hence, the question of the accuracy needed for astrophysical modeling is crucial. Recent comparisons between theory and experiment show that such a level of agreement (10-20\%) is reached for the best possible calculations but this kind of agreement is obtained only for a limited number of systems that has been the object of intensive studies. The availability of experimental results is crucial to this respect  and allows to check the theoretical calculations as disentangling the respective effects of the PES and dynamical approximation is not commonly addressed. 

\paragraph{Is such an accuracy enough for astrophysical purposes?}
Collisional excitation rate coefficients as well as Einstein emission coefficients are introduced in radiative transfer models aimed at computing emission probabilities
of the various molecular species. The escape probability formalism introduced by Sobolev \cite{sobolev:60} allows to
reduce significantly the   time necessary to solve the corresponding non linear equations coupling statistical equilibrium equations and radiation source terms. This method is equivalent to the LVG  method and may be superseded by more sophisticated methods such as the Monte-Carlo method or  a direct resolution of the integro-differential equations of radiative transfer via the short characteristic method \cite{daniel:13}.
The recently computed collisional excitation rate coefficients described in Section \ref{sec:col} have been tested against previous results in
astrophysical excitation models in fruitful interdisciplinary collaborations. On may emphasize the most significant 
achievements performed for CS \cite{Lique:06CS}, SO \cite{lique:06}, {\NNHp} \cite{daniel:06}, HCN / HNC \cite{Sarrasin:10,Dumouchel:11,benabdallah:12}, {\NHHH} \cite{maret09}, H$_2$CO\cite{Troscompt:09}  and last but not least  {\HHO} \cite{Daniel:12}.
The availability of those data also allowed to overcome previous simplistic LTE approximations involving excitation temperature assumptions such as for NO \cite{lique:09}.
Sensitivity of these excitation models to collisional rate coefficients is manifest and an order of magnitude difference in the collisional excitation rate coefficient induces typically a factor of 2 to 5 in the abundance determinations  from our own experience. Systematic analyses, such as those performed for chemical models \cite{wakelam:10} would be helpful. 
Then, we may answer that an accuracy  of 30--50\% in the collisional excitation rate coefficients  is tolerable for  astrophysical modeling purposes. 
A closely related point concerns the question if collisional rates involving Helium can be used to infer those for para-H$_2$($j=0$), through an adequate scaling factor involving the reduced mass. As the potential surfaces are usually more shallow when {\HH} is involved, significant differences 
may arise, as in the case of \NHHH.  Those discrepancies are thought to decrease with increasing temperatures. However, {\HH}($j=2$) becomes populated around 500K and He is not anymore a good template of \HH. 
Similarly, collisional excitation rate coefficients involving para-H$_2$($j=0$) are not representative of those with ortho-H$_2$, J=1  as the spherical harmonic expansion of the PES entails new terms leading possible significant differences. Such effects have been shown to reach a factor larger than 3   for  H$_2$O. 
What may have not be realized is the requirement of completeness of the collisional data sets.  As an example, {\NHHH} has been detected up to its (18,18) metastable level \cite{wilson:06}.  This level is only quenched via collisions for which not any, even approximate value, is available. For these specific levels, the availability  of collisional values with even an accuracy of a factor 2 (typically what we could expect from calculations with He as a model for H$_2$ for heavy molecule) would be a major advance. 
An additional point relies to other possible excitation mechanisms and, as an example, we refer to the recent finding on CH$^+$ that overlooking radiative excitation  and chemical pumping \cite{zanchet:13}   leads to overestimating density by about one order of magnitude \cite{godard:13}.
As far as radiative and chemical effects are concerned, we emphasize again that such effects may be introduced
only in well defined environments, when collisional effects are reasonably well understood. 
The  relevant data  consist of accurate transition frequencies and associate oscillator strengths for radiative pumping and  temperature dependent state to state reaction rate coefficients for chemical pumping.  

\paragraph{What is the future of collisional excitation studies?}
If we take the point of view of astrophysicists, we stress that a significant number of collisional rates is available for most of the currently detected molecules, as displayed in Tables \ref{tab1} and \ref{tab2}. Extension to higher temperatures and to a larger number of rotationally excited levels is desirable
and sometimes even critical, as for example for ammonia. Some related computations are indeed in progress.
Extrapolation to higher (lower) temperatures is addressed in Ref. \citen{schoier:05}  where specific analytic extrapolation formulae are suggested. An artificial neural network has also been recently investigated \cite{neufeld:10} as a tool for estimating 
rate coefficients for the collisional excitation of molecules. The danger of  extrapolation is however emphasized
by Flower and Pineau des For{\^e}ts  \cite{Flower:12extr} for a number of molecules (\HH, CO, SiO, o-\HHO, o-NH$_3$ and CH$_3$OH) and we endorse these warnings on an even larger scale.
Despite all previous positive statements, some collisional data are still badly missing such as metal cyanide compounds, linear polyatomic carbon chains C$_n$H, SiC$_n$, .... 
Oxygen hydride ions (OH$^+$, \HHOp, \HHHOp)  should also be considered as  target molecules for collisional studies since they play  a crucial role in ion-molecule chemistry and high metastable levels of {\HHHOp} have been detected which cannot be understood within the present knowledge.
From the fundamental molecular physics point of view,  several features deserve still more investigation. The balance and/or competition between reactive and non reactive channels is an open issue for a number of case studies and remains to be addressed both experimentally and theoretically. 
The {\HHHp} +  {\HH} collisional system is a good example of joined theoretical \cite{Hugo:09,crabtree:11a,gomez:12} and
experimental \cite{gerlich:06,crabtree:11b} efforts to disentangle non reactive, proton hoping, proton exchange 
collisions but the present status is not yet completely satisfactory and more work is welcome. {\HHHp} has also been detected in highly excited metastable levels  which can only decay via collisions, which are presently not well accounted for \cite{oka:04}. In a study on the H + {\DD} system, Lique et al. \cite{lique:12D2} found that  the reactive channel  may be neglected in the rotational excitation theoretical calculations in the presence of a potential barrier. Such an approximation should be tested against other benchmark systems. 
Reactive molecular ions such as  CO$^+$, OH$^+$, {\HHOp}, CH$^+$, .... are additional examples of
systems where  both reactive and non reactive 
channels are at work, with a rather attractive PES.  Additional complexity is brought by the open shell structure of 
 these ions (except CH$^+$).

Studies of vibrational collisional excitation of low energy bending modes of polyatomics have received little interest until now, both experimentally and theoretically. The theoretical approach implies
the determination of angular dependence of the PES and subsequent fitting and relevant dynamical treatments which are not available in open source programs, as is the case for rotational excitation and/or stretching mode excitation. Such data are nevertheless of astrophysical interest as excited bending modes of HCN, C$_2$H$_2$, C$_2$H$_4$ are detected in circumstellar envelopes and warm astrophysical environments.

Complex organic molecules are an additional challenge for theoreticians and experimentalists and their presence
is manifest not only in the complex Galactic Center Sgr B2 region \cite{loomis:13, zaleski:13} but surprisingly also in low mass star forming region such as IRAS +16293 - 2422 \cite{caux:11}
 and pre-stellar cores such as L1689N \cite{bacmann:12} and B1 \cite{cernicharo:12}. Exceptional care is required in the identification of these species which involve a huge number of closely spaced levels and transitions present over a large range of frequencies. As far as observations are concerned, their analysis involves essentially rotational diagrams as not any relevant data is available.  Whereas LTE conditions are probably fulfilled under hot core conditions, this is probably not true in the cold pre-stellar cores where information about collisional rates is badly missing. We finally want to emphasize the new perspectives raised by the studies allowing to probe state dependent chemistry
in close connection with spectroscopic information. 

\acknowledgement
We acknowledge the CNRS national program  ''Physique et Chimie du Milieu Interstellaire''  for supporting such an interdisciplinary  research field  and  all our present and past collaborators who allowed us to perform part of the work that has been reviewed in this paper. ER thanks the Agence Nationale de la Recherche (ANR, France) for financial support under contract 09-BLAN-020901. FL is very grateful to Nicole Feautrier, Annie Spielfiedel, Jacek K{\l}os and Millard Alexander for sharing their knowledge on scattering calculations. 

\paragraph{Biography}
Evelyne Roueff was first student and then lecturer in Physics at the Ecole Normale Sup\'erieure de Jeunes Filles before getting an Astronomer position at the Observatoire de Paris in 1982.
She prepared her thesis in Molecular Physics under the supervision of H. van Regemorter where she theoretically studied  pressure broadening of sodium atoms in a bath of Helium and atomic Hydrogen. She then focused her research towards interstellar applications and computed, amongst different processes, fine structure collisional excitation of important atoms and ions by atomic Hydrogen, radiative association reaction  between C$^+$ and H and the theoretical accurate electronic spectrum of molecular Hydrogen, which is widely used not only for astrophysical but also in planetary and laboratory applications. She is now involved in astrochemical modeling tightly coupled to observations  and has been part of the team detecting, for the first time, fully deuterated ammonia and thioformaldehyde as well as  several new isomers of isocyanic acid.
She has been deeply involved in the origin and in the management of the interdisciplinary program PCMI (Physique et Chimie du Milieu Interstellaire), which was funded by CNRS (France) and fostered fruitull collaborations between astrophysicists, molecular physicists and chemists in France.
She obtained the Deslandres Price of the Acad\'emie fran\c{c}aise in 1999 with G. Pineau des For\^ets
for their work on bistability in the interstellar chemical modeling studies.

\begin{figure}
\begin{center}
\includegraphics[width=5. cm]{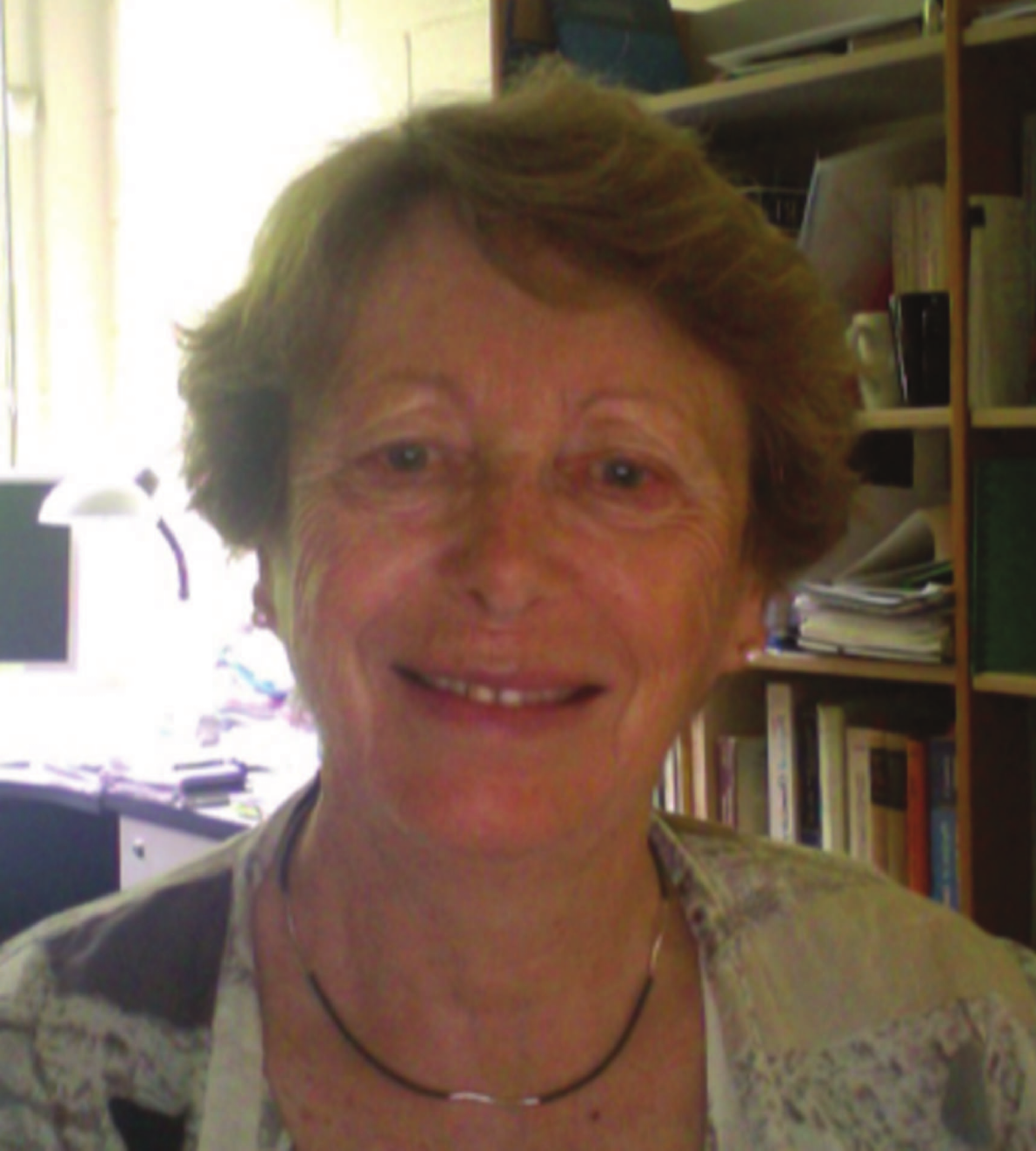}
\end{center}
\end{figure}

\paragraph{Biography}

Fran\c{c}ois Lique (born in 1980, in France) studied physics at the Universit\'e Pierre et Marie Curie, Paris (France) and received his Ph.D. in 2006, from the same university, under the supervision of Nicole Feautrier and Annie Spielfiedel. His dissertation focused on the collisional excitation of interstellar species, from the theoretical modeling to astrophysical applications. During his Ph.D, he spent several months in Madrid (Spain) working on astrophysical modeling. As a postdoc, he joined Millard Alexander's group at the University of Maryland (USA) to work on inelastic and reactive collisions implying open-shell molecules. In 2008, he obtained a lecturer position at the Laboratoire Ondes et Milieux Complexes of the University of Le Havre (France). In 2010, he defended his Habilitation. His current research focuses on the study of physical and chemical processes of astrophysical interest in close collaboration with experimentalists and astrophysicists.

\begin{figure}
\begin{center}
\includegraphics[width=5. cm]{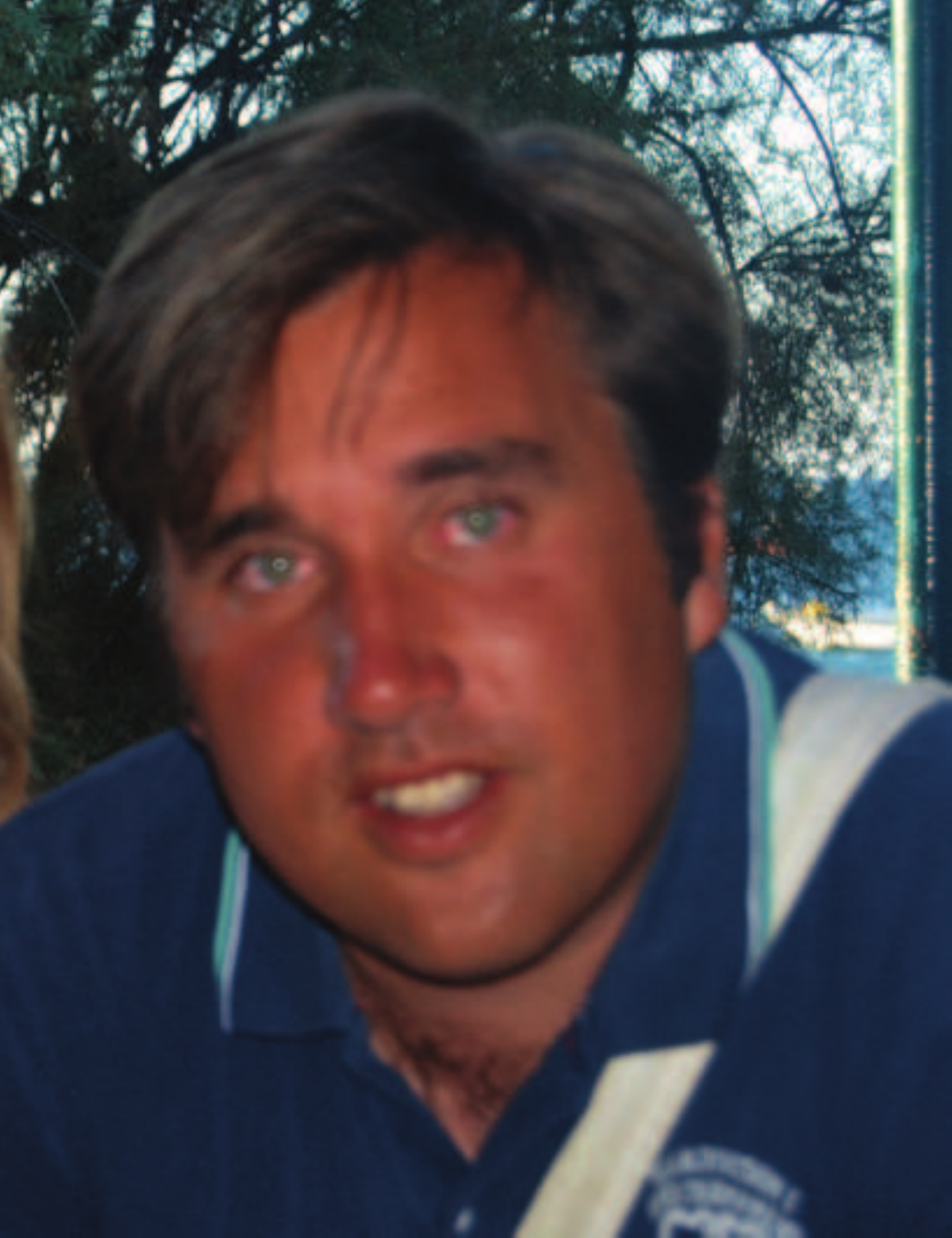}
\end{center}
\end{figure}



\begin{mcitethebibliography}{457}
\providecommand*\natexlab[1]{#1}
\providecommand*\mciteSetBstSublistMode[1]{}
\providecommand*\mciteSetBstMaxWidthForm[2]{}
\providecommand*\mciteBstWouldAddEndPuncttrue
  {\def\EndOfBibitem{\unskip.}}
\providecommand*\mciteBstWouldAddEndPunctfalse
  {\let\EndOfBibitem\relax}
\providecommand*\mciteSetBstMidEndSepPunct[3]{}
\providecommand*\mciteSetBstSublistLabelBeginEnd[3]{}
\providecommand*\EndOfBibitem{}
\mciteSetBstSublistMode{f}
\mciteSetBstMaxWidthForm{subitem}{(\alph{mcitesubitemcount})}
\mciteSetBstSublistLabelBeginEnd
  {\mcitemaxwidthsubitemform\space}
  {\relax}
  {\relax}

\bibitem[Lequeux et~al.(2005)Lequeux, Falgarone, and Ryter]{lequeux:05}
Lequeux,~J.; Falgarone,~E.; Ryter,~C. \emph{The Interstellar Medium}; Astronomy
  and Astrophysics Library; Springer: Berlin, 2005\relax
\mciteBstWouldAddEndPuncttrue
\mciteSetBstMidEndSepPunct{\mcitedefaultmidpunct}
{\mcitedefaultendpunct}{\mcitedefaultseppunct}\relax
\EndOfBibitem
\bibitem[{Bernes}(1979)]{bernes:79}
{Bernes},~C. \emph{\aap} \textbf{1979}, \emph{73}, 67\relax
\mciteBstWouldAddEndPuncttrue
\mciteSetBstMidEndSepPunct{\mcitedefaultmidpunct}
{\mcitedefaultendpunct}{\mcitedefaultseppunct}\relax
\EndOfBibitem
\bibitem[{van der Tak} et~al.(2007){van der Tak}, {Black}, {Sch{\"o}ier},
  {Jansen}, and {van Dishoeck}]{vandertak:07}
{van der Tak},~F.~F.~S.; {Black},~J.~H.; {Sch{\"o}ier},~F.~L.; {Jansen},~D.~J.;
  {van Dishoeck},~E.~F. \emph{\aap} \textbf{2007}, \emph{468}, 627\relax
\mciteBstWouldAddEndPuncttrue
\mciteSetBstMidEndSepPunct{\mcitedefaultmidpunct}
{\mcitedefaultendpunct}{\mcitedefaultseppunct}\relax
\EndOfBibitem
\bibitem[{Goldsmith} and {Langer}(1999){Goldsmith}, and {Langer}]{goldsmith:99}
{Goldsmith},~P.~F.; {Langer},~W.~D. \emph{\apj} \textbf{1999}, \emph{517},
  209\relax
\mciteBstWouldAddEndPuncttrue
\mciteSetBstMidEndSepPunct{\mcitedefaultmidpunct}
{\mcitedefaultendpunct}{\mcitedefaultseppunct}\relax
\EndOfBibitem
\bibitem[{Elitzur}(1994)]{elitzur:94}
{Elitzur},~M. \emph{\apj} \textbf{1994}, \emph{422}, 751\relax
\mciteBstWouldAddEndPuncttrue
\mciteSetBstMidEndSepPunct{\mcitedefaultmidpunct}
{\mcitedefaultendpunct}{\mcitedefaultseppunct}\relax
\EndOfBibitem
\bibitem[{Daniel} and {Cernicharo}(2013){Daniel}, and {Cernicharo}]{daniel:13}
{Daniel},~F.; {Cernicharo},~J. \emph{\aap} \textbf{2013}, \emph{553}, A70\relax
\mciteBstWouldAddEndPuncttrue
\mciteSetBstMidEndSepPunct{\mcitedefaultmidpunct}
{\mcitedefaultendpunct}{\mcitedefaultseppunct}\relax
\EndOfBibitem
\bibitem[{Jensen} et~al.(2010){Jensen}, {Snow}, {Sonneborn}, and
  {Rachford}]{jensen:10}
{Jensen},~A.~G.; {Snow},~T.~P.; {Sonneborn},~G.; {Rachford},~B.~L. \emph{\apj}
  \textbf{2010}, \emph{711}, 1236\relax
\mciteBstWouldAddEndPuncttrue
\mciteSetBstMidEndSepPunct{\mcitedefaultmidpunct}
{\mcitedefaultendpunct}{\mcitedefaultseppunct}\relax
\EndOfBibitem
\bibitem[{Neufeld}(2012)]{neufeld:12}
{Neufeld},~D.~A. \emph{\apj} \textbf{2012}, \emph{749}, 125\relax
\mciteBstWouldAddEndPuncttrue
\mciteSetBstMidEndSepPunct{\mcitedefaultmidpunct}
{\mcitedefaultendpunct}{\mcitedefaultseppunct}\relax
\EndOfBibitem
\bibitem[cdm(2005)]{cdms}
\url{http://www.astro.uni-koeln.de/cdms/}, 2005\relax
\mciteBstWouldAddEndPuncttrue
\mciteSetBstMidEndSepPunct{\mcitedefaultmidpunct}
{\mcitedefaultendpunct}{\mcitedefaultseppunct}\relax
\EndOfBibitem
\bibitem[{M{\"u}ller} et~al.(2001){M{\"u}ller}, {Thorwirth}, {Roth}, and
  {Winnewisser}]{muller:01}
{M{\"u}ller},~H.~S.~P.; {Thorwirth},~S.; {Roth},~D.~A.; {Winnewisser},~G.
  \emph{\aap} \textbf{2001}, \emph{370}, L49\relax
\mciteBstWouldAddEndPuncttrue
\mciteSetBstMidEndSepPunct{\mcitedefaultmidpunct}
{\mcitedefaultendpunct}{\mcitedefaultseppunct}\relax
\EndOfBibitem
\bibitem[{M{\"u}ller} et~al.(2005){M{\"u}ller}, {Schl{\"o}der}, {Stutzki}, and
  {Winnewisser}]{muller:05}
{M{\"u}ller},~H.~S.~P.; {Schl{\"o}der},~F.; {Stutzki},~J.; {Winnewisser},~G.
  \emph{J. Mol. Struct.} \textbf{2005}, \emph{742}, 215\relax
\mciteBstWouldAddEndPuncttrue
\mciteSetBstMidEndSepPunct{\mcitedefaultmidpunct}
{\mcitedefaultendpunct}{\mcitedefaultseppunct}\relax
\EndOfBibitem
\bibitem[{Dumouchel} et~al.(2012){Dumouchel}, {K{\l}os}, {Tobo{\l}a},
  {Bacmann}, {Maret}, {Hily-Blant}, {Faure}, and {Lique}]{dumouchel:12NH}
{Dumouchel},~F.; {K{\l}os},~J.; {Tobo{\l}a},~R.; {Bacmann},~A.; {Maret},~S.;
  {Hily-Blant},~P.; {Faure},~A.; {Lique},~F. \emph{\jcp} \textbf{2012},
  \emph{137}, 114306\relax
\mciteBstWouldAddEndPuncttrue
\mciteSetBstMidEndSepPunct{\mcitedefaultmidpunct}
{\mcitedefaultendpunct}{\mcitedefaultseppunct}\relax
\EndOfBibitem
\bibitem[lam(2005)]{lamda}
\url{http://home.strw.leidenuniv.nl/~moldata/}, 2005\relax
\mciteBstWouldAddEndPuncttrue
\mciteSetBstMidEndSepPunct{\mcitedefaultmidpunct}
{\mcitedefaultendpunct}{\mcitedefaultseppunct}\relax
\EndOfBibitem
\bibitem[{Sch{\"o}ier} et~al.(2005){Sch{\"o}ier}, {van der Tak}, {van
  Dishoeck}, and {Black}]{schoier:05}
{Sch{\"o}ier},~F.~L.; {van der Tak},~F.~F.~S.; {van Dishoeck},~E.~F.;
  {Black},~J.~H. \emph{\aap} \textbf{2005}, \emph{432}, 369\relax
\mciteBstWouldAddEndPuncttrue
\mciteSetBstMidEndSepPunct{\mcitedefaultmidpunct}
{\mcitedefaultendpunct}{\mcitedefaultseppunct}\relax
\EndOfBibitem
\bibitem[bas(2012)]{basecol}
\url{http://basecol.obspm.fr/}, 2012\relax
\mciteBstWouldAddEndPuncttrue
\mciteSetBstMidEndSepPunct{\mcitedefaultmidpunct}
{\mcitedefaultendpunct}{\mcitedefaultseppunct}\relax
\EndOfBibitem
\bibitem[{Dubernet, M.-L.} et~al.(2013){Dubernet, M.-L.}, {Alexander, M. H.},
  {Ba, Y. A.}, {Balakrishnan, N.}, {Balan\c{c}a, C.}, {Ceccarelli, C.},
  {Cernicharo, J.}, {Daniel, F.}, {Dayou, F.}, {Doronin, M.}, {Dumouchel, F.},
  {Faure, A.}, {Feautrier, N.}, {Flower, D. R.}, {Grosjean, A.}, {Halvick, P.},
  {Klos, J.}, {Lique, F.}, {McBane, G. C.}, {Marinakis, S.}, {Moreau, N.},
  {Moszynski, R.}, {Neufeld, D. A.}, {Roueff, E.}, {Schilke, P.}, {Spielfiedel,
  A.}, {Stancil, P. C.}, {Stoecklin, T.}, {Tennyson, J.}, {Yang, B.},
  {Vasserot, A.-M.}, and {Wiesenfeld, L.}]{Dubernet:13}
{Dubernet, M.-L.}, et~al.  \emph{\aap} \textbf{2013}, \emph{553}, A50\relax
\mciteBstWouldAddEndPuncttrue
\mciteSetBstMidEndSepPunct{\mcitedefaultmidpunct}
{\mcitedefaultendpunct}{\mcitedefaultseppunct}\relax
\EndOfBibitem
\bibitem[Bernstein(1979)]{bernstein1979atom}
Bernstein,~R. \emph{Atom-molecule collision theory: a guide for the
  experimentalist}; Physics of atoms and molecules; Plenum Press: New York,
  1979\relax
\mciteBstWouldAddEndPuncttrue
\mciteSetBstMidEndSepPunct{\mcitedefaultmidpunct}
{\mcitedefaultendpunct}{\mcitedefaultseppunct}\relax
\EndOfBibitem
\bibitem[Flower(2007)]{Flower:07book}
Flower,~D. \emph{Molecular collisions in the interstellar medium.}; Cambridge
  astrophysics; Cambridge University Press: Cambridge, 2007; Vol.~42\relax
\mciteBstWouldAddEndPuncttrue
\mciteSetBstMidEndSepPunct{\mcitedefaultmidpunct}
{\mcitedefaultendpunct}{\mcitedefaultseppunct}\relax
\EndOfBibitem
\bibitem[Paterson et~al.(2012)Paterson, Costen, and McKendrick]{Paterson:12}
Paterson,~G.; Costen,~M.~L.; McKendrick,~K.~G. \emph{Int. Rev. in Phys. Chem.}
  \textbf{2012}, \emph{31}, 69\relax
\mciteBstWouldAddEndPuncttrue
\mciteSetBstMidEndSepPunct{\mcitedefaultmidpunct}
{\mcitedefaultendpunct}{\mcitedefaultseppunct}\relax
\EndOfBibitem
\bibitem[{Green} and {Thaddeus}(1974){Green}, and {Thaddeus}]{green74}
{Green},~S.; {Thaddeus},~P. \emph{\apj} \textbf{1974}, \emph{191}, 653\relax
\mciteBstWouldAddEndPuncttrue
\mciteSetBstMidEndSepPunct{\mcitedefaultmidpunct}
{\mcitedefaultendpunct}{\mcitedefaultseppunct}\relax
\EndOfBibitem
\bibitem[{Lique} et~al.(2008){Lique}, {Tobo{\l}a}, {K{\l}os}, {Feautrier},
  {Spielfiedel}, {Vincent}, {Cha{\l}asi{\'n}ski}, and {Alexander}]{lique08b}
{Lique},~F.; {Tobo{\l}a},~R.; {K{\l}os},~J.; {Feautrier},~N.;
  {Spielfiedel},~A.; {Vincent},~L.~F.~M.; {Cha{\l}asi{\'n}ski},~G.;
  {Alexander},~M.~H. \emph{\aap} \textbf{2008}, \emph{478}, 567\relax
\mciteBstWouldAddEndPuncttrue
\mciteSetBstMidEndSepPunct{\mcitedefaultmidpunct}
{\mcitedefaultendpunct}{\mcitedefaultseppunct}\relax
\EndOfBibitem
\bibitem[{Arthurs} and {Dalgarno}(1960){Arthurs}, and {Dalgarno}]{arthurs:60}
{Arthurs},~A.~M.; {Dalgarno},~A. \emph{Proc. R. Soc. A} \textbf{1960},
  \emph{256}, 540\relax
\mciteBstWouldAddEndPuncttrue
\mciteSetBstMidEndSepPunct{\mcitedefaultmidpunct}
{\mcitedefaultendpunct}{\mcitedefaultseppunct}\relax
\EndOfBibitem
\bibitem[{Townes} and {Cheung}(1969){Townes}, and {Cheung}]{Townes:69}
{Townes},~C.~H.; {Cheung},~A.~C. \emph{\apjl} \textbf{1969}, \emph{157},
  L103\relax
\mciteBstWouldAddEndPuncttrue
\mciteSetBstMidEndSepPunct{\mcitedefaultmidpunct}
{\mcitedefaultendpunct}{\mcitedefaultseppunct}\relax
\EndOfBibitem
\bibitem[{Augustin} and {Miller}(1974){Augustin}, and {Miller}]{Augustin:74}
{Augustin},~S.~D.; {Miller},~W.~H. \emph{\jcp} \textbf{1974}, \emph{61},
  3155\relax
\mciteBstWouldAddEndPuncttrue
\mciteSetBstMidEndSepPunct{\mcitedefaultmidpunct}
{\mcitedefaultendpunct}{\mcitedefaultseppunct}\relax
\EndOfBibitem
\bibitem[{Green} and {Thaddeus}(1976){Green}, and {Thaddeus}]{green:76}
{Green},~S.; {Thaddeus},~P. \emph{\apj} \textbf{1976}, \emph{205}, 766\relax
\mciteBstWouldAddEndPuncttrue
\mciteSetBstMidEndSepPunct{\mcitedefaultmidpunct}
{\mcitedefaultendpunct}{\mcitedefaultseppunct}\relax
\EndOfBibitem
\bibitem[{Watson} et~al.(1980){Watson}, {Elitzur}, and {Bieniek}]{Watson:80}
{Watson},~W.~D.; {Elitzur},~M.; {Bieniek},~R.~J. \emph{\apj} \textbf{1980},
  \emph{240}, 547\relax
\mciteBstWouldAddEndPuncttrue
\mciteSetBstMidEndSepPunct{\mcitedefaultmidpunct}
{\mcitedefaultendpunct}{\mcitedefaultseppunct}\relax
\EndOfBibitem
\bibitem[{Chu} and {Dalgarno}(1975){Chu}, and {Dalgarno}]{Chu:75}
{Chu},~S.-I.; {Dalgarno},~A. \emph{Proc. R. Soc. A} \textbf{1975}, \emph{342},
  191\relax
\mciteBstWouldAddEndPuncttrue
\mciteSetBstMidEndSepPunct{\mcitedefaultmidpunct}
{\mcitedefaultendpunct}{\mcitedefaultseppunct}\relax
\EndOfBibitem
\bibitem[{Lepp} et~al.(1995){Lepp}, {Buch}, and {Dalgarno}]{Lepp:95}
{Lepp},~S.; {Buch},~V.; {Dalgarno},~A. \emph{\apjs} \textbf{1995}, \emph{98},
  345\relax
\mciteBstWouldAddEndPuncttrue
\mciteSetBstMidEndSepPunct{\mcitedefaultmidpunct}
{\mcitedefaultendpunct}{\mcitedefaultseppunct}\relax
\EndOfBibitem
\bibitem[{Green} and {Chapman}(1978){Green}, and {Chapman}]{green78}
{Green},~S.; {Chapman},~S. \emph{\apjs} \textbf{1978}, \emph{37}, 169\relax
\mciteBstWouldAddEndPuncttrue
\mciteSetBstMidEndSepPunct{\mcitedefaultmidpunct}
{\mcitedefaultendpunct}{\mcitedefaultseppunct}\relax
\EndOfBibitem
\bibitem[{Bieniek} and {Green}(1983){Bieniek}, and {Green}]{bieniek83}
{Bieniek},~R.~J.; {Green},~S. \emph{\apjl} \textbf{1983}, \emph{265}, L29\relax
\mciteBstWouldAddEndPuncttrue
\mciteSetBstMidEndSepPunct{\mcitedefaultmidpunct}
{\mcitedefaultendpunct}{\mcitedefaultseppunct}\relax
\EndOfBibitem
\bibitem[{Green}(1991)]{green91}
{Green},~S. \emph{\apjs} \textbf{1991}, \emph{76}, 979\relax
\mciteBstWouldAddEndPuncttrue
\mciteSetBstMidEndSepPunct{\mcitedefaultmidpunct}
{\mcitedefaultendpunct}{\mcitedefaultseppunct}\relax
\EndOfBibitem
\bibitem[Green et~al.(1993)Green, Maluendes, and McLean]{green93}
Green,~S.; Maluendes,~S.; McLean,~A.~D. \emph{\apjs} \textbf{1993}, \emph{85},
  181\relax
\mciteBstWouldAddEndPuncttrue
\mciteSetBstMidEndSepPunct{\mcitedefaultmidpunct}
{\mcitedefaultendpunct}{\mcitedefaultseppunct}\relax
\EndOfBibitem
\bibitem[{Green}(1994)]{green94}
{Green},~S. \emph{\apj} \textbf{1994}, \emph{434}, 188\relax
\mciteBstWouldAddEndPuncttrue
\mciteSetBstMidEndSepPunct{\mcitedefaultmidpunct}
{\mcitedefaultendpunct}{\mcitedefaultseppunct}\relax
\EndOfBibitem
\bibitem[{Neufeld} and {Green}(1994){Neufeld}, and {Green}]{Neufeld:94}
{Neufeld},~D.~A.; {Green},~S. \emph{\apj} \textbf{1994}, \emph{432}, 158\relax
\mciteBstWouldAddEndPuncttrue
\mciteSetBstMidEndSepPunct{\mcitedefaultmidpunct}
{\mcitedefaultendpunct}{\mcitedefaultseppunct}\relax
\EndOfBibitem
\bibitem[Dewangan and Flower(1981)Dewangan, and Flower]{flower1:81}
Dewangan,~D.; Flower,~D. \emph{J. Phys. B At. Mol. Opt. Phys.} \textbf{1981},
  \emph{14}, 2179\relax
\mciteBstWouldAddEndPuncttrue
\mciteSetBstMidEndSepPunct{\mcitedefaultmidpunct}
{\mcitedefaultendpunct}{\mcitedefaultseppunct}\relax
\EndOfBibitem
\bibitem[Dewangan and Flower(1983)Dewangan, and Flower]{flower2:83}
Dewangan,~D.; Flower,~D. \emph{J. Phys. B At. Mol. Opt. Phys.} \textbf{1983},
  \emph{16}, 2157\relax
\mciteBstWouldAddEndPuncttrue
\mciteSetBstMidEndSepPunct{\mcitedefaultmidpunct}
{\mcitedefaultendpunct}{\mcitedefaultseppunct}\relax
\EndOfBibitem
\bibitem[{Danby} et~al.(1986){Danby}, {Flower}, {Kochanski}, {Kurdi}, and
  {Valiron}]{danby86}
{Danby},~G.; {Flower},~D.~R.; {Kochanski},~E.; {Kurdi},~L.; {Valiron},~P.
  \emph{J. Phys. B At. Mol. Opt. Phys.} \textbf{1986}, \emph{19}, 2891\relax
\mciteBstWouldAddEndPuncttrue
\mciteSetBstMidEndSepPunct{\mcitedefaultmidpunct}
{\mcitedefaultendpunct}{\mcitedefaultseppunct}\relax
\EndOfBibitem
\bibitem[{Danby} et~al.(1988){Danby}, {Flower}, {Valiron}, {Schilke}, and
  {Walmsley}]{danby88}
{Danby},~G.; {Flower},~D.~R.; {Valiron},~P.; {Schilke},~P.; {Walmsley},~C.~M.
  \emph{\mnras} \textbf{1988}, \emph{235}, 229\relax
\mciteBstWouldAddEndPuncttrue
\mciteSetBstMidEndSepPunct{\mcitedefaultmidpunct}
{\mcitedefaultendpunct}{\mcitedefaultseppunct}\relax
\EndOfBibitem
\bibitem[{Jaquet} et~al.(1992){Jaquet}, {Staemmler}, {Smith}, and
  {Flower}]{jaquet92}
{Jaquet},~R.; {Staemmler},~V.; {Smith},~M.~D.; {Flower},~D.~R. \emph{J. Phys. B
  At. Mol. Opt. Phys.} \textbf{1992}, \emph{25}, 285\relax
\mciteBstWouldAddEndPuncttrue
\mciteSetBstMidEndSepPunct{\mcitedefaultmidpunct}
{\mcitedefaultendpunct}{\mcitedefaultseppunct}\relax
\EndOfBibitem
\bibitem[{Flower} and {Roueff}(1999){Flower}, and {Roueff}]{Flower:99a}
{Flower},~D.~R.; {Roueff},~E. \emph{J. Phys. B} \textbf{1999}, \emph{32},
  3399\relax
\mciteBstWouldAddEndPuncttrue
\mciteSetBstMidEndSepPunct{\mcitedefaultmidpunct}
{\mcitedefaultendpunct}{\mcitedefaultseppunct}\relax
\EndOfBibitem
\bibitem[{Monteiro}(1984)]{monteiro84}
{Monteiro},~T. \emph{\mnras} \textbf{1984}, \emph{210}, 1\relax
\mciteBstWouldAddEndPuncttrue
\mciteSetBstMidEndSepPunct{\mcitedefaultmidpunct}
{\mcitedefaultendpunct}{\mcitedefaultseppunct}\relax
\EndOfBibitem
\bibitem[{Monteiro} and {Stutzki}(1986){Monteiro}, and {Stutzki}]{monteiro86}
{Monteiro},~T.~S.; {Stutzki},~J. \emph{\mnras} \textbf{1986}, \emph{221},
  33P\relax
\mciteBstWouldAddEndPuncttrue
\mciteSetBstMidEndSepPunct{\mcitedefaultmidpunct}
{\mcitedefaultendpunct}{\mcitedefaultseppunct}\relax
\EndOfBibitem
\bibitem[{Balakrishnan} et~al.(1999){Balakrishnan}, {Vieira}, {Babb},
  {Dalgarno}, {Forrey}, and {Lepp}]{Balakrishnan:99}
{Balakrishnan},~N.; {Vieira},~M.; {Babb},~J.~F.; {Dalgarno},~A.;
  {Forrey},~R.~C.; {Lepp},~S. \emph{\apj} \textbf{1999}, \emph{524}, 1122\relax
\mciteBstWouldAddEndPuncttrue
\mciteSetBstMidEndSepPunct{\mcitedefaultmidpunct}
{\mcitedefaultendpunct}{\mcitedefaultseppunct}\relax
\EndOfBibitem
\bibitem[{Qu{\'e}m{\'e}ner} and {Balakrishnan}(2009){Qu{\'e}m{\'e}ner}, and
  {Balakrishnan}]{Quemener:09}
{Qu{\'e}m{\'e}ner},~G.; {Balakrishnan},~N. \emph{\jcp} \textbf{2009},
  \emph{130}, 114303\relax
\mciteBstWouldAddEndPuncttrue
\mciteSetBstMidEndSepPunct{\mcitedefaultmidpunct}
{\mcitedefaultendpunct}{\mcitedefaultseppunct}\relax
\EndOfBibitem
\bibitem[{Lique} et~al.(2012){Lique}, {Honvault}, and {Faure}]{lique:12:H2}
{Lique},~F.; {Honvault},~P.; {Faure},~A. \emph{\jcp} \textbf{2012}, \emph{137},
  154303\relax
\mciteBstWouldAddEndPuncttrue
\mciteSetBstMidEndSepPunct{\mcitedefaultmidpunct}
{\mcitedefaultendpunct}{\mcitedefaultseppunct}\relax
\EndOfBibitem
\bibitem[{Roueff} and {Zeippen}(2000){Roueff}, and {Zeippen}]{Roueff:00}
{Roueff},~E.; {Zeippen},~C.~J. \emph{\aaps} \textbf{2000}, \emph{142},
  475\relax
\mciteBstWouldAddEndPuncttrue
\mciteSetBstMidEndSepPunct{\mcitedefaultmidpunct}
{\mcitedefaultendpunct}{\mcitedefaultseppunct}\relax
\EndOfBibitem
\bibitem[{Flower} and {Roueff}(1999){Flower}, and {Roueff}]{Flower:99c}
{Flower},~D.~R.; {Roueff},~E. \emph{\mnras} \textbf{1999}, \emph{309},
  833\relax
\mciteBstWouldAddEndPuncttrue
\mciteSetBstMidEndSepPunct{\mcitedefaultmidpunct}
{\mcitedefaultendpunct}{\mcitedefaultseppunct}\relax
\EndOfBibitem
\bibitem[{Lique} and {Faure}(2012){Lique}, and {Faure}]{lique:12D2}
{Lique},~F.; {Faure},~A. \emph{\jcp} \textbf{2012}, \emph{136}, 031101\relax
\mciteBstWouldAddEndPuncttrue
\mciteSetBstMidEndSepPunct{\mcitedefaultmidpunct}
{\mcitedefaultendpunct}{\mcitedefaultseppunct}\relax
\EndOfBibitem
\bibitem[Najar et~al.(2008)Najar, Abdallah, Jaidane, and Lakhdar]{najar:08}
Najar,~F.; Abdallah,~D.~B.; Jaidane,~N.; Lakhdar,~Z.~B. \emph{Chem. Phys.
  Lett.} \textbf{2008}, \emph{460}, 31\relax
\mciteBstWouldAddEndPuncttrue
\mciteSetBstMidEndSepPunct{\mcitedefaultmidpunct}
{\mcitedefaultendpunct}{\mcitedefaultseppunct}\relax
\EndOfBibitem
\bibitem[{Najar} et~al.(2009){Najar}, {Ben Abdallah}, {Jaidane}, {Ben Lakhdar},
  {Chambaud}, and {Hochlaf}]{najar:09}
{Najar},~F.; {Ben Abdallah},~D.; {Jaidane},~N.; {Ben Lakhdar},~Z.;
  {Chambaud},~G.; {Hochlaf},~M. \emph{\jcp} \textbf{2009}, \emph{130},
  204305\relax
\mciteBstWouldAddEndPuncttrue
\mciteSetBstMidEndSepPunct{\mcitedefaultmidpunct}
{\mcitedefaultendpunct}{\mcitedefaultseppunct}\relax
\EndOfBibitem
\bibitem[{Cecchi-Pestellini} et~al.(2002){Cecchi-Pestellini}, {Bodo},
  {Balakrishnan}, and {Dalgarno}]{Cecchi:02}
{Cecchi-Pestellini},~C.; {Bodo},~E.; {Balakrishnan},~N.; {Dalgarno},~A.
  \emph{\apj} \textbf{2002}, \emph{571}, 1015\relax
\mciteBstWouldAddEndPuncttrue
\mciteSetBstMidEndSepPunct{\mcitedefaultmidpunct}
{\mcitedefaultendpunct}{\mcitedefaultseppunct}\relax
\EndOfBibitem
\bibitem[{Flower}(2012)]{Flower:12CO}
{Flower},~D.~R. \emph{\mnras} \textbf{2012}, \emph{425}, 1350\relax
\mciteBstWouldAddEndPuncttrue
\mciteSetBstMidEndSepPunct{\mcitedefaultmidpunct}
{\mcitedefaultendpunct}{\mcitedefaultseppunct}\relax
\EndOfBibitem
\bibitem[{Shepler} et~al.(2007){Shepler}, {Yang}, {Dhilip Kumar}, {Stancil},
  {Bowman}, {Balakrishnan}, {Zhang}, {Bodo}, and {Dalgarno}]{Shepler:07}
{Shepler},~B.~C.; {Yang},~B.~H.; {Dhilip Kumar},~T.~J.; {Stancil},~P.~C.;
  {Bowman},~J.~M.; {Balakrishnan},~N.; {Zhang},~P.; {Bodo},~E.; {Dalgarno},~A.
  \emph{\aap} \textbf{2007}, \emph{475}, L15\relax
\mciteBstWouldAddEndPuncttrue
\mciteSetBstMidEndSepPunct{\mcitedefaultmidpunct}
{\mcitedefaultendpunct}{\mcitedefaultseppunct}\relax
\EndOfBibitem
\bibitem[{Lique} et~al.(2006){Lique}, {Spielfiedel}, and
  {Cernicharo}]{Lique:06CS}
{Lique},~F.; {Spielfiedel},~A.; {Cernicharo},~J. \emph{\aap} \textbf{2006},
  \emph{451}, 1125\relax
\mciteBstWouldAddEndPuncttrue
\mciteSetBstMidEndSepPunct{\mcitedefaultmidpunct}
{\mcitedefaultendpunct}{\mcitedefaultseppunct}\relax
\EndOfBibitem
\bibitem[Lique and Spielfiedel(2007)Lique, and Spielfiedel]{Lique:07}
Lique,~F.; Spielfiedel,~A. \emph{\aap} \textbf{2007}, \emph{462}, 1179\relax
\mciteBstWouldAddEndPuncttrue
\mciteSetBstMidEndSepPunct{\mcitedefaultmidpunct}
{\mcitedefaultendpunct}{\mcitedefaultseppunct}\relax
\EndOfBibitem
\bibitem[Dumouchel et~al.(2010)Dumouchel, Faure, and Lique]{Dumouchel:10}
Dumouchel,~F.; Faure,~A.; Lique,~F. \emph{\mnras} \textbf{2010}, \emph{406},
  2488\relax
\mciteBstWouldAddEndPuncttrue
\mciteSetBstMidEndSepPunct{\mcitedefaultmidpunct}
{\mcitedefaultendpunct}{\mcitedefaultseppunct}\relax
\EndOfBibitem
\bibitem[{Ben Abdallah} et~al.(2012){Ben Abdallah}, {Najar}, {Jaidane},
  {Dumouchel}, and {Lique}]{benabdallah:12}
{Ben Abdallah},~D.; {Najar},~F.; {Jaidane},~N.; {Dumouchel},~F.; {Lique},~F.
  \emph{\mnras} \textbf{2012}, \emph{419}, 2441\relax
\mciteBstWouldAddEndPuncttrue
\mciteSetBstMidEndSepPunct{\mcitedefaultmidpunct}
{\mcitedefaultendpunct}{\mcitedefaultseppunct}\relax
\EndOfBibitem
\bibitem[{Dumouchel} et~al.(2011){Dumouchel}, {K{\l}os}, and
  {Lique}]{Dumouchel:11}
{Dumouchel},~F.; {K{\l}os},~J.; {Lique},~F. \emph{Phys. Chem. Chem. Phys. (Inc.
  Faraday Trans.)} \textbf{2011}, \emph{13}, 8204\relax
\mciteBstWouldAddEndPuncttrue
\mciteSetBstMidEndSepPunct{\mcitedefaultmidpunct}
{\mcitedefaultendpunct}{\mcitedefaultseppunct}\relax
\EndOfBibitem
\bibitem[{Lanza} and {Lique}(2012){Lanza}, and {Lique}]{Lanza:12}
{Lanza},~M.; {Lique},~F. \emph{\mnras} \textbf{2012}, \emph{424}, 1261\relax
\mciteBstWouldAddEndPuncttrue
\mciteSetBstMidEndSepPunct{\mcitedefaultmidpunct}
{\mcitedefaultendpunct}{\mcitedefaultseppunct}\relax
\EndOfBibitem
\bibitem[{Reese} et~al.(2005){Reese}, {Stoecklin}, {Voronin}, and
  {Rayez}]{Reese:05}
{Reese},~C.; {Stoecklin},~T.; {Voronin},~A.; {Rayez},~J.~C. \emph{\aap}
  \textbf{2005}, \emph{430}, 1139\relax
\mciteBstWouldAddEndPuncttrue
\mciteSetBstMidEndSepPunct{\mcitedefaultmidpunct}
{\mcitedefaultendpunct}{\mcitedefaultseppunct}\relax
\EndOfBibitem
\bibitem[{Guillon} and {Stoecklin}(2012){Guillon}, and {Stoecklin}]{Guillon:12}
{Guillon},~G.; {Stoecklin},~T. \emph{\mnras} \textbf{2012}, \emph{420},
  579\relax
\mciteBstWouldAddEndPuncttrue
\mciteSetBstMidEndSepPunct{\mcitedefaultmidpunct}
{\mcitedefaultendpunct}{\mcitedefaultseppunct}\relax
\EndOfBibitem
\bibitem[{Gotoum} et~al.(2012){Gotoum}, {Hammami}, {Owono Owono}, and
  {Jaidane}]{gotoum:12}
{Gotoum},~N.; {Hammami},~K.; {Owono Owono},~L.~C.; {Jaidane},~N.-E.
  \emph{Astrophys. Sp. Science} \textbf{2012}, \emph{337}, 553\relax
\mciteBstWouldAddEndPuncttrue
\mciteSetBstMidEndSepPunct{\mcitedefaultmidpunct}
{\mcitedefaultendpunct}{\mcitedefaultseppunct}\relax
\EndOfBibitem
\bibitem[{Tobo{\l}a} et~al.(2007){Tobo{\l}a}, {K{\l}os}, {Lique},
  {Cha{\l}asi{\'n}ski}, and {Alexander}]{Tobola:07}
{Tobo{\l}a},~R.; {K{\l}os},~J.; {Lique},~F.; {Cha{\l}asi{\'n}ski},~G.;
  {Alexander},~M.~H. \emph{\aap} \textbf{2007}, \emph{468}, 1123\relax
\mciteBstWouldAddEndPuncttrue
\mciteSetBstMidEndSepPunct{\mcitedefaultmidpunct}
{\mcitedefaultendpunct}{\mcitedefaultseppunct}\relax
\EndOfBibitem
\bibitem[{Wernli} et~al.(2007){Wernli}, {Wiesenfeld}, {Faure}, and
  {Valiron}]{Wernli:07}
{Wernli},~M.; {Wiesenfeld},~L.; {Faure},~A.; {Valiron},~P. \emph{\aap}
  \textbf{2007}, \emph{464}, 1147\relax
\mciteBstWouldAddEndPuncttrue
\mciteSetBstMidEndSepPunct{\mcitedefaultmidpunct}
{\mcitedefaultendpunct}{\mcitedefaultseppunct}\relax
\EndOfBibitem
\bibitem[{Wernli} et~al.(2007){Wernli}, {Wiesenfeld}, {Faure}, and
  {Valiron}]{Wernli:07a}
{Wernli},~M.; {Wiesenfeld},~L.; {Faure},~A.; {Valiron},~P. \emph{\aap}
  \textbf{2007}, \emph{475}, 391\relax
\mciteBstWouldAddEndPuncttrue
\mciteSetBstMidEndSepPunct{\mcitedefaultmidpunct}
{\mcitedefaultendpunct}{\mcitedefaultseppunct}\relax
\EndOfBibitem
\bibitem[Hammami et~al.(2008)Hammami, Owono, Jaidane, and Lakhdar]{Hammami2008}
Hammami,~K.; Owono,~L.~O.; Jaidane,~N.; Lakhdar,~Z.~B. \emph{J. Mol. Struct.:
  THEOCHEM} \textbf{2008}, \emph{860}, 45\relax
\mciteBstWouldAddEndPuncttrue
\mciteSetBstMidEndSepPunct{\mcitedefaultmidpunct}
{\mcitedefaultendpunct}{\mcitedefaultseppunct}\relax
\EndOfBibitem
\bibitem[{Hammami} et~al.(2008){Hammami}, {Nkem}, {Owono Owono}, {Jaidane}, and
  {Ben Lakhdar}]{Hammami:08}
{Hammami},~K.; {Nkem},~C.; {Owono Owono},~L.~C.; {Jaidane},~N.; {Ben
  Lakhdar},~Z. \emph{\jcp} \textbf{2008}, \emph{129}, 204305\relax
\mciteBstWouldAddEndPuncttrue
\mciteSetBstMidEndSepPunct{\mcitedefaultmidpunct}
{\mcitedefaultendpunct}{\mcitedefaultseppunct}\relax
\EndOfBibitem
\bibitem[{Vincent} et~al.(2007){Vincent}, {Spielfiedel}, and
  {Lique}]{Vincent:07}
{Vincent},~L.~F.~M.; {Spielfiedel},~A.; {Lique},~F. \emph{\aap} \textbf{2007},
  \emph{472}, 1037\relax
\mciteBstWouldAddEndPuncttrue
\mciteSetBstMidEndSepPunct{\mcitedefaultmidpunct}
{\mcitedefaultendpunct}{\mcitedefaultseppunct}\relax
\EndOfBibitem
\bibitem[{Tobo{\l}a} et~al.(2008){Tobo{\l}a}, {Lique}, {K{\l}os}, and
  {Cha{\l}asi{\'n}ski}]{Tobola:08}
{Tobo{\l}a},~R.; {Lique},~F.; {K{\l}os},~J.; {Cha{\l}asi{\'n}ski},~G. \emph{J.
  Phys. B At. Mol. Opt. Phys.} \textbf{2008}, \emph{41}, 155702\relax
\mciteBstWouldAddEndPuncttrue
\mciteSetBstMidEndSepPunct{\mcitedefaultmidpunct}
{\mcitedefaultendpunct}{\mcitedefaultseppunct}\relax
\EndOfBibitem
\bibitem[{K{\l}os} and {Lique}(2008){K{\l}os}, and {Lique}]{klos08SiS}
{K{\l}os},~J.; {Lique},~F. \emph{\mnras} \textbf{2008}, \emph{390}, 239\relax
\mciteBstWouldAddEndPuncttrue
\mciteSetBstMidEndSepPunct{\mcitedefaultmidpunct}
{\mcitedefaultendpunct}{\mcitedefaultseppunct}\relax
\EndOfBibitem
\bibitem[{Ben Abdallah} et~al.(2008){Ben Abdallah}, {Hammami}, {Najar},
  {Jaidane}, {Ben Lakhdar}, {Senent}, {Chambaud}, and
  {Hochlaf}]{Benabdallah:08}
{Ben Abdallah},~D.; {Hammami},~K.; {Najar},~F.; {Jaidane},~N.; {Ben
  Lakhdar},~Z.; {Senent},~M.~L.; {Chambaud},~G.; {Hochlaf},~M. \emph{\apj}
  \textbf{2008}, \emph{686}, 379\relax
\mciteBstWouldAddEndPuncttrue
\mciteSetBstMidEndSepPunct{\mcitedefaultmidpunct}
{\mcitedefaultendpunct}{\mcitedefaultseppunct}\relax
\EndOfBibitem
\bibitem[{Turpin} et~al.(2010){Turpin}, {Stoecklin}, and {Voronin}]{Turpin:10}
{Turpin},~F.; {Stoecklin},~T.; {Voronin},~A. \emph{\aap} \textbf{2010},
  \emph{511}, A28\relax
\mciteBstWouldAddEndPuncttrue
\mciteSetBstMidEndSepPunct{\mcitedefaultmidpunct}
{\mcitedefaultendpunct}{\mcitedefaultseppunct}\relax
\EndOfBibitem
\bibitem[Nkem et~al.(2009)Nkem, Hammami, Manga, Owono, Jaidane, and
  Lakhdar]{Nkem:09}
Nkem,~C.; Hammami,~K.; Manga,~A.; Owono,~L.~O.; Jaidane,~N.; Lakhdar,~Z.~B.
  \emph{J. Mol. Struct.: THEOCHEM} \textbf{2009}, \emph{901}, 220\relax
\mciteBstWouldAddEndPuncttrue
\mciteSetBstMidEndSepPunct{\mcitedefaultmidpunct}
{\mcitedefaultendpunct}{\mcitedefaultseppunct}\relax
\EndOfBibitem
\bibitem[{Buffa} et~al.(2009){Buffa}, {Dore}, and {Meuwly}]{Buffa:09}
{Buffa},~G.; {Dore},~L.; {Meuwly},~M. \emph{\mnras} \textbf{2009}, \emph{397},
  1909\relax
\mciteBstWouldAddEndPuncttrue
\mciteSetBstMidEndSepPunct{\mcitedefaultmidpunct}
{\mcitedefaultendpunct}{\mcitedefaultseppunct}\relax
\EndOfBibitem
\bibitem[{Flower}(1999)]{Flower:HCOp}
{Flower},~D.~R. \emph{\mnras} \textbf{1999}, \emph{305}, 651\relax
\mciteBstWouldAddEndPuncttrue
\mciteSetBstMidEndSepPunct{\mcitedefaultmidpunct}
{\mcitedefaultendpunct}{\mcitedefaultseppunct}\relax
\EndOfBibitem
\bibitem[{Buffa}(2012)]{Buffa:12}
{Buffa},~G. \emph{\mnras} \textbf{2012}, \emph{421}, 719\relax
\mciteBstWouldAddEndPuncttrue
\mciteSetBstMidEndSepPunct{\mcitedefaultmidpunct}
{\mcitedefaultendpunct}{\mcitedefaultseppunct}\relax
\EndOfBibitem
\bibitem[{Pagani} et~al.(2012){Pagani}, {Bourgoin}, and {Lique}]{Pagani:12}
{Pagani},~L.; {Bourgoin},~A.; {Lique},~F. \emph{\aap} \textbf{2012},
  \emph{548}, L4\relax
\mciteBstWouldAddEndPuncttrue
\mciteSetBstMidEndSepPunct{\mcitedefaultmidpunct}
{\mcitedefaultendpunct}{\mcitedefaultseppunct}\relax
\EndOfBibitem
\bibitem[{Andersson} et~al.(2008){Andersson}, {Barinovs}, and
  {Nyman}]{andersson:08}
{Andersson},~S.; {Barinovs},~{\c G}.; {Nyman},~G. \emph{\apj} \textbf{2008},
  \emph{678}, 1042\relax
\mciteBstWouldAddEndPuncttrue
\mciteSetBstMidEndSepPunct{\mcitedefaultmidpunct}
{\mcitedefaultendpunct}{\mcitedefaultseppunct}\relax
\EndOfBibitem
\bibitem[{K{\l}os} and {Lique}(2011){K{\l}os}, and {Lique}]{Klos:11}
{K{\l}os},~J.; {Lique},~F. \emph{\mnras} \textbf{2011}, \emph{418}, 271\relax
\mciteBstWouldAddEndPuncttrue
\mciteSetBstMidEndSepPunct{\mcitedefaultmidpunct}
{\mcitedefaultendpunct}{\mcitedefaultseppunct}\relax
\EndOfBibitem
\bibitem[{Dumouchel} et~al.(2012){Dumouchel}, {Spielfiedel}, {Senent}, and
  {Feautrier}]{dumouchel:12}
{Dumouchel},~F.; {Spielfiedel},~A.; {Senent},~M.~L.; {Feautrier},~N.
  \emph{Chem. Phys. Lett.} \textbf{2012}, \emph{533}, 6\relax
\mciteBstWouldAddEndPuncttrue
\mciteSetBstMidEndSepPunct{\mcitedefaultmidpunct}
{\mcitedefaultendpunct}{\mcitedefaultseppunct}\relax
\EndOfBibitem
\bibitem[{Lique} et~al.(2010){Lique}, {Spielfiedel}, {Feautrier}, {Schneider},
  {K{\l}os}, and {Alexander}]{lique:10CN}
{Lique},~F.; {Spielfiedel},~A.; {Feautrier},~N.; {Schneider},~I.~F.;
  {K{\l}os},~J.; {Alexander},~M.~H. \emph{\jcp} \textbf{2010}, \emph{132},
  024303\relax
\mciteBstWouldAddEndPuncttrue
\mciteSetBstMidEndSepPunct{\mcitedefaultmidpunct}
{\mcitedefaultendpunct}{\mcitedefaultseppunct}\relax
\EndOfBibitem
\bibitem[{Lique} and {K{\l}os}(2011){Lique}, and {K{\l}os}]{lique:11CN}
{Lique},~F.; {K{\l}os},~J. \emph{\mnras} \textbf{2011}, \emph{413}, L20\relax
\mciteBstWouldAddEndPuncttrue
\mciteSetBstMidEndSepPunct{\mcitedefaultmidpunct}
{\mcitedefaultendpunct}{\mcitedefaultseppunct}\relax
\EndOfBibitem
\bibitem[{Kalugina} et~al.(2012){Kalugina}, {Lique}, and {K{\l}os}]{kalugina12}
{Kalugina},~Y.; {Lique},~F.; {K{\l}os},~J. \emph{\mnras} \textbf{2012},
  \emph{422}, 812\relax
\mciteBstWouldAddEndPuncttrue
\mciteSetBstMidEndSepPunct{\mcitedefaultmidpunct}
{\mcitedefaultendpunct}{\mcitedefaultseppunct}\relax
\EndOfBibitem
\bibitem[{Spielfiedel} et~al.(2012){Spielfiedel}, {Feautrier}, {Najar}, {Ben
  Abdallah}, {Dayou}, {Senent}, and {Lique}]{Spielfiedel:12}
{Spielfiedel},~A.; {Feautrier},~N.; {Najar},~F.; {Ben Abdallah},~D.;
  {Dayou},~F.; {Senent},~M.~L.; {Lique},~F. \emph{\mnras} \textbf{2012},
  \emph{421}, 1891\relax
\mciteBstWouldAddEndPuncttrue
\mciteSetBstMidEndSepPunct{\mcitedefaultmidpunct}
{\mcitedefaultendpunct}{\mcitedefaultseppunct}\relax
\EndOfBibitem
\bibitem[{Lique}(2010)]{lique:10}
{Lique},~F. \emph{\jcp} \textbf{2010}, \emph{132}, 044311\relax
\mciteBstWouldAddEndPuncttrue
\mciteSetBstMidEndSepPunct{\mcitedefaultmidpunct}
{\mcitedefaultendpunct}{\mcitedefaultseppunct}\relax
\EndOfBibitem
\bibitem[Kalugina et~al.(2012)Kalugina, Denis~Alpizar, Stoecklin, and
  Lique]{Kalugina:12O2}
Kalugina,~Y.; Denis~Alpizar,~O.; Stoecklin,~T.; Lique,~F. \emph{Phys. Chem.
  Chem. Phys. (Inc. Faraday Trans.)} \textbf{2012}, \emph{14}, 16458\relax
\mciteBstWouldAddEndPuncttrue
\mciteSetBstMidEndSepPunct{\mcitedefaultmidpunct}
{\mcitedefaultendpunct}{\mcitedefaultseppunct}\relax
\EndOfBibitem
\bibitem[Lique et~al.(2005)Lique, Spielfiedel, Dubernet, and
  Feautrier]{lique:05}
Lique,~F.; Spielfiedel,~A.; Dubernet,~M.~L.; Feautrier,~N. \emph{\jcp}
  \textbf{2005}, \emph{123}, 134316\relax
\mciteBstWouldAddEndPuncttrue
\mciteSetBstMidEndSepPunct{\mcitedefaultmidpunct}
{\mcitedefaultendpunct}{\mcitedefaultseppunct}\relax
\EndOfBibitem
\bibitem[{Lique} et~al.(2006){Lique}, {Dubernet}, {Spielfiedel}, and
  {Feautrier}]{Lique:06SO}
{Lique},~F.; {Dubernet},~M.-L.; {Spielfiedel},~A.; {Feautrier},~N. \emph{\aap}
  \textbf{2006}, \emph{450}, 399\relax
\mciteBstWouldAddEndPuncttrue
\mciteSetBstMidEndSepPunct{\mcitedefaultmidpunct}
{\mcitedefaultendpunct}{\mcitedefaultseppunct}\relax
\EndOfBibitem
\bibitem[{Lique} et~al.(2006){Lique}, {Spielfiedel}, {Dhont}, and
  {Feautrier}]{Lique:06SO_2}
{Lique},~F.; {Spielfiedel},~A.; {Dhont},~G.; {Feautrier},~N. \emph{\aap}
  \textbf{2006}, \emph{458}, 331\relax
\mciteBstWouldAddEndPuncttrue
\mciteSetBstMidEndSepPunct{\mcitedefaultmidpunct}
{\mcitedefaultendpunct}{\mcitedefaultseppunct}\relax
\EndOfBibitem
\bibitem[{Lique} et~al.(2007){Lique}, {Senent}, {Spielfiedel}, and
  {Feautrier}]{lique:07b}
{Lique},~F.; {Senent},~M.; {Spielfiedel},~A.; {Feautrier},~N. \emph{\jcp}
  \textbf{2007}, \emph{126}, 164312\relax
\mciteBstWouldAddEndPuncttrue
\mciteSetBstMidEndSepPunct{\mcitedefaultmidpunct}
{\mcitedefaultendpunct}{\mcitedefaultseppunct}\relax
\EndOfBibitem
\bibitem[{Tobo{\l}a} et~al.(2011){Tobo{\l}a}, {Dumouchel}, {K{\l}os}, and
  {Lique}]{tobola:11}
{Tobo{\l}a},~R.; {Dumouchel},~F.; {K{\l}os},~J.; {Lique},~F. \emph{\jcp}
  \textbf{2011}, \emph{134}, 024305\relax
\mciteBstWouldAddEndPuncttrue
\mciteSetBstMidEndSepPunct{\mcitedefaultmidpunct}
{\mcitedefaultendpunct}{\mcitedefaultseppunct}\relax
\EndOfBibitem
\bibitem[{Lique} et~al.(2010){Lique}, {K{\l}os}, and {Hochlaf}]{Lique:10C4}
{Lique},~F.; {K{\l}os},~J.; {Hochlaf},~M. \emph{Phys. Chem. Chem. Phys. (Inc.
  Faraday Trans.)} \textbf{2010}, \emph{12}, 15672\relax
\mciteBstWouldAddEndPuncttrue
\mciteSetBstMidEndSepPunct{\mcitedefaultmidpunct}
{\mcitedefaultendpunct}{\mcitedefaultseppunct}\relax
\EndOfBibitem
\bibitem[K{\l}os et~al.(2007)K{\l}os, Lique, and Alexander]{klos:07OH}
K{\l}os,~J.; Lique,~F.; Alexander,~M.~H. \emph{Chem. Phys. Lett.}
  \textbf{2007}, \emph{445}, 12\relax
\mciteBstWouldAddEndPuncttrue
\mciteSetBstMidEndSepPunct{\mcitedefaultmidpunct}
{\mcitedefaultendpunct}{\mcitedefaultseppunct}\relax
\EndOfBibitem
\bibitem[{Offer} et~al.(1994){Offer}, {van Hemert}, and {van
  Dishoeck}]{offer:94}
{Offer},~A.~R.; {van Hemert},~M.~C.; {van Dishoeck},~E.~F. \emph{\jcp}
  \textbf{1994}, \emph{100}, 362\relax
\mciteBstWouldAddEndPuncttrue
\mciteSetBstMidEndSepPunct{\mcitedefaultmidpunct}
{\mcitedefaultendpunct}{\mcitedefaultseppunct}\relax
\EndOfBibitem
\bibitem[{K{\l}os} et~al.(2009){K{\l}os}, {Lique}, and {Alexander}]{klos:09}
{K{\l}os},~J.; {Lique},~F.; {Alexander},~M.~H. \emph{Chem. Phys. Lett.}
  \textbf{2009}, \emph{476}, 135\relax
\mciteBstWouldAddEndPuncttrue
\mciteSetBstMidEndSepPunct{\mcitedefaultmidpunct}
{\mcitedefaultendpunct}{\mcitedefaultseppunct}\relax
\EndOfBibitem
\bibitem[K{\l}os et~al.(2008)K{\l}os, Lique, and Alexander]{klos:08NO}
K{\l}os,~J.; Lique,~F.; Alexander,~M.~H. \emph{Chem. Phys. Lett.}
  \textbf{2008}, \emph{455}, 1\relax
\mciteBstWouldAddEndPuncttrue
\mciteSetBstMidEndSepPunct{\mcitedefaultmidpunct}
{\mcitedefaultendpunct}{\mcitedefaultseppunct}\relax
\EndOfBibitem
\bibitem[Lique et~al.(2009)Lique, van~der Tak, K{\l}os, Bulthuis, and
  Alexander]{lique:09}
Lique,~F.; van~der Tak,~F. F.~S.; K{\l}os,~J.; Bulthuis,~J.; Alexander,~M.~H.
  \emph{\aap} \textbf{2009}, \emph{493}, 557\relax
\mciteBstWouldAddEndPuncttrue
\mciteSetBstMidEndSepPunct{\mcitedefaultmidpunct}
{\mcitedefaultendpunct}{\mcitedefaultseppunct}\relax
\EndOfBibitem
\bibitem[{Yang} and {Stancil}(2007){Yang}, and {Stancil}]{yang:07}
{Yang},~B.; {Stancil},~P.~C. \emph{\jcp} \textbf{2007}, \emph{126},
  154306\relax
\mciteBstWouldAddEndPuncttrue
\mciteSetBstMidEndSepPunct{\mcitedefaultmidpunct}
{\mcitedefaultendpunct}{\mcitedefaultseppunct}\relax
\EndOfBibitem
\bibitem[{Yang} et~al.(2013){Yang}, {Nagao}, {Satomi}, {Kimura}, and
  {Stancil}]{yang:13h2o}
{Yang},~B.; {Nagao},~M.; {Satomi},~W.; {Kimura},~M.; {Stancil},~P.~C.
  \emph{\apj} \textbf{2013}, \emph{765}, 77\relax
\mciteBstWouldAddEndPuncttrue
\mciteSetBstMidEndSepPunct{\mcitedefaultmidpunct}
{\mcitedefaultendpunct}{\mcitedefaultseppunct}\relax
\EndOfBibitem
\bibitem[{Dubernet} et~al.(2009){Dubernet}, {Daniel}, {Grosjean}, and
  {Lin}]{Dubernet:09}
{Dubernet},~M.-L.; {Daniel},~F.; {Grosjean},~A.; {Lin},~C.~Y. \emph{\aap}
  \textbf{2009}, \emph{497}, 911\relax
\mciteBstWouldAddEndPuncttrue
\mciteSetBstMidEndSepPunct{\mcitedefaultmidpunct}
{\mcitedefaultendpunct}{\mcitedefaultseppunct}\relax
\EndOfBibitem
\bibitem[{Daniel} et~al.(2010){Daniel}, {Dubernet}, {Pacaud}, and
  {Grosjean}]{Daniel:10}
{Daniel},~F.; {Dubernet},~M.-L.; {Pacaud},~F.; {Grosjean},~A. \emph{\aap}
  \textbf{2010}, \emph{517}, A13\relax
\mciteBstWouldAddEndPuncttrue
\mciteSetBstMidEndSepPunct{\mcitedefaultmidpunct}
{\mcitedefaultendpunct}{\mcitedefaultseppunct}\relax
\EndOfBibitem
\bibitem[{Daniel} et~al.(2011){Daniel}, {Dubernet}, and {Grosjean}]{Daniel:11}
{Daniel},~F.; {Dubernet},~M.-L.; {Grosjean},~A. \emph{\aap} \textbf{2011},
  \emph{536}, A76\relax
\mciteBstWouldAddEndPuncttrue
\mciteSetBstMidEndSepPunct{\mcitedefaultmidpunct}
{\mcitedefaultendpunct}{\mcitedefaultseppunct}\relax
\EndOfBibitem
\bibitem[{Faure} et~al.(2012){Faure}, {Wiesenfeld}, {Scribano}, and
  {Ceccarelli}]{Faure:12}
{Faure},~A.; {Wiesenfeld},~L.; {Scribano},~Y.; {Ceccarelli},~C. \emph{\mnras}
  \textbf{2012}, \emph{420}, 699\relax
\mciteBstWouldAddEndPuncttrue
\mciteSetBstMidEndSepPunct{\mcitedefaultmidpunct}
{\mcitedefaultendpunct}{\mcitedefaultseppunct}\relax
\EndOfBibitem
\bibitem[{Troscompt} et~al.(2009){Troscompt}, {Faure}, {Wiesenfeld},
  {Ceccarelli}, and {Valiron}]{Troscompt:09}
{Troscompt},~N.; {Faure},~A.; {Wiesenfeld},~L.; {Ceccarelli},~C.; {Valiron},~P.
  \emph{\aap} \textbf{2009}, \emph{493}, 687\relax
\mciteBstWouldAddEndPuncttrue
\mciteSetBstMidEndSepPunct{\mcitedefaultmidpunct}
{\mcitedefaultendpunct}{\mcitedefaultseppunct}\relax
\EndOfBibitem
\bibitem[{Machin} and {Roueff}(2005){Machin}, and {Roueff}]{machin:05}
{Machin},~L.; {Roueff},~E. \emph{J. Phys. B At. Mol. Opt. Phys.} \textbf{2005},
  \emph{38}, 1519\relax
\mciteBstWouldAddEndPuncttrue
\mciteSetBstMidEndSepPunct{\mcitedefaultmidpunct}
{\mcitedefaultendpunct}{\mcitedefaultseppunct}\relax
\EndOfBibitem
\bibitem[{Yang} and {Stancil}(2008){Yang}, and {Stancil}]{yang:08}
{Yang},~B.~H.; {Stancil},~P.~C. \emph{Eur. Phys. J. D} \textbf{2008},
  \emph{47}, 351\relax
\mciteBstWouldAddEndPuncttrue
\mciteSetBstMidEndSepPunct{\mcitedefaultmidpunct}
{\mcitedefaultendpunct}{\mcitedefaultseppunct}\relax
\EndOfBibitem
\bibitem[{Maret} et~al.(2009){Maret}, {Faure}, {Scifoni}, and
  {Wiesenfeld}]{maret09}
{Maret},~S.; {Faure},~A.; {Scifoni},~E.; {Wiesenfeld},~L. \emph{\mnras}
  \textbf{2009}, \emph{399}, 425\relax
\mciteBstWouldAddEndPuncttrue
\mciteSetBstMidEndSepPunct{\mcitedefaultmidpunct}
{\mcitedefaultendpunct}{\mcitedefaultseppunct}\relax
\EndOfBibitem
\bibitem[{Machin} and {Roueff}(2006){Machin}, and {Roueff}]{Machin:06}
{Machin},~L.; {Roueff},~E. \emph{\aap} \textbf{2006}, \emph{460}, 953\relax
\mciteBstWouldAddEndPuncttrue
\mciteSetBstMidEndSepPunct{\mcitedefaultmidpunct}
{\mcitedefaultendpunct}{\mcitedefaultseppunct}\relax
\EndOfBibitem
\bibitem[{Machin} and {Roueff}(2007){Machin}, and {Roueff}]{Machin:07}
{Machin},~L.; {Roueff},~E. \emph{\aap} \textbf{2007}, \emph{465}, 647\relax
\mciteBstWouldAddEndPuncttrue
\mciteSetBstMidEndSepPunct{\mcitedefaultmidpunct}
{\mcitedefaultendpunct}{\mcitedefaultseppunct}\relax
\EndOfBibitem
\bibitem[{Wiesenfeld} et~al.(2011){Wiesenfeld}, {Scifoni}, {Faure}, and
  {Roueff}]{wiesenfeld11}
{Wiesenfeld},~L.; {Scifoni},~E.; {Faure},~A.; {Roueff},~E. \emph{\mnras}
  \textbf{2011}, \emph{413}, 509\relax
\mciteBstWouldAddEndPuncttrue
\mciteSetBstMidEndSepPunct{\mcitedefaultmidpunct}
{\mcitedefaultendpunct}{\mcitedefaultseppunct}\relax
\EndOfBibitem
\bibitem[{Chandra} and {Kegel}(2000){Chandra}, and {Kegel}]{Chandra:00}
{Chandra},~S.; {Kegel},~W.~H. \emph{\aaps} \textbf{2000}, \emph{142}, 113\relax
\mciteBstWouldAddEndPuncttrue
\mciteSetBstMidEndSepPunct{\mcitedefaultmidpunct}
{\mcitedefaultendpunct}{\mcitedefaultseppunct}\relax
\EndOfBibitem
\bibitem[{Rabli} and {Flower}(2010){Rabli}, and {Flower}]{Rabli:10a}
{Rabli},~D.; {Flower},~D.~R. \emph{\mnras} \textbf{2010}, \emph{403},
  2033\relax
\mciteBstWouldAddEndPuncttrue
\mciteSetBstMidEndSepPunct{\mcitedefaultmidpunct}
{\mcitedefaultendpunct}{\mcitedefaultseppunct}\relax
\EndOfBibitem
\bibitem[{Rabli} and {Flower}(2011){Rabli}, and {Flower}]{Rabli:11}
{Rabli},~D.; {Flower},~D.~R. \emph{\mnras} \textbf{2011}, \emph{411},
  2093\relax
\mciteBstWouldAddEndPuncttrue
\mciteSetBstMidEndSepPunct{\mcitedefaultmidpunct}
{\mcitedefaultendpunct}{\mcitedefaultseppunct}\relax
\EndOfBibitem
\bibitem[{Rabli} and {Flower}(2010){Rabli}, and {Flower}]{Rabli:10b}
{Rabli},~D.; {Flower},~D.~R. \emph{\mnras} \textbf{2010}, \emph{406}, 95\relax
\mciteBstWouldAddEndPuncttrue
\mciteSetBstMidEndSepPunct{\mcitedefaultmidpunct}
{\mcitedefaultendpunct}{\mcitedefaultseppunct}\relax
\EndOfBibitem
\bibitem[{Green}(1986)]{Green:86CH3CN}
{Green},~S. \emph{\apj} \textbf{1986}, \emph{309}, 331\relax
\mciteBstWouldAddEndPuncttrue
\mciteSetBstMidEndSepPunct{\mcitedefaultmidpunct}
{\mcitedefaultendpunct}{\mcitedefaultseppunct}\relax
\EndOfBibitem
\bibitem[{Faure} et~al.(2011){Faure}, {Szalewicz}, and {Wiesenfeld}]{Faure:11}
{Faure},~A.; {Szalewicz},~K.; {Wiesenfeld},~L. \emph{\jcp} \textbf{2011},
  \emph{135}, 024301\relax
\mciteBstWouldAddEndPuncttrue
\mciteSetBstMidEndSepPunct{\mcitedefaultmidpunct}
{\mcitedefaultendpunct}{\mcitedefaultseppunct}\relax
\EndOfBibitem
\bibitem[Green(1986)]{HNCO}
Green,~S. \url{http://data.giss.nasa.gov/mcrates/\#hnco/}, 1986\relax
\mciteBstWouldAddEndPuncttrue
\mciteSetBstMidEndSepPunct{\mcitedefaultmidpunct}
{\mcitedefaultendpunct}{\mcitedefaultseppunct}\relax
\EndOfBibitem
\bibitem[{Ma} et~al.(2012){Ma}, {Dagdigian}, and {Alexander}]{Ma:12}
{Ma},~L.; {Dagdigian},~P.~J.; {Alexander},~M.~H. \emph{\jcp} \textbf{2012},
  \emph{136}, 224306\relax
\mciteBstWouldAddEndPuncttrue
\mciteSetBstMidEndSepPunct{\mcitedefaultmidpunct}
{\mcitedefaultendpunct}{\mcitedefaultseppunct}\relax
\EndOfBibitem
\bibitem[Dagdigian and Alexander(2011)Dagdigian, and Alexander]{Dagdigian:11}
Dagdigian,~P.; Alexander,~M.~H. \emph{\jcp} \textbf{2011}, \emph{135},
  064306\relax
\mciteBstWouldAddEndPuncttrue
\mciteSetBstMidEndSepPunct{\mcitedefaultmidpunct}
{\mcitedefaultendpunct}{\mcitedefaultseppunct}\relax
\EndOfBibitem
\bibitem[{Hugo} et~al.(2009){Hugo}, {Asvany}, and {Schlemmer}]{Hugo:09}
{Hugo},~E.; {Asvany},~O.; {Schlemmer},~S. \emph{\jcp} \textbf{2009},
  \emph{130}, 164302\relax
\mciteBstWouldAddEndPuncttrue
\mciteSetBstMidEndSepPunct{\mcitedefaultmidpunct}
{\mcitedefaultendpunct}{\mcitedefaultseppunct}\relax
\EndOfBibitem
\bibitem[{Hammami} et~al.(2007){Hammami}, {Lique}, {Ja{\"i}dane}, {Ben
  Lakhdar}, {Spielfiedel}, and {Feautrier}]{Hammami:07}
{Hammami},~K.; {Lique},~F.; {Ja{\"i}dane},~N.; {Ben Lakhdar},~Z.;
  {Spielfiedel},~A.; {Feautrier},~N. \emph{\aap} \textbf{2007}, \emph{462},
  789\relax
\mciteBstWouldAddEndPuncttrue
\mciteSetBstMidEndSepPunct{\mcitedefaultmidpunct}
{\mcitedefaultendpunct}{\mcitedefaultseppunct}\relax
\EndOfBibitem
\bibitem[{Carty} et~al.(2004){Carty}, {Goddard}, {Sims}, and {Smith}]{Carty:04}
{Carty},~D.; {Goddard},~A.; {Sims},~I.~R.; {Smith},~I.~W.~M. \emph{\jcp}
  \textbf{2004}, \emph{121}, 4671\relax
\mciteBstWouldAddEndPuncttrue
\mciteSetBstMidEndSepPunct{\mcitedefaultmidpunct}
{\mcitedefaultendpunct}{\mcitedefaultseppunct}\relax
\EndOfBibitem
\bibitem[Werner et~al.(2012)Werner, Knowles, Knizia, Manby, and
  Sch{\"u}tz]{Werner:12}
Werner,~H.-J.; Knowles,~P.~J.; Knizia,~G.; Manby,~F.~R.; Sch{\"u}tz,~M.
  \emph{Wiley Interdisciplinary Reviews: Computational Molecular Science}
  \textbf{2012}, \emph{2}, 242\relax
\mciteBstWouldAddEndPuncttrue
\mciteSetBstMidEndSepPunct{\mcitedefaultmidpunct}
{\mcitedefaultendpunct}{\mcitedefaultseppunct}\relax
\EndOfBibitem
\bibitem[{Hampel} et~al.(1992){Hampel}, {Peterson}, and {Werner}]{Hampel92}
{Hampel},~C.; {Peterson},~K.~A.; {Werner},~H.-J. \emph{Chem. Phys. Lett.}
  \textbf{1992}, \emph{190}, 1\relax
\mciteBstWouldAddEndPuncttrue
\mciteSetBstMidEndSepPunct{\mcitedefaultmidpunct}
{\mcitedefaultendpunct}{\mcitedefaultseppunct}\relax
\EndOfBibitem
\bibitem[Knowles et~al.(1993)Knowles, Hampel, and Werner]{knowles:93}
Knowles,~P.~J.; Hampel,~C.; Werner,~H.-J. \emph{\jcp} \textbf{1993}, \emph{99},
  5219\relax
\mciteBstWouldAddEndPuncttrue
\mciteSetBstMidEndSepPunct{\mcitedefaultmidpunct}
{\mcitedefaultendpunct}{\mcitedefaultseppunct}\relax
\EndOfBibitem
\bibitem[Knowles et~al.(2000)Knowles, Hampel, and Werner]{knowles:00}
Knowles,~P.~J.; Hampel,~C.; Werner,~H.-J. \emph{\jcp} \textbf{2000},
  \emph{112}, 3106\relax
\mciteBstWouldAddEndPuncttrue
\mciteSetBstMidEndSepPunct{\mcitedefaultmidpunct}
{\mcitedefaultendpunct}{\mcitedefaultseppunct}\relax
\EndOfBibitem
\bibitem[{Werner}(1996)]{Werner:96}
{Werner},~H.-J. \emph{Mol. Phys.} \textbf{1996}, \emph{89}, 645\relax
\mciteBstWouldAddEndPuncttrue
\mciteSetBstMidEndSepPunct{\mcitedefaultmidpunct}
{\mcitedefaultendpunct}{\mcitedefaultseppunct}\relax
\EndOfBibitem
\bibitem[{Celani} and {Werner}(2000){Celani}, and {Werner}]{celani:00}
{Celani},~P.; {Werner},~H.-J. \emph{\jcp} \textbf{2000}, \emph{112}, 5546\relax
\mciteBstWouldAddEndPuncttrue
\mciteSetBstMidEndSepPunct{\mcitedefaultmidpunct}
{\mcitedefaultendpunct}{\mcitedefaultseppunct}\relax
\EndOfBibitem
\bibitem[Jeziorski et~al.(1994)Jeziorski, Moszynski, and
  Szalewicz]{Jeziorski:94}
Jeziorski,~B.; Moszynski,~R.; Szalewicz,~K. \emph{Chem. Rev.} \textbf{1994},
  \emph{94}, 1887\relax
\mciteBstWouldAddEndPuncttrue
\mciteSetBstMidEndSepPunct{\mcitedefaultmidpunct}
{\mcitedefaultendpunct}{\mcitedefaultseppunct}\relax
\EndOfBibitem
\bibitem[Werner and Knowles(1988)Werner, and Knowles]{Werner:88}
Werner,~H.-J.; Knowles,~P.~J. \emph{\jcp} \textbf{1988}, \emph{89}, 5803\relax
\mciteBstWouldAddEndPuncttrue
\mciteSetBstMidEndSepPunct{\mcitedefaultmidpunct}
{\mcitedefaultendpunct}{\mcitedefaultseppunct}\relax
\EndOfBibitem
\bibitem[{Knowles}(1988)]{knowles:88}
{Knowles},~P. \emph{Chem. Phys. Lett.} \textbf{1988}, \emph{145}, 514\relax
\mciteBstWouldAddEndPuncttrue
\mciteSetBstMidEndSepPunct{\mcitedefaultmidpunct}
{\mcitedefaultendpunct}{\mcitedefaultseppunct}\relax
\EndOfBibitem
\bibitem[{Adler} et~al.(2007){Adler}, {Knizia}, and {Werner}]{adler:07}
{Adler},~T.~B.; {Knizia},~G.; {Werner},~H. \emph{\jcp} \textbf{2007},
  \emph{127}, 221106\relax
\mciteBstWouldAddEndPuncttrue
\mciteSetBstMidEndSepPunct{\mcitedefaultmidpunct}
{\mcitedefaultendpunct}{\mcitedefaultseppunct}\relax
\EndOfBibitem
\bibitem[{Knizia} et~al.(2009){Knizia}, {Adler}, and {Werner}]{Knizia:09}
{Knizia},~G.; {Adler},~T.~B.; {Werner},~H. \emph{\jcp} \textbf{2009},
  \emph{130}, 054104\relax
\mciteBstWouldAddEndPuncttrue
\mciteSetBstMidEndSepPunct{\mcitedefaultmidpunct}
{\mcitedefaultendpunct}{\mcitedefaultseppunct}\relax
\EndOfBibitem
\bibitem[Kendall et~al.(1992)Kendall, Dunning, and Harrison]{kendall:92}
Kendall,~R.~A.; Dunning,~T.~H.; Harrison,~R.~J. \emph{\jcp} \textbf{1992},
  \emph{96}, 6796\relax
\mciteBstWouldAddEndPuncttrue
\mciteSetBstMidEndSepPunct{\mcitedefaultmidpunct}
{\mcitedefaultendpunct}{\mcitedefaultseppunct}\relax
\EndOfBibitem
\bibitem[Woon and Dunning{, Jr.}(1993)Woon, and Dunning{, Jr.}]{woon:93}
Woon,~D.~E.; Dunning{, Jr.},~T.~H. \emph{\jcp} \textbf{1993}, \emph{98},
  1358\relax
\mciteBstWouldAddEndPuncttrue
\mciteSetBstMidEndSepPunct{\mcitedefaultmidpunct}
{\mcitedefaultendpunct}{\mcitedefaultseppunct}\relax
\EndOfBibitem
\bibitem[Woon and Dunning(1994)Woon, and Dunning]{woon:94}
Woon,~D.~E.; Dunning,~T.~H. \emph{\jcp} \textbf{1994}, \emph{100}, 2975\relax
\mciteBstWouldAddEndPuncttrue
\mciteSetBstMidEndSepPunct{\mcitedefaultmidpunct}
{\mcitedefaultendpunct}{\mcitedefaultseppunct}\relax
\EndOfBibitem
\bibitem[Tao and Pan(1992)Tao, and Pan]{Tao:92}
Tao,~F.~M.; Pan,~Y.~K. \emph{\jcp} \textbf{1992}, \emph{97}, 4989\relax
\mciteBstWouldAddEndPuncttrue
\mciteSetBstMidEndSepPunct{\mcitedefaultmidpunct}
{\mcitedefaultendpunct}{\mcitedefaultseppunct}\relax
\EndOfBibitem
\bibitem[mol()]{molpro}
MOLPRO is a package of {\em ab initio} programs written by H.-J. Werner and P.
  J. Knowles, with contributions from R. D. Amos, A. Berning, D. L.Cooper, M.
  J. O. Deegan, A. J. Dobbyn, F. Eckert, C. Hampel, T. Leininger, R. Lindh, A.
  W. Lloyd, W. Meyer, M. E. Mura, A. Nickla\ss, P.Palmieri, K. Peterson, R.
  Pitzer, P. Pulay, G. Rauhut, M. Sch{\"u}tz, H. Stoll, A. J. Stone and T.
  Thorsteinsson.\relax
\mciteBstWouldAddEndPunctfalse
\mciteSetBstMidEndSepPunct{\mcitedefaultmidpunct}
{}{\mcitedefaultseppunct}\relax
\EndOfBibitem
\bibitem[Frisch et~al.()Frisch, Trucks, Schlegel, Scuseria, Robb, Cheeseman,
  Scalmani, Barone, Mennucci, Petersson, Nakatsuji, Caricato, Li, Hratchian,
  Izmaylov, Bloino, Zheng, Sonnenberg, Hada, Ehara, Toyota, Fukuda, Hasegawa,
  Ishida, Nakajima, Honda, Kitao, Nakai, Vreven, Montgomery, Peralta, Ogliaro,
  Bearpark, Heyd, Brothers, Kudin, Staroverov, Kobayashi, Normand,
  Raghavachari, Rendell, Burant, Iyengar, Tomasi, Cossi, Rega, Millam, Klene,
  Knox, Cross, Bakken, Adamo, Jaramillo, Gomperts, Stratmann, Yazyev, Austin,
  Cammi, Pomelli, Ochterski, Martin, Morokuma, Zakrzewski, Voth, Salvador,
  Dannenberg, Dapprich, Daniels, Farkas, Foresman, Ortiz, Cioslowski, and
  Fox]{g09}
Frisch,~M.~J. et~al.  Gaussian~09 {R}evision {A}.1. Gaussian Inc. Wallingford
  CT 2009\relax
\mciteBstWouldAddEndPuncttrue
\mciteSetBstMidEndSepPunct{\mcitedefaultmidpunct}
{\mcitedefaultendpunct}{\mcitedefaultseppunct}\relax
\EndOfBibitem
\bibitem[Bukowski et~al.(2008)Bukowski, Cencek, Jankowski, Jeziorska,
  Jeziorski, Kucharski, Lotrich, Misquitta, {Moszy\'{n}ski}, Patkowski,
  Podeszwa, Rybak, Szalewicz, Williams, Wheatley, Wormer, and
  {\.Z}uchowski]{SAPT2008}
Bukowski,~R.; Cencek,~W.; Jankowski,~P.; Jeziorska,~M.; Jeziorski,~B.;
  Kucharski,~S.~A.; Lotrich,~V.~F.; Misquitta,~A.~J.; {Moszy\'{n}ski},~R.;
  Patkowski,~K.; Podeszwa,~R.; Rybak,~S.; Szalewicz,~K.; Williams,~H.~L.;
  Wheatley,~R.~J.; Wormer,~P. E.~S.; {\.Z}uchowski,~P.~S. 2008; {SAPT2008}: An
  {\em Ab Initio} Program for Many-Body Symmetry-Adapted Perturbation Theory
  Calculations of Intermolecular Interaction Energies\relax
\mciteBstWouldAddEndPuncttrue
\mciteSetBstMidEndSepPunct{\mcitedefaultmidpunct}
{\mcitedefaultendpunct}{\mcitedefaultseppunct}\relax
\EndOfBibitem
\bibitem[{Green}(1975)]{green:75}
{Green},~S. \emph{\jcp} \textbf{1975}, \emph{62}, 2271\relax
\mciteBstWouldAddEndPuncttrue
\mciteSetBstMidEndSepPunct{\mcitedefaultmidpunct}
{\mcitedefaultendpunct}{\mcitedefaultseppunct}\relax
\EndOfBibitem
\bibitem[Ho and Rabitz(1996)Ho, and Rabitz]{ho:96}
Ho,~T.-S.; Rabitz,~H. \emph{\jcp} \textbf{1996}, \emph{104}, 2584\relax
\mciteBstWouldAddEndPuncttrue
\mciteSetBstMidEndSepPunct{\mcitedefaultmidpunct}
{\mcitedefaultendpunct}{\mcitedefaultseppunct}\relax
\EndOfBibitem
\bibitem[{Valiron} et~al.(2008){Valiron}, {Wernli}, {Faure}, {Wiesenfeld},
  {Rist}, {Ked{\v z}uch}, and {Noga}]{valiron:08}
{Valiron},~P.; {Wernli},~M.; {Faure},~A.; {Wiesenfeld},~L.; {Rist},~C.; {Ked{\v
  z}uch},~S.; {Noga},~J. \emph{\jcp} \textbf{2008}, \emph{129}, 134306\relax
\mciteBstWouldAddEndPuncttrue
\mciteSetBstMidEndSepPunct{\mcitedefaultmidpunct}
{\mcitedefaultendpunct}{\mcitedefaultseppunct}\relax
\EndOfBibitem
\bibitem[Stone(1996)]{Stone:96}
Stone,~A.~J. \emph{The Theory of Intermolecular forces}; Clarendon Press:
  Oxford, 1996\relax
\mciteBstWouldAddEndPuncttrue
\mciteSetBstMidEndSepPunct{\mcitedefaultmidpunct}
{\mcitedefaultendpunct}{\mcitedefaultseppunct}\relax
\EndOfBibitem
\bibitem[Hutson and Green(1994)Hutson, and Green]{molscat:94}
Hutson,~J.~M.; Green,~S. 1994; {\sc molscat} computer code, version 14 (1994),
  distributed by Collaborative Computational Project No. 6 of the Engineering
  and Physical Sciences Research Council (UK)\relax
\mciteBstWouldAddEndPuncttrue
\mciteSetBstMidEndSepPunct{\mcitedefaultmidpunct}
{\mcitedefaultendpunct}{\mcitedefaultseppunct}\relax
\EndOfBibitem
\bibitem[{Flower} et~al.(2000){Flower}, {Bourhis}, and {Launay}]{molcol}
{Flower},~D.~R.; {Bourhis},~G.; {Launay},~J.-M. \emph{Comp. Phys. Comm.}
  \textbf{2000}, \emph{131}, 187\relax
\mciteBstWouldAddEndPuncttrue
\mciteSetBstMidEndSepPunct{\mcitedefaultmidpunct}
{\mcitedefaultendpunct}{\mcitedefaultseppunct}\relax
\EndOfBibitem
\bibitem[hib()]{hibridon}
The HIBRIDON package was written by M. H. Alexander, D. E. Manolopoulos, H.-J.
  Werner, and B. Follmeg, with contributions by P. F. Vohralik, D. Lemoine, G.
  Corey, R. Gordon, B. Johnson, T. Orlikowski, A. Berning, A. Degli-Esposti, C.
  Rist, P. Dagdigian, B. Pouilly, G. van der Sanden, M. Yang, F. de Weerd, S.
  Gregurick, J. K{\l}os and F. Lique,
  http://www2.chem.umd.edu/groups/alexander/\relax
\mciteBstWouldAddEndPuncttrue
\mciteSetBstMidEndSepPunct{\mcitedefaultmidpunct}
{\mcitedefaultendpunct}{\mcitedefaultseppunct}\relax
\EndOfBibitem
\bibitem[{Lique} and {K{\l}os}(2008){Lique}, and {K{\l}os}]{lique08SiS}
{Lique},~F.; {K{\l}os},~J. \emph{\jcp} \textbf{2008}, \emph{128}, 034306\relax
\mciteBstWouldAddEndPuncttrue
\mciteSetBstMidEndSepPunct{\mcitedefaultmidpunct}
{\mcitedefaultendpunct}{\mcitedefaultseppunct}\relax
\EndOfBibitem
\bibitem[{Green}(1976)]{Green:76top}
{Green},~S. \emph{\jcp} \textbf{1976}, \emph{64}, 3463\relax
\mciteBstWouldAddEndPuncttrue
\mciteSetBstMidEndSepPunct{\mcitedefaultmidpunct}
{\mcitedefaultendpunct}{\mcitedefaultseppunct}\relax
\EndOfBibitem
\bibitem[{Green}(1979)]{Green:79top}
{Green},~S. \emph{\jcp} \textbf{1979}, \emph{70}, 816\relax
\mciteBstWouldAddEndPuncttrue
\mciteSetBstMidEndSepPunct{\mcitedefaultmidpunct}
{\mcitedefaultendpunct}{\mcitedefaultseppunct}\relax
\EndOfBibitem
\bibitem[{Garrison} et~al.(1976){Garrison}, {Lester}, and
  {Miller}]{Garrison:76}
{Garrison},~B.~J.; {Lester},~W.~A.,~Jr.; {Miller},~W.~H. \emph{\jcp}
  \textbf{1976}, \emph{65}, 2193\relax
\mciteBstWouldAddEndPuncttrue
\mciteSetBstMidEndSepPunct{\mcitedefaultmidpunct}
{\mcitedefaultendpunct}{\mcitedefaultseppunct}\relax
\EndOfBibitem
\bibitem[{Phillips} et~al.(1995){Phillips}, {Maluendes}, and
  {Green}]{Phillips:95}
{Phillips},~T.~R.; {Maluendes},~S.; {Green},~S. \emph{\jcp} \textbf{1995},
  \emph{102}, 6024\relax
\mciteBstWouldAddEndPuncttrue
\mciteSetBstMidEndSepPunct{\mcitedefaultmidpunct}
{\mcitedefaultendpunct}{\mcitedefaultseppunct}\relax
\EndOfBibitem
\bibitem[{Rist} et~al.(1993){Rist}, {Alexander}, and {Valiron}]{Rist:93}
{Rist},~C.; {Alexander},~M.~H.; {Valiron},~P. \emph{\jcp} \textbf{1993},
  \emph{98}, 4662\relax
\mciteBstWouldAddEndPuncttrue
\mciteSetBstMidEndSepPunct{\mcitedefaultmidpunct}
{\mcitedefaultendpunct}{\mcitedefaultseppunct}\relax
\EndOfBibitem
\bibitem[Alexander(1982)]{alexander1:82}
Alexander,~M. \emph{\jcp} \textbf{1982}, \emph{76}, 3637\relax
\mciteBstWouldAddEndPuncttrue
\mciteSetBstMidEndSepPunct{\mcitedefaultmidpunct}
{\mcitedefaultendpunct}{\mcitedefaultseppunct}\relax
\EndOfBibitem
\bibitem[Alexander(1982)]{alexander2:82}
Alexander,~M. \emph{\jcp} \textbf{1982}, \emph{76}, 5974\relax
\mciteBstWouldAddEndPuncttrue
\mciteSetBstMidEndSepPunct{\mcitedefaultmidpunct}
{\mcitedefaultendpunct}{\mcitedefaultseppunct}\relax
\EndOfBibitem
\bibitem[Herzberg(1950)]{herzberg}
Herzberg,~G. \emph{Molecular Spectra and Molecular Structure, Vol. 1: Spectra
  of Diatomic Molecules}; Van Nostrand: New York, 1950\relax
\mciteBstWouldAddEndPuncttrue
\mciteSetBstMidEndSepPunct{\mcitedefaultmidpunct}
{\mcitedefaultendpunct}{\mcitedefaultseppunct}\relax
\EndOfBibitem
\bibitem[{Alexander} and {Stolte}(2000){Alexander}, and {Stolte}]{alexander:00}
{Alexander},~M.~H.; {Stolte},~S. \emph{\jcp} \textbf{2000}, \emph{112},
  8017\relax
\mciteBstWouldAddEndPuncttrue
\mciteSetBstMidEndSepPunct{\mcitedefaultmidpunct}
{\mcitedefaultendpunct}{\mcitedefaultseppunct}\relax
\EndOfBibitem
\bibitem[Alexander(1985)]{alexander:85}
Alexander,~M.~H. \emph{Chem. Phys.} \textbf{1985}, \emph{92}, 337\relax
\mciteBstWouldAddEndPuncttrue
\mciteSetBstMidEndSepPunct{\mcitedefaultmidpunct}
{\mcitedefaultendpunct}{\mcitedefaultseppunct}\relax
\EndOfBibitem
\bibitem[Corey and McCourt(1983)Corey, and McCourt]{corey:83}
Corey,~G.~C.; McCourt,~F.~R. \emph{J. Phys. Chem} \textbf{1983}, \emph{87},
  2723\relax
\mciteBstWouldAddEndPuncttrue
\mciteSetBstMidEndSepPunct{\mcitedefaultmidpunct}
{\mcitedefaultendpunct}{\mcitedefaultseppunct}\relax
\EndOfBibitem
\bibitem[{Offer} and {van Dishoeck}(1992){Offer}, and {van Dishoeck}]{offer:92}
{Offer},~A.~R.; {van Dishoeck},~E.~F. \emph{\mnras} \textbf{1992}, \emph{257},
  377\relax
\mciteBstWouldAddEndPuncttrue
\mciteSetBstMidEndSepPunct{\mcitedefaultmidpunct}
{\mcitedefaultendpunct}{\mcitedefaultseppunct}\relax
\EndOfBibitem
\bibitem[{Faure} and {Lique}(2012){Faure}, and {Lique}]{Faure:12HFS}
{Faure},~A.; {Lique},~F. \emph{\mnras} \textbf{2012}, \emph{425}, 740\relax
\mciteBstWouldAddEndPuncttrue
\mciteSetBstMidEndSepPunct{\mcitedefaultmidpunct}
{\mcitedefaultendpunct}{\mcitedefaultseppunct}\relax
\EndOfBibitem
\bibitem[{Stutzki} and {Winnewisser}(1985){Stutzki}, and
  {Winnewisser}]{Stutzki:85}
{Stutzki},~J.; {Winnewisser},~G. \emph{\aap} \textbf{1985}, \emph{144}, 1\relax
\mciteBstWouldAddEndPuncttrue
\mciteSetBstMidEndSepPunct{\mcitedefaultmidpunct}
{\mcitedefaultendpunct}{\mcitedefaultseppunct}\relax
\EndOfBibitem
\bibitem[{Alexander} and {Dagdigian}(1985){Alexander}, and
  {Dagdigian}]{alexander:85hyp}
{Alexander},~M.~H.; {Dagdigian},~P.~J. \emph{\jcp} \textbf{1985}, \emph{83},
  2191\relax
\mciteBstWouldAddEndPuncttrue
\mciteSetBstMidEndSepPunct{\mcitedefaultmidpunct}
{\mcitedefaultendpunct}{\mcitedefaultseppunct}\relax
\EndOfBibitem
\bibitem[{Daniel} et~al.(2004){Daniel}, {Dubernet}, and {Meuwly}]{Daniel:04}
{Daniel},~F.; {Dubernet},~M.-L.; {Meuwly},~M. \emph{\jcp} \textbf{2004},
  \emph{121}, 4540\relax
\mciteBstWouldAddEndPuncttrue
\mciteSetBstMidEndSepPunct{\mcitedefaultmidpunct}
{\mcitedefaultendpunct}{\mcitedefaultseppunct}\relax
\EndOfBibitem
\bibitem[{Guilloteau} and {Baudry}(1981){Guilloteau}, and
  {Baudry}]{Guilloteau:81}
{Guilloteau},~S.; {Baudry},~A. \emph{\aap} \textbf{1981}, \emph{97}, 213\relax
\mciteBstWouldAddEndPuncttrue
\mciteSetBstMidEndSepPunct{\mcitedefaultmidpunct}
{\mcitedefaultendpunct}{\mcitedefaultseppunct}\relax
\EndOfBibitem
\bibitem[{Keto} and {Rybicki}(2010){Keto}, and {Rybicki}]{keto:10}
{Keto},~E.; {Rybicki},~G. \emph{\apj} \textbf{2010}, \emph{716}, 1315\relax
\mciteBstWouldAddEndPuncttrue
\mciteSetBstMidEndSepPunct{\mcitedefaultmidpunct}
{\mcitedefaultendpunct}{\mcitedefaultseppunct}\relax
\EndOfBibitem
\bibitem[{Ag{\'u}ndez} et~al.(2012){Ag{\'u}ndez}, {Fonfr{\'{\i}}a},
  {Cernicharo}, {Kahane}, {Daniel}, and {Gu{\'e}lin}]{agundez:12}
{Ag{\'u}ndez},~M.; {Fonfr{\'{\i}}a},~J.~P.; {Cernicharo},~J.; {Kahane},~C.;
  {Daniel},~F.; {Gu{\'e}lin},~M. \emph{\aap} \textbf{2012}, \emph{543},
  A48\relax
\mciteBstWouldAddEndPuncttrue
\mciteSetBstMidEndSepPunct{\mcitedefaultmidpunct}
{\mcitedefaultendpunct}{\mcitedefaultseppunct}\relax
\EndOfBibitem
\bibitem[{Cernicharo} et~al.(1999){Cernicharo}, {Yamamura},
  {Gonz{\'a}lez-Alfonso}, {de Jong}, {Heras}, {Escribano}, and
  {Ortigoso}]{Cernicharo:99}
{Cernicharo},~J.; {Yamamura},~I.; {Gonz{\'a}lez-Alfonso},~E.; {de Jong},~T.;
  {Heras},~A.; {Escribano},~R.; {Ortigoso},~J. \emph{\apjl} \textbf{1999},
  \emph{526}, L41\relax
\mciteBstWouldAddEndPuncttrue
\mciteSetBstMidEndSepPunct{\mcitedefaultmidpunct}
{\mcitedefaultendpunct}{\mcitedefaultseppunct}\relax
\EndOfBibitem
\bibitem[{Highberger} et~al.(2000){Highberger}, {Apponi}, {Bieging}, {Ziurys},
  and {Mangum}]{Highberger:00}
{Highberger},~J.~L.; {Apponi},~A.~J.; {Bieging},~J.~H.; {Ziurys},~L.~M.;
  {Mangum},~J.~G. \emph{\apj} \textbf{2000}, \emph{544}, 881\relax
\mciteBstWouldAddEndPuncttrue
\mciteSetBstMidEndSepPunct{\mcitedefaultmidpunct}
{\mcitedefaultendpunct}{\mcitedefaultseppunct}\relax
\EndOfBibitem
\bibitem[{Krems}(2002)]{krems:02}
{Krems},~R.~V. \emph{\jcp} \textbf{2002}, \emph{116}, 4517\relax
\mciteBstWouldAddEndPuncttrue
\mciteSetBstMidEndSepPunct{\mcitedefaultmidpunct}
{\mcitedefaultendpunct}{\mcitedefaultseppunct}\relax
\EndOfBibitem
\bibitem[{Flower}(2000)]{Flower:00}
{Flower},~D.~R. \emph{J. Phys. B At. Mol. Opt. Phys.} \textbf{2000}, \emph{33},
  L193\relax
\mciteBstWouldAddEndPuncttrue
\mciteSetBstMidEndSepPunct{\mcitedefaultmidpunct}
{\mcitedefaultendpunct}{\mcitedefaultseppunct}\relax
\EndOfBibitem
\bibitem[{Parker} and {Pack}(1978){Parker}, and {Pack}]{parker:78}
{Parker},~G.~A.; {Pack},~R.~T. \emph{\jcp} \textbf{1978}, \emph{68}, 1585\relax
\mciteBstWouldAddEndPuncttrue
\mciteSetBstMidEndSepPunct{\mcitedefaultmidpunct}
{\mcitedefaultendpunct}{\mcitedefaultseppunct}\relax
\EndOfBibitem
\bibitem[{Stoecklin} et~al.(2013){Stoecklin}, {Denis-Alpizar}, {Halvick}, and
  {Dubernet}]{stoecklin:13}
{Stoecklin},~T.; {Denis-Alpizar},~O.; {Halvick},~P.; {Dubernet},~M.-L.
  \emph{ArXiv e-prints} \textbf{2013}, \relax
\mciteBstWouldAddEndPunctfalse
\mciteSetBstMidEndSepPunct{\mcitedefaultmidpunct}
{}{\mcitedefaultseppunct}\relax
\EndOfBibitem
\bibitem[{Abrines} and {Percival}(1966){Abrines}, and {Percival}]{Abrines:66}
{Abrines},~R.; {Percival},~I.~C. \emph{Proc. Phys. Soc., London} \textbf{1966},
  \emph{88}, 861\relax
\mciteBstWouldAddEndPuncttrue
\mciteSetBstMidEndSepPunct{\mcitedefaultmidpunct}
{\mcitedefaultendpunct}{\mcitedefaultseppunct}\relax
\EndOfBibitem
\bibitem[{Bonnet} and {Rayez}(2004){Bonnet}, and {Rayez}]{Bonnet:04}
{Bonnet},~L.; {Rayez},~J. \emph{Chem. Phys. Lett.} \textbf{2004}, \emph{397},
  106\relax
\mciteBstWouldAddEndPuncttrue
\mciteSetBstMidEndSepPunct{\mcitedefaultmidpunct}
{\mcitedefaultendpunct}{\mcitedefaultseppunct}\relax
\EndOfBibitem
\bibitem[{Mandy}(2009)]{Mandy:09}
{Mandy},~M.~E. \emph{Chem. Phys.} \textbf{2009}, \emph{365}, 1\relax
\mciteBstWouldAddEndPuncttrue
\mciteSetBstMidEndSepPunct{\mcitedefaultmidpunct}
{\mcitedefaultendpunct}{\mcitedefaultseppunct}\relax
\EndOfBibitem
\bibitem[Schiffman and Chandler(1995)Schiffman, and Chandler]{Schiffman:95}
Schiffman,~A.; Chandler,~D.~W. \emph{Int. Rev. in Phys. Chem.} \textbf{1995},
  \emph{14}, 371\relax
\mciteBstWouldAddEndPuncttrue
\mciteSetBstMidEndSepPunct{\mcitedefaultmidpunct}
{\mcitedefaultendpunct}{\mcitedefaultseppunct}\relax
\EndOfBibitem
\bibitem[{Dagdigian}(1996)]{Dagdigian:96}
{Dagdigian},~P.~J. In \emph{{Experimental Studies of Rotationally Inelastic
  State-Resolved Collisions of Small Molecular Free Radicals}}; Liu,~K.,
  Wagner,~A., Eds.; Adv. Ser. Phys. Chem.; World Scientific Publishing Company,
  1996; Vol.~6; p 315\relax
\mciteBstWouldAddEndPuncttrue
\mciteSetBstMidEndSepPunct{\mcitedefaultmidpunct}
{\mcitedefaultendpunct}{\mcitedefaultseppunct}\relax
\EndOfBibitem
\bibitem[{Smith}(2011)]{Smith:11}
{Smith},~I.~W.~M. \emph{\araa} \textbf{2011}, \emph{49}, 29\relax
\mciteBstWouldAddEndPuncttrue
\mciteSetBstMidEndSepPunct{\mcitedefaultmidpunct}
{\mcitedefaultendpunct}{\mcitedefaultseppunct}\relax
\EndOfBibitem
\bibitem[{Daly} and {Oka}(1970){Daly}, and {Oka}]{daly:70}
{Daly},~P.~W.; {Oka},~T. \emph{\jcp} \textbf{1970}, \emph{53}, 3272\relax
\mciteBstWouldAddEndPuncttrue
\mciteSetBstMidEndSepPunct{\mcitedefaultmidpunct}
{\mcitedefaultendpunct}{\mcitedefaultseppunct}\relax
\EndOfBibitem
\bibitem[{Br{\'e}chignac} et~al.(1980){Br{\'e}chignac}, {Picard-Bersellini},
  {Charneau}, and {Launay}]{Brechignac:80}
{Br{\'e}chignac},~P.; {Picard-Bersellini},~A.; {Charneau},~N.; {Launay},~J.~M.
  \emph{Chem. Phys.} \textbf{1980}, \emph{53}, 165\relax
\mciteBstWouldAddEndPuncttrue
\mciteSetBstMidEndSepPunct{\mcitedefaultmidpunct}
{\mcitedefaultendpunct}{\mcitedefaultseppunct}\relax
\EndOfBibitem
\bibitem[{Orr}(1995)]{Orr:95}
{Orr},~B. \emph{Chem. Phys.} \textbf{1995}, \emph{190}, 261\relax
\mciteBstWouldAddEndPuncttrue
\mciteSetBstMidEndSepPunct{\mcitedefaultmidpunct}
{\mcitedefaultendpunct}{\mcitedefaultseppunct}\relax
\EndOfBibitem
\bibitem[{James} et~al.(1998){James}, {Sims}, {Smith}, {Alexander}, and
  {Yang}]{James:98}
{James},~P.~L.; {Sims},~I.~R.; {Smith},~I.~W.~M.; {Alexander},~M.~H.;
  {Yang},~M. \emph{\jcp} \textbf{1998}, \emph{109}, 3882\relax
\mciteBstWouldAddEndPuncttrue
\mciteSetBstMidEndSepPunct{\mcitedefaultmidpunct}
{\mcitedefaultendpunct}{\mcitedefaultseppunct}\relax
\EndOfBibitem
\bibitem[{De Lucia} and {Green}(1988){De Lucia}, and {Green}]{Delucia:88}
{De Lucia},~F.; {Green},~S. \emph{J. Mol. Struct.} \textbf{1988}, \emph{190},
  435\relax
\mciteBstWouldAddEndPuncttrue
\mciteSetBstMidEndSepPunct{\mcitedefaultmidpunct}
{\mcitedefaultendpunct}{\mcitedefaultseppunct}\relax
\EndOfBibitem
\bibitem[{Willey} et~al.(1989){Willey}, {Bittner}, and {De Lucia}]{Willey:89}
{Willey},~D.~R.; {Bittner},~D.~N.; {De Lucia},~F.~C. \emph{Mol. Phys.}
  \textbf{1989}, \emph{67}, 455\relax
\mciteBstWouldAddEndPuncttrue
\mciteSetBstMidEndSepPunct{\mcitedefaultmidpunct}
{\mcitedefaultendpunct}{\mcitedefaultseppunct}\relax
\EndOfBibitem
\bibitem[{Beaky} et~al.(1996){Beaky}, {Goyette}, and {de Lucia}]{Beaky:96}
{Beaky},~M.~M.; {Goyette},~T.~M.; {de Lucia},~F.~C. \emph{\jcp} \textbf{1996},
  \emph{105}, 3994\relax
\mciteBstWouldAddEndPuncttrue
\mciteSetBstMidEndSepPunct{\mcitedefaultmidpunct}
{\mcitedefaultendpunct}{\mcitedefaultseppunct}\relax
\EndOfBibitem
\bibitem[{Willey} et~al.(2000){Willey}, {Timlin}, {Deramo}, {Pondillo},
  {Wesolek}, and {Wig}]{Willey:00}
{Willey},~D.~R.; {Timlin},~R.~E.; {Deramo},~M.; {Pondillo},~P.~L.;
  {Wesolek},~D.~M.; {Wig},~R.~W. \emph{\jcp} \textbf{2000}, \emph{113},
  611\relax
\mciteBstWouldAddEndPuncttrue
\mciteSetBstMidEndSepPunct{\mcitedefaultmidpunct}
{\mcitedefaultendpunct}{\mcitedefaultseppunct}\relax
\EndOfBibitem
\bibitem[{Ben-Reuven}(1966)]{benreuven:66}
{Ben-Reuven},~A. \emph{Phys. Rev.} \textbf{1966}, \emph{145}, 7\relax
\mciteBstWouldAddEndPuncttrue
\mciteSetBstMidEndSepPunct{\mcitedefaultmidpunct}
{\mcitedefaultendpunct}{\mcitedefaultseppunct}\relax
\EndOfBibitem
\bibitem[{Baranger}(1958)]{Baranger:58}
{Baranger},~M. \emph{Phys. Rev.} \textbf{1958}, \emph{112}, 855\relax
\mciteBstWouldAddEndPuncttrue
\mciteSetBstMidEndSepPunct{\mcitedefaultmidpunct}
{\mcitedefaultendpunct}{\mcitedefaultseppunct}\relax
\EndOfBibitem
\bibitem[{Ter Meulen}(1997)]{meulen:97}
{Ter Meulen},~J.~J. In \emph{IAU Symposium}; {van Dishoeck},~E.~F., Ed.; IAU
  Symposium; 1997; Vol. 178; p 241\relax
\mciteBstWouldAddEndPuncttrue
\mciteSetBstMidEndSepPunct{\mcitedefaultmidpunct}
{\mcitedefaultendpunct}{\mcitedefaultseppunct}\relax
\EndOfBibitem
\bibitem[{Yang} et~al.(2010){Yang}, {Sarma}, {Ter Meulen}, {Parker}, {McBane},
  {Wiesenfeld}, {Faure}, {Scribano}, and {Feautrier}]{Yang:10}
{Yang},~C.-H.; {Sarma},~G.; {Ter Meulen},~J.~J.; {Parker},~D.~H.;
  {McBane},~G.~C.; {Wiesenfeld},~L.; {Faure},~A.; {Scribano},~Y.;
  {Feautrier},~N. \emph{\jcp} \textbf{2010}, \emph{133}, 131103\relax
\mciteBstWouldAddEndPuncttrue
\mciteSetBstMidEndSepPunct{\mcitedefaultmidpunct}
{\mcitedefaultendpunct}{\mcitedefaultseppunct}\relax
\EndOfBibitem
\bibitem[{Yang} et~al.(2011){Yang}, {Sarma}, {Parker}, {Ter Meulen}, and
  {Wiesenfeld}]{Yang:11}
{Yang},~C.-H.; {Sarma},~G.; {Parker},~D.~H.; {Ter Meulen},~J.~J.;
  {Wiesenfeld},~L. \emph{\jcp} \textbf{2011}, \emph{134}, 204308\relax
\mciteBstWouldAddEndPuncttrue
\mciteSetBstMidEndSepPunct{\mcitedefaultmidpunct}
{\mcitedefaultendpunct}{\mcitedefaultseppunct}\relax
\EndOfBibitem
\bibitem[{Chefdeville} et~al.(2012){Chefdeville}, {Stoecklin}, {Bergeat},
  {Hickson}, {Naulin}, and {Costes}]{Chefdeville:12}
{Chefdeville},~S.; {Stoecklin},~T.; {Bergeat},~A.; {Hickson},~K.~M.;
  {Naulin},~C.; {Costes},~M. \emph{Phys. Rev. Lett.} \textbf{2012}, \emph{109},
  023201\relax
\mciteBstWouldAddEndPuncttrue
\mciteSetBstMidEndSepPunct{\mcitedefaultmidpunct}
{\mcitedefaultendpunct}{\mcitedefaultseppunct}\relax
\EndOfBibitem
\bibitem[{Tejeda} et~al.(2008){Tejeda}, {Thibault}, {Fern{\'a}ndez}, and
  {Montero}]{Tejeda:08}
{Tejeda},~G.; {Thibault},~F.; {Fern{\'a}ndez},~J.~M.; {Montero},~S. \emph{\jcp}
  \textbf{2008}, \emph{128}, 224308\relax
\mciteBstWouldAddEndPuncttrue
\mciteSetBstMidEndSepPunct{\mcitedefaultmidpunct}
{\mcitedefaultendpunct}{\mcitedefaultseppunct}\relax
\EndOfBibitem
\bibitem[{Flower} and {Roueff}(1998){Flower}, and {Roueff}]{Flower:98b}
{Flower},~D.~R.; {Roueff},~E. \emph{J. Phys. B At. Mol. Opt. Phys.}
  \textbf{1998}, \emph{31}, 2935\relax
\mciteBstWouldAddEndPuncttrue
\mciteSetBstMidEndSepPunct{\mcitedefaultmidpunct}
{\mcitedefaultendpunct}{\mcitedefaultseppunct}\relax
\EndOfBibitem
\bibitem[{Muchnick} and {Russek}(1994){Muchnick}, and {Russek}]{Muchnick:94}
{Muchnick},~P.; {Russek},~A. \emph{\jcp} \textbf{1994}, \emph{100}, 4336\relax
\mciteBstWouldAddEndPuncttrue
\mciteSetBstMidEndSepPunct{\mcitedefaultmidpunct}
{\mcitedefaultendpunct}{\mcitedefaultseppunct}\relax
\EndOfBibitem
\bibitem[{Boothroyd} et~al.(2003){Boothroyd}, {Martin}, and
  {Peterson}]{boothroyd:03}
{Boothroyd},~A.~I.; {Martin},~P.~G.; {Peterson},~M.~R. \emph{\jcp}
  \textbf{2003}, \emph{119}, 3187\relax
\mciteBstWouldAddEndPuncttrue
\mciteSetBstMidEndSepPunct{\mcitedefaultmidpunct}
{\mcitedefaultendpunct}{\mcitedefaultseppunct}\relax
\EndOfBibitem
\bibitem[{Boothroyd} et~al.(1991){Boothroyd}, {Dove}, {Keogh}, {Martin}, and
  {Peterson}]{Boothroyd:91H4}
{Boothroyd},~A.~I.; {Dove},~J.~E.; {Keogh},~W.~J.; {Martin},~P.~G.;
  {Peterson},~M.~R. \emph{\jcp} \textbf{1991}, \emph{95}, 4331\relax
\mciteBstWouldAddEndPuncttrue
\mciteSetBstMidEndSepPunct{\mcitedefaultmidpunct}
{\mcitedefaultendpunct}{\mcitedefaultseppunct}\relax
\EndOfBibitem
\bibitem[{Aguado} et~al.(1994){Aguado}, {Su{\'a}rez}, and
  {Paniagua}]{Aguado:94}
{Aguado},~A.; {Su{\'a}rez},~C.; {Paniagua},~M. \emph{\jcp} \textbf{1994},
  \emph{101}, 4004\relax
\mciteBstWouldAddEndPuncttrue
\mciteSetBstMidEndSepPunct{\mcitedefaultmidpunct}
{\mcitedefaultendpunct}{\mcitedefaultseppunct}\relax
\EndOfBibitem
\bibitem[{Boothroyd} et~al.(2002){Boothroyd}, {Martin}, {Keogh}, and
  {Peterson}]{Boothroyd:02}
{Boothroyd},~A.~I.; {Martin},~P.~G.; {Keogh},~W.~J.; {Peterson},~M.~J.
  \emph{\jcp} \textbf{2002}, \emph{116}, 666\relax
\mciteBstWouldAddEndPuncttrue
\mciteSetBstMidEndSepPunct{\mcitedefaultmidpunct}
{\mcitedefaultendpunct}{\mcitedefaultseppunct}\relax
\EndOfBibitem
\bibitem[{Hinde}(2008)]{Hinde:08}
{Hinde},~R.~J. \emph{\jcp} \textbf{2008}, \emph{128}, 154308\relax
\mciteBstWouldAddEndPuncttrue
\mciteSetBstMidEndSepPunct{\mcitedefaultmidpunct}
{\mcitedefaultendpunct}{\mcitedefaultseppunct}\relax
\EndOfBibitem
\bibitem[{Danby} et~al.(1987){Danby}, {Flower}, and {Monteiro}]{Danby:87}
{Danby},~G.; {Flower},~D.~R.; {Monteiro},~T.~S. \emph{\mnras} \textbf{1987},
  \emph{226}, 739\relax
\mciteBstWouldAddEndPuncttrue
\mciteSetBstMidEndSepPunct{\mcitedefaultmidpunct}
{\mcitedefaultendpunct}{\mcitedefaultseppunct}\relax
\EndOfBibitem
\bibitem[{Flower}(1998)]{Flower:98}
{Flower},~D.~R. \emph{\mnras} \textbf{1998}, \emph{297}, 334\relax
\mciteBstWouldAddEndPuncttrue
\mciteSetBstMidEndSepPunct{\mcitedefaultmidpunct}
{\mcitedefaultendpunct}{\mcitedefaultseppunct}\relax
\EndOfBibitem
\bibitem[Lin and Guo(2003)Lin, and Guo]{Lin:03}
Lin,~S.; Guo,~H. \emph{Chem. Phys.} \textbf{2003}, \emph{289}, 191\relax
\mciteBstWouldAddEndPuncttrue
\mciteSetBstMidEndSepPunct{\mcitedefaultmidpunct}
{\mcitedefaultendpunct}{\mcitedefaultseppunct}\relax
\EndOfBibitem
\bibitem[{Otto} et~al.(2008){Otto}, {Gatti}, and {Meyer}]{Otto:08}
{Otto},~F.; {Gatti},~F.; {Meyer},~H.-D. \emph{\jcp} \textbf{2008}, \emph{128},
  064305\relax
\mciteBstWouldAddEndPuncttrue
\mciteSetBstMidEndSepPunct{\mcitedefaultmidpunct}
{\mcitedefaultendpunct}{\mcitedefaultseppunct}\relax
\EndOfBibitem
\bibitem[{Zenevich} and {Billing}(1999){Zenevich}, and {Billing}]{Zenevich:99}
{Zenevich},~V.~A.; {Billing},~G.~D. \emph{\jcp} \textbf{1999}, \emph{111},
  2401\relax
\mciteBstWouldAddEndPuncttrue
\mciteSetBstMidEndSepPunct{\mcitedefaultmidpunct}
{\mcitedefaultendpunct}{\mcitedefaultseppunct}\relax
\EndOfBibitem
\bibitem[{dos Santos} et~al.(2011){dos Santos}, {Balakrishnan}, {Lepp},
  {Qu{\'e}m{\'e}ner}, {Forrey}, {Hinde}, and {Stancil}]{santos:11}
{dos Santos},~S.~F.; {Balakrishnan},~N.; {Lepp},~S.; {Qu{\'e}m{\'e}ner},~G.;
  {Forrey},~R.~C.; {Hinde},~R.~J.; {Stancil},~P.~C. \emph{\jcp} \textbf{2011},
  \emph{134}, 214303\relax
\mciteBstWouldAddEndPuncttrue
\mciteSetBstMidEndSepPunct{\mcitedefaultmidpunct}
{\mcitedefaultendpunct}{\mcitedefaultseppunct}\relax
\EndOfBibitem
\bibitem[{dos Santos} et~al.(2013){dos Santos}, {Balakrishnan}, {Forrey}, and
  {Stancil}]{santos:13}
{dos Santos},~S.~F.; {Balakrishnan},~N.; {Forrey},~R.~C.; {Stancil},~P.~C.
  \emph{\jcp} \textbf{2013}, \emph{138}, 104302\relax
\mciteBstWouldAddEndPuncttrue
\mciteSetBstMidEndSepPunct{\mcitedefaultmidpunct}
{\mcitedefaultendpunct}{\mcitedefaultseppunct}\relax
\EndOfBibitem
\bibitem[{Mat{\'e}} et~al.(2005){Mat{\'e}}, {Thibault}, {Tejeda},
  {Fern{\'a}ndez}, and {Montero}]{Mate:05}
{Mat{\'e}},~B.; {Thibault},~F.; {Tejeda},~G.; {Fern{\'a}ndez},~J.~M.;
  {Montero},~S. \emph{\jcp} \textbf{2005}, \emph{122}, 064313\relax
\mciteBstWouldAddEndPuncttrue
\mciteSetBstMidEndSepPunct{\mcitedefaultmidpunct}
{\mcitedefaultendpunct}{\mcitedefaultseppunct}\relax
\EndOfBibitem
\bibitem[{Audibert} et~al.(1975){Audibert}, {Vilaseca}, {Lukasik}, and
  {Ducuing}]{Audibert:75}
{Audibert},~M.-M.; {Vilaseca},~R.; {Lukasik},~J.; {Ducuing},~J. \emph{Chem.
  Phys. Lett.} \textbf{1975}, \emph{31}, 232\relax
\mciteBstWouldAddEndPuncttrue
\mciteSetBstMidEndSepPunct{\mcitedefaultmidpunct}
{\mcitedefaultendpunct}{\mcitedefaultseppunct}\relax
\EndOfBibitem
\bibitem[{Boothroyd} et~al.(1991){Boothroyd}, {Martin}, {Keogh}, and
  {Peterson}]{Boothroyd:91}
{Boothroyd},~A.~I.; {Martin},~P.~G.; {Keogh},~W.~J.; {Peterson},~M.~R.
  \emph{\jcp} \textbf{1991}, \emph{95}, 4343\relax
\mciteBstWouldAddEndPuncttrue
\mciteSetBstMidEndSepPunct{\mcitedefaultmidpunct}
{\mcitedefaultendpunct}{\mcitedefaultseppunct}\relax
\EndOfBibitem
\bibitem[{Mielke} et~al.(2002){Mielke}, {Garrett}, and {Peterson}]{Mielke02}
{Mielke},~S.~L.; {Garrett},~B.~C.; {Peterson},~K.~A. \emph{\jcp} \textbf{2002},
  \emph{116}, 4142\relax
\mciteBstWouldAddEndPuncttrue
\mciteSetBstMidEndSepPunct{\mcitedefaultmidpunct}
{\mcitedefaultendpunct}{\mcitedefaultseppunct}\relax
\EndOfBibitem
\bibitem[{Wrathmall} et~al.(2007){Wrathmall}, {Gusdorf}, and
  {Flower}]{wrathmall:07}
{Wrathmall},~S.~A.; {Gusdorf},~A.; {Flower},~D.~R. \emph{\mnras} \textbf{2007},
  \emph{382}, 133\relax
\mciteBstWouldAddEndPuncttrue
\mciteSetBstMidEndSepPunct{\mcitedefaultmidpunct}
{\mcitedefaultendpunct}{\mcitedefaultseppunct}\relax
\EndOfBibitem
\bibitem[{Le Bourlot} et~al.(1999){Le Bourlot}, {Pineau des For{\^e}ts}, and
  {Flower}]{lebourlot:99}
{Le Bourlot},~J.; {Pineau des For{\^e}ts},~G.; {Flower},~D.~R. \emph{\mnras}
  \textbf{1999}, \emph{305}, 802\relax
\mciteBstWouldAddEndPuncttrue
\mciteSetBstMidEndSepPunct{\mcitedefaultmidpunct}
{\mcitedefaultendpunct}{\mcitedefaultseppunct}\relax
\EndOfBibitem
\bibitem[{Flower} et~al.(2000){Flower}, {Le Bourlot}, {Pineau des For{\^e}ts},
  and {Roueff}]{flower:00b}
{Flower},~D.~R.; {Le Bourlot},~J.; {Pineau des For{\^e}ts},~G.; {Roueff},~E.
  \emph{\mnras} \textbf{2000}, \emph{314}, 753\relax
\mciteBstWouldAddEndPuncttrue
\mciteSetBstMidEndSepPunct{\mcitedefaultmidpunct}
{\mcitedefaultendpunct}{\mcitedefaultseppunct}\relax
\EndOfBibitem
\bibitem[{Wrathmall} and {Flower}(2006){Wrathmall}, and {Flower}]{wrathmall:06}
{Wrathmall},~S.~A.; {Flower},~D.~R. \emph{J. Phys. B At. Mol. Opt. Phys.}
  \textbf{2006}, \emph{39}, L249\relax
\mciteBstWouldAddEndPuncttrue
\mciteSetBstMidEndSepPunct{\mcitedefaultmidpunct}
{\mcitedefaultendpunct}{\mcitedefaultseppunct}\relax
\EndOfBibitem
\bibitem[{Flower}(1999)]{Flower:99}
{Flower},~D.~R. \emph{J. Phys. B} \textbf{1999}, \emph{32}, 1755\relax
\mciteBstWouldAddEndPuncttrue
\mciteSetBstMidEndSepPunct{\mcitedefaultmidpunct}
{\mcitedefaultendpunct}{\mcitedefaultseppunct}\relax
\EndOfBibitem
\bibitem[{Schulz} and {Le Roy}(1965){Schulz}, and {Le Roy}]{Schulz:65}
{Schulz},~W.~R.; {Le Roy},~D.~J. \emph{\jcp} \textbf{1965}, \emph{42},
  3869\relax
\mciteBstWouldAddEndPuncttrue
\mciteSetBstMidEndSepPunct{\mcitedefaultmidpunct}
{\mcitedefaultendpunct}{\mcitedefaultseppunct}\relax
\EndOfBibitem
\bibitem[{Aoiz} et~al.(2005){Aoiz}, {Ba{\~n}ares}, and {Herrero}]{Aoiz05}
{Aoiz},~F.~J.; {Ba{\~n}ares},~L.; {Herrero},~V.~J. \emph{Int. Rev. in Phys.
  Chem.} \textbf{2005}, \emph{24}, 119\relax
\mciteBstWouldAddEndPuncttrue
\mciteSetBstMidEndSepPunct{\mcitedefaultmidpunct}
{\mcitedefaultendpunct}{\mcitedefaultseppunct}\relax
\EndOfBibitem
\bibitem[{Cernicharo} et~al.(1996){Cernicharo}, {Barlow}, {Gonzalez-Alfonso},
  {Cox}, {Clegg}, {Nguyen-Q-Rieu}, {Omont}, {Guelin}, {Liu}, {Sylvester},
  {Lim}, {Griffin}, {Swinyard}, {Unger}, {Ade}, {Baluteau}, {Caux}, {Cohen},
  {Emery}, {Fischer}, {Furniss}, {Glencross}, {Greenhouse}, {Gry}, {Joubert},
  {Lorenzetti}, {Nisini}, {Orfei}, {Pequignot}, {Saraceno}, {Serra}, {Skinner},
  {Smith}, {Towlson}, {Walker}, {Armand}, {Burgdorf}, {Ewart}, {di Giorgio},
  {Molinari}, {Price}, {Sidher}, {Texier}, and {Trams}]{cernicharo:96}
{Cernicharo},~J. et~al.  \emph{\aap} \textbf{1996}, \emph{315}, L201\relax
\mciteBstWouldAddEndPuncttrue
\mciteSetBstMidEndSepPunct{\mcitedefaultmidpunct}
{\mcitedefaultendpunct}{\mcitedefaultseppunct}\relax
\EndOfBibitem
\bibitem[{Manoj} et~al.(2013){Manoj}, {Watson}, {Neufeld}, {Megeath}, {Vavrek},
  {Yu}, {Visser}, {Bergin}, {Fischer}, {Tobin}, {Stutz}, {Ali}, {Wilson}, {Di
  Francesco}, {Osorio}, {Maret}, and {Poteet}]{manoj:13}
{Manoj},~P.; {Watson},~D.~M.; {Neufeld},~D.~A.; {Megeath},~S.~T.; {Vavrek},~R.;
  {Yu},~V.; {Visser},~R.; {Bergin},~E.~A.; {Fischer},~W.~J.; {Tobin},~J.~J.;
  {Stutz},~A.~M.; {Ali},~B.; {Wilson},~T.~L.; {Di Francesco},~J.; {Osorio},~M.;
  {Maret},~S.; {Poteet},~C.~A. \emph{\apj} \textbf{2013}, \emph{763}, 83\relax
\mciteBstWouldAddEndPuncttrue
\mciteSetBstMidEndSepPunct{\mcitedefaultmidpunct}
{\mcitedefaultendpunct}{\mcitedefaultseppunct}\relax
\EndOfBibitem
\bibitem[{San Jos{\'e}-Garc{\'{\i}}a} et~al.(2013){San Jos{\'e}-Garc{\'{\i}}a},
  {Mottram}, {Kristensen}, {van Dishoeck}, {Y{\i}ld{\i}z}, {van der Tak},
  {Herpin}, {Visser}, {McCoey}, {Wyrowski}, {Braine}, and
  {Johnstone}]{garcia:13}
{San Jos{\'e}-Garc{\'{\i}}a},~I.; {Mottram},~J.~C.; {Kristensen},~L.~E.; {van
  Dishoeck},~E.~F.; {Y{\i}ld{\i}z},~U.~A.; {van der Tak},~F.~F.~S.;
  {Herpin},~F.; {Visser},~R.; {McCoey},~C.; {Wyrowski},~F.; {Braine},~J.;
  {Johnstone},~D. \emph{\aap} \textbf{2013}, \emph{553}, A125\relax
\mciteBstWouldAddEndPuncttrue
\mciteSetBstMidEndSepPunct{\mcitedefaultmidpunct}
{\mcitedefaultendpunct}{\mcitedefaultseppunct}\relax
\EndOfBibitem
\bibitem[Heijmen et~al.(1997)Heijmen, Moszynski, Wormer, and {van der
  Avoird}]{Heijmen:97b}
Heijmen,~T. G.~A.; Moszynski,~R.; Wormer,~P. E.~S.; {van der Avoird},~A.
  \emph{\jcp} \textbf{1997}, \emph{107}, 9921\relax
\mciteBstWouldAddEndPuncttrue
\mciteSetBstMidEndSepPunct{\mcitedefaultmidpunct}
{\mcitedefaultendpunct}{\mcitedefaultseppunct}\relax
\EndOfBibitem
\bibitem[{Antonova} et~al.(1999){Antonova}, {Lin}, {Tsakotellis}, and
  {McBane}]{Antonova:99}
{Antonova},~S.; {Lin},~A.; {Tsakotellis},~A.~P.; {McBane},~G.~C. \emph{\jcp}
  \textbf{1999}, \emph{110}, 2384\relax
\mciteBstWouldAddEndPuncttrue
\mciteSetBstMidEndSepPunct{\mcitedefaultmidpunct}
{\mcitedefaultendpunct}{\mcitedefaultseppunct}\relax
\EndOfBibitem
\bibitem[{Smith} et~al.(2004){Smith}, {Hostutler}, {Hager}, {Heaven}, and
  {McBane}]{Smith:04}
{Smith},~T.~C.; {Hostutler},~D.~A.; {Hager},~G.~D.; {Heaven},~M.~C.;
  {McBane},~G.~C. \emph{\jcp} \textbf{2004}, \emph{120}, 2285\relax
\mciteBstWouldAddEndPuncttrue
\mciteSetBstMidEndSepPunct{\mcitedefaultmidpunct}
{\mcitedefaultendpunct}{\mcitedefaultseppunct}\relax
\EndOfBibitem
\bibitem[{Yang} et~al.(2005){Yang}, {Stancil}, {Balakrishnan}, and
  {Forrey}]{Yang:05}
{Yang},~B.; {Stancil},~P.~C.; {Balakrishnan},~N.; {Forrey},~R.~C. \emph{\jcp}
  \textbf{2005}, \emph{123}, 134326\relax
\mciteBstWouldAddEndPuncttrue
\mciteSetBstMidEndSepPunct{\mcitedefaultmidpunct}
{\mcitedefaultendpunct}{\mcitedefaultseppunct}\relax
\EndOfBibitem
\bibitem[{Kobayashi} et~al.(2000){Kobayashi}, {Amos}, {Reid}, {Quiney}, and
  {Simpson}]{kobayashi:00}
{Kobayashi},~R.; {Amos},~R.~D.; {Reid},~J.~P.; {Quiney},~H.~M.;
  {Simpson},~C.~J.~S.~M. \emph{Mol. Phys.} \textbf{2000}, \emph{98}, 1995\relax
\mciteBstWouldAddEndPuncttrue
\mciteSetBstMidEndSepPunct{\mcitedefaultmidpunct}
{\mcitedefaultendpunct}{\mcitedefaultseppunct}\relax
\EndOfBibitem
\bibitem[{Wickham-Jones} et~al.(1987){Wickham-Jones}, {Williams}, and
  {Simpson}]{jones:87}
{Wickham-Jones},~C.~T.; {Williams},~H.~T.; {Simpson},~C.~J.~S.~M. \emph{\jcp}
  \textbf{1987}, \emph{87}, 5294\relax
\mciteBstWouldAddEndPuncttrue
\mciteSetBstMidEndSepPunct{\mcitedefaultmidpunct}
{\mcitedefaultendpunct}{\mcitedefaultseppunct}\relax
\EndOfBibitem
\bibitem[{Peterson} and {McBane}(2005){Peterson}, and {McBane}]{peterson:05}
{Peterson},~K.~A.; {McBane},~G.~C. \emph{\jcp} \textbf{2005}, \emph{123},
  084314\relax
\mciteBstWouldAddEndPuncttrue
\mciteSetBstMidEndSepPunct{\mcitedefaultmidpunct}
{\mcitedefaultendpunct}{\mcitedefaultseppunct}\relax
\EndOfBibitem
\bibitem[{Jankowski} and {Szalewicz}(1998){Jankowski}, and
  {Szalewicz}]{Jankowski:98}
{Jankowski},~P.; {Szalewicz},~K. \emph{\jcp} \textbf{1998}, \emph{108},
  3554\relax
\mciteBstWouldAddEndPuncttrue
\mciteSetBstMidEndSepPunct{\mcitedefaultmidpunct}
{\mcitedefaultendpunct}{\mcitedefaultseppunct}\relax
\EndOfBibitem
\bibitem[{McKellar}(1998)]{McKellar:98}
{McKellar},~A.~R.~W. \emph{\jcp} \textbf{1998}, \emph{108}, 1811\relax
\mciteBstWouldAddEndPuncttrue
\mciteSetBstMidEndSepPunct{\mcitedefaultmidpunct}
{\mcitedefaultendpunct}{\mcitedefaultseppunct}\relax
\EndOfBibitem
\bibitem[{Flower}(2001)]{Flower:01}
{Flower},~D.~R. \emph{J. Phys. B At. Mol. Opt. Phys.} \textbf{2001}, \emph{34},
  2731\relax
\mciteBstWouldAddEndPuncttrue
\mciteSetBstMidEndSepPunct{\mcitedefaultmidpunct}
{\mcitedefaultendpunct}{\mcitedefaultseppunct}\relax
\EndOfBibitem
\bibitem[{Mengel} et~al.(2001){Mengel}, {de Lucia}, and {Herbst}]{Mengel:01}
{Mengel},~M.; {de Lucia},~F.~C.; {Herbst},~E. \emph{Can. J. Phys.}
  \textbf{2001}, \emph{79}, 589\relax
\mciteBstWouldAddEndPuncttrue
\mciteSetBstMidEndSepPunct{\mcitedefaultmidpunct}
{\mcitedefaultendpunct}{\mcitedefaultseppunct}\relax
\EndOfBibitem
\bibitem[{Schinke} et~al.(1985){Schinke}, {Engel}, {Buck}, {Meyer}, and
  {Diercksen}]{Schinke:85}
{Schinke},~R.; {Engel},~V.; {Buck},~U.; {Meyer},~H.; {Diercksen},~G.~H.~F.
  \emph{\apj} \textbf{1985}, \emph{299}, 939\relax
\mciteBstWouldAddEndPuncttrue
\mciteSetBstMidEndSepPunct{\mcitedefaultmidpunct}
{\mcitedefaultendpunct}{\mcitedefaultseppunct}\relax
\EndOfBibitem
\bibitem[Jankowski and Szalewicz(2005)Jankowski, and Szalewicz]{Jankowski:05}
Jankowski,~P.; Szalewicz,~K. \emph{\jcp} \textbf{2005}, \emph{123},
  104301\relax
\mciteBstWouldAddEndPuncttrue
\mciteSetBstMidEndSepPunct{\mcitedefaultmidpunct}
{\mcitedefaultendpunct}{\mcitedefaultseppunct}\relax
\EndOfBibitem
\bibitem[{Wernli} et~al.(2006){Wernli}, {Valiron}, {Faure}, {Wiesenfeld},
  {Jankowski}, and {Szalewicz}]{wernli06}
{Wernli},~M.; {Valiron},~P.; {Faure},~A.; {Wiesenfeld},~L.; {Jankowski},~P.;
  {Szalewicz},~K. \emph{\aap} \textbf{2006}, \emph{446}, 367\relax
\mciteBstWouldAddEndPuncttrue
\mciteSetBstMidEndSepPunct{\mcitedefaultmidpunct}
{\mcitedefaultendpunct}{\mcitedefaultseppunct}\relax
\EndOfBibitem
\bibitem[{Yang} et~al.(2010){Yang}, {Stancil}, {Balakrishnan}, and
  {Forrey}]{yang:10b}
{Yang},~B.; {Stancil},~P.~C.; {Balakrishnan},~N.; {Forrey},~R.~C. \emph{\apj}
  \textbf{2010}, \emph{718}, 1062\relax
\mciteBstWouldAddEndPuncttrue
\mciteSetBstMidEndSepPunct{\mcitedefaultmidpunct}
{\mcitedefaultendpunct}{\mcitedefaultseppunct}\relax
\EndOfBibitem
\bibitem[{Bowman} et~al.(1986){Bowman}, {Bittman}, and {Harding}]{Bowman:86}
{Bowman},~J.~M.; {Bittman},~J.~S.; {Harding},~L.~B. \emph{\jcp} \textbf{1986},
  \emph{85}, 911\relax
\mciteBstWouldAddEndPuncttrue
\mciteSetBstMidEndSepPunct{\mcitedefaultmidpunct}
{\mcitedefaultendpunct}{\mcitedefaultseppunct}\relax
\EndOfBibitem
\bibitem[{Keller} et~al.(1996){Keller}, {Floethmann}, {Dobbyn}, {Schinke},
  {Werner}, {Bauer}, and {Rosmus}]{Keller:96}
{Keller},~H.-M.; {Floethmann},~H.; {Dobbyn},~A.~J.; {Schinke},~R.;
  {Werner},~H.-J.; {Bauer},~C.; {Rosmus},~P. \emph{\jcp} \textbf{1996},
  \emph{105}, 4983\relax
\mciteBstWouldAddEndPuncttrue
\mciteSetBstMidEndSepPunct{\mcitedefaultmidpunct}
{\mcitedefaultendpunct}{\mcitedefaultseppunct}\relax
\EndOfBibitem
\bibitem[Zanchet et~al.(2006)Zanchet, Bussery-Honvault, and
  Honvault]{Zanchet:06}
Zanchet,~A.; Bussery-Honvault,~B.; Honvault,~P. \emph{J. Phys. Chem. A}
  \textbf{2006}, \emph{110}, 12017\relax
\mciteBstWouldAddEndPuncttrue
\mciteSetBstMidEndSepPunct{\mcitedefaultmidpunct}
{\mcitedefaultendpunct}{\mcitedefaultseppunct}\relax
\EndOfBibitem
\bibitem[{Balakrishnan} et~al.(2002){Balakrishnan}, {Yan}, and
  {Dalgarno}]{Balakrishnan:02}
{Balakrishnan},~N.; {Yan},~M.; {Dalgarno},~A. \emph{\apj} \textbf{2002},
  \emph{568}, 443\relax
\mciteBstWouldAddEndPuncttrue
\mciteSetBstMidEndSepPunct{\mcitedefaultmidpunct}
{\mcitedefaultendpunct}{\mcitedefaultseppunct}\relax
\EndOfBibitem
\bibitem[{Yang} et~al.(2013){Yang}, {Stancil}, {Balakrishnan}, {Forrey}, and
  {Bowman}]{yang:13co}
{Yang},~B.; {Stancil},~P.~C.; {Balakrishnan},~N.; {Forrey},~R.~C.;
  {Bowman},~J.~M. \emph{\apj} \textbf{2013}, \emph{771}, 49\relax
\mciteBstWouldAddEndPuncttrue
\mciteSetBstMidEndSepPunct{\mcitedefaultmidpunct}
{\mcitedefaultendpunct}{\mcitedefaultseppunct}\relax
\EndOfBibitem
\bibitem[{Cheung} et~al.(1969){Cheung}, {Rank}, {Townes}, {Thornton}, and
  {Welch}]{cheung:69}
{Cheung},~A.~C.; {Rank},~D.~M.; {Townes},~C.~H.; {Thornton},~D.~D.;
  {Welch},~W.~J. \emph{\nat} \textbf{1969}, \emph{221}, 626\relax
\mciteBstWouldAddEndPuncttrue
\mciteSetBstMidEndSepPunct{\mcitedefaultmidpunct}
{\mcitedefaultendpunct}{\mcitedefaultseppunct}\relax
\EndOfBibitem
\bibitem[{Benedettini} et~al.(2002){Benedettini}, {Viti}, {Giannini}, {Nisini},
  {Goldsmith}, and {Saraceno}]{benedettini:02}
{Benedettini},~M.; {Viti},~S.; {Giannini},~T.; {Nisini},~B.;
  {Goldsmith},~P.~F.; {Saraceno},~P. \emph{\aap} \textbf{2002}, \emph{395},
  657\relax
\mciteBstWouldAddEndPuncttrue
\mciteSetBstMidEndSepPunct{\mcitedefaultmidpunct}
{\mcitedefaultendpunct}{\mcitedefaultseppunct}\relax
\EndOfBibitem
\bibitem[{Green}(1980)]{Green:80}
{Green},~S. \emph{\apjs} \textbf{1980}, \emph{42}, 103\relax
\mciteBstWouldAddEndPuncttrue
\mciteSetBstMidEndSepPunct{\mcitedefaultmidpunct}
{\mcitedefaultendpunct}{\mcitedefaultseppunct}\relax
\EndOfBibitem
\bibitem[{Hodges} et~al.(2002){Hodges}, {Wheatley}, and {Harvey}]{hodges:02}
{Hodges},~M.~P.; {Wheatley},~R.~J.; {Harvey},~A.~H. \emph{\jcp} \textbf{2002},
  \emph{116}, 1397\relax
\mciteBstWouldAddEndPuncttrue
\mciteSetBstMidEndSepPunct{\mcitedefaultmidpunct}
{\mcitedefaultendpunct}{\mcitedefaultseppunct}\relax
\EndOfBibitem
\bibitem[Patkowski et~al.(2002)Patkowski, Morona, Moszynski, Jeziorski, and
  Szalewicz]{patkowski:02}
Patkowski,~K.; Morona,~T.; Moszynski,~R.; Jeziorski,~B.; Szalewicz,~K. \emph{J.
  Mol. Struct.: THEOCHEM} \textbf{2002}, \emph{591}, 231\relax
\mciteBstWouldAddEndPuncttrue
\mciteSetBstMidEndSepPunct{\mcitedefaultmidpunct}
{\mcitedefaultendpunct}{\mcitedefaultseppunct}\relax
\EndOfBibitem
\bibitem[{Calderoni}(2003)]{calderoni:03}
{Calderoni},~G. \emph{Chem. Phys. Lett.} \textbf{2003}, \emph{370}, 233\relax
\mciteBstWouldAddEndPuncttrue
\mciteSetBstMidEndSepPunct{\mcitedefaultmidpunct}
{\mcitedefaultendpunct}{\mcitedefaultseppunct}\relax
\EndOfBibitem
\bibitem[{Phillips} et~al.(1996){Phillips}, {Maluendes}, and
  {Green}]{Phillips:96}
{Phillips},~T.~R.; {Maluendes},~S.; {Green},~S. \emph{\apjs} \textbf{1996},
  \emph{107}, 467\relax
\mciteBstWouldAddEndPuncttrue
\mciteSetBstMidEndSepPunct{\mcitedefaultmidpunct}
{\mcitedefaultendpunct}{\mcitedefaultseppunct}\relax
\EndOfBibitem
\bibitem[{Phillips} et~al.(1994){Phillips}, {Maluendes}, {McLean}, and
  {Green}]{Phillips:94}
{Phillips},~T.~R.; {Maluendes},~S.; {McLean},~A.~D.; {Green},~S. \emph{\jcp}
  \textbf{1994}, \emph{101}, 5824\relax
\mciteBstWouldAddEndPuncttrue
\mciteSetBstMidEndSepPunct{\mcitedefaultmidpunct}
{\mcitedefaultendpunct}{\mcitedefaultseppunct}\relax
\EndOfBibitem
\bibitem[{Dubernet} and {Grosjean}(2002){Dubernet}, and
  {Grosjean}]{Dubernet:02}
{Dubernet},~M.-L.; {Grosjean},~A. \emph{\aap} \textbf{2002}, \emph{390},
  793\relax
\mciteBstWouldAddEndPuncttrue
\mciteSetBstMidEndSepPunct{\mcitedefaultmidpunct}
{\mcitedefaultendpunct}{\mcitedefaultseppunct}\relax
\EndOfBibitem
\bibitem[{Grosjean} et~al.(2003){Grosjean}, {Dubernet}, and
  {Ceccarelli}]{Dubernet:03}
{Grosjean},~A.; {Dubernet},~M.-L.; {Ceccarelli},~C. \emph{\aap} \textbf{2003},
  \emph{408}, 1197\relax
\mciteBstWouldAddEndPuncttrue
\mciteSetBstMidEndSepPunct{\mcitedefaultmidpunct}
{\mcitedefaultendpunct}{\mcitedefaultseppunct}\relax
\EndOfBibitem
\bibitem[{Faure} et~al.(2005){Faure}, {Wiesenfeld}, {Wernli}, and
  {Valiron}]{Faure:05}
{Faure},~A.; {Wiesenfeld},~L.; {Wernli},~M.; {Valiron},~P. \emph{\jcp}
  \textbf{2005}, \emph{123}, 104309\relax
\mciteBstWouldAddEndPuncttrue
\mciteSetBstMidEndSepPunct{\mcitedefaultmidpunct}
{\mcitedefaultendpunct}{\mcitedefaultseppunct}\relax
\EndOfBibitem
\bibitem[{Dubernet} et~al.(2006){Dubernet}, {Daniel}, {Grosjean}, {Faure},
  {Valiron}, {Wernli}, {Wiesenfeld}, {Rist}, {Noga}, and
  {Tennyson}]{Dubernet:06}
{Dubernet},~M.-L.; {Daniel},~F.; {Grosjean},~A.; {Faure},~A.; {Valiron},~P.;
  {Wernli},~M.; {Wiesenfeld},~L.; {Rist},~C.; {Noga},~J.; {Tennyson},~J.
  \emph{\aap} \textbf{2006}, \emph{460}, 323\relax
\mciteBstWouldAddEndPuncttrue
\mciteSetBstMidEndSepPunct{\mcitedefaultmidpunct}
{\mcitedefaultendpunct}{\mcitedefaultseppunct}\relax
\EndOfBibitem
\bibitem[{Faure} et~al.(2007){Faure}, {Crimier}, {Ceccarelli}, {Valiron},
  {Wiesenfeld}, and {Dubernet}]{Faure:07}
{Faure},~A.; {Crimier},~N.; {Ceccarelli},~C.; {Valiron},~P.; {Wiesenfeld},~L.;
  {Dubernet},~M.~L. \emph{\aap} \textbf{2007}, \emph{472}, 1029\relax
\mciteBstWouldAddEndPuncttrue
\mciteSetBstMidEndSepPunct{\mcitedefaultmidpunct}
{\mcitedefaultendpunct}{\mcitedefaultseppunct}\relax
\EndOfBibitem
\bibitem[{Faure} and {Josselin}(2008){Faure}, and {Josselin}]{Faure:08}
{Faure},~A.; {Josselin},~E. \emph{\aap} \textbf{2008}, \emph{492}, 257\relax
\mciteBstWouldAddEndPuncttrue
\mciteSetBstMidEndSepPunct{\mcitedefaultmidpunct}
{\mcitedefaultendpunct}{\mcitedefaultseppunct}\relax
\EndOfBibitem
\bibitem[{van der Avoird} and {Nesbitt}(2011){van der Avoird}, and
  {Nesbitt}]{Avoird:11}
{van der Avoird},~A.; {Nesbitt},~D.~J. \emph{\jcp} \textbf{2011}, \emph{134},
  044314\relax
\mciteBstWouldAddEndPuncttrue
\mciteSetBstMidEndSepPunct{\mcitedefaultmidpunct}
{\mcitedefaultendpunct}{\mcitedefaultseppunct}\relax
\EndOfBibitem
\bibitem[{Weida} and {Nesbitt}(1999){Weida}, and {Nesbitt}]{Weida:99}
{Weida},~M.~J.; {Nesbitt},~D.~J. \emph{\jcp} \textbf{1999}, \emph{110},
  156\relax
\mciteBstWouldAddEndPuncttrue
\mciteSetBstMidEndSepPunct{\mcitedefaultmidpunct}
{\mcitedefaultendpunct}{\mcitedefaultseppunct}\relax
\EndOfBibitem
\bibitem[{Daniel} et~al.(2012){Daniel}, {Goicoechea}, {Cernicharo}, {Dubernet},
  and {Faure}]{Daniel:12}
{Daniel},~F.; {Goicoechea},~J.~R.; {Cernicharo},~J.; {Dubernet},~M.-L.;
  {Faure},~A. \emph{\aap} \textbf{2012}, \emph{547}, A81\relax
\mciteBstWouldAddEndPuncttrue
\mciteSetBstMidEndSepPunct{\mcitedefaultmidpunct}
{\mcitedefaultendpunct}{\mcitedefaultseppunct}\relax
\EndOfBibitem
\bibitem[{McKellar}(1940)]{mckellar:40}
{McKellar},~A. \emph{\pasp} \textbf{1940}, \emph{52}, 187\relax
\mciteBstWouldAddEndPuncttrue
\mciteSetBstMidEndSepPunct{\mcitedefaultmidpunct}
{\mcitedefaultendpunct}{\mcitedefaultseppunct}\relax
\EndOfBibitem
\bibitem[{Adams}(1941)]{adams:41}
{Adams},~W.~S. \emph{\apj} \textbf{1941}, \emph{93}, 11\relax
\mciteBstWouldAddEndPuncttrue
\mciteSetBstMidEndSepPunct{\mcitedefaultmidpunct}
{\mcitedefaultendpunct}{\mcitedefaultseppunct}\relax
\EndOfBibitem
\bibitem[{Fuente} et~al.(1995){Fuente}, {Martin-Pintado}, and
  {Gaume}]{fuente:95}
{Fuente},~A.; {Martin-Pintado},~J.; {Gaume},~R. \emph{\apjl} \textbf{1995},
  \emph{442}, L33\relax
\mciteBstWouldAddEndPuncttrue
\mciteSetBstMidEndSepPunct{\mcitedefaultmidpunct}
{\mcitedefaultendpunct}{\mcitedefaultseppunct}\relax
\EndOfBibitem
\bibitem[{Hakobian} and {Crutcher}(2011){Hakobian}, and
  {Crutcher}]{Hakobian:11}
{Hakobian},~N.~S.; {Crutcher},~R.~M. \emph{\apj} \textbf{2011}, \emph{733},
  6\relax
\mciteBstWouldAddEndPuncttrue
\mciteSetBstMidEndSepPunct{\mcitedefaultmidpunct}
{\mcitedefaultendpunct}{\mcitedefaultseppunct}\relax
\EndOfBibitem
\bibitem[{Hily-Blant} et~al.(2008){Hily-Blant}, {Walmsley}, {Pineau Des
  For{\^e}ts}, and {Flower}]{hilyblant:08}
{Hily-Blant},~P.; {Walmsley},~M.; {Pineau Des For{\^e}ts},~G.; {Flower},~D.
  \emph{\aap} \textbf{2008}, \emph{480}, L5\relax
\mciteBstWouldAddEndPuncttrue
\mciteSetBstMidEndSepPunct{\mcitedefaultmidpunct}
{\mcitedefaultendpunct}{\mcitedefaultseppunct}\relax
\EndOfBibitem
\bibitem[{Truong-Bach} et~al.(1987){Truong-Bach}, {Nguyen-Q-Rieu}, {Omont},
  {Olofsson}, and {Johansson}]{Truong87}
{Truong-Bach},; {Nguyen-Q-Rieu},; {Omont},~A.; {Olofsson},~H.;
  {Johansson},~L.~E.~B. \emph{\aap} \textbf{1987}, \emph{176}, 285\relax
\mciteBstWouldAddEndPuncttrue
\mciteSetBstMidEndSepPunct{\mcitedefaultmidpunct}
{\mcitedefaultendpunct}{\mcitedefaultseppunct}\relax
\EndOfBibitem
\bibitem[Fei et~al.(1994)Fei, Lambert, Carrington, Filseth, Sadowski, and
  Dugan]{fei:94}
Fei,~R.; Lambert,~H.~M.; Carrington,~T.; Filseth,~S.~V.; Sadowski,~C.~M.;
  Dugan,~C.~H. \emph{\jcp} \textbf{1994}, \emph{100}, 1190\relax
\mciteBstWouldAddEndPuncttrue
\mciteSetBstMidEndSepPunct{\mcitedefaultmidpunct}
{\mcitedefaultendpunct}{\mcitedefaultseppunct}\relax
\EndOfBibitem
\bibitem[{Walmsley} et~al.(1982){Walmsley}, {Churchwell}, {Nash}, and
  {Fitzpatrick}]{Walmsley:82}
{Walmsley},~C.~M.; {Churchwell},~E.; {Nash},~A.; {Fitzpatrick},~E. \emph{\apjl}
  \textbf{1982}, \emph{258}, L75\relax
\mciteBstWouldAddEndPuncttrue
\mciteSetBstMidEndSepPunct{\mcitedefaultmidpunct}
{\mcitedefaultendpunct}{\mcitedefaultseppunct}\relax
\EndOfBibitem
\bibitem[{Turner} et~al.(1997){Turner}, {Pirogov}, and {Minh}]{Turner:97}
{Turner},~B.~E.; {Pirogov},~L.; {Minh},~Y.~C. \emph{\apj} \textbf{1997},
  \emph{483}, 235\relax
\mciteBstWouldAddEndPuncttrue
\mciteSetBstMidEndSepPunct{\mcitedefaultmidpunct}
{\mcitedefaultendpunct}{\mcitedefaultseppunct}\relax
\EndOfBibitem
\bibitem[{Hirota} et~al.(1998){Hirota}, {Yamamoto}, {Mikami}, and
  {Ohishi}]{hirota98}
{Hirota},~T.; {Yamamoto},~S.; {Mikami},~H.; {Ohishi},~M. \emph{\apj}
  \textbf{1998}, \emph{503}, 717\relax
\mciteBstWouldAddEndPuncttrue
\mciteSetBstMidEndSepPunct{\mcitedefaultmidpunct}
{\mcitedefaultendpunct}{\mcitedefaultseppunct}\relax
\EndOfBibitem
\bibitem[{Schilke} et~al.(1992){Schilke}, {Walmsley}, {Pineau Des Forets},
  {Roueff}, {Flower}, and {Guilloteau}]{schilke92}
{Schilke},~P.; {Walmsley},~C.~M.; {Pineau Des Forets},~G.; {Roueff},~E.;
  {Flower},~D.~R.; {Guilloteau},~S. \emph{\aap} \textbf{1992}, \emph{256},
  595\relax
\mciteBstWouldAddEndPuncttrue
\mciteSetBstMidEndSepPunct{\mcitedefaultmidpunct}
{\mcitedefaultendpunct}{\mcitedefaultseppunct}\relax
\EndOfBibitem
\bibitem[{Talbi} et~al.(1996){Talbi}, {Ellinger}, and {Herbst}]{talbi:96}
{Talbi},~D.; {Ellinger},~Y.; {Herbst},~E. \emph{\aap} \textbf{1996},
  \emph{314}, 688\relax
\mciteBstWouldAddEndPuncttrue
\mciteSetBstMidEndSepPunct{\mcitedefaultmidpunct}
{\mcitedefaultendpunct}{\mcitedefaultseppunct}\relax
\EndOfBibitem
\bibitem[{Sarrasin} et~al.(2010){Sarrasin}, {Abdallah}, {Wernli}, {Faure},
  {Cernicharo}, and {Lique}]{Sarrasin:10}
{Sarrasin},~E.; {Abdallah},~D.~B.; {Wernli},~M.; {Faure},~A.; {Cernicharo},~J.;
  {Lique},~F. \emph{\mnras} \textbf{2010}, \emph{404}, 518\relax
\mciteBstWouldAddEndPuncttrue
\mciteSetBstMidEndSepPunct{\mcitedefaultmidpunct}
{\mcitedefaultendpunct}{\mcitedefaultseppunct}\relax
\EndOfBibitem
\bibitem[{Dayou} and {Balan{\c c}a}(2006){Dayou}, and {Balan{\c c}a}]{Dayou:06}
{Dayou},~F.; {Balan{\c c}a},~C. \emph{\aap} \textbf{2006}, \emph{459},
  297\relax
\mciteBstWouldAddEndPuncttrue
\mciteSetBstMidEndSepPunct{\mcitedefaultmidpunct}
{\mcitedefaultendpunct}{\mcitedefaultseppunct}\relax
\EndOfBibitem
\bibitem[{Turner} et~al.(1992){Turner}, {Chan}, {Green}, and
  {Lubowich}]{turner:92}
{Turner},~B.~E.; {Chan},~K.-W.; {Green},~S.; {Lubowich},~D.~A. \emph{\apj}
  \textbf{1992}, \emph{399}, 114\relax
\mciteBstWouldAddEndPuncttrue
\mciteSetBstMidEndSepPunct{\mcitedefaultmidpunct}
{\mcitedefaultendpunct}{\mcitedefaultseppunct}\relax
\EndOfBibitem
\bibitem[Corey et~al.(1986)Corey, Alexander, and Schaefer]{corey:86}
Corey,~G.~C.; Alexander,~M.~H.; Schaefer,~J. \emph{\jcp} \textbf{1986},
  \emph{85}, 2726\relax
\mciteBstWouldAddEndPuncttrue
\mciteSetBstMidEndSepPunct{\mcitedefaultmidpunct}
{\mcitedefaultendpunct}{\mcitedefaultseppunct}\relax
\EndOfBibitem
\bibitem[Lique et~al.(2006)Lique, Cernicharo, and Cox]{lique:06}
Lique,~F.; Cernicharo,~J.; Cox,~P. \emph{Astrophys. J.} \textbf{2006},
  \emph{653}, 1342\relax
\mciteBstWouldAddEndPuncttrue
\mciteSetBstMidEndSepPunct{\mcitedefaultmidpunct}
{\mcitedefaultendpunct}{\mcitedefaultseppunct}\relax
\EndOfBibitem
\bibitem[{Spielfiedel} et~al.(2009){Spielfiedel}, {Senent}, {Dayou}, {Balan{\c
  c}a}, {Cressiot-Vincent}, {Faure}, {Wiesenfeld}, and
  {Feautrier}]{spielfiedel:09}
{Spielfiedel},~A.; {Senent},~M.-L.; {Dayou},~F.; {Balan{\c c}a},~C.;
  {Cressiot-Vincent},~L.; {Faure},~A.; {Wiesenfeld},~L.; {Feautrier},~N.
  \emph{\jcp} \textbf{2009}, \emph{131}, 014305\relax
\mciteBstWouldAddEndPuncttrue
\mciteSetBstMidEndSepPunct{\mcitedefaultmidpunct}
{\mcitedefaultendpunct}{\mcitedefaultseppunct}\relax
\EndOfBibitem
\bibitem[{Cernicharo} et~al.(2011){Cernicharo}, {Spielfiedel}, {Balan{\c c}a},
  {Dayou}, {Senent}, {Feautrier}, {Faure}, {Cressiot-Vincent}, {Wiesenfeld},
  and {Pardo}]{Cernicharo:11}
{Cernicharo},~J.; {Spielfiedel},~A.; {Balan{\c c}a},~C.; {Dayou},~F.;
  {Senent},~M.-L.; {Feautrier},~N.; {Faure},~A.; {Cressiot-Vincent},~L.;
  {Wiesenfeld},~L.; {Pardo},~J.~R. \emph{\aap} \textbf{2011}, \emph{531},
  A103\relax
\mciteBstWouldAddEndPuncttrue
\mciteSetBstMidEndSepPunct{\mcitedefaultmidpunct}
{\mcitedefaultendpunct}{\mcitedefaultseppunct}\relax
\EndOfBibitem
\bibitem[{Green}(1995)]{green95}
{Green},~S. \emph{\apjs} \textbf{1995}, \emph{100}, 213\relax
\mciteBstWouldAddEndPuncttrue
\mciteSetBstMidEndSepPunct{\mcitedefaultmidpunct}
{\mcitedefaultendpunct}{\mcitedefaultseppunct}\relax
\EndOfBibitem
\bibitem[{Cheung} et~al.(1968){Cheung}, {Rank}, {Townes}, {Thornton}, and
  {Welch}]{Cheung:68}
{Cheung},~A.~C.; {Rank},~D.~M.; {Townes},~C.~H.; {Thornton},~D.~D.;
  {Welch},~W.~J. \emph{Phys. Rev. Lett.} \textbf{1968}, \emph{21}, 1701\relax
\mciteBstWouldAddEndPuncttrue
\mciteSetBstMidEndSepPunct{\mcitedefaultmidpunct}
{\mcitedefaultendpunct}{\mcitedefaultseppunct}\relax
\EndOfBibitem
\bibitem[{Persson} et~al.(2010){Persson}, {Black}, {Cernicharo}, {Goicoechea},
  {Hassel}, {Herbst}, {Gerin}, {de Luca}, {Bell}, {Coutens}, {Falgarone},
  {Goldsmith}, {Gupta}, {Ka{\'z}mierczak}, {Lis}, {Mookerjea}, {Neufeld},
  {Pearson}, {Phillips}, {Sonnentrucker}, {Stutzki}, {Vastel}, {Yu},
  {Boulanger}, {Dartois}, {Encrenaz}, {Geballe}, {Giesen}, {Godard}, {Gry},
  {Hennebelle}, {Hily-Blant}, {Joblin}, {Ko{\l}os}, {Kre{\l}owski},
  {Mart{\'{\i}}n-Pintado}, {Menten}, {Monje}, {Perault}, {Plume}, {Salez},
  {Schlemmer}, {Schmidt}, {Teyssier}, {P{\'e}ron}, {Cais}, {Gaufre}, {Cros},
  {Ravera}, {Morris}, {Lord}, and {Planesas}]{persson:10}
{Persson},~C.~M. et~al.  \emph{\aap} \textbf{2010}, \emph{521}, L45\relax
\mciteBstWouldAddEndPuncttrue
\mciteSetBstMidEndSepPunct{\mcitedefaultmidpunct}
{\mcitedefaultendpunct}{\mcitedefaultseppunct}\relax
\EndOfBibitem
\bibitem[{Hily-Blant} et~al.(2010){Hily-Blant}, {Maret}, {Bacmann},
  {Bottinelli}, {Parise}, {Caux}, {Faure}, {Bergin}, {Blake}, {Castets},
  {Ceccarelli}, {Cernicharo}, {Coutens}, {Crimier}, {Demyk}, {Dominik},
  {Gerin}, {Hennebelle}, {Henning}, {Kahane}, {Klotz}, {Melnick}, {Pagani},
  {Schilke}, {Vastel}, {Wakelam}, {Walters}, {Baudry}, {Bell}, {Benedettini},
  {Boogert}, {Cabrit}, {Caselli}, {Codella}, {Comito}, {Encrenaz}, {Falgarone},
  {Fuente}, {Goldsmith}, {Helmich}, {Herbst}, {Jacq}, {Kama}, {Langer},
  {Lefloch}, {Lis}, {Lord}, {Lorenzani}, {Neufeld}, {Nisini}, {Pacheco},
  {Phillips}, {Salez}, {Saraceno}, {Schuster}, {Tielens}, {van der Tak}, {van
  der Wiel}, {Viti}, {Wyrowski}, and {Yorke}]{hily10}
{Hily-Blant},~P. et~al.  \emph{\aap} \textbf{2010}, \emph{521}, L52\relax
\mciteBstWouldAddEndPuncttrue
\mciteSetBstMidEndSepPunct{\mcitedefaultmidpunct}
{\mcitedefaultendpunct}{\mcitedefaultseppunct}\relax
\EndOfBibitem
\bibitem[{Meyer} and {Roth}(1991){Meyer}, and {Roth}]{meyer:91}
{Meyer},~D.~M.; {Roth},~K.~C. \emph{\apjl} \textbf{1991}, \emph{376}, L49\relax
\mciteBstWouldAddEndPuncttrue
\mciteSetBstMidEndSepPunct{\mcitedefaultmidpunct}
{\mcitedefaultendpunct}{\mcitedefaultseppunct}\relax
\EndOfBibitem
\bibitem[{van Dishoeck} et~al.(1993){van Dishoeck}, {Jansen}, {Schilke}, and
  {Phillips}]{vandishoeck:93}
{van Dishoeck},~E.~F.; {Jansen},~D.~J.; {Schilke},~P.; {Phillips},~T.~G.
  \emph{\apjl} \textbf{1993}, \emph{416}, L83\relax
\mciteBstWouldAddEndPuncttrue
\mciteSetBstMidEndSepPunct{\mcitedefaultmidpunct}
{\mcitedefaultendpunct}{\mcitedefaultseppunct}\relax
\EndOfBibitem
\bibitem[{Willey} et~al.(2002){Willey}, {Timlin}, {Merlin}, {Sowa}, and
  {Wesolek}]{willey:02}
{Willey},~D.~R.; {Timlin},~R.~E.,~Jr.; {Merlin},~J.~M.; {Sowa},~M.~M.;
  {Wesolek},~D.~M. \emph{\apjs} \textbf{2002}, \emph{139}, 191\relax
\mciteBstWouldAddEndPuncttrue
\mciteSetBstMidEndSepPunct{\mcitedefaultmidpunct}
{\mcitedefaultendpunct}{\mcitedefaultseppunct}\relax
\EndOfBibitem
\bibitem[{Hodges} and {Wheatley}(2001){Hodges}, and {Wheatley}]{Hodges:01}
{Hodges},~M.~P.; {Wheatley},~R.~J. \emph{\jcp} \textbf{2001}, \emph{114},
  8836\relax
\mciteBstWouldAddEndPuncttrue
\mciteSetBstMidEndSepPunct{\mcitedefaultmidpunct}
{\mcitedefaultendpunct}{\mcitedefaultseppunct}\relax
\EndOfBibitem
\bibitem[Schleipen and {ter Meulen}(1991)Schleipen, and {ter
  Meulen}]{schleipen:91}
Schleipen,~J.; {ter Meulen},~J.~J. \emph{Chem. Phys.} \textbf{1991},
  \emph{156}, 479\relax
\mciteBstWouldAddEndPuncttrue
\mciteSetBstMidEndSepPunct{\mcitedefaultmidpunct}
{\mcitedefaultendpunct}{\mcitedefaultseppunct}\relax
\EndOfBibitem
\bibitem[Rinnenthal and Gericke(2002)Rinnenthal, and
  Gericke]{rinnenthaltully:02}
Rinnenthal,~J.~L.; Gericke,~K.-H. \emph{\jcp} \textbf{2002}, \emph{116},
  9776\relax
\mciteBstWouldAddEndPuncttrue
\mciteSetBstMidEndSepPunct{\mcitedefaultmidpunct}
{\mcitedefaultendpunct}{\mcitedefaultseppunct}\relax
\EndOfBibitem
\bibitem[{Larsson} et~al.(2007){Larsson}, {Liseau}, {Pagani}, {Bergman},
  {Bernath}, {Biver}, {Black}, {Booth}, {Buat}, {Crovisier}, {Curry},
  {Dahlgren}, {Encrenaz}, {Falgarone}, {Feldman}, {Fich}, {Flor{\'e}n},
  {Fredrixon}, {Frisk}, {Gahm}, {Gerin}, {Hagstr{\"o}m}, {Harju}, {Hasegawa},
  {Hjalmarson}, {Johansson}, {Justtanont}, {Klotz}, {Kyr{\"o}l{\"a}}, {Kwok},
  {Lecacheux}, {Liljestr{\"o}m}, {Llewellyn}, {Lundin}, {M{\'e}gie},
  {Mitchell}, {Murtagh}, {Nordh}, {Nyman}, {Olberg}, {Olofsson}, {Olofsson},
  {Olofsson}, {Persson}, {Plume}, {Rickman}, {Ristorcelli}, {Rydbeck},
  {Sandqvist}, {Sch{\'e}ele}, {Serra}, {Torchinsky}, {Tothill}, {Volk},
  {Wiklind}, {Wilson}, {Winnberg}, and {Witt}]{larsson:07}
{Larsson},~B. et~al.  \emph{\aap} \textbf{2007}, \emph{466}, 999\relax
\mciteBstWouldAddEndPuncttrue
\mciteSetBstMidEndSepPunct{\mcitedefaultmidpunct}
{\mcitedefaultendpunct}{\mcitedefaultseppunct}\relax
\EndOfBibitem
\bibitem[{Goldsmith} et~al.(2011){Goldsmith}, {Liseau}, {Bell}, {Black},
  {Chen}, {Hollenbach}, {Kaufman}, {Li}, {Lis}, {Melnick}, {Neufeld}, {Pagani},
  {Snell}, {Benz}, {Bergin}, {Bruderer}, {Caselli}, {Caux}, {Encrenaz},
  {Falgarone}, {Gerin}, {Goicoechea}, {Hjalmarson}, {Larsson}, {Le Bourlot},
  {Le Petit}, {De Luca}, {Nagy}, {Roueff}, {Sandqvist}, {van der Tak}, {van
  Dishoeck}, {Vastel}, {Viti}, and {Y{\i}ld{\i}z}]{goldsmith:11}
{Goldsmith},~P.~F. et~al.  \emph{\apj} \textbf{2011}, \emph{737}, 96\relax
\mciteBstWouldAddEndPuncttrue
\mciteSetBstMidEndSepPunct{\mcitedefaultmidpunct}
{\mcitedefaultendpunct}{\mcitedefaultseppunct}\relax
\EndOfBibitem
\bibitem[{Liseau} et~al.(2012){Liseau}, {Goldsmith}, {Larsson}, {Pagani},
  {Bergman}, {Le Bourlot}, {Bell}, {Benz}, {Bergin}, {Bjerkeli}, {Black},
  {Bruderer}, {Caselli}, {Caux}, {Chen}, {de Luca}, {Encrenaz}, {Falgarone},
  {Gerin}, {Goicoechea}, {Hjalmarson}, {Hollenbach}, {Justtanont}, {Kaufman},
  {Le Petit}, {Li}, {Lis}, {Melnick}, {Nagy}, {Olofsson}, {Olofsson}, {Roueff},
  {Sandqvist}, {Snell}, {van der Tak}, {van Dishoeck}, {Vastel}, {Viti}, and
  {Y{\i}ld{\i}z}]{liseau:12}
{Liseau},~R. et~al.  \emph{\aap} \textbf{2012}, \emph{541}, A73\relax
\mciteBstWouldAddEndPuncttrue
\mciteSetBstMidEndSepPunct{\mcitedefaultmidpunct}
{\mcitedefaultendpunct}{\mcitedefaultseppunct}\relax
\EndOfBibitem
\bibitem[Carty et~al.(2006)Carty, Goddard, K{\"o}hler, Sims, and
  Smith]{carty:06}
Carty,~D.; Goddard,~A.; K{\"o}hler,~S. P.~K.; Sims,~I.~R.; Smith,~I. W.~M.
  \emph{J. Phys. Chem. A} \textbf{2006}, \emph{110}, 3101\relax
\mciteBstWouldAddEndPuncttrue
\mciteSetBstMidEndSepPunct{\mcitedefaultmidpunct}
{\mcitedefaultendpunct}{\mcitedefaultseppunct}\relax
\EndOfBibitem
\bibitem[{Lique} et~al.(2009){Lique}, {Jorfi}, {Honvault}, {Halvick}, {Lin},
  {Guo}, {Xie}, {Dagdigian}, {K{\l}os}, and {Alexander}]{Lique:09O2}
{Lique},~F.; {Jorfi},~M.; {Honvault},~P.; {Halvick},~P.; {Lin},~S.~Y.;
  {Guo},~H.; {Xie},~D.~Q.; {Dagdigian},~P.~J.; {K{\l}os},~J.;
  {Alexander},~M.~H. \emph{\jcp} \textbf{2009}, \emph{131}, 221104\relax
\mciteBstWouldAddEndPuncttrue
\mciteSetBstMidEndSepPunct{\mcitedefaultmidpunct}
{\mcitedefaultendpunct}{\mcitedefaultseppunct}\relax
\EndOfBibitem
\bibitem[{Liszt} and {Turner}(1978){Liszt}, and {Turner}]{liszt:78}
{Liszt},~H.~S.; {Turner},~B.~E. \emph{\apjl} \textbf{1978}, \emph{224},
  L73\relax
\mciteBstWouldAddEndPuncttrue
\mciteSetBstMidEndSepPunct{\mcitedefaultmidpunct}
{\mcitedefaultendpunct}{\mcitedefaultseppunct}\relax
\EndOfBibitem
\bibitem[{Gerin} et~al.(1992){Gerin}, {Viala}, {Pauzat}, and
  {Ellinger}]{gerin:92}
{Gerin},~M.; {Viala},~Y.; {Pauzat},~F.; {Ellinger},~Y. \emph{\aap}
  \textbf{1992}, \emph{266}, 463\relax
\mciteBstWouldAddEndPuncttrue
\mciteSetBstMidEndSepPunct{\mcitedefaultmidpunct}
{\mcitedefaultendpunct}{\mcitedefaultseppunct}\relax
\EndOfBibitem
\bibitem[{Dewangan} et~al.(1987){Dewangan}, {Flower}, and
  {Alexander}]{dewangan:87}
{Dewangan},~D.~P.; {Flower},~D.~R.; {Alexander},~M.~H. \emph{\mnras}
  \textbf{1987}, \emph{226}, 505\relax
\mciteBstWouldAddEndPuncttrue
\mciteSetBstMidEndSepPunct{\mcitedefaultmidpunct}
{\mcitedefaultendpunct}{\mcitedefaultseppunct}\relax
\EndOfBibitem
\bibitem[{Schreel} et~al.(1993){Schreel}, {Schleipen}, {Eppink}, and {Ter
  Meulen}]{Schreel:93}
{Schreel},~K.; {Schleipen},~J.; {Eppink},~A.; {Ter Meulen},~J.~J. \emph{\jcp}
  \textbf{1993}, \emph{99}, 8713\relax
\mciteBstWouldAddEndPuncttrue
\mciteSetBstMidEndSepPunct{\mcitedefaultmidpunct}
{\mcitedefaultendpunct}{\mcitedefaultseppunct}\relax
\EndOfBibitem
\bibitem[{Kirste} et~al.(2010){Kirste}, {Scharfenberg}, {K{\l}os}, {Lique},
  {Alexander}, {Meijer}, and {van de Meerakker}]{Kirste:10}
{Kirste},~M.; {Scharfenberg},~L.; {K{\l}os},~J.; {Lique},~F.;
  {Alexander},~M.~H.; {Meijer},~G.; {van de Meerakker},~S.~Y.~T. \emph{\pra}
  \textbf{2010}, \emph{82}, 042717\relax
\mciteBstWouldAddEndPuncttrue
\mciteSetBstMidEndSepPunct{\mcitedefaultmidpunct}
{\mcitedefaultendpunct}{\mcitedefaultseppunct}\relax
\EndOfBibitem
\bibitem[{Brown} et~al.(1975){Brown}, {Hougen}, {Huber}, {Johns}, {Kopp},
  {Lefebvre-Brion}, {Merer}, {Ramsay}, {Rostas}, and {Zare}]{brown:75}
{Brown},~J.~M.; {Hougen},~J.~T.; {Huber},~K.-P.; {Johns},~J.~W.~C.; {Kopp},~I.;
  {Lefebvre-Brion},~H.; {Merer},~A.~J.; {Ramsay},~D.~A.; {Rostas},~J.;
  {Zare},~R.~N. \emph{J. Mol. Spect.} \textbf{1975}, \emph{55}, 500\relax
\mciteBstWouldAddEndPuncttrue
\mciteSetBstMidEndSepPunct{\mcitedefaultmidpunct}
{\mcitedefaultendpunct}{\mcitedefaultseppunct}\relax
\EndOfBibitem
\bibitem[{Guelin} et~al.(1978){Guelin}, {Green}, and {Thaddeus}]{guelin:78}
{Guelin},~M.; {Green},~S.; {Thaddeus},~P. \emph{\apjl} \textbf{1978},
  \emph{224}, L27\relax
\mciteBstWouldAddEndPuncttrue
\mciteSetBstMidEndSepPunct{\mcitedefaultmidpunct}
{\mcitedefaultendpunct}{\mcitedefaultseppunct}\relax
\EndOfBibitem
\bibitem[{Cernicharo} et~al.(1991){Cernicharo}, {Gottlieb}, {Guelin},
  {Killian}, {Paubert}, {Thaddeus}, and {Vrtilek}]{cernicharo:91a}
{Cernicharo},~J.; {Gottlieb},~C.~A.; {Guelin},~M.; {Killian},~T.~C.;
  {Paubert},~G.; {Thaddeus},~P.; {Vrtilek},~J.~M. \emph{\apjl} \textbf{1991},
  \emph{368}, L39\relax
\mciteBstWouldAddEndPuncttrue
\mciteSetBstMidEndSepPunct{\mcitedefaultmidpunct}
{\mcitedefaultendpunct}{\mcitedefaultseppunct}\relax
\EndOfBibitem
\bibitem[{Cernicharo} et~al.(1991){Cernicharo}, {Gottlieb}, {Guelin},
  {Killian}, {Thaddeus}, and {Vrtilek}]{cernicharo:91b}
{Cernicharo},~J.; {Gottlieb},~C.~A.; {Guelin},~M.; {Killian},~T.~C.;
  {Thaddeus},~P.; {Vrtilek},~J.~M. \emph{\apjl} \textbf{1991}, \emph{368},
  L43\relax
\mciteBstWouldAddEndPuncttrue
\mciteSetBstMidEndSepPunct{\mcitedefaultmidpunct}
{\mcitedefaultendpunct}{\mcitedefaultseppunct}\relax
\EndOfBibitem
\bibitem[{Guelin} et~al.(1997){Guelin}, {Cernicharo}, {Travers}, {McCarthy},
  {Gottlieb}, {Thaddeus}, {Ohishi}, {Saito}, and {Yamamoto}]{guelin:97}
{Guelin},~M.; {Cernicharo},~J.; {Travers},~M.~J.; {McCarthy},~M.~C.;
  {Gottlieb},~C.~A.; {Thaddeus},~P.; {Ohishi},~M.; {Saito},~S.; {Yamamoto},~S.
  \emph{\aap} \textbf{1997}, \emph{317}, L1\relax
\mciteBstWouldAddEndPuncttrue
\mciteSetBstMidEndSepPunct{\mcitedefaultmidpunct}
{\mcitedefaultendpunct}{\mcitedefaultseppunct}\relax
\EndOfBibitem
\bibitem[{{\'A}d{\'a}mkovics} et~al.(2003){{\'A}d{\'a}mkovics}, {Blake}, and
  {McCall}]{adamkovics:03}
{{\'A}d{\'a}mkovics},~M.; {Blake},~G.~A.; {McCall},~B.~J. \emph{\apj}
  \textbf{2003}, \emph{595}, 235\relax
\mciteBstWouldAddEndPuncttrue
\mciteSetBstMidEndSepPunct{\mcitedefaultmidpunct}
{\mcitedefaultendpunct}{\mcitedefaultseppunct}\relax
\EndOfBibitem
\bibitem[{Iglesias-Groth}(2011)]{iglesias:11}
{Iglesias-Groth},~S. \emph{\mnras} \textbf{2011}, \emph{411}, 1857\relax
\mciteBstWouldAddEndPuncttrue
\mciteSetBstMidEndSepPunct{\mcitedefaultmidpunct}
{\mcitedefaultendpunct}{\mcitedefaultseppunct}\relax
\EndOfBibitem
\bibitem[{Welty} et~al.(2013){Welty}, {Howk}, {Lehner}, and {Black}]{welty:13}
{Welty},~D.~E.; {Howk},~J.~C.; {Lehner},~N.; {Black},~J.~H. \emph{\mnras}
  \textbf{2013}, \emph{428}, 1107\relax
\mciteBstWouldAddEndPuncttrue
\mciteSetBstMidEndSepPunct{\mcitedefaultmidpunct}
{\mcitedefaultendpunct}{\mcitedefaultseppunct}\relax
\EndOfBibitem
\bibitem[{Wang} et~al.(2009){Wang}, {Kuan}, {Liu}, {Huang}, and
  {Charnley}]{wang:09}
{Wang},~K.-S.; {Kuan},~Y.-J.; {Liu},~S.-Y.; {Huang},~H.-C.; {Charnley},~S.~B.
  In \emph{Bioastronomy 2007: Molecules, Microbes and Extraterrestrial Life};
  {Meech},~K.~J., {Keane},~J.~V., {Mumma},~M.~J., {Siefert},~J.~L.,
  {Werthimer},~D.~J., Eds.; Astron. Soc. Pacific Conf. Ser.; 2009; Vol. 420;
  p~49\relax
\mciteBstWouldAddEndPuncttrue
\mciteSetBstMidEndSepPunct{\mcitedefaultmidpunct}
{\mcitedefaultendpunct}{\mcitedefaultseppunct}\relax
\EndOfBibitem
\bibitem[{Sakai} et~al.(2012){Sakai}, {Ceccarelli}, {Bottinelli}, {Sakai}, and
  {Yamamoto}]{sakai:12}
{Sakai},~N.; {Ceccarelli},~C.; {Bottinelli},~S.; {Sakai},~T.; {Yamamoto},~S.
  \emph{\apj} \textbf{2012}, \emph{754}, 70\relax
\mciteBstWouldAddEndPuncttrue
\mciteSetBstMidEndSepPunct{\mcitedefaultmidpunct}
{\mcitedefaultendpunct}{\mcitedefaultseppunct}\relax
\EndOfBibitem
\bibitem[{Caux} et~al.(2011){Caux}, {Kahane}, {Castets}, {Coutens},
  {Ceccarelli}, {Bacmann}, {Bisschop}, {Bottinelli}, {Comito}, {Helmich},
  {Lefloch}, {Parise}, {Schilke}, {Tielens}, {van Dishoeck}, {Vastel},
  {Wakelam}, and {Walters}]{caux:11}
{Caux},~E.; {Kahane},~C.; {Castets},~A.; {Coutens},~A.; {Ceccarelli},~C.;
  {Bacmann},~A.; {Bisschop},~S.; {Bottinelli},~S.; {Comito},~C.;
  {Helmich},~F.~P.; {Lefloch},~B.; {Parise},~B.; {Schilke},~P.;
  {Tielens},~A.~G.~G.~M.; {van Dishoeck},~E.; {Vastel},~C.; {Wakelam},~V.;
  {Walters},~A. \emph{\aap} \textbf{2011}, \emph{532}, A23\relax
\mciteBstWouldAddEndPuncttrue
\mciteSetBstMidEndSepPunct{\mcitedefaultmidpunct}
{\mcitedefaultendpunct}{\mcitedefaultseppunct}\relax
\EndOfBibitem
\bibitem[{De Lucia}(2006)]{delucia:06}
{De Lucia},~F. {Exterminating the weeds in the astronomical garden}. Complex
  Molecules in Space: Present Status and Prospects with ALMA. 2006\relax
\mciteBstWouldAddEndPuncttrue
\mciteSetBstMidEndSepPunct{\mcitedefaultmidpunct}
{\mcitedefaultendpunct}{\mcitedefaultseppunct}\relax
\EndOfBibitem
\bibitem[{Palmer} et~al.(1969){Palmer}, {Zuckerman}, {Buhl}, and
  {Snyder}]{palmer:69}
{Palmer},~P.; {Zuckerman},~B.; {Buhl},~D.; {Snyder},~L.~E. \emph{\apjl}
  \textbf{1969}, \emph{156}, L147\relax
\mciteBstWouldAddEndPuncttrue
\mciteSetBstMidEndSepPunct{\mcitedefaultmidpunct}
{\mcitedefaultendpunct}{\mcitedefaultseppunct}\relax
\EndOfBibitem
\bibitem[{Troscompt} et~al.(2009){Troscompt}, {Faure}, {Maret}, {Ceccarelli},
  {Hily-Blant}, and {Wiesenfeld}]{Troscompt:09b}
{Troscompt},~N.; {Faure},~A.; {Maret},~S.; {Ceccarelli},~C.; {Hily-Blant},~P.;
  {Wiesenfeld},~L. \emph{\aap} \textbf{2009}, \emph{506}, 1243\relax
\mciteBstWouldAddEndPuncttrue
\mciteSetBstMidEndSepPunct{\mcitedefaultmidpunct}
{\mcitedefaultendpunct}{\mcitedefaultseppunct}\relax
\EndOfBibitem
\bibitem[{Pottage} et~al.(2001){Pottage}, {Flower}, and {Davis}]{pottage:01}
{Pottage},~J.~T.; {Flower},~D.~R.; {Davis},~S.~L. \emph{J. Phys. B At. Mol.
  Opt. Phys.} \textbf{2001}, \emph{34}, 3313\relax
\mciteBstWouldAddEndPuncttrue
\mciteSetBstMidEndSepPunct{\mcitedefaultmidpunct}
{\mcitedefaultendpunct}{\mcitedefaultseppunct}\relax
\EndOfBibitem
\bibitem[{Pottage} et~al.(2002){Pottage}, {Flower}, and {Davis}]{pottage:02}
{Pottage},~J.~T.; {Flower},~D.~R.; {Davis},~S.~L. \emph{J. Phys. B At. Mol.
  Opt. Phys.} \textbf{2002}, \emph{35}, 2541\relax
\mciteBstWouldAddEndPuncttrue
\mciteSetBstMidEndSepPunct{\mcitedefaultmidpunct}
{\mcitedefaultendpunct}{\mcitedefaultseppunct}\relax
\EndOfBibitem
\bibitem[{Pottage} et~al.(2004){Pottage}, {Flower}, and {Davis}]{pottage:04}
{Pottage},~J.~T.; {Flower},~D.~R.; {Davis},~S.~L. \emph{J. Phys. B At. Mol.
  Opt. Phys.} \textbf{2004}, \emph{37}, 165\relax
\mciteBstWouldAddEndPuncttrue
\mciteSetBstMidEndSepPunct{\mcitedefaultmidpunct}
{\mcitedefaultendpunct}{\mcitedefaultseppunct}\relax
\EndOfBibitem
\bibitem[{Pottage} et~al.(2004){Pottage}, {Flower}, and {Davis}]{pottage:04a}
{Pottage},~J.~T.; {Flower},~D.~R.; {Davis},~S.~L. \emph{\mnras} \textbf{2004},
  \emph{352}, 39\relax
\mciteBstWouldAddEndPuncttrue
\mciteSetBstMidEndSepPunct{\mcitedefaultmidpunct}
{\mcitedefaultendpunct}{\mcitedefaultseppunct}\relax
\EndOfBibitem
\bibitem[{Flower} et~al.(2010){Flower}, {Pineau Des For{\^e}ts}, and
  {Rabli}]{flower:10}
{Flower},~D.~R.; {Pineau Des For{\^e}ts},~G.; {Rabli},~D. \emph{\mnras}
  \textbf{2010}, \emph{409}, 29\relax
\mciteBstWouldAddEndPuncttrue
\mciteSetBstMidEndSepPunct{\mcitedefaultmidpunct}
{\mcitedefaultendpunct}{\mcitedefaultseppunct}\relax
\EndOfBibitem
\bibitem[{Flower} and {Pineau des For{\^e}ts}(2012){Flower}, and {Pineau des
  For{\^e}ts}]{Flower:12extr}
{Flower},~D.~R.; {Pineau des For{\^e}ts},~G. \emph{\mnras} \textbf{2012},
  \emph{421}, 2786\relax
\mciteBstWouldAddEndPuncttrue
\mciteSetBstMidEndSepPunct{\mcitedefaultmidpunct}
{\mcitedefaultendpunct}{\mcitedefaultseppunct}\relax
\EndOfBibitem
\bibitem[{Voronkov} et~al.(2010){Voronkov}, {Caswell}, {Britton}, {Green},
  {Sobolev}, and {Ellingsen}]{voronkov:10}
{Voronkov},~M.~A.; {Caswell},~J.~L.; {Britton},~T.~R.; {Green},~J.~A.;
  {Sobolev},~A.~M.; {Ellingsen},~S.~P. \emph{\mnras} \textbf{2010}, \emph{408},
  133\relax
\mciteBstWouldAddEndPuncttrue
\mciteSetBstMidEndSepPunct{\mcitedefaultmidpunct}
{\mcitedefaultendpunct}{\mcitedefaultseppunct}\relax
\EndOfBibitem
\bibitem[{Walmsley} et~al.(1988){Walmsley}, {Batrla}, {Matthews}, and
  {Menten}]{walmsley:88}
{Walmsley},~C.~M.; {Batrla},~W.; {Matthews},~H.~E.; {Menten},~K.~M. \emph{\aap}
  \textbf{1988}, \emph{197}, 271\relax
\mciteBstWouldAddEndPuncttrue
\mciteSetBstMidEndSepPunct{\mcitedefaultmidpunct}
{\mcitedefaultendpunct}{\mcitedefaultseppunct}\relax
\EndOfBibitem
\bibitem[{Herbst} et~al.(1977){Herbst}, {Green}, {Thaddeus}, and
  {Klemperer}]{herbst:77}
{Herbst},~E.; {Green},~S.; {Thaddeus},~P.; {Klemperer},~W. \emph{\apj}
  \textbf{1977}, \emph{215}, 503\relax
\mciteBstWouldAddEndPuncttrue
\mciteSetBstMidEndSepPunct{\mcitedefaultmidpunct}
{\mcitedefaultendpunct}{\mcitedefaultseppunct}\relax
\EndOfBibitem
\bibitem[{Hammami} et~al.(2009){Hammami}, {Owono Owono}, and
  {St{\"a}uber}]{Hammami:09}
{Hammami},~K.; {Owono Owono},~L.~C.; {St{\"a}uber},~P. \emph{\aap}
  \textbf{2009}, \emph{507}, 1083\relax
\mciteBstWouldAddEndPuncttrue
\mciteSetBstMidEndSepPunct{\mcitedefaultmidpunct}
{\mcitedefaultendpunct}{\mcitedefaultseppunct}\relax
\EndOfBibitem
\bibitem[{Buhl} and {Snyder}(1973){Buhl}, and {Snyder}]{buhl:73}
{Buhl},~D.; {Snyder},~L.~E. \emph{\apj} \textbf{1973}, \emph{180}, 791\relax
\mciteBstWouldAddEndPuncttrue
\mciteSetBstMidEndSepPunct{\mcitedefaultmidpunct}
{\mcitedefaultendpunct}{\mcitedefaultseppunct}\relax
\EndOfBibitem
\bibitem[{Herbst} and {Klemperer}(1974){Herbst}, and {Klemperer}]{herbst:74}
{Herbst},~E.; {Klemperer},~W. \emph{\apj} \textbf{1974}, \emph{188}, 255\relax
\mciteBstWouldAddEndPuncttrue
\mciteSetBstMidEndSepPunct{\mcitedefaultmidpunct}
{\mcitedefaultendpunct}{\mcitedefaultseppunct}\relax
\EndOfBibitem
\bibitem[{Kraemer} and {Diercksen}(1976){Kraemer}, and {Diercksen}]{Kraemer:76}
{Kraemer},~W.~P.; {Diercksen},~G.~H.~F. \emph{\apjl} \textbf{1976}, \emph{205},
  L97\relax
\mciteBstWouldAddEndPuncttrue
\mciteSetBstMidEndSepPunct{\mcitedefaultmidpunct}
{\mcitedefaultendpunct}{\mcitedefaultseppunct}\relax
\EndOfBibitem
\bibitem[{Monteiro}(1985)]{monteiro:85}
{Monteiro},~T.~S. \emph{\mnras} \textbf{1985}, \emph{214}, 419\relax
\mciteBstWouldAddEndPuncttrue
\mciteSetBstMidEndSepPunct{\mcitedefaultmidpunct}
{\mcitedefaultendpunct}{\mcitedefaultseppunct}\relax
\EndOfBibitem
\bibitem[{Green}(1975)]{Green:75N2Hp}
{Green},~S. \emph{\apj} \textbf{1975}, \emph{201}, 366\relax
\mciteBstWouldAddEndPuncttrue
\mciteSetBstMidEndSepPunct{\mcitedefaultmidpunct}
{\mcitedefaultendpunct}{\mcitedefaultseppunct}\relax
\EndOfBibitem
\bibitem[{Daniel} et~al.(2005){Daniel}, {Dubernet}, {Meuwly}, {Cernicharo}, and
  {Pagani}]{Daniel:05}
{Daniel},~F.; {Dubernet},~M.-L.; {Meuwly},~M.; {Cernicharo},~J.; {Pagani},~L.
  \emph{\mnras} \textbf{2005}, \emph{363}, 1083\relax
\mciteBstWouldAddEndPuncttrue
\mciteSetBstMidEndSepPunct{\mcitedefaultmidpunct}
{\mcitedefaultendpunct}{\mcitedefaultseppunct}\relax
\EndOfBibitem
\bibitem[{Geballe} and {Oka}(1996){Geballe}, and {Oka}]{geballe:96}
{Geballe},~T.~R.; {Oka},~T. \emph{\nat} \textbf{1996}, \emph{384}, 334\relax
\mciteBstWouldAddEndPuncttrue
\mciteSetBstMidEndSepPunct{\mcitedefaultmidpunct}
{\mcitedefaultendpunct}{\mcitedefaultseppunct}\relax
\EndOfBibitem
\bibitem[{McCall} et~al.(1998){McCall}, {Geballe}, {Hinkle}, and
  {Oka}]{mccall:98}
{McCall},~B.~J.; {Geballe},~T.~R.; {Hinkle},~K.~H.; {Oka},~T. \emph{Science}
  \textbf{1998}, \emph{279}, 1910\relax
\mciteBstWouldAddEndPuncttrue
\mciteSetBstMidEndSepPunct{\mcitedefaultmidpunct}
{\mcitedefaultendpunct}{\mcitedefaultseppunct}\relax
\EndOfBibitem
\bibitem[{McCall} et~al.(1999){McCall}, {Geballe}, {Hinkle}, and
  {Oka}]{mccall:99}
{McCall},~B.~J.; {Geballe},~T.~R.; {Hinkle},~K.~H.; {Oka},~T. \emph{\apj}
  \textbf{1999}, \emph{522}, 338\relax
\mciteBstWouldAddEndPuncttrue
\mciteSetBstMidEndSepPunct{\mcitedefaultmidpunct}
{\mcitedefaultendpunct}{\mcitedefaultseppunct}\relax
\EndOfBibitem
\bibitem[{Geballe}(2006)]{geballe:06}
{Geballe},~T.~R. \emph{Phil. Trans. R. Soc., A} \textbf{2006}, \emph{364},
  3035\relax
\mciteBstWouldAddEndPuncttrue
\mciteSetBstMidEndSepPunct{\mcitedefaultmidpunct}
{\mcitedefaultendpunct}{\mcitedefaultseppunct}\relax
\EndOfBibitem
\bibitem[{Geballe}(2012)]{geballe:12}
{Geballe},~T.~R. \emph{Phil. Trans. R. Soc., A} \textbf{2012}, \emph{370},
  5151\relax
\mciteBstWouldAddEndPuncttrue
\mciteSetBstMidEndSepPunct{\mcitedefaultmidpunct}
{\mcitedefaultendpunct}{\mcitedefaultseppunct}\relax
\EndOfBibitem
\bibitem[{Oka} and {Epp}(2004){Oka}, and {Epp}]{oka:04}
{Oka},~T.; {Epp},~E. \emph{\apj} \textbf{2004}, \emph{613}, 349\relax
\mciteBstWouldAddEndPuncttrue
\mciteSetBstMidEndSepPunct{\mcitedefaultmidpunct}
{\mcitedefaultendpunct}{\mcitedefaultseppunct}\relax
\EndOfBibitem
\bibitem[{Xie} et~al.(2005){Xie}, {Braams}, and {Bowman}]{xie:05}
{Xie},~Z.; {Braams},~B.~J.; {Bowman},~J.~M. \emph{\jcp} \textbf{2005},
  \emph{122}, 224307\relax
\mciteBstWouldAddEndPuncttrue
\mciteSetBstMidEndSepPunct{\mcitedefaultmidpunct}
{\mcitedefaultendpunct}{\mcitedefaultseppunct}\relax
\EndOfBibitem
\bibitem[{Aguado} et~al.(2010){Aguado}, {Barrag{\'a}n}, {Prosmiti},
  {Delgado-Barrio}, {Villarreal}, and {Roncero}]{aguado:10}
{Aguado},~A.; {Barrag{\'a}n},~P.; {Prosmiti},~R.; {Delgado-Barrio},~G.;
  {Villarreal},~P.; {Roncero},~O. \emph{\jcp} \textbf{2010}, \emph{133},
  024306\relax
\mciteBstWouldAddEndPuncttrue
\mciteSetBstMidEndSepPunct{\mcitedefaultmidpunct}
{\mcitedefaultendpunct}{\mcitedefaultseppunct}\relax
\EndOfBibitem
\bibitem[{G{\'o}mez-Carrasco} et~al.(2012){G{\'o}mez-Carrasco},
  {Gonz{\'a}lez-S{\'a}nchez}, {Aguado}, {Sanz-Sanz}, {Zanchet}, and
  {Roncero}]{gomez:12}
{G{\'o}mez-Carrasco},~S.; {Gonz{\'a}lez-S{\'a}nchez},~L.; {Aguado},~A.;
  {Sanz-Sanz},~C.; {Zanchet},~A.; {Roncero},~O. \emph{\jcp} \textbf{2012},
  \emph{137}, 094303\relax
\mciteBstWouldAddEndPuncttrue
\mciteSetBstMidEndSepPunct{\mcitedefaultmidpunct}
{\mcitedefaultendpunct}{\mcitedefaultseppunct}\relax
\EndOfBibitem
\bibitem[{Park} and {Light}(2007){Park}, and {Light}]{park:07}
{Park},~K.; {Light},~J.~C. \emph{\jcp} \textbf{2007}, \emph{126}, 044305\relax
\mciteBstWouldAddEndPuncttrue
\mciteSetBstMidEndSepPunct{\mcitedefaultmidpunct}
{\mcitedefaultendpunct}{\mcitedefaultseppunct}\relax
\EndOfBibitem
\bibitem[{Ag{\'u}ndez} et~al.(2010){Ag{\'u}ndez}, {Cernicharo}, {Gu{\'e}lin},
  {Kahane}, {Roueff}, {K{\l}os}, {Aoiz}, {Lique}, {Marcelino}, {Goicoechea},
  {Gonz{\'a}lez Garc{\'{\i}}a}, {Gottlieb}, {McCarthy}, and
  {Thaddeus}]{Agundez:10}
{Ag{\'u}ndez},~M.; {Cernicharo},~J.; {Gu{\'e}lin},~M.; {Kahane},~C.;
  {Roueff},~E.; {K{\l}os},~J.; {Aoiz},~F.~J.; {Lique},~F.; {Marcelino},~N.;
  {Goicoechea},~J.~R.; {Gonz{\'a}lez Garc{\'{\i}}a},~M.; {Gottlieb},~C.~A.;
  {McCarthy},~M.~C.; {Thaddeus},~P. \emph{\aap} \textbf{2010}, \emph{517},
  L2\relax
\mciteBstWouldAddEndPuncttrue
\mciteSetBstMidEndSepPunct{\mcitedefaultmidpunct}
{\mcitedefaultendpunct}{\mcitedefaultseppunct}\relax
\EndOfBibitem
\bibitem[{Faure} et~al.(2004){Faure}, {Gorfinkiel}, and {Tennyson}]{Faure:04}
{Faure},~A.; {Gorfinkiel},~J.~D.; {Tennyson},~J. \emph{\mnras} \textbf{2004},
  \emph{347}, 323\relax
\mciteBstWouldAddEndPuncttrue
\mciteSetBstMidEndSepPunct{\mcitedefaultmidpunct}
{\mcitedefaultendpunct}{\mcitedefaultseppunct}\relax
\EndOfBibitem
\bibitem[{Daniel} et~al.(2007){Daniel}, {Cernicharo}, {Roueff}, {Gerin}, and
  {Dubernet}]{Daniel:07}
{Daniel},~F.; {Cernicharo},~J.; {Roueff},~E.; {Gerin},~M.; {Dubernet},~M.~L.
  \emph{\apj} \textbf{2007}, \emph{667}, 980\relax
\mciteBstWouldAddEndPuncttrue
\mciteSetBstMidEndSepPunct{\mcitedefaultmidpunct}
{\mcitedefaultendpunct}{\mcitedefaultseppunct}\relax
\EndOfBibitem
\bibitem[{Scribano} et~al.(2010){Scribano}, {Faure}, and
  {Wiesenfeld}]{scribano10}
{Scribano},~Y.; {Faure},~A.; {Wiesenfeld},~L. \emph{\jcp} \textbf{2010},
  \emph{133}, 231105\relax
\mciteBstWouldAddEndPuncttrue
\mciteSetBstMidEndSepPunct{\mcitedefaultmidpunct}
{\mcitedefaultendpunct}{\mcitedefaultseppunct}\relax
\EndOfBibitem
\bibitem[{Draine}(1978)]{draine:78}
{Draine},~B.~T. \emph{\apjs} \textbf{1978}, \emph{36}, 595\relax
\mciteBstWouldAddEndPuncttrue
\mciteSetBstMidEndSepPunct{\mcitedefaultmidpunct}
{\mcitedefaultendpunct}{\mcitedefaultseppunct}\relax
\EndOfBibitem
\bibitem[{Mathis} et~al.(1983){Mathis}, {Mezger}, and {Panagia}]{mathis:83}
{Mathis},~J.~S.; {Mezger},~P.~G.; {Panagia},~N. \emph{\aap} \textbf{1983},
  \emph{128}, 212\relax
\mciteBstWouldAddEndPuncttrue
\mciteSetBstMidEndSepPunct{\mcitedefaultmidpunct}
{\mcitedefaultendpunct}{\mcitedefaultseppunct}\relax
\EndOfBibitem
\bibitem[Lefebvre-Brion and Field(1986)Lefebvre-Brion, and Field]{lefebvre:86}
Lefebvre-Brion,~H.; Field,~R.~W. \emph{Perturbations in the spectra of diatomic
  molecules}; Academic Press: New York, 1986\relax
\mciteBstWouldAddEndPuncttrue
\mciteSetBstMidEndSepPunct{\mcitedefaultmidpunct}
{\mcitedefaultendpunct}{\mcitedefaultseppunct}\relax
\EndOfBibitem
\bibitem[Lefebvre-Brion and Field(2006)Lefebvre-Brion, and Field]{lefebvre:04}
Lefebvre-Brion,~H.; Field,~R.~W. \emph{The spectra and dynamics of diatomic
  molecules}; Elsevier: Amsterdam, 2006\relax
\mciteBstWouldAddEndPuncttrue
\mciteSetBstMidEndSepPunct{\mcitedefaultmidpunct}
{\mcitedefaultendpunct}{\mcitedefaultseppunct}\relax
\EndOfBibitem
\bibitem[Herzberg(1989)]{herzberg:89}
Herzberg,~G. \emph{Molecular Spectra and Molecular Structure, Vol. 1: Spectra
  of Diatomic Molecules}; Krieger Publishing Company: Malabar, 1989\relax
\mciteBstWouldAddEndPuncttrue
\mciteSetBstMidEndSepPunct{\mcitedefaultmidpunct}
{\mcitedefaultendpunct}{\mcitedefaultseppunct}\relax
\EndOfBibitem
\bibitem[Herzberg(1950)]{herzberg3}
Herzberg,~G. \emph{Molecular Spectra and Molecular Structure, Vol. 3:
  Electronic Spectra and Electronic Structure of Polyatomic Molecules}; Van
  Nostrand: New York, 1950; p 757\relax
\mciteBstWouldAddEndPuncttrue
\mciteSetBstMidEndSepPunct{\mcitedefaultmidpunct}
{\mcitedefaultendpunct}{\mcitedefaultseppunct}\relax
\EndOfBibitem
\bibitem[{Hsu} and {Smith}(1977){Hsu}, and {Smith}]{hsu:77}
{Hsu},~D.~K.; {Smith},~W.~H. \emph{Spectrosc. Lett.} \textbf{1977}, \emph{10},
  181\relax
\mciteBstWouldAddEndPuncttrue
\mciteSetBstMidEndSepPunct{\mcitedefaultmidpunct}
{\mcitedefaultendpunct}{\mcitedefaultseppunct}\relax
\EndOfBibitem
\bibitem[{Kuznetsova}(1987)]{kuznetsova:87}
{Kuznetsova},~L.~A. \emph{Spectrosc. Lett.} \textbf{1987}, \emph{20}, 665\relax
\mciteBstWouldAddEndPuncttrue
\mciteSetBstMidEndSepPunct{\mcitedefaultmidpunct}
{\mcitedefaultendpunct}{\mcitedefaultseppunct}\relax
\EndOfBibitem
\bibitem[{Reddy} et~al.(2003){Reddy}, {Nazeer Ahammed}, {Rama Gopal}, and {Baba
  Basha}]{reddy:03}
{Reddy},~R.~R.; {Nazeer Ahammed},~Y.; {Rama Gopal},~K.; {Baba Basha},~D.
  \emph{Astrophys. Sp. Science} \textbf{2003}, \emph{286}, 419\relax
\mciteBstWouldAddEndPuncttrue
\mciteSetBstMidEndSepPunct{\mcitedefaultmidpunct}
{\mcitedefaultendpunct}{\mcitedefaultseppunct}\relax
\EndOfBibitem
\bibitem[jpl(1998)]{jpl}
\url{http://spec.jpl.nasa.gov/}, 1998\relax
\mciteBstWouldAddEndPuncttrue
\mciteSetBstMidEndSepPunct{\mcitedefaultmidpunct}
{\mcitedefaultendpunct}{\mcitedefaultseppunct}\relax
\EndOfBibitem
\bibitem[{Pickett} et~al.(1998){Pickett}, {Poynter}, {Cohen}, {Delitsky},
  {Pearson}, and {M{\"u}ller}]{Pickett:98}
{Pickett},~H.~M.; {Poynter},~R.~L.; {Cohen},~E.~A.; {Delitsky},~M.~L.;
  {Pearson},~J.~C.; {M{\"u}ller},~H.~S.~P. \emph{\jqsrt} \textbf{1998},
  \emph{60}, 883--890\relax
\mciteBstWouldAddEndPuncttrue
\mciteSetBstMidEndSepPunct{\mcitedefaultmidpunct}
{\mcitedefaultendpunct}{\mcitedefaultseppunct}\relax
\EndOfBibitem
\bibitem[hit()]{hitran}
\url{http://www.cfa.Harvard.edu/HITRAN/}\relax
\mciteBstWouldAddEndPuncttrue
\mciteSetBstMidEndSepPunct{\mcitedefaultmidpunct}
{\mcitedefaultendpunct}{\mcitedefaultseppunct}\relax
\EndOfBibitem
\bibitem[Rothman et~al.(2013)Rothman, Gordon, Babikov, Barbe, Benner, Bernath,
  Birk, Bizzocchi, Boudon, Brown, Campargue, Chance, Cohen, Coudert, Devi,
  Drouin, Fayt, Flaud, Gamache, Harrison, Hartmann, Hill, Hodges, Jacquemart,
  Jolly, Lamouroux, Roy, Li, Long, Lyulin, Mackie, Massie, Mikhailenko, M?ller,
  Naumenko, Nikitin, Orphal, Perevalov, Perrin, Polovtseva, Richard, Smith,
  Starikova, Sung, Tashkun, Tennyson, Toon, Tyuterev, and Wagner]{Rothman:13}
Rothman,~L. et~al.  \emph{\jqsrt} \textbf{2013}, --\relax
\mciteBstWouldAddEndPuncttrue
\mciteSetBstMidEndSepPunct{\mcitedefaultmidpunct}
{\mcitedefaultendpunct}{\mcitedefaultseppunct}\relax
\EndOfBibitem
\bibitem[gei()]{geisa}
\url{http://ether.ipsl.jussieu.fr/etherTypo/?id=950}\relax
\mciteBstWouldAddEndPuncttrue
\mciteSetBstMidEndSepPunct{\mcitedefaultmidpunct}
{\mcitedefaultendpunct}{\mcitedefaultseppunct}\relax
\EndOfBibitem
\bibitem[Jacquinet-Husson et~al.(2011)Jacquinet-Husson, Crepeau, Armante,
  Boutammine, Ch?din, Scott, Crevoisier, Capelle, Boone, Poulet-Crovisier,
  Barbe, Campargue, Benner, Benilan, B?zard, Boudon, Brown, Coudert, Coustenis,
  Dana, Devi, Fally, Fayt, Flaud, Goldman, Herman, Harris, Jacquemart, Jolly,
  Kleiner, Kleinb?hl, Kwabia-Tchana, Lavrentieva, Lacome, Xu, Lyulin, Mandin,
  Maki, Mikhailenko, Miller, Mishina, Moazzen-Ahmadi, M?ller, Nikitin, Orphal,
  Perevalov, Perrin, Petkie, Predoi-Cross, Rinsland, Remedios, Rotger, Smith,
  Sung, Tashkun, Tennyson, Toth, Vandaele, and Auwera]{Jacquinet:11}
Jacquinet-Husson,~N. et~al.  \emph{\jqsrt} \textbf{2011}, \emph{112}, 2395 --
  2445\relax
\mciteBstWouldAddEndPuncttrue
\mciteSetBstMidEndSepPunct{\mcitedefaultmidpunct}
{\mcitedefaultendpunct}{\mcitedefaultseppunct}\relax
\EndOfBibitem
\bibitem[{Fox}(2005)]{fox:05}
{Fox},~J.~L. \emph{J. Phys. Conf. Ser.} \textbf{2005}, \emph{4}, 32\relax
\mciteBstWouldAddEndPuncttrue
\mciteSetBstMidEndSepPunct{\mcitedefaultmidpunct}
{\mcitedefaultendpunct}{\mcitedefaultseppunct}\relax
\EndOfBibitem
\bibitem[{Fox}(2012)]{fox:12}
{Fox},~J.~L. \emph{Icarus} \textbf{2012}, \emph{221}, 787\relax
\mciteBstWouldAddEndPuncttrue
\mciteSetBstMidEndSepPunct{\mcitedefaultmidpunct}
{\mcitedefaultendpunct}{\mcitedefaultseppunct}\relax
\EndOfBibitem
\bibitem[{Fray} et~al.(2005){Fray}, {B{\'e}nilan}, {Cottin}, {Gazeau}, and
  {Crovisier}]{fray:05}
{Fray},~N.; {B{\'e}nilan},~Y.; {Cottin},~H.; {Gazeau},~M.-C.; {Crovisier},~J.
  \emph{Plan. Sp. Science} \textbf{2005}, \emph{53}, 1243\relax
\mciteBstWouldAddEndPuncttrue
\mciteSetBstMidEndSepPunct{\mcitedefaultmidpunct}
{\mcitedefaultendpunct}{\mcitedefaultseppunct}\relax
\EndOfBibitem
\bibitem[Lindqvist et~al.(2000)Lindqvist, Schoier, Lucas, and
  Olofsson]{lindqvist:00}
Lindqvist,~M.; Schoier,~F.; Lucas,~R.; Olofsson,~H. \emph{\aap} \textbf{2000},
  \emph{361}, 1036\relax
\mciteBstWouldAddEndPuncttrue
\mciteSetBstMidEndSepPunct{\mcitedefaultmidpunct}
{\mcitedefaultendpunct}{\mcitedefaultseppunct}\relax
\EndOfBibitem
\bibitem[Cook et~al.(2000)Cook, Langford, Ashfold, and Dixon]{cook:00}
Cook,~P.~A.; Langford,~S.~R.; Ashfold,~M. N.~R.; Dixon,~R.~N. \emph{\jcp}
  \textbf{2000}, \emph{113}, 994\relax
\mciteBstWouldAddEndPuncttrue
\mciteSetBstMidEndSepPunct{\mcitedefaultmidpunct}
{\mcitedefaultendpunct}{\mcitedefaultseppunct}\relax
\EndOfBibitem
\bibitem[{Varju} et~al.(2011){Varju}, {Hejduk}, {Dohnal}, {J{\'{\i}}lek},
  {Kotr{\'{\i}}k}, {Pla{\v s}il}, {Gerlich}, and {Glos{\'{\i}}k}]{varju:11}
{Varju},~J.; {Hejduk},~M.; {Dohnal},~P.; {J{\'{\i}}lek},~M.;
  {Kotr{\'{\i}}k},~T.; {Pla{\v s}il},~R.; {Gerlich},~D.; {Glos{\'{\i}}k},~J.
  \emph{Phys. Rev. Lett.} \textbf{2011}, \emph{106}, 203201\relax
\mciteBstWouldAddEndPuncttrue
\mciteSetBstMidEndSepPunct{\mcitedefaultmidpunct}
{\mcitedefaultendpunct}{\mcitedefaultseppunct}\relax
\EndOfBibitem
\bibitem[{Dohnal} et~al.(2012){Dohnal}, {Hejduk}, {Varju}, {Rubovi{\v c}},
  {Rou{\v c}ka}, {Kotr{\'{\i}}k}, {Pla{\v s}il}, {Glos{\'{\i}}k}, and
  {Johnsen}]{dohnal:12}
{Dohnal},~P.; {Hejduk},~M.; {Varju},~J.; {Rubovi{\v c}},~P.; {Rou{\v c}ka},~{\v
  S}.; {Kotr{\'{\i}}k},~T.; {Pla{\v s}il},~R.; {Glos{\'{\i}}k},~J.;
  {Johnsen},~R. \emph{\jcp} \textbf{2012}, \emph{136}, 244304\relax
\mciteBstWouldAddEndPuncttrue
\mciteSetBstMidEndSepPunct{\mcitedefaultmidpunct}
{\mcitedefaultendpunct}{\mcitedefaultseppunct}\relax
\EndOfBibitem
\bibitem[{Pagani} et~al.(2009){Pagani}, {Vastel}, {Hugo}, {Kokoouline},
  {Greene}, {Bacmann}, {Bayet}, {Ceccarelli}, {Peng}, and
  {Schlemmer}]{pagani:09}
{Pagani},~L.; {Vastel},~C.; {Hugo},~E.; {Kokoouline},~V.; {Greene},~C.~H.;
  {Bacmann},~A.; {Bayet},~E.; {Ceccarelli},~C.; {Peng},~R.; {Schlemmer},~S.
  \emph{\aap} \textbf{2009}, \emph{494}, 623\relax
\mciteBstWouldAddEndPuncttrue
\mciteSetBstMidEndSepPunct{\mcitedefaultmidpunct}
{\mcitedefaultendpunct}{\mcitedefaultseppunct}\relax
\EndOfBibitem
\bibitem[Che et~al.(2007)Che, Ren, Wang, Dong, Dai, Wang, Zhang, Yang, Li,
  Werner, Lique, and Alexander]{che:07}
Che,~L.; Ren,~Z.; Wang,~X.; Dong,~W.; Dai,~D.; Wang,~X.; Zhang,~D.~H.;
  Yang,~X.; Li,~G.; Werner,~H.-J.; Lique,~F.; Alexander,~M.~H. \emph{Science}
  \textbf{2007}, \emph{317}, 1061\relax
\mciteBstWouldAddEndPuncttrue
\mciteSetBstMidEndSepPunct{\mcitedefaultmidpunct}
{\mcitedefaultendpunct}{\mcitedefaultseppunct}\relax
\EndOfBibitem
\bibitem[Lique et~al.(2008)Lique, Alexander, Li, Werner, Nizkorodov, Harper,
  and Nesbitt]{lique:08FH2}
Lique,~F.; Alexander,~M.~H.; Li,~G.; Werner,~H.-J.; Nizkorodov,~S.~A.;
  Harper,~W.~W.; Nesbitt,~D.~J. \emph{J. Chem. Phys.} \textbf{2008},
  \emph{128}, 084313\relax
\mciteBstWouldAddEndPuncttrue
\mciteSetBstMidEndSepPunct{\mcitedefaultmidpunct}
{\mcitedefaultendpunct}{\mcitedefaultseppunct}\relax
\EndOfBibitem
\bibitem[{Wang} et~al.(2008){Wang}, {Dong}, {Xiao}, {Che}, {Ren}, {Dai},
  {Wang}, {Casavecchia}, {Yang}, {Jiang}, {Xie}, {Sun}, {Lee}, {Zhang},
  {Werner}, and {Alexander}]{Wang:08}
{Wang},~X.; {Dong},~W.; {Xiao},~C.; {Che},~L.; {Ren},~Z.; {Dai},~D.;
  {Wang},~X.; {Casavecchia},~P.; {Yang},~X.; {Jiang},~B.; {Xie},~D.; {Sun},~Z.;
  {Lee},~S.; {Zhang},~D.~H.; {Werner},~H.; {Alexander},~M.~H. \emph{Science}
  \textbf{2008}, \emph{322}, 573\relax
\mciteBstWouldAddEndPuncttrue
\mciteSetBstMidEndSepPunct{\mcitedefaultmidpunct}
{\mcitedefaultendpunct}{\mcitedefaultseppunct}\relax
\EndOfBibitem
\bibitem[Zhou et~al.(0)Zhou, Jiang, Xie, and Guo]{zhou:13}
Zhou,~L.; Jiang,~B.; Xie,~D.; Guo,~H. \emph{J. Phys. Chem. A} \textbf{0},
  \emph{0}, null\relax
\mciteBstWouldAddEndPuncttrue
\mciteSetBstMidEndSepPunct{\mcitedefaultmidpunct}
{\mcitedefaultendpunct}{\mcitedefaultseppunct}\relax
\EndOfBibitem
\bibitem[{Ma} et~al.(2012){Ma}, {Zhu}, {Guo}, and {Yarkony}]{ma:12nh3}
{Ma},~J.; {Zhu},~X.; {Guo},~H.; {Yarkony},~D.~R. \emph{\jcp} \textbf{2012},
  \emph{137}, 220000\relax
\mciteBstWouldAddEndPuncttrue
\mciteSetBstMidEndSepPunct{\mcitedefaultmidpunct}
{\mcitedefaultendpunct}{\mcitedefaultseppunct}\relax
\EndOfBibitem
\bibitem[{Spitzer}(1976)]{spitzer:76}
{Spitzer},~L. \emph{\qjras} \textbf{1976}, \emph{17}, 97\relax
\mciteBstWouldAddEndPuncttrue
\mciteSetBstMidEndSepPunct{\mcitedefaultmidpunct}
{\mcitedefaultendpunct}{\mcitedefaultseppunct}\relax
\EndOfBibitem
\bibitem[{Snow} and {McCall}(2006){Snow}, and {McCall}]{snow:06}
{Snow},~T.~P.; {McCall},~B.~J. \emph{\araa} \textbf{2006}, \emph{44}, 367\relax
\mciteBstWouldAddEndPuncttrue
\mciteSetBstMidEndSepPunct{\mcitedefaultmidpunct}
{\mcitedefaultendpunct}{\mcitedefaultseppunct}\relax
\EndOfBibitem
\bibitem[{Wolniewicz} et~al.(1998){Wolniewicz}, {Simbotin}, and
  {Dalgarno}]{wolniewicz:98}
{Wolniewicz},~L.; {Simbotin},~I.; {Dalgarno},~A. \emph{\apjs} \textbf{1998},
  \emph{115}, 293\relax
\mciteBstWouldAddEndPuncttrue
\mciteSetBstMidEndSepPunct{\mcitedefaultmidpunct}
{\mcitedefaultendpunct}{\mcitedefaultseppunct}\relax
\EndOfBibitem
\bibitem[{Lacour} et~al.(2005){Lacour}, {Ziskin}, {H{\'e}brard}, {Oliveira},
  {Andr{\'e}}, {Ferlet}, and {Vidal-Madjar}]{lacour:05}
{Lacour},~S.; {Ziskin},~V.; {H{\'e}brard},~G.; {Oliveira},~C.;
  {Andr{\'e}},~M.~K.; {Ferlet},~R.; {Vidal-Madjar},~A. \emph{\apj}
  \textbf{2005}, \emph{627}, 251\relax
\mciteBstWouldAddEndPuncttrue
\mciteSetBstMidEndSepPunct{\mcitedefaultmidpunct}
{\mcitedefaultendpunct}{\mcitedefaultseppunct}\relax
\EndOfBibitem
\bibitem[{Ingalls} et~al.(2011){Ingalls}, {Bania}, {Boulanger}, {Draine},
  {Falgarone}, and {Hily-Blant}]{ingalls:11}
{Ingalls},~J.~G.; {Bania},~T.~M.; {Boulanger},~F.; {Draine},~B.~T.;
  {Falgarone},~E.; {Hily-Blant},~P. \emph{\apj} \textbf{2011}, \emph{743},
  174\relax
\mciteBstWouldAddEndPuncttrue
\mciteSetBstMidEndSepPunct{\mcitedefaultmidpunct}
{\mcitedefaultendpunct}{\mcitedefaultseppunct}\relax
\EndOfBibitem
\bibitem[{Federman} et~al.(1995){Federman}, {Cardell}, {van Dishoeck},
  {Lambert}, and {Black}]{federman:95}
{Federman},~S.~R.; {Cardell},~J.~A.; {van Dishoeck},~E.~F.; {Lambert},~D.~L.;
  {Black},~J.~H. \emph{\apj} \textbf{1995}, \emph{445}, 325\relax
\mciteBstWouldAddEndPuncttrue
\mciteSetBstMidEndSepPunct{\mcitedefaultmidpunct}
{\mcitedefaultendpunct}{\mcitedefaultseppunct}\relax
\EndOfBibitem
\bibitem[{Meyer} et~al.(2001){Meyer}, {Lauroesch}, {Sofia}, {Draine}, and
  {Bertoldi}]{meyer:01}
{Meyer},~D.~M.; {Lauroesch},~J.~T.; {Sofia},~U.~J.; {Draine},~B.~T.;
  {Bertoldi},~F. \emph{\apjl} \textbf{2001}, \emph{553}, L59\relax
\mciteBstWouldAddEndPuncttrue
\mciteSetBstMidEndSepPunct{\mcitedefaultmidpunct}
{\mcitedefaultendpunct}{\mcitedefaultseppunct}\relax
\EndOfBibitem
\bibitem[{Gnaci{\'n}ski}(2011)]{gnacinski:11}
{Gnaci{\'n}ski},~P. \emph{\aap} \textbf{2011}, \emph{532}, A122\relax
\mciteBstWouldAddEndPuncttrue
\mciteSetBstMidEndSepPunct{\mcitedefaultmidpunct}
{\mcitedefaultendpunct}{\mcitedefaultseppunct}\relax
\EndOfBibitem
\bibitem[{Gnaci{\'n}ski}(2013)]{gnacinski:13}
{Gnaci{\'n}ski},~P. \emph{\aap} \textbf{2013}, \emph{549}, A37\relax
\mciteBstWouldAddEndPuncttrue
\mciteSetBstMidEndSepPunct{\mcitedefaultmidpunct}
{\mcitedefaultendpunct}{\mcitedefaultseppunct}\relax
\EndOfBibitem
\bibitem[{Abgrall} et~al.(1993){Abgrall}, {Roueff}, {Launay}, {Roncin}, and
  {Subtil}]{abgrall:93}
{Abgrall},~H.; {Roueff},~E.; {Launay},~F.; {Roncin},~J.~Y.; {Subtil},~J.~L.
  \emph{J. Mol. Spect.} \textbf{1993}, \emph{157}, 512\relax
\mciteBstWouldAddEndPuncttrue
\mciteSetBstMidEndSepPunct{\mcitedefaultmidpunct}
{\mcitedefaultendpunct}{\mcitedefaultseppunct}\relax
\EndOfBibitem
\bibitem[{Abgrall} et~al.(1993){Abgrall}, {Roueff}, {Launay}, {Roncin}, and
  {Subtil}]{abgrall:93a}
{Abgrall},~H.; {Roueff},~E.; {Launay},~F.; {Roncin},~J.~Y.; {Subtil},~J.~L.
  \emph{\aaps} \textbf{1993}, \emph{101}, 273\relax
\mciteBstWouldAddEndPuncttrue
\mciteSetBstMidEndSepPunct{\mcitedefaultmidpunct}
{\mcitedefaultendpunct}{\mcitedefaultseppunct}\relax
\EndOfBibitem
\bibitem[{Abgrall} et~al.(1993){Abgrall}, {Roueff}, {Launay}, {Roncin}, and
  {Subtil}]{abgrall:93b}
{Abgrall},~H.; {Roueff},~E.; {Launay},~F.; {Roncin},~J.~Y.; {Subtil},~J.~L.
  \emph{\aaps} \textbf{1993}, \emph{101}, 323\relax
\mciteBstWouldAddEndPuncttrue
\mciteSetBstMidEndSepPunct{\mcitedefaultmidpunct}
{\mcitedefaultendpunct}{\mcitedefaultseppunct}\relax
\EndOfBibitem
\bibitem[{Roncin} and {Launay}(1995){Roncin}, and {Launay}]{roncin:95}
{Roncin},~J.-Y.; {Launay},~F. In \emph{Laboratory and Astronomical High
  Resolution Spectra}; {Sauval},~A.~J., {Blomme},~R., {Grevesse},~N., Eds.;
  Astron. Soc. Pacific Conf. Ser.; 1995; Vol.~81; p 310\relax
\mciteBstWouldAddEndPuncttrue
\mciteSetBstMidEndSepPunct{\mcitedefaultmidpunct}
{\mcitedefaultendpunct}{\mcitedefaultseppunct}\relax
\EndOfBibitem
\bibitem[Liu et~al.(1999)Liu, Kolessov, Partin, Bezel, and Wittig]{liu:99}
Liu,~K.; Kolessov,~A.; Partin,~J.~W.; Bezel,~I.; Wittig,~C. \emph{Chem. Phys.
  Lett.} \textbf{1999}, \emph{299}, 374\relax
\mciteBstWouldAddEndPuncttrue
\mciteSetBstMidEndSepPunct{\mcitedefaultmidpunct}
{\mcitedefaultendpunct}{\mcitedefaultseppunct}\relax
\EndOfBibitem
\bibitem[{Liu} et~al.(2003){Liu}, {Shemansky}, {Abgrall}, {Roueff}, {Ahmed},
  and {Ajello}]{liu:03}
{Liu},~X.; {Shemansky},~D.~E.; {Abgrall},~H.; {Roueff},~E.; {Ahmed},~S.~M.;
  {Ajello},~J.~M. \emph{J. Phys. B At. Mol. Opt. Phys.} \textbf{2003},
  \emph{36}, 173\relax
\mciteBstWouldAddEndPuncttrue
\mciteSetBstMidEndSepPunct{\mcitedefaultmidpunct}
{\mcitedefaultendpunct}{\mcitedefaultseppunct}\relax
\EndOfBibitem
\bibitem[{Dabrowski}(1984)]{dabrowski:84}
{Dabrowski},~I. \emph{Can. J. Phys.} \textbf{1984}, \emph{62}, 1639\relax
\mciteBstWouldAddEndPuncttrue
\mciteSetBstMidEndSepPunct{\mcitedefaultmidpunct}
{\mcitedefaultendpunct}{\mcitedefaultseppunct}\relax
\EndOfBibitem
\bibitem[{Abgrall} et~al.(2000){Abgrall}, {Roueff}, and {Drira}]{abgrall:00}
{Abgrall},~H.; {Roueff},~E.; {Drira},~I. \emph{\aaps} \textbf{2000},
  \emph{141}, 297\relax
\mciteBstWouldAddEndPuncttrue
\mciteSetBstMidEndSepPunct{\mcitedefaultmidpunct}
{\mcitedefaultendpunct}{\mcitedefaultseppunct}\relax
\EndOfBibitem
\bibitem[{Jonin} et~al.(2000){Jonin}, {Liu}, {Ajello}, {James}, and
  {Abgrall}]{jonin:00}
{Jonin},~C.; {Liu},~X.; {Ajello},~J.~M.; {James},~G.~K.; {Abgrall},~H.
  \emph{\apjs} \textbf{2000}, \emph{129}, 247\relax
\mciteBstWouldAddEndPuncttrue
\mciteSetBstMidEndSepPunct{\mcitedefaultmidpunct}
{\mcitedefaultendpunct}{\mcitedefaultseppunct}\relax
\EndOfBibitem
\bibitem[{Black} and {van Dishoeck}(1987){Black}, and {van Dishoeck}]{black:87}
{Black},~J.~H.; {van Dishoeck},~E.~F. \emph{\apj} \textbf{1987}, \emph{322},
  412\relax
\mciteBstWouldAddEndPuncttrue
\mciteSetBstMidEndSepPunct{\mcitedefaultmidpunct}
{\mcitedefaultendpunct}{\mcitedefaultseppunct}\relax
\EndOfBibitem
\bibitem[{R{\"o}llig} et~al.(2007){R{\"o}llig}, {Abel}, {Bell}, {Bensch},
  {Black}, {Ferland}, {Jonkheid}, {Kamp}, {Kaufman}, {Le Bourlot}, {Le Petit},
  {Meijerink}, {Morata}, {Ossenkopf}, {Roueff}, {Shaw}, {Spaans}, {Sternberg},
  {Stutzki}, {Thi}, {van Dishoeck}, {van Hoof}, {Viti}, and
  {Wolfire}]{roellig:07}
{R{\"o}llig},~M. et~al.  \emph{\aap} \textbf{2007}, \emph{467}, 187\relax
\mciteBstWouldAddEndPuncttrue
\mciteSetBstMidEndSepPunct{\mcitedefaultmidpunct}
{\mcitedefaultendpunct}{\mcitedefaultseppunct}\relax
\EndOfBibitem
\bibitem[{Gough} et~al.(1996){Gough}, {Schermann}, {Pichou}, {Landau}, {Cadez},
  and {Hall}]{gough:96}
{Gough},~S.; {Schermann},~C.; {Pichou},~F.; {Landau},~M.; {Cadez},~I.;
  {Hall},~R.~I. \emph{\aap} \textbf{1996}, \emph{305}, 687\relax
\mciteBstWouldAddEndPuncttrue
\mciteSetBstMidEndSepPunct{\mcitedefaultmidpunct}
{\mcitedefaultendpunct}{\mcitedefaultseppunct}\relax
\EndOfBibitem
\bibitem[{Creighan} et~al.(2006){Creighan}, {Perry}, and {Price}]{creighan:06}
{Creighan},~S.~C.; {Perry},~J.~S.~A.; {Price},~S.~D. \emph{\jcp} \textbf{2006},
  \emph{124}, 114701\relax
\mciteBstWouldAddEndPuncttrue
\mciteSetBstMidEndSepPunct{\mcitedefaultmidpunct}
{\mcitedefaultendpunct}{\mcitedefaultseppunct}\relax
\EndOfBibitem
\bibitem[{Lemaire} et~al.(2010){Lemaire}, {Vidali}, {Baouche}, {Chehrouri},
  {Chaabouni}, and {Mokrane}]{lemaire:10}
{Lemaire},~J.~L.; {Vidali},~G.; {Baouche},~S.; {Chehrouri},~M.;
  {Chaabouni},~H.; {Mokrane},~H. \emph{\apjl} \textbf{2010}, \emph{725},
  L156\relax
\mciteBstWouldAddEndPuncttrue
\mciteSetBstMidEndSepPunct{\mcitedefaultmidpunct}
{\mcitedefaultendpunct}{\mcitedefaultseppunct}\relax
\EndOfBibitem
\bibitem[{Sizun} et~al.(2010){Sizun}, {Bachellerie}, {Aguillon}, and
  {Sidis}]{sizun:10}
{Sizun},~M.; {Bachellerie},~D.; {Aguillon},~F.; {Sidis},~V. \emph{Chem. Phys.
  Lett.} \textbf{2010}, \emph{498}, 32\relax
\mciteBstWouldAddEndPuncttrue
\mciteSetBstMidEndSepPunct{\mcitedefaultmidpunct}
{\mcitedefaultendpunct}{\mcitedefaultseppunct}\relax
\EndOfBibitem
\bibitem[{Islam} et~al.(2010){Islam}, {Cecchi-Pestellini}, {Viti}, and
  {Casu}]{islam:10}
{Islam},~F.; {Cecchi-Pestellini},~C.; {Viti},~S.; {Casu},~S. \emph{\apj}
  \textbf{2010}, \emph{725}, 1111\relax
\mciteBstWouldAddEndPuncttrue
\mciteSetBstMidEndSepPunct{\mcitedefaultmidpunct}
{\mcitedefaultendpunct}{\mcitedefaultseppunct}\relax
\EndOfBibitem
\bibitem[{Le Bourlot} et~al.(2012){Le Bourlot}, {Le Petit}, {Pinto}, {Roueff},
  and {Roy}]{lebourlot:12}
{Le Bourlot},~J.; {Le Petit},~F.; {Pinto},~C.; {Roueff},~E.; {Roy},~F.
  \emph{\aap} \textbf{2012}, \emph{541}, A76\relax
\mciteBstWouldAddEndPuncttrue
\mciteSetBstMidEndSepPunct{\mcitedefaultmidpunct}
{\mcitedefaultendpunct}{\mcitedefaultseppunct}\relax
\EndOfBibitem
\bibitem[{Burton} et~al.(2002){Burton}, {Londish}, and {Brand}]{burton:02}
{Burton},~M.~G.; {Londish},~D.; {Brand},~P.~W.~J.~L. \emph{\mnras}
  \textbf{2002}, \emph{333}, 721\relax
\mciteBstWouldAddEndPuncttrue
\mciteSetBstMidEndSepPunct{\mcitedefaultmidpunct}
{\mcitedefaultendpunct}{\mcitedefaultseppunct}\relax
\EndOfBibitem
\bibitem[{Congiu} et~al.(2009){Congiu}, {Matar}, {Kristensen}, {Dulieu}, and
  {Lemaire}]{congiu:09}
{Congiu},~E.; {Matar},~E.; {Kristensen},~L.~E.; {Dulieu},~F.; {Lemaire},~J.~L.
  \emph{\mnras} \textbf{2009}, \emph{397}, L96\relax
\mciteBstWouldAddEndPuncttrue
\mciteSetBstMidEndSepPunct{\mcitedefaultmidpunct}
{\mcitedefaultendpunct}{\mcitedefaultseppunct}\relax
\EndOfBibitem
\bibitem[{Solomon} et~al.(1987){Solomon}, {Rivolo}, {Barrett}, and
  {Yahil}]{solomon:87}
{Solomon},~P.~M.; {Rivolo},~A.~R.; {Barrett},~J.; {Yahil},~A. \emph{\apj}
  \textbf{1987}, \emph{319}, 730\relax
\mciteBstWouldAddEndPuncttrue
\mciteSetBstMidEndSepPunct{\mcitedefaultmidpunct}
{\mcitedefaultendpunct}{\mcitedefaultseppunct}\relax
\EndOfBibitem
\bibitem[{Bell} et~al.(2006){Bell}, {Roueff}, {Viti}, and {Williams}]{bell:06}
{Bell},~T.~A.; {Roueff},~E.; {Viti},~S.; {Williams},~D.~A. \emph{\mnras}
  \textbf{2006}, \emph{371}, 1865\relax
\mciteBstWouldAddEndPuncttrue
\mciteSetBstMidEndSepPunct{\mcitedefaultmidpunct}
{\mcitedefaultendpunct}{\mcitedefaultseppunct}\relax
\EndOfBibitem
\bibitem[{Pineda} et~al.(2008){Pineda}, {Caselli}, and {Goodman}]{pineda:08}
{Pineda},~J.~E.; {Caselli},~P.; {Goodman},~A.~A. \emph{\apj} \textbf{2008},
  \emph{679}, 481\relax
\mciteBstWouldAddEndPuncttrue
\mciteSetBstMidEndSepPunct{\mcitedefaultmidpunct}
{\mcitedefaultendpunct}{\mcitedefaultseppunct}\relax
\EndOfBibitem
\bibitem[{Bolatto} et~al.(2013){Bolatto}, {Wolfire}, and {Leroy}]{bolatto:13}
{Bolatto},~A.~D.; {Wolfire},~M.; {Leroy},~A.~K. \emph{ArXiv e-prints}
  \textbf{2013}, \relax
\mciteBstWouldAddEndPunctfalse
\mciteSetBstMidEndSepPunct{\mcitedefaultmidpunct}
{}{\mcitedefaultseppunct}\relax
\EndOfBibitem
\bibitem[{Letzelter} et~al.(1987){Letzelter}, {Eidelsberg}, {Rostas}, {Breton},
  and {Thieblemont}]{letzelter:87}
{Letzelter},~C.; {Eidelsberg},~M.; {Rostas},~F.; {Breton},~J.;
  {Thieblemont},~B. \emph{Chem. Phys.} \textbf{1987}, \emph{114}, 273\relax
\mciteBstWouldAddEndPuncttrue
\mciteSetBstMidEndSepPunct{\mcitedefaultmidpunct}
{\mcitedefaultendpunct}{\mcitedefaultseppunct}\relax
\EndOfBibitem
\bibitem[{Tchang-Brillet} et~al.(1992){Tchang-Brillet}, {Julienne}, {Robbe},
  {Letzelter}, and {Rostas}]{tchang:92}
{Tchang-Brillet},~W.-U.~L.; {Julienne},~P.~S.; {Robbe},~J.-M.; {Letzelter},~C.;
  {Rostas},~F. \emph{\jcp} \textbf{1992}, \emph{96}, 6735\relax
\mciteBstWouldAddEndPuncttrue
\mciteSetBstMidEndSepPunct{\mcitedefaultmidpunct}
{\mcitedefaultendpunct}{\mcitedefaultseppunct}\relax
\EndOfBibitem
\bibitem[{Eidelsberg} et~al.(1991){Eidelsberg}, {Viala}, {Rostas}, and
  {Benayoun}]{eidelsberg:91}
{Eidelsberg},~M.; {Viala},~Y.; {Rostas},~F.; {Benayoun},~J.~J. \emph{\aaps}
  \textbf{1991}, \emph{90}, 231\relax
\mciteBstWouldAddEndPuncttrue
\mciteSetBstMidEndSepPunct{\mcitedefaultmidpunct}
{\mcitedefaultendpunct}{\mcitedefaultseppunct}\relax
\EndOfBibitem
\bibitem[{Cacciani} et~al.(1995){Cacciani}, {Hogervorst}, and
  {Ubachs}]{cacciani:95}
{Cacciani},~P.; {Hogervorst},~W.; {Ubachs},~W. \emph{\jcp} \textbf{1995},
  \emph{102}, 8308\relax
\mciteBstWouldAddEndPuncttrue
\mciteSetBstMidEndSepPunct{\mcitedefaultmidpunct}
{\mcitedefaultendpunct}{\mcitedefaultseppunct}\relax
\EndOfBibitem
\bibitem[{Eidelsberg} et~al.(2006){Eidelsberg}, {Sheffer}, {Federman},
  {Lemaire}, {Fillion}, {Rostas}, and {Ruiz}]{eidelsberg:06}
{Eidelsberg},~M.; {Sheffer},~Y.; {Federman},~S.~R.; {Lemaire},~J.~L.;
  {Fillion},~J.~H.; {Rostas},~F.; {Ruiz},~J. \emph{\apj} \textbf{2006},
  \emph{647}, 1543\relax
\mciteBstWouldAddEndPuncttrue
\mciteSetBstMidEndSepPunct{\mcitedefaultmidpunct}
{\mcitedefaultendpunct}{\mcitedefaultseppunct}\relax
\EndOfBibitem
\bibitem[{Chakraborty} et~al.(2008){Chakraborty}, {Ahmed}, {Jackson}, and
  {Thiemens}]{chakraborty:08}
{Chakraborty},~S.; {Ahmed},~M.; {Jackson},~T.~L.; {Thiemens},~M.~H.
  \emph{Science} \textbf{2008}, \emph{321}, 1328\relax
\mciteBstWouldAddEndPuncttrue
\mciteSetBstMidEndSepPunct{\mcitedefaultmidpunct}
{\mcitedefaultendpunct}{\mcitedefaultseppunct}\relax
\EndOfBibitem
\bibitem[{Eidelsberg} et~al.(2012){Eidelsberg}, {Lemaire}, {Federman}, {Stark},
  {Heays}, {Sheffer}, {Gavilan}, {Fillion}, {Rostas}, {Lyons}, {Smith}, {de
  Oliveira}, {Joyeux}, {Roudjane}, and {Nahon}]{eidelsberg:12}
{Eidelsberg},~M.; {Lemaire},~J.~L.; {Federman},~S.~R.; {Stark},~G.;
  {Heays},~A.~N.; {Sheffer},~Y.; {Gavilan},~L.; {Fillion},~J.-H.; {Rostas},~F.;
  {Lyons},~J.~R.; {Smith},~P.~L.; {de Oliveira},~N.; {Joyeux},~D.;
  {Roudjane},~M.; {Nahon},~L. \emph{\aap} \textbf{2012}, \emph{543}, A69\relax
\mciteBstWouldAddEndPuncttrue
\mciteSetBstMidEndSepPunct{\mcitedefaultmidpunct}
{\mcitedefaultendpunct}{\mcitedefaultseppunct}\relax
\EndOfBibitem
\bibitem[{Le Bourlot} et~al.(1993){Le Bourlot}, {Pineau Des Forets}, {Roueff},
  and {Flower}]{lebourlot:93}
{Le Bourlot},~J.; {Pineau Des Forets},~G.; {Roueff},~E.; {Flower},~D.~R.
  \emph{\aap} \textbf{1993}, \emph{267}, 233\relax
\mciteBstWouldAddEndPuncttrue
\mciteSetBstMidEndSepPunct{\mcitedefaultmidpunct}
{\mcitedefaultendpunct}{\mcitedefaultseppunct}\relax
\EndOfBibitem
\bibitem[{Visser} et~al.(2009){Visser}, {van Dishoeck}, and {Black}]{visser:09}
{Visser},~R.; {van Dishoeck},~E.~F.; {Black},~J.~H. \emph{\aap} \textbf{2009},
  \emph{503}, 323\relax
\mciteBstWouldAddEndPuncttrue
\mciteSetBstMidEndSepPunct{\mcitedefaultmidpunct}
{\mcitedefaultendpunct}{\mcitedefaultseppunct}\relax
\EndOfBibitem
\bibitem[{R{\"o}llig} and {Ossenkopf}(2013){R{\"o}llig}, and
  {Ossenkopf}]{roellig:13}
{R{\"o}llig},~M.; {Ossenkopf},~V. \emph{\aap} \textbf{2013}, \emph{550},
  A56\relax
\mciteBstWouldAddEndPuncttrue
\mciteSetBstMidEndSepPunct{\mcitedefaultmidpunct}
{\mcitedefaultendpunct}{\mcitedefaultseppunct}\relax
\EndOfBibitem
\bibitem[{Bulut} et~al.(2009){Bulut}, {Zanchet}, {Honvault},
  {Bussery-Honvault}, and {Ba{\~n}ares}]{bulut:09}
{Bulut},~N.; {Zanchet},~A.; {Honvault},~P.; {Bussery-Honvault},~B.;
  {Ba{\~n}ares},~L. \emph{\jcp} \textbf{2009}, \emph{130}, 194303\relax
\mciteBstWouldAddEndPuncttrue
\mciteSetBstMidEndSepPunct{\mcitedefaultmidpunct}
{\mcitedefaultendpunct}{\mcitedefaultseppunct}\relax
\EndOfBibitem
\bibitem[{Tappe} et~al.(2008){Tappe}, {Lada}, {Black}, and {Muench}]{tappe:08}
{Tappe},~A.; {Lada},~C.~J.; {Black},~J.~H.; {Muench},~A.~A. \emph{\apjl}
  \textbf{2008}, \emph{680}, L117\relax
\mciteBstWouldAddEndPuncttrue
\mciteSetBstMidEndSepPunct{\mcitedefaultmidpunct}
{\mcitedefaultendpunct}{\mcitedefaultseppunct}\relax
\EndOfBibitem
\bibitem[{van Harrevelt} and {van Hemert}(2000){van Harrevelt}, and {van
  Hemert}]{vanharrevelt:00}
{van Harrevelt},~R.; {van Hemert},~M.~C. \emph{\jcp} \textbf{2000}, \emph{112},
  5787\relax
\mciteBstWouldAddEndPuncttrue
\mciteSetBstMidEndSepPunct{\mcitedefaultmidpunct}
{\mcitedefaultendpunct}{\mcitedefaultseppunct}\relax
\EndOfBibitem
\bibitem[{Harich} et~al.(2000){Harich}, {Hwang}, {Yang}, {Lin}, {Yang}, and
  {Dixon}]{Harich:00}
{Harich},~S.~A.; {Hwang},~D.~W.~H.; {Yang},~X.; {Lin},~J.~J.; {Yang},~X.;
  {Dixon},~R.~N. \emph{\jcp} \textbf{2000}, \emph{113}, 10073\relax
\mciteBstWouldAddEndPuncttrue
\mciteSetBstMidEndSepPunct{\mcitedefaultmidpunct}
{\mcitedefaultendpunct}{\mcitedefaultseppunct}\relax
\EndOfBibitem
\bibitem[{Najita} et~al.(2010){Najita}, {Carr}, {Strom}, {Watson}, {Pascucci},
  {Hollenbach}, {Gorti}, and {Keller}]{najita:10}
{Najita},~J.~R.; {Carr},~J.~S.; {Strom},~S.~E.; {Watson},~D.~M.;
  {Pascucci},~I.; {Hollenbach},~D.; {Gorti},~U.; {Keller},~L. \emph{\apj}
  \textbf{2010}, \emph{712}, 274\relax
\mciteBstWouldAddEndPuncttrue
\mciteSetBstMidEndSepPunct{\mcitedefaultmidpunct}
{\mcitedefaultendpunct}{\mcitedefaultseppunct}\relax
\EndOfBibitem
\bibitem[{Watson} et~al.(2007){Watson}, {Bohac}, {Hull}, {Forrest}, {Furlan},
  {Najita}, {Calvet}, {D'Alessio}, {Hartmann}, {Sargent}, {Green}, {Kim}, and
  {Houck}]{watson:07}
{Watson},~D.~M.; {Bohac},~C.~J.; {Hull},~C.; {Forrest},~W.~J.; {Furlan},~E.;
  {Najita},~J.; {Calvet},~N.; {D'Alessio},~P.; {Hartmann},~L.; {Sargent},~B.;
  {Green},~J.~D.; {Kim},~K.~H.; {Houck},~J.~R. \emph{\nat} \textbf{2007},
  \emph{448}, 1026\relax
\mciteBstWouldAddEndPuncttrue
\mciteSetBstMidEndSepPunct{\mcitedefaultmidpunct}
{\mcitedefaultendpunct}{\mcitedefaultseppunct}\relax
\EndOfBibitem
\bibitem[{Herczeg} et~al.(2012){Herczeg}, {Karska}, {Bruderer}, {Kristensen},
  {van Dishoeck}, {J{\o}rgensen}, {Visser}, {Wampfler}, {Bergin},
  {Y{\i}ld{\i}z}, {Pontoppidan}, and {Gracia-Carpio}]{herczeg:12}
{Herczeg},~G.~J.; {Karska},~A.; {Bruderer},~S.; {Kristensen},~L.~E.; {van
  Dishoeck},~E.~F.; {J{\o}rgensen},~J.~K.; {Visser},~R.; {Wampfler},~S.~F.;
  {Bergin},~E.~A.; {Y{\i}ld{\i}z},~U.~A.; {Pontoppidan},~K.~M.;
  {Gracia-Carpio},~J. \emph{\aap} \textbf{2012}, \emph{540}, A84\relax
\mciteBstWouldAddEndPuncttrue
\mciteSetBstMidEndSepPunct{\mcitedefaultmidpunct}
{\mcitedefaultendpunct}{\mcitedefaultseppunct}\relax
\EndOfBibitem
\bibitem[{Lis} et~al.(2012){Lis}, {Schilke}, {Bergin}, {Emprechtinger}, and the
  HEXOS~Team]{lis:12}
{Lis},~D.; {Schilke},~P.; {Bergin},~E.; {Emprechtinger},~M.; the HEXOS~Team,
  \emph{Phil. Trans. R. Soc., A} \textbf{2012}, \emph{370}, 5162\relax
\mciteBstWouldAddEndPuncttrue
\mciteSetBstMidEndSepPunct{\mcitedefaultmidpunct}
{\mcitedefaultendpunct}{\mcitedefaultseppunct}\relax
\EndOfBibitem
\bibitem[{Wilson} et~al.(2006){Wilson}, {Henkel}, and
  {H{\"u}ttemeister}]{wilson:06}
{Wilson},~T.~L.; {Henkel},~C.; {H{\"u}ttemeister},~S. \emph{\aap}
  \textbf{2006}, \emph{460}, 533\relax
\mciteBstWouldAddEndPuncttrue
\mciteSetBstMidEndSepPunct{\mcitedefaultmidpunct}
{\mcitedefaultendpunct}{\mcitedefaultseppunct}\relax
\EndOfBibitem
\bibitem[{Plasil} et~al.(2011){Plasil}, {Mehner}, {Dohnal}, {Kotrik}, {Glosik},
  and {Gerlich}]{plasil:11}
{Plasil},~R.; {Mehner},~T.; {Dohnal},~P.; {Kotrik},~T.; {Glosik},~J.;
  {Gerlich},~D. \emph{\apj} \textbf{2011}, \emph{737}, 60\relax
\mciteBstWouldAddEndPuncttrue
\mciteSetBstMidEndSepPunct{\mcitedefaultmidpunct}
{\mcitedefaultendpunct}{\mcitedefaultseppunct}\relax
\EndOfBibitem
\bibitem[{Pineau des Forets} et~al.(1986){Pineau des Forets}, {Flower},
  {Hartquist}, and {Dalgarno}]{forets:86}
{Pineau des Forets},~G.; {Flower},~D.~R.; {Hartquist},~T.~W.; {Dalgarno},~A.
  \emph{\mnras} \textbf{1986}, \emph{220}, 801\relax
\mciteBstWouldAddEndPuncttrue
\mciteSetBstMidEndSepPunct{\mcitedefaultmidpunct}
{\mcitedefaultendpunct}{\mcitedefaultseppunct}\relax
\EndOfBibitem
\bibitem[{Joulain} et~al.(1998){Joulain}, {Falgarone}, {Pineau des Forets}, and
  {Flower}]{joulain:98}
{Joulain},~K.; {Falgarone},~E.; {Pineau des Forets},~G.; {Flower},~D.
  \emph{\aap} \textbf{1998}, \emph{340}, 241\relax
\mciteBstWouldAddEndPuncttrue
\mciteSetBstMidEndSepPunct{\mcitedefaultmidpunct}
{\mcitedefaultendpunct}{\mcitedefaultseppunct}\relax
\EndOfBibitem
\bibitem[{Zsarg{\'o}} and {Federman}(2003){Zsarg{\'o}}, and
  {Federman}]{zsargo:03}
{Zsarg{\'o}},~J.; {Federman},~S.~R. \emph{\apj} \textbf{2003}, \emph{589},
  319\relax
\mciteBstWouldAddEndPuncttrue
\mciteSetBstMidEndSepPunct{\mcitedefaultmidpunct}
{\mcitedefaultendpunct}{\mcitedefaultseppunct}\relax
\EndOfBibitem
\bibitem[{Godard} et~al.(2009){Godard}, {Falgarone}, and {Pineau Des
  For{\^e}ts}]{godard:09}
{Godard},~B.; {Falgarone},~E.; {Pineau Des For{\^e}ts},~G. \emph{\aap}
  \textbf{2009}, \emph{495}, 847\relax
\mciteBstWouldAddEndPuncttrue
\mciteSetBstMidEndSepPunct{\mcitedefaultmidpunct}
{\mcitedefaultendpunct}{\mcitedefaultseppunct}\relax
\EndOfBibitem
\bibitem[{Ag{\'u}ndez} et~al.(2010){Ag{\'u}ndez}, {Goicoechea}, {Cernicharo},
  {Faure}, and {Roueff}]{agundez:10b}
{Ag{\'u}ndez},~M.; {Goicoechea},~J.~R.; {Cernicharo},~J.; {Faure},~A.;
  {Roueff},~E. \emph{\apj} \textbf{2010}, \emph{713}, 662\relax
\mciteBstWouldAddEndPuncttrue
\mciteSetBstMidEndSepPunct{\mcitedefaultmidpunct}
{\mcitedefaultendpunct}{\mcitedefaultseppunct}\relax
\EndOfBibitem
\bibitem[{Godard} and {Cernicharo}(2013){Godard}, and {Cernicharo}]{godard:13}
{Godard},~B.; {Cernicharo},~J. \emph{\aap} \textbf{2013}, \emph{550}, A8\relax
\mciteBstWouldAddEndPuncttrue
\mciteSetBstMidEndSepPunct{\mcitedefaultmidpunct}
{\mcitedefaultendpunct}{\mcitedefaultseppunct}\relax
\EndOfBibitem
\bibitem[{Field} and {Hitchcock}(1966){Field}, and {Hitchcock}]{field:66}
{Field},~G.~B.; {Hitchcock},~J.~L. \emph{\apj} \textbf{1966}, \emph{146},
  1\relax
\mciteBstWouldAddEndPuncttrue
\mciteSetBstMidEndSepPunct{\mcitedefaultmidpunct}
{\mcitedefaultendpunct}{\mcitedefaultseppunct}\relax
\EndOfBibitem
\bibitem[{Penzias} and {Wilson}(1965){Penzias}, and {Wilson}]{penzias:65}
{Penzias},~A.~A.; {Wilson},~R.~W. \emph{\apj} \textbf{1965}, \emph{142},
  419\relax
\mciteBstWouldAddEndPuncttrue
\mciteSetBstMidEndSepPunct{\mcitedefaultmidpunct}
{\mcitedefaultendpunct}{\mcitedefaultseppunct}\relax
\EndOfBibitem
\bibitem[{Crane} et~al.(1986){Crane}, {Hegyi}, {Mandolesi}, and
  {Danks}]{crane:86}
{Crane},~P.; {Hegyi},~D.~J.; {Mandolesi},~N.; {Danks},~A.~C. \emph{\apj}
  \textbf{1986}, \emph{309}, 822\relax
\mciteBstWouldAddEndPuncttrue
\mciteSetBstMidEndSepPunct{\mcitedefaultmidpunct}
{\mcitedefaultendpunct}{\mcitedefaultseppunct}\relax
\EndOfBibitem
\bibitem[{Meyer} et~al.(1989){Meyer}, {Roth}, and {Hawkins}]{meyer:89c}
{Meyer},~D.~M.; {Roth},~K.~C.; {Hawkins},~I. \emph{\apjl} \textbf{1989},
  \emph{343}, L1\relax
\mciteBstWouldAddEndPuncttrue
\mciteSetBstMidEndSepPunct{\mcitedefaultmidpunct}
{\mcitedefaultendpunct}{\mcitedefaultseppunct}\relax
\EndOfBibitem
\bibitem[{Kaiser} and {Wright}(1990){Kaiser}, and {Wright}]{kaiser:90}
{Kaiser},~M.~E.; {Wright},~E.~L. \emph{\apjl} \textbf{1990}, \emph{356},
  L1\relax
\mciteBstWouldAddEndPuncttrue
\mciteSetBstMidEndSepPunct{\mcitedefaultmidpunct}
{\mcitedefaultendpunct}{\mcitedefaultseppunct}\relax
\EndOfBibitem
\bibitem[{Black} and {van Dishoeck}(1991){Black}, and {van Dishoeck}]{black:91}
{Black},~J.~H.; {van Dishoeck},~E.~F. \emph{\apjl} \textbf{1991}, \emph{369},
  L9\relax
\mciteBstWouldAddEndPuncttrue
\mciteSetBstMidEndSepPunct{\mcitedefaultmidpunct}
{\mcitedefaultendpunct}{\mcitedefaultseppunct}\relax
\EndOfBibitem
\bibitem[{Ritchey} et~al.(2011){Ritchey}, {Federman}, and
  {Lambert}]{ritchey:11}
{Ritchey},~A.~M.; {Federman},~S.~R.; {Lambert},~D.~L. \emph{\apj}
  \textbf{2011}, \emph{728}, 36\relax
\mciteBstWouldAddEndPuncttrue
\mciteSetBstMidEndSepPunct{\mcitedefaultmidpunct}
{\mcitedefaultendpunct}{\mcitedefaultseppunct}\relax
\EndOfBibitem
\bibitem[{Kre{\l}owski} et~al.(2012){Kre{\l}owski}, {Galazutdinov}, and
  {Gnaci{\'n}ski}]{krelowski:12}
{Kre{\l}owski},~J.; {Galazutdinov},~G.; {Gnaci{\'n}ski},~P. \emph{Astron.
  Nachr.} \textbf{2012}, \emph{333}, 627\relax
\mciteBstWouldAddEndPuncttrue
\mciteSetBstMidEndSepPunct{\mcitedefaultmidpunct}
{\mcitedefaultendpunct}{\mcitedefaultseppunct}\relax
\EndOfBibitem
\bibitem[{Fixsen}(2009)]{fixsen:09}
{Fixsen},~D.~J. \emph{\apj} \textbf{2009}, \emph{707}, 916\relax
\mciteBstWouldAddEndPuncttrue
\mciteSetBstMidEndSepPunct{\mcitedefaultmidpunct}
{\mcitedefaultendpunct}{\mcitedefaultseppunct}\relax
\EndOfBibitem
\bibitem[{Leach}(2012)]{leach:12}
{Leach},~S. \emph{\mnras} \textbf{2012}, \emph{421}, 1325\relax
\mciteBstWouldAddEndPuncttrue
\mciteSetBstMidEndSepPunct{\mcitedefaultmidpunct}
{\mcitedefaultendpunct}{\mcitedefaultseppunct}\relax
\EndOfBibitem
\bibitem[{Noterdaeme} et~al.(2008){Noterdaeme}, {Ledoux}, {Petitjean}, and
  {Srianand}]{noterdaeme:08}
{Noterdaeme},~P.; {Ledoux},~C.; {Petitjean},~P.; {Srianand},~R. \emph{\aap}
  \textbf{2008}, \emph{481}, 327\relax
\mciteBstWouldAddEndPuncttrue
\mciteSetBstMidEndSepPunct{\mcitedefaultmidpunct}
{\mcitedefaultendpunct}{\mcitedefaultseppunct}\relax
\EndOfBibitem
\bibitem[{Srianand} et~al.(2008){Srianand}, {Noterdaeme}, {Ledoux}, and
  {Petitjean}]{srianand:08}
{Srianand},~R.; {Noterdaeme},~P.; {Ledoux},~C.; {Petitjean},~P. \emph{\aap}
  \textbf{2008}, \emph{482}, L39\relax
\mciteBstWouldAddEndPuncttrue
\mciteSetBstMidEndSepPunct{\mcitedefaultmidpunct}
{\mcitedefaultendpunct}{\mcitedefaultseppunct}\relax
\EndOfBibitem
\bibitem[{Noterdaeme} et~al.(2009){Noterdaeme}, {Ledoux}, {Srianand},
  {Petitjean}, and {Lopez}]{noterdaeme:09}
{Noterdaeme},~P.; {Ledoux},~C.; {Srianand},~R.; {Petitjean},~P.; {Lopez},~S.
  \emph{\aap} \textbf{2009}, \emph{503}, 765\relax
\mciteBstWouldAddEndPuncttrue
\mciteSetBstMidEndSepPunct{\mcitedefaultmidpunct}
{\mcitedefaultendpunct}{\mcitedefaultseppunct}\relax
\EndOfBibitem
\bibitem[{Noterdaeme} et~al.(2010){Noterdaeme}, {Petitjean}, {Ledoux},
  {L{\'o}pez}, {Srianand}, and {Vergani}]{noterdaeme:10}
{Noterdaeme},~P.; {Petitjean},~P.; {Ledoux},~C.; {L{\'o}pez},~S.;
  {Srianand},~R.; {Vergani},~S.~D. \emph{\aap} \textbf{2010}, \emph{523},
  A80\relax
\mciteBstWouldAddEndPuncttrue
\mciteSetBstMidEndSepPunct{\mcitedefaultmidpunct}
{\mcitedefaultendpunct}{\mcitedefaultseppunct}\relax
\EndOfBibitem
\bibitem[{Noterdaeme} et~al.(2011){Noterdaeme}, {Petitjean}, {Srianand},
  {Ledoux}, and {L{\'o}pez}]{noterdaeme:11}
{Noterdaeme},~P.; {Petitjean},~P.; {Srianand},~R.; {Ledoux},~C.;
  {L{\'o}pez},~S. \emph{\aap} \textbf{2011}, \emph{526}, L7\relax
\mciteBstWouldAddEndPuncttrue
\mciteSetBstMidEndSepPunct{\mcitedefaultmidpunct}
{\mcitedefaultendpunct}{\mcitedefaultseppunct}\relax
\EndOfBibitem
\bibitem[{Muller} et~al.(2013){Muller}, {Beelen}, {Black}, {Curran},
  {Horellou}, {Aalto}, {Combes}, {Gu{\'e}lin}, and {Henkel}]{muller:13}
{Muller},~S.; {Beelen},~A.; {Black},~J.~H.; {Curran},~S.~J.; {Horellou},~C.;
  {Aalto},~S.; {Combes},~F.; {Gu{\'e}lin},~M.; {Henkel},~C. \emph{\aap}
  \textbf{2013}, \emph{551}, A109\relax
\mciteBstWouldAddEndPuncttrue
\mciteSetBstMidEndSepPunct{\mcitedefaultmidpunct}
{\mcitedefaultendpunct}{\mcitedefaultseppunct}\relax
\EndOfBibitem
\bibitem[Sobolev(1960)]{sobolev:60}
Sobolev,~V. \emph{Moving envelopes of stars}; Harvard books on astronomy;
  Harvard University Press: Cambridge, 1960\relax
\mciteBstWouldAddEndPuncttrue
\mciteSetBstMidEndSepPunct{\mcitedefaultmidpunct}
{\mcitedefaultendpunct}{\mcitedefaultseppunct}\relax
\EndOfBibitem
\bibitem[{Daniel} et~al.(2006){Daniel}, {Cernicharo}, and
  {Dubernet}]{daniel:06}
{Daniel},~F.; {Cernicharo},~J.; {Dubernet},~M. \emph{\apj} \textbf{2006},
  \emph{648}, 461\relax
\mciteBstWouldAddEndPuncttrue
\mciteSetBstMidEndSepPunct{\mcitedefaultmidpunct}
{\mcitedefaultendpunct}{\mcitedefaultseppunct}\relax
\EndOfBibitem
\bibitem[{Wakelam} et~al.(2010){Wakelam}, {Herbst}, {Le Bourlot}, {Hersant},
  {Selsis}, and {Guilloteau}]{wakelam:10}
{Wakelam},~V.; {Herbst},~E.; {Le Bourlot},~J.; {Hersant},~F.; {Selsis},~F.;
  {Guilloteau},~S. \emph{\aap} \textbf{2010}, \emph{517}, A21\relax
\mciteBstWouldAddEndPuncttrue
\mciteSetBstMidEndSepPunct{\mcitedefaultmidpunct}
{\mcitedefaultendpunct}{\mcitedefaultseppunct}\relax
\EndOfBibitem
\bibitem[{Zanchet} et~al.(2013){Zanchet}, {Godard}, {Bulut}, {Roncero},
  {Halvick}, and {Cernicharo}]{zanchet:13}
{Zanchet},~A.; {Godard},~B.; {Bulut},~N.; {Roncero},~O.; {Halvick},~P.;
  {Cernicharo},~J. \emph{\apj} \textbf{2013}, \emph{766}, 80\relax
\mciteBstWouldAddEndPuncttrue
\mciteSetBstMidEndSepPunct{\mcitedefaultmidpunct}
{\mcitedefaultendpunct}{\mcitedefaultseppunct}\relax
\EndOfBibitem
\bibitem[{Neufeld}(2010)]{neufeld:10}
{Neufeld},~D.~A. \emph{\apj} \textbf{2010}, \emph{708}, 635\relax
\mciteBstWouldAddEndPuncttrue
\mciteSetBstMidEndSepPunct{\mcitedefaultmidpunct}
{\mcitedefaultendpunct}{\mcitedefaultseppunct}\relax
\EndOfBibitem
\bibitem[{Crabtree} et~al.(2011){Crabtree}, {Tom}, and {McCall}]{crabtree:11a}
{Crabtree},~K.~N.; {Tom},~B.~A.; {McCall},~B.~J. \emph{\jcp} \textbf{2011},
  \emph{134}, 194310\relax
\mciteBstWouldAddEndPuncttrue
\mciteSetBstMidEndSepPunct{\mcitedefaultmidpunct}
{\mcitedefaultendpunct}{\mcitedefaultseppunct}\relax
\EndOfBibitem
\bibitem[{Gerlich} et~al.(2006){Gerlich}, {Windisch}, {Hlavenka}, and {et
  al.}]{gerlich:06}
{Gerlich},~D.; {Windisch},~F.; {Hlavenka},~P.; {et al.}, \emph{Phil. Trans. R.
  Soc., A} \textbf{2006}, \emph{364}, 3007\relax
\mciteBstWouldAddEndPuncttrue
\mciteSetBstMidEndSepPunct{\mcitedefaultmidpunct}
{\mcitedefaultendpunct}{\mcitedefaultseppunct}\relax
\EndOfBibitem
\bibitem[{Crabtree} et~al.(2011){Crabtree}, {Kauffman}, {Tom}, {Be{\c c}ka},
  {McGuire}, and {McCall}]{crabtree:11b}
{Crabtree},~K.~N.; {Kauffman},~C.~A.; {Tom},~B.~A.; {Be{\c c}ka},~E.;
  {McGuire},~B.~A.; {McCall},~B.~J. \emph{\jcp} \textbf{2011}, \emph{134},
  194311\relax
\mciteBstWouldAddEndPuncttrue
\mciteSetBstMidEndSepPunct{\mcitedefaultmidpunct}
{\mcitedefaultendpunct}{\mcitedefaultseppunct}\relax
\EndOfBibitem
\bibitem[{Loomis} et~al.(2013){Loomis}, {Zaleski}, {Steber}, {Neill}, {Muckle},
  {Harris}, {Hollis}, {Jewell}, {Lattanzi}, {Lovas}, {Martinez}, {McCarthy},
  {Remijan}, {Pate}, and {Corby}]{loomis:13}
{Loomis},~R.~A.; {Zaleski},~D.~P.; {Steber},~A.~L.; {Neill},~J.~L.;
  {Muckle},~M.~T.; {Harris},~B.~J.; {Hollis},~J.~M.; {Jewell},~P.~R.;
  {Lattanzi},~V.; {Lovas},~F.~J.; {Martinez},~O.,~Jr.; {McCarthy},~M.~C.;
  {Remijan},~A.~J.; {Pate},~B.~H.; {Corby},~J.~F. \emph{\apjl} \textbf{2013},
  \emph{765}, L9\relax
\mciteBstWouldAddEndPuncttrue
\mciteSetBstMidEndSepPunct{\mcitedefaultmidpunct}
{\mcitedefaultendpunct}{\mcitedefaultseppunct}\relax
\EndOfBibitem
\bibitem[{Zaleski} et~al.(2013){Zaleski}, {Seifert}, {Steber}, {Muckle},
  {Loomis}, {Corby}, {Martinez}, {Crabtree}, {Jewell}, {Hollis}, {Lovas},
  {Vasquez}, {Nyiramahirwe}, {Sciortino}, {Johnson}, {McCarthy}, {Remijan}, and
  {Pate}]{zaleski:13}
{Zaleski},~D.~P.; {Seifert},~N.~A.; {Steber},~A.~L.; {Muckle},~M.~T.;
  {Loomis},~R.~A.; {Corby},~J.~F.; {Martinez},~O.,~Jr.; {Crabtree},~K.~N.;
  {Jewell},~P.~R.; {Hollis},~J.~M.; {Lovas},~F.~J.; {Vasquez},~D.;
  {Nyiramahirwe},~J.; {Sciortino},~N.; {Johnson},~K.; {McCarthy},~M.~C.;
  {Remijan},~A.~J.; {Pate},~B.~H. \emph{\apjl} \textbf{2013}, \emph{765},
  L10\relax
\mciteBstWouldAddEndPuncttrue
\mciteSetBstMidEndSepPunct{\mcitedefaultmidpunct}
{\mcitedefaultendpunct}{\mcitedefaultseppunct}\relax
\EndOfBibitem
\bibitem[{Bacmann} et~al.(2012){Bacmann}, {Taquet}, {Faure}, {Kahane}, and
  {Ceccarelli}]{bacmann:12}
{Bacmann},~A.; {Taquet},~V.; {Faure},~A.; {Kahane},~C.; {Ceccarelli},~C.
  \emph{\aap} \textbf{2012}, \emph{541}, L12\relax
\mciteBstWouldAddEndPuncttrue
\mciteSetBstMidEndSepPunct{\mcitedefaultmidpunct}
{\mcitedefaultendpunct}{\mcitedefaultseppunct}\relax
\EndOfBibitem
\bibitem[{Cernicharo} et~al.(2012){Cernicharo}, {Marcelino}, {Roueff}, {Gerin},
  {Jim{\'e}nez-Escobar}, and {Mu{\~n}oz Caro}]{cernicharo:12}
{Cernicharo},~J.; {Marcelino},~N.; {Roueff},~E.; {Gerin},~M.;
  {Jim{\'e}nez-Escobar},~A.; {Mu{\~n}oz Caro},~G.~M. \emph{\apjl}
  \textbf{2012}, \emph{759}, L43\relax
\mciteBstWouldAddEndPuncttrue
\mciteSetBstMidEndSepPunct{\mcitedefaultmidpunct}
{\mcitedefaultendpunct}{\mcitedefaultseppunct}\relax
\EndOfBibitem
\end{mcitethebibliography}

\providecommand*\mcitethebibliography{\thebibliography}
\csname @ifundefined\endcsname{endmcitethebibliography}
  {\let\endmcitethebibliography\endthebibliography}{}

\end{document}